\newif{\ifarxiv}
\newif{\ifdraft}
\newif{\ifremarks}

\arxivtrue 
\remarkstrue 

\ifremarks
\newcommand{\remarktb}[1]{{\renewcommand{\bfdefault}{b}{\color[RGB]{0,150,0}{\textbf{#1}}}}}
\newcommand{\remarktf}[1]{{\renewcommand{\bfdefault}{b}{\color[RGB]{150,0,150}{\textbf{#1}}}}}
\newcommand{\remarkjc}[1]{{\renewcommand{\bfdefault}{b}{\color[RGB]{0,0,150}{\textbf{#1}}}}}
\newcommand{\remarkpv}[1]{{\renewcommand{\bfdefault}{b}{\color[RGB]{150,0,0}{\textbf{#1}}}}}
\fi
\providecommand{\remarktb}[1]{\ignorespaces}
\providecommand{\remarktf}[1]{\ignorespaces}
\providecommand{\remarkjc}[1]{\ignorespaces}
\providecommand{\remarkpv}[1]{\ignorespaces}

\documentclass[12pt,a4paper]{article}
\ifdraft\else\fi
\pdfoutput=1
\usepackage{amsmath,amssymb,mathtools}
\usepackage[nosort]{cite}
\usepackage[bulletsep]{collref} 
\ifdraft\usepackage{showkeys}\fi 
\usepackage[bookmarks=true,ocgcolorlinks=true]{hyperref} 
\usepackage{graphicx}
\usepackage[usenames,dvipsnames,svgnames,table]{xcolor}
\usepackage[normalem]{ulem}
\usepackage{booktabs} 
\usepackage{calc}
\usepackage{listings}

\definecolor{mypurple}{rgb}{0.6,0,0.8}

\usepackage{lmodern}
\usepackage[T1]{fontenc}
\usepackage{microtype}

\newcommand{\HexagonPaper}{Basso:2015zoa}
\newcommand{\Hexagonalizationone}{Fleury:2016ykk}

\newcommand{\Hexagonalizationtwo}{Fleury:2017eph}

\usepackage[left=2.5cm,right=2.5cm,top=2.8cm,bottom=2.8cm]{geometry}

\allowdisplaybreaks

\makeatletter
\renewcommand*\l@section[2]{%
  \ifnum \c@tocdepth >\z@
    \addpenalty\@secpenalty
    \addvspace{0.8em \@plus\p@}%
    \setlength\@tempdima{1.5em}%
    \begingroup
      \parindent \z@ \rightskip \@pnumwidth
      \parfillskip -\@pnumwidth
      \leavevmode \bfseries
      \advance\leftskip\@tempdima
      \hskip -\leftskip
      #1\nobreak\hfil \nobreak\hb@xt@\@pnumwidth{\hss #2}\par
    \endgroup
  \fi}
\makeatother

\usepackage{enumitem}

\let\oldbfseries=\bfseries
\let\oldmdseries=\mdseries
\let\oldnormalfont=\normalfont
\renewcommand{\bfseries}{\oldbfseries\boldmath}
\renewcommand{\mdseries}{\oldmdseries\unboldmath}
\renewcommand{\normalfont}{\oldnormalfont\unboldmath}

\usepackage{fixmath}

\usepackage{empheq}
\newcommand{\nn}{\nonumber}

\newenvironment{myeqnarray}{\arraycolsep0pt\begin{eqnarray}}{\end{eqnarray}\ignorespacesafterend}
\newenvironment{myeqnarray*}{\arraycolsep0pt\begin{eqnarray*}}{\end{eqnarray*}\ignorespacesafterend}

\newcommand{\beq}{\begin{equation}}
\newcommand{\eeq}{\end{equation}}
\def\[{\begin{equation}}
\def\]{\end{equation}}
\def\<{\begin{myeqnarray}}
\def\>{\end{myeqnarray}}

\numberwithin{equation}{section}

\def\etal.{et\penalty50\ al.}
\usepackage{xspace}
\newcommand*{\eg}{e.\,g.\@\xspace}
\newcommand*{\ie}{i.\,e.\@\xspace}
\makeatletter\newcommand*{\etc}{%
    \@ifnextchar{.}%
        {etc}%
        {etc.\@\xspace}%
}\makeatother

\makeatletter
\newlength{\apb@width}
\newcommand{\autoparbox}[2][c]{\settowidth{\apb@width}{#2}\parbox[#1]{\apb@width}{#2}}
\newcommand{\includegraphicsbox}[2][]{\autoparbox{\includegraphics[#1]{#2}}}
\makeatother
\usepackage[font=small,labelfont=bf,width=0.9\textwidth]{caption}

\providecommand{\hypersetup}[1]{}
\providecommand{\hyperref}[2][]{#2}
\providecommand{\texorpdfstring}[2]{#1}
\providecommand{\pdfbookmark}[3][]{}

\hypersetup{plainpages=false}
\hypersetup{pdfpagemode=UseOutlines}
\hypersetup{bookmarksnumbered=true}
\hypersetup{bookmarksopen=true}
\hypersetup{pdfstartview=FitH}
\hypersetup{citecolor=[rgb]{0 .4 0}}
\hypersetup{urlcolor=[rgb]{.4 0 0}}
\hypersetup{linkcolor=[rgb]{0 0 .4}}
\hypersetup{citebordercolor={.6 .9 .6}}
\hypersetup{urlbordercolor={.7 .8 1}}
\hypersetup{linkbordercolor={1 .7 .7}}
\hypersetup{pdfborder={0 0 .5}}

\newcommand{\namedref}[2]{\hyperref[#2]{#1~\ref*{#2}}}
\newcommand{\secref}[1]{\namedref{Section}{#1}}
\newcommand{\appref}[1]{\namedref{Appendix}{#1}}
\newcommand{\tabref}[1]{\namedref{Table}{#1}}
\newcommand{\figref}[1]{\namedref{Figure}{#1}}

\makeatletter
\let\@myabstract\@empty
\let\@keywords\@empty
\let\@subject\@empty
\providecommand{\affiliation}[1]{\gdef\@affiliation{#1}}
\providecommand{\myabstract}[1]{\gdef\@myabstract{#1}}
\providecommand{\keywords}[1]{\gdef\@keywords{#1}}
\providecommand{\subject}[1]{\gdef\@subject{#1}}
\def\thetitle{\@title}
\def\theauthor{\@author}
\def\theaffiliation{\@affiliation}
\def\theabstract{\@myabstract}
\def\thesubject{\@subject}
\def\thedate{\@date}
\def\thekeywords{\@keywords}
\makeatother
\AtBeginDocument{
\hypersetup{pdftitle={\thetitle}}%
\hypersetup{pdfauthor={\theauthor}}%
\hypersetup{pdfsubject={\thesubject}}%
\hypersetup{pdfkeywords={\thekeywords}}%
}

\newcommand{\sfrac}[2]{{\textstyle\frac{#1}{#2}}}
\newcommand{\half}{\sfrac{1}{2}}

\newcommand{\order}[1]{\mathcal{O}(#1)}

\newcommand{\superN}{\mathcal{N}}
\newcommand{\gym}{g_{\scriptscriptstyle\mathrm{YM}}}

\newcommand{\Real}{\mathbb{R}}
\newcommand{\Integers}{\mathbb{Z}}

\newcommand{\dd}{\mathrm{d}}
\newcommand{\Nc}{N\subrm{c}}
\newcommand{\Csphere}{{}^\bullet\kern-1.2pt C}
\newcommand{\Ctorus}{{}^\circ\kern-1.2pt C}
\newcommand{\polygon}{\mathop{\texttt{polygon}}}

\DeclareMathOperator{\tr}{tr}
\DeclareMathOperator{\aut}{Aut}
\DeclareMathOperator{\Li}{Li}

\DeclareMathOperator{\Disk}{Disk}

\newcommand{\op}[1]{\mathcal{#1}}
\newcommand{\suprm}[1]{^{\text{#1}}}
\newcommand{\subrm}[1]{_{\text{#1}}}
\newcommand{\alg}[1]{\mathfrak{#1}}
\newcommand{\grp}[1]{\mathrm{#1}}

\newcommand{\mathematica}{\textsc{Mathematica}\@\xspace}
\newcommand{\form}{\textsc{Form}\@\xspace}

\newcommand{\brk}[1]{(#1)}
\newcommand{\lrbrk}[1]{\left(#1\right)}
\newcommand{\bigbrk}[1]{\bigl(#1\bigr)}
\newcommand{\Bigbrk}[1]{\Bigl(#1\Bigr)}

\newcommand{\biggbrk}[1]{\biggl(#1\biggr)}
\newcommand{\biggbrkopen}{\biggl(}
\newcommand{\biggbrkclose}{\biggr)}

\newcommand{\lrsbrk}[1]{\left[#1\right]}
\newcommand{\bigsbrk}[1]{\bigl[#1\bigr]}
\newcommand{\Bigsbrk}[1]{\Bigl[#1\Bigr]}
\newcommand{\biggsbrk}[1]{\biggl[#1\biggr]}
\newcommand{\Biggsbrk}[1]{\Biggl[#1\Biggr]}
\newcommand{\brc}[1]{\{#1\}}
\newcommand{\lrbrc}[1]{\left\{#1\right\}}
\newcommand{\bigbrc}[1]{\bigl\{#1\bigr\}}
\newcommand{\Bigbrc}[1]{\Bigl\{#1\Bigr\}}
\newcommand{\biggbrc}[1]{\biggl\{#1\biggr\}}
\newcommand{\Biggbrc}[1]{\Biggl\{#1\Biggr\}}

\newcommand{\comm}[2]{[#1,#2]}

\newcommand{\vev}[1]{\langle#1\rangle}

\newcommand{\abs}[1]{|#1|}

\ifx\href\asklfhas\newcommand{\href}[2]{#2}\fi

\title{Handling Handles. Part II.\texorpdfstring{\\}{} Stratification and Data Analysis}

\author{%
T. Bargheer\texorpdfstring{$^{a,b,c}$}{},
J. Caetano\texorpdfstring{$^{d}$}{},
T. Fleury\texorpdfstring{$^{d,e}$}{},
S. Komatsu\texorpdfstring{$^{f}$}{},
P. Vieira\texorpdfstring{$^{g,h}$}{}}

\subject{Mathematical Physics}


\begin{document}

\pdfbookmark[1]{Title Page}{title}

\thispagestyle{empty}
\setcounter{page}{0}

\mbox{}
\vfill

\begin{center}

{\Large\textbf{\thetitle}\par}

\bigskip

\bigskip

\textsc{\theauthor}

\bigskip

\begingroup
\footnotesize\itshape

$^{a}$Institut f\"ur Theoretische Physik,
Leibniz Universit\"at Hannover,
\\
Appelstra{\ss}e 2, 30167 Hannover, Germany

\medskip

$^{b}$DESY Theory Group, DESY Hamburg,
Notkestra\ss e 85, D-22603 Hamburg, Germany

\medskip

$^{c}$Kavli Institute for Theoretical Physics,
University of California
Santa Barbara, CA 93106, USA

\medskip

$^{d}$Laboratoire de Physique Th\'{e}orique
de l'\'{E}cole Normale Sup\'{e}rieure de Paris,
CNRS, ENS \& PSL Research University, UPMC \& Sorbonne Universit\'{e}s,
75005 Paris, France.

\medskip

$^{e}$International Institute of Physics,
Universidade Federal do Rio Grande do Norte,
\\
Campus Universitario, Lagoa Nova, Natal-RN 59078-970, Brazil

\medskip

$^{f}$School of Natural Sciences, Institute for Advanced Study,
Einstein Drive, Princeton, NJ 08540, USA

\medskip

$^{g}$Perimeter Institute for Theoretical Physics,
31 Caroline St N Waterloo, Ontario N2L 2Y5, Canada

\medskip

$^{h}$Instituto de F\'isica Te\'orica, UNESP - Univ. Estadual Paulista,
ICTP South American Institute for Fundamental Research,
Rua Dr. Bento Teobaldo Ferraz 271, 01140-070, S\~ao Paulo, SP, Brasil

\endgroup

\bigskip

\newcommand{\email}[1]{\href{mailto:#1}{#1}}

{\small\ttfamily
\email{till.bargheer@desy.de},
\email{joao.caetanus@gmail.com},
\email{tsi.fleury@gmail.com},
\email{shota.komadze@gmail.com},
\email{pedrogvieira@gmail.com}}
\par

\bigskip
\bigskip

\textbf{Abstract}\vspace{3mm}

\begin{minipage}{12.5cm}
In a previous work~\cite{Bargheer:2017nne},
we proposed an integrability setup for computing
non-planar
corrections to correlation functions
in $\superN=4$ super Yang--Mills theory at any value of the coupling constant.
The procedure consists of drawing all possible tree-level
graphs on a Riemann surface of given genus, completing each graph
to a triangulation, inserting a hexagon form factor into each face,
and summing over a complete set of states on each edge of the
triangulation.
The summation over graphs can be interpreted as a quantization of the
string moduli space integration.
The quantization requires a careful treatment of the moduli space boundaries,
which is realized by subtracting degenerate Riemann
surfaces; this procedure is called stratification.
In this work, we precisely formulate our proposal
and perform
several perturbative checks.
These checks require hitherto unknown multi-particle mirror
contributions at one loop, which we also compute.
\end{minipage}

\end{center}

\vfill

\newpage

\pdfbookmark[1]{\contentsname}{contents}
\tableofcontents

\newpage

\section{Introduction}

\begin{figure}[t]
\centering
\includegraphics[angle=-90,width=\textwidth]{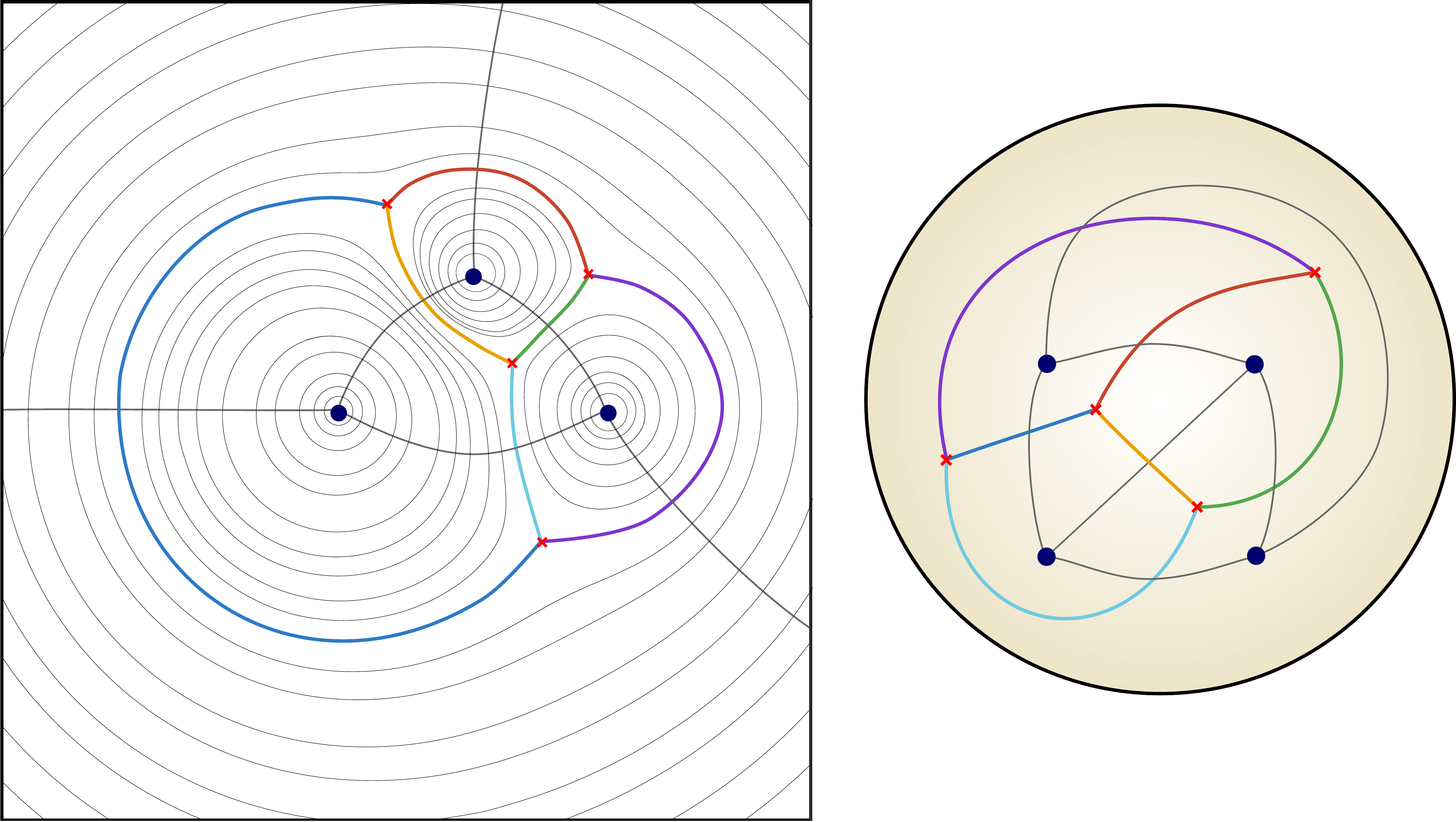}
\caption{In the left figure, four strings radiated from four
worldsheet punctures interact (blue dots, with the fourth puncture located at
infinity). The punctures are encircled by contour lines of constant
worldsheet time (black). Critical lines (colorful) divide
the contour lines encircling different punctures. The critical lines
intersect at the red crosses, and thus define a
cubic graph. The dual graph is shown in gray. On the right,
we represent a standard tree-level field-theory four-point
graph with the topology of a sphere (gray), as well as its dual graph (colorful).}
\label{fig:4ptStrebel}
\end{figure}

Like in any perturbative string theory, closed string amplitudes in
$\grp{AdS}_5\times\grp{S}^5$ superstring theory are given by integrations over
the moduli space of Riemann surfaces of various genus. Like in any
large-$\Nc$ gauge theory, correlation functions of local single-trace
gauge-invariant operators in $\superN=4$ SYM theory are given by sums over
double-line Feynman (ribbon) graphs of various genus. By virtue of the AdS/CFT
duality, these two quantities ought to
be the same. Clearly, to better understand the nature of holography,
it is crucial to understand how the sum over graphs connects to the
integration over the string moduli.

Our proposal in~\cite{Bargheer:2017nne} provides one realization. It can
be motivated as a finite-coupling extension of a very nice proposal
by Razamat~\cite{Razamat:2008zr}, built up on the works of Gopakumar
\etal.~\cite{Gopakumar:2003ns, Gopakumar:2004qb,Gopakumar:2004ys,
Gopakumar:2005fx,Aharony:2006th,Aharony:2007fs}, which in turn relied on beautiful
classical mathematics by Strebel~\cite{StrebelBook,MulasePenkava:1998a}, where an
isomorphism between the space of metric ribbon graphs and moduli spaces of
Riemann surfaces was first understood.%
\footnote{The present work is a continuation of the hexagonalization proposal for planar
correlation functions~\cite{Fleury:2016ykk,Eden:2016xvg} (see
also~\cite{Eden:2017ozn,Fleury:2017eph}), which
was an extension of the three-point function hexagon
construction~\cite{Basso:2015zoa}, which in turn was strongly inspired
by numerous
weak-coupling~\cite{Escobedo:2010xs,Escobedo:2011xw,Gromov:2011jh,Gromov:2012vu,Gromov:2012uv,Caetano:2014gwa}
and strong-coupling~\cite{Kazama:2011cp,Kazama:2012is,Kazama:2013qsa}
studies. It was these weak- and strong-coupling
mathematical structures -- only available due to
integrability -- which were the most important hints in arriving at
our proposal~\cite{Bargheer:2017nne}.}

Let us briefly describe some of these ideas. \figref{fig:4ptStrebel}
is a very inspiring example, so let us explain a few of its features.
The figure describes four strings interacting at tree level,
\ie a four-punctured sphere (in the figure, one of the punctures is
at infinity). The black lines are sections of the incoming strings.
Close to each puncture, the string world-sheet behaves as a normal
single string, so here the black lines are simple circles. They are
the \emph{lines of constant $\tau$} for each string.
These lines of constant $\tau$ need to fit
together into a global picture, as shown in the figure. Note
that there are four special points, the red crosses, which can be
connected along critical lines (the colorful lines), across which we
``jump from one string to another''.
These critical lines define a graph. There is also a dual
graph, drawn in gray.%
\footnote{In this example, both the graph and its dual graph are
cubic graphs, but this is not necessarily true in general.}
This construction creates a map between the moduli space of a
four-punctured Riemann sphere and a class of graphs, as anticipated above.

These cartoons can be made mathematically rigorous. For each
punctured Riemann surface, there is a unique quadratic differential
$\phi$, called the Strebel differential,
with fixed residues at each puncture, which decomposes the surface
into disk-like regions -- the faces delimited by the colorful
lines~\cite{StrebelBook,MulasePenkava:1998a} (see the appendices
in~\cite{Razamat:2008zr} for a beautiful review).
The
red crosses are
the zeros of the Strebel differential. The line integrals between these critical
points, \ie the integrals along the colorful lines are real,
and thus define a (positive) length for
each line of the graph.
In this way the graph becomes a \emph{metric} graph.
(The sum over the lengths of the critical lines that
encircle a puncture equals the residue of the
Strebel differential at that puncture by contour integral arguments.)
By construction, the critical lines emanating from each zero have a
definite ordering around that zero. This ordering can equivalently be
achieved by promoting each line to a ``ribbon'' by giving it a
non-zero width; for this reason the relevant graphs are called \emph{metric
ribbon graphs}.
Conversely, fixing a graph topology and assigning a length to each
edge uniquely fixes the Strebel differential and thus a point in the
moduli space.

Such metric ribbon graphs, like the one on the right of~\figref{fig:4ptStrebel},
also arise at zero coupling in the dual gauge theory. There,
the number associated to each line is nothing but
the number of propagators connecting two operators along that line.
These numbers are thus integers in this case, as emphasized
in~\cite{Razamat:2008zr}.
Note that the total number of lines getting out
of a given operator is fixed, which is the gauge-theory counterpart of
the above contour integral argument.

As such, it is very tempting to propose that we fix the residue
of the Strebel differential at each puncture to be equal to the
number of fields%
\footnote{The ``number of fields'' is inherently a weak-coupling
concept, which could be replaced by \eg the total R-charge of the operator.}
inside the trace of the dual operator.%
\footnote{Note that until now the value of the residue remained
arbitrary. Indeed, the map between the space of metric ribbon graphs $\Gamma_{n,g}$ and
the moduli space of Riemann surfaces $\mathcal{M}_{n,g}$ conveniently
contains a factor of $\mathbb{R}_+^n$ as
$\mathcal{M}_{n,g} \times \mathbb{R}_+^n \simeq \Gamma_{n,g}$,
so we
can think of the space of metric ribbon graphs as a fibration over the Riemann
surface moduli space. Fixing the residues of the Strebel differential
to the natural gauge-theory values simply amounts to picking a section
of this fibration.}
Then there is a discrete subset of points within the string moduli space
where those integer residues are split into integer subsets, which
define a valid gauge-theory ribbon graph. By our weak-coupling analysis,
it seems that the string path integral is localizing at these
points. Note that the graphs defined by the Strebel differential
change as we move in the string moduli space, and that all free gauge-theory
graphs nicely show up when doing so, such that the map is truly
complete. The jump from one graph to another is mathematically very
similar to the wall-crossing phenomenon within the space of $4$d
$\superN=2$ theories~\cite{Gaiotto:2008cd,Gaiotto:2009hg}.

What about finite coupling? Here it is where the hexagons come
in. The gray lines in~\figref{fig:4ptStrebel} typically define a triangulation of the
Riemann surface (since the colored dual graph is a cubic graph). The
triangular faces become hexagons once we
blow up all punctures into small circles, such that small extra
segments get inserted into all triangle vertices, effectively
converting all triangles into hexagons.
In order to glue together these hexagons, we insert a complete basis of (open
mirror) string states at each of the gray lines. The sum over these
complete bases of states can be thought of as exploring the vicinity of
each discrete point in the moduli space, thus covering the full string
path integral.

For correlation functions of more/fewer operators, and/or different
worldsheet genus, the picture is very similar. What changes, of course, is the number
of zeros of the Strebel differential,%
\footnote{The zeros of the Strebel differential may vary in degree.
The number of zeros equals the number
of faces of the (dual) graph, whereas the sum of their degrees equals the number
of hexagons.}
that is the number of hexagon
operators we should glue together. In the example above, we had four
red crosses, that is four hexagons. This number is very easy to
understand. Topologically, a four-point function can be thought of as
gluing together two pairs of pants, and each pair of pants is the union of
two hexagons.
To obtain a genus~$g$ correlation function of $n$
closed strings, we would glue together $2n+4g-4$ hexagons. We ought to
glue all these hexagons together and sum over a complete basis of
mirror states on each gluing line. Each
hexagon has three such mirror lines, as illustrated
in~\figref{fig:4ptStrebel}, and each line is shared by two hexagons, so
there will be a $(3n+6g-6)$-fold sum over mirror
states.%
\footnote{Note that we should also sum over the lengths
associated to the gluing lines. These lines always connect two physical
operators, with the $n$ constraints that the sum of lengths leaving each
puncture equals the length (charge) of the corresponding physical operator,
such that one ends up with a $(2n+6g-6)$-dimensional sum, which is the
appropriate dimension of the string moduli space. For instance, for
$n=4$ and $g=0$ we have a two-fold sum, which matches nicely with the
two real parameters of the complex position of the fourth
puncture on the sphere, once the other three positions are fixed.}
This is admissibly a hard task, but, until now, there is no
alternative for studying correlation functions at finite coupling and
genus in this gauge theory. So this is the best we have thus far.%
\footnote{Of course, there are simplifying limits. In perturbation
theory, most of these sums collapse, since it is costly to create and
annihilate mirror particles. Hence, the hexagonalization procedure often
becomes quite efficient, see \eg~\cite{Basso:2015eqa}. At strong
coupling, the sums sometimes exponentiate and can be resummed, see
\eg~\cite{Jiang:2016ulr}. And for very large operators, the various lengths
that have to be traversed by mirror states as we glue together two hexagons are often
very large, projecting the state sum to the lowest-energy states,
thus also simplifying the computations greatly, as in~\cite{Bargheer:2017nne}.}

For higher genus -- \ie as we venture into the non-planar regime --
there is a final and very important ingredient called the
\emph{stratification}, which appeared already in the context of matrix
models~\cite{Chekov:1992ns,Chekhov:1995cq,Anselmi:1994em}, and
which gives the name to this paper. It can be motivated from
gauge theory as well as from string theory considerations. From the
gauge theory viewpoint, it is clear that simply drawing all tree-level graphs of a
given genus, and dressing them by hexagons and mirror states cannot be the full
story: As we go to higher loops in 't~Hooft coupling, there will be
handles formed by purely virtual processes, which are not present at lower
orders. So including only genus-$g$ tree-level graphs misses
some contributions. One naive idea would be to include -- at a given
genus -- all graphs which can be drawn on surfaces of that genus or
less. But this would be no good either, as it would vastly over-count contributions.
The stratification procedure explained in this paper prescribes precisely
which contributions have to be added or subtracted, so that -- we hope --
everything works out. From a string theory perspective, this stratification
originates in the boundaries of the moduli space. We can have tori,
for example, degenerating into spheres, and to properly avoid missing
(or double-counting) such degenerate contributions, we need to
carefully understand what to sum over. In more conventional string
perturbation theory, we are used to continuous integrations over the
moduli space, where such degenerate contributions typically amount to
measure-zero sets, which we can ignore. But here -- as emphasized above and already proposed in~\cite{Razamat:2008zr}
-- the sum is rather a discrete one, hence
missing or adding particular terms matters.

All in all, our final proposal can be summarized in equation~\eqref{eq:mainformula}
below, where the seemingly innocuous~$\mathcal{S}$ operation is the
stratification procedure, which is
further motivated and made precise below, see \eg~\eqref{eq:stratification4}
for a taste of what it ends up looking
like.

In the end, all this is a plausible yet conjectural picture. Clearly,
many checks are crucial to validate this proposal, and to iron out its
details. A most obvious test is to carry out the hexagonalization and
stratification procedure to study the first non-planar quantum
correction to a gauge-theory four-point correlation function, and
to compare the result with available perturbative data. That is what this
paper is about.

\section{Developing the Proposal}
\label{sec:developing-proposal}

In the following, we introduce our main formula and explain its
ingredients in~\secref{sec:main-formula}. In the
subsequent~\secref{sec:polygonization}, we explain the summation over
graphs at the example of a four-point function on the torus.
\secref{sec:stratification} and \secref{sec:subtractions} are devoted
to the effects of stratification.

\subsection{The Main Formula}
\label{sec:main-formula}

Recall that in a general large-$\Nc$ gauge theory with adjoint matter,
each Feynman diagram is assigned a genus by promoting all propagators
to double-lines (pairs of fundamental color lines). At each
single-trace operator insertion, the color trace induces a definite
ordering of the attached (double) lines. By this ordering, the color
lines of the resulting double-line graph form well-defined closed
loops. Assigning an oriented disk (face) to each of these color loops,
we obtain an oriented compact surface. The genus of the graph (Wick
contraction) is the genus of this surface. Counting powers of $\Nc$
and $\gym^2$ for propagators (${\sim}\gym^2$), vertices (${\sim}1/\gym^2$), and
faces (${\sim}\Nc$), taking into account that every operator insertion adds
a boundary component to the surface, absorbing one power of $\Nc$ into
the 't~Hooft coupling $\lambda=\gym^2\Nc$, and using the formula for the Euler
characteristic, we arrive at the well-known genus
expansion formula~\cite{'tHooft:1974jz} for connected correlators of
(canonically normalized) single-trace operators~$\op{O}_i$:
\begin{equation}
\vev{\op{O}_1\dotsc\op{O}_n}
=\frac{1}{\Nc^{n-2}}\sum_{g=0}^\infty \frac{1}{\Nc^{2g}}\,\mathcal{G}_{1,\dotsc,n}^{(g)}(\lambda)
\,,\qquad
\lambda=\gym^2\Nc
\,.
\label{eq:genusexpansion}
\end{equation}
Here, $\mathcal{G}_{1,\dotsc,n}^{(g)}(\lambda)$ is the correlator
restricted to genus-$g$ contributions. Via the AdS/CFT duality, the
surface defined by Feynman diagrams at large $\Nc$ becomes the
worldsheet of the dual string with $n$ vertex operator insertions.

The purpose of this paper is to give a concrete and explicit realization
of the general large-$\Nc$ genus expansion
formula~\eqref{eq:genusexpansion} for the case of $\superN=4$ super
Yang--Mills theory. The proposed formula is based on the integrability
of the (gauge/worldsheet/string) theory, and should be valid at any order in
the 't~Hooft coupling constant $\lambda$. The general formula reads
\begin{align}
\vev{\op{Q}_1\dotsc \op{Q}_n}
=\frac{\prod_{i=1}^{n}\sqrt{k_i}}{\Nc^{n-2}}\,{\mathcal{S}}\circ \sum_{\Gamma\in\mathbold{\Gamma}}\frac{1}{\Nc^{2g(\Gamma)}}
\lrsbrk{\prod_{b\in\mathbold{b}(\Gamma_{\!\triangle})}d_b^{\ell_b}\int_{\mathbold{M}_b}\dd\psi_b \, \mathcal{W}(\psi_b)}
\prod_{a=1}^{2n+4g(\Gamma)-4}\mathcal{H}_a
\,.
\label{eq:mainformula}
\end{align}
Let us explain the ingredients:
The operators $\op{Q}_i$ we consider are half-BPS operators, which are
characterized by a position~$x_i$, an internal polarization
$\alpha_i$, and a weight $k_i$,
\begin{equation}
\op{Q}_i=\op{Q}(\alpha_i,x_i,k_i)
=\tr\lrbrk{\brk{\alpha_i\cdot\Phi(x_i)}^{k_i}}
\,,\qquad
\alpha_i^2=0
\,.
\label{eq:BPSop}
\end{equation}
Here, $\Phi=\brk{\Phi_1,\dotsc,\Phi_6}$ are the six real scalar fields of
$\superN=4$ super Yang--Mills theory, and $\alpha$ is a
six-dimensional null vector.
We start with the set $\mathbold{\Gamma}$ of all Wick contractions of
the $n$ operators \emph{in the free theory}. Each Wick contraction
defines a graph, whose edges are the propagators. We will use the
terms ``graph'' and ``Wick contraction'' interchangeably. By the procedure
described above, we can associate a compact oriented surface to each
Wick contraction, and thereby define the genus $g(\Gamma)$ of any
given graph $\Gamma$. Importantly, the edges emanating from each
operator have a definite ordering around that operator due to the
color trace in~\eqref{eq:BPSop}.%
\footnote{Graphs with this ordering property are called \emph{ribbon graphs}.}

Next, we promote each graph $\Gamma$ to a
triangulation $\Gamma_\triangle$ in two steps: First, we identify
(``glue together'') all homotopically equivalent (that is, parallel
and non-crossing) lines of the original graph $\Gamma$. The resulting graph is called a
\emph{skeleton graph}. We can assign a ``width'' to each line of the
skeleton graph, which equals the number of lines (propagators) that
have been identified. Each line of the skeleton graph is called a
\emph{bridge} $b$, and the width of the line is conventionally called the
\emph{bridge length} $\ell_b$. There is a propagator factor
$d_b^{\ell_b}$ for each bridge. By definition, each face of a skeleton graph is
bounded by three or more bridges. In a second step, we subdivide faces
that are bounded by $(m>3)$ bridges into triangles by inserting $(m-3)$
further \emph{zero-length bridges} (ZLBs). Using the formula for the Euler
characteristic, one finds that the fully triangulated graph
$\Gamma_\triangle$ has $2n+4g(\Gamma)-4$ faces.

For each bridge $b$ of the triangulated skeleton graph
$\Gamma_\triangle$, we
integrate over a complete set of states $\psi_b$ living on that bridge,
and we insert a weight factor $\mathcal{W}(\psi_b)$. The weight factor
measures the charges of the state $\psi_b$ under a
superconformal transformation that relates the two adjacent
triangular faces; it thus
depends on both the cross ratios of the four neighboring vertices, and
on the labels of the state $\psi_b$.
The worldsheet theory on each bridge is a ``mirror theory'' which is
obtained from the physical worldsheet theory by an analytic continuation via a
double-Wick rotation. States in this theory are
composed of magnons with definite rapidities $u_i\in\Real$ and bound
state indices $a_i\in\Integers_{\geq1}$. A complete set of states is
given by all Bethe states, where each Bethe state is characterized by
the number $m$ of magnons, their rapidities $\brc{u_1,\dotsc,u_m}$,
their bound state indices $\brc{a_1,\dotsc,a_m}$, and their
$\alg{su}(2|2)^2$ flavor labels $(A,\dot A)$. The integration
over the space $\mathbold{M}_b$ of mirror states hence expands to
\begin{equation}
\int_{\mathbold{M}_b}\dd\psi_b
=\sum_{m=0}^\infty\prod_{i=1}^m\sum_{a_i=1}^\infty\sum_{A_i,\dot A_i}\int_{u_i=-\infty}^\infty
\dd u_i\,\mu_{a_i}(u_i)\,e^{-\tilde{E}_{a_i}(u_i)\,\ell_b}
\,,
\end{equation}
where $\mu_{a_i}(u_i)$ is a measure factor, $\tilde{E}$ is the mirror
energy, $\ell_b$ is the length of the bridge~$b$, and the exponential is
a Boltzmann factor for the propagation of the mirror particles across
the bridge.

Finally, each face $a$ of the triangulated skeleton
graph~$\Gamma_\triangle$ carries one hexagon form
factor~$\mathcal{H}_a$, which accounts for the interactions among the
three physical operators $\op{Q}_i$, $\op{Q}_j$, $\op{Q}_k$ as well as the mirror
states on the three edges $b_1$, $b_2$, $b_3$ adjacent to the face. It
is therefore a function of all this data:
\begin{equation}
\mathcal{H}_a
=\mathcal{H}_a(x_i,\alpha_i,x_j,\alpha_j,x_k,\alpha_k;\psi_{b_1},\psi_{b_2},\psi_{b_3})
\,.
\end{equation}
The hexagon form factor is a worldsheet branch-point twist operator that
inserts an excess angle of $\pi$ on the worldsheet. It has been
introduced in~\cite{Basso:2015zoa} for the purpose of computing planar
three-point functions, and has later been applied to compute planar
four-point~\cite{Fleury:2016ykk,Eden:2016xvg} and five-point
functions~\cite{Fleury:2017eph}. Our formula~\eqref{eq:mainformula} is
an extension and generalization of these works to the non-planar
regime. Notably, all ingredients of the formula~\eqref{eq:mainformula}
(measures $\mu_a(u)$, mirror energies $\tilde{E}_a(u)$, and hexagon
form factors $\mathcal{H}$) are known as exact functions of the
coupling $\lambda$, and hence the formula should be valid at finite
coupling.%
\footnote{Of course it is still a sum over infinitely many mirror
states, and as such cannot be evaluated exactly in general. What one
can hope for is that it admits high-loop or even exact expansions in
specific limits. This is the focus of upcoming
work~\cite{TillPedroFrankToAppear,TillPedroFrankVascoWorkInProgress}.}
The hexagon form factors are given in terms of the Beisert
S-matrix~\cite{Beisert:2005tm}, the dressing
phase~\cite{Beisert:2006ib}, as well as analytic continuations among
the three physical and the three mirror theories on the perimeter of the hexagon~\cite{Basso:2015zoa}.

Unlike the general genus expansion~\eqref{eq:genusexpansion}, the
formula~\eqref{eq:mainformula} nicely separates the combinatorial sum
over graphs and topologies from the coupling dependence, since the sum
over graphs only runs over Wick contractions of the \emph{free} theory. At
any fixed genus, the list of contributing graphs can be constructed
once and for all. The dependence on the coupling $\lambda$ sits purely
in the dynamics of the integrable hexagonal patches of worldsheet
$\mathcal{H}$ and their gluing properties.

Finally, we have the very important \emph{stratification} operation indicated
by the operator~${\mathcal{S}}$ in~\eqref{eq:mainformula}. The basic
idea already anticipated in the introduction is that the sum over
graphs mimics the integration over the string moduli space, which
contains boundaries. At those boundaries, it is crucial to avoid
missing or over-counting contributions,
specially in a discrete sum as we have here.\footnote{In moduli space
integrations, this issue can sometimes be glossed over, since the
boundaries are immaterial measure-zero subsets; this is
definitely not the case in our sums.} Despite its innocuous
appearance, it is perhaps the most non-trivial aspect of this paper
and is discussed in great detail below; the curious reader can give a
quick peek at~\eqref{eq:stratification4} below.

In the remainder of this paper, we will flesh out the details of the
formula~\eqref{eq:mainformula}, test it against known perturbative
data at genus one, and use it to make a few higher-loop predictions.

\subsection{Polygonization and Hexagonalization}
\label{sec:polygonization}

The combinatorial part of the 
prescription is to sum
over planar contractions of $n$ operators on a surface with given
genus. We refer to this step as the \emph{polygonization}.
This task can be split into three steps: (1) construct all
inequivalent skeleton graphs with $n$ vertices on the given surface (excluding
edges that connect a vertex to itself), (2) sum over all inequivalent
labelings of the vertices and identify each labeled vertex with one of
the operators, and (3) for each labeled skeleton graph, sum over all possible
distributions of propagators on the edges (bridges) of the graph that is
compatible with the choice of operators, such that each edge carries
at least one propagator.

\paragraph{Maximal Graphs on the Torus.}

In the following, we will construct all inequivalent graphs with four
vertices on the torus. To begin, we classify all graphs with a maximal
number of edges. All other graphs (including those with genus zero)
will be obtained from these ``maximal'' graphs by deleting edges. The
maximal number of edges of a graph with four vertices on the
torus is $12$. Graphs with $12$ edges cut the torus into $8$
triangles. For some maximal graphs, the
number of edges drops to $11$ or $10$, such graphs include squares
involving only two of the four vertices.
Once we blow up the operator insertions to
finite-size circles, all triangles will become hexagons, all squares
will become octagons, and more generally all $n$-gons will become $2n$-gons.

We classify all possible maximal graphs by first putting
only two operators on the torus, and by listing all inequivalent ways to
contract those two operators. This results in a torus cut into some
faces by the bridges among the two operators. Subsequently, we insert
two more operators in all possible ways, and add as many bridges as
possible. We end up with the $16$
inequivalent graphs shown in~\tabref{tab:maxgraphs}.
\begin{table}
\centering
\begin{tabular}{ccccc}
\includegraphicsbox{FigTorusCase11} &
\includegraphicsbox{FigTorusCase121} &
\includegraphicsbox{FigTorusCase122} &
\includegraphicsbox{FigTorusCase13}%
\rule[-1.02cm]{0pt}{1ex}%
\\
1.1 &
1.2.1 &
1.2.2 &
1.3
\\
\includegraphicsbox{FigTorusCase141} &
\includegraphicsbox{FigTorusCase142} &
\includegraphicsbox{FigTorusCase151} &
\includegraphicsbox{FigTorusCase152}
\\
1.4.1 &
1.4.2 &
1.5.1 &
1.5.2
\\
\includegraphicsbox{FigTorusCase153} &
\includegraphicsbox{FigTorusCase16} &
\includegraphicsbox{FigTorusCase211} &
\includegraphicsbox{FigTorusCase212}
\\
1.5.3 &
1.6 &
2.1.1 &
2.1.2
\\
\includegraphicsbox{FigTorusCase213} &
\includegraphicsbox{FigTorusCase22} &
\includegraphicsbox{FigTorusCase31} &
\includegraphicsbox{FigTorusCase32}
\\
2.1.3 &
2.2 &
3.1 &
3.2
\end{tabular}
\caption{Inequivalent maximal graphs on the torus.}
\label{tab:maxgraphs}
\end{table}
Let us explain how we arrive at this classification: Two operators on
the torus can be connected by at most four bridges. It is useful to
draw such a configuration as follows:
\begin{equation}
\includegraphicsbox{FigTorus24}
\;,
\label{eq:torus2op4br}
\end{equation}
where the box represents the torus, with opposing edges identified.
The four bridges cut the torus into two octagons. Placing one further
operator into each octagon and adding all possible bridges gives case~1.1 in~\tabref{tab:maxgraphs}.
When both further operators are placed in the same octagon, there are
two inequivalent ways to distribute the bridges, these are the
cases~1.2.1 and~1.2.2 (here, the fundamental domain of the torus has
been shifted to put the initial octagon in the center). Since each edge in general represents multiple
propagators, we also need to consider cases where the two further
operators are placed \emph{inside} the bridges
of~\eqref{eq:torus2op4br}. Placing one operator in one of the bridges
and the other operator into one of the octagons gives case~1.3
in~\tabref{tab:maxgraphs}. Placing both operators in separate bridges
gives cases~1.4.1 and~1.4.2. Placing both operators into the
same bridge yields cases~1.5.1, 1.5.2, and~1.5.3. Finally, placing
the third operator inside one of the octagons and the fourth operator into
one of the bridges attached to the third operator results in case~1.6.

Next, we need to consider cases where no two operators are connected
by more than three bridges (otherwise we would end up with one of the
previous cases). Again we start by only putting two operators on the
torus. Connecting them by three bridges cuts the torus into one big
dodecagon, which we can depict in two useful ways:
\begin{equation}
\includegraphicsbox{FigTorus23}
\;=\;
\includegraphicsbox{FigTorus23hex}
\;.
\label{eq:torus2op3br}
\end{equation}
In the right figure, opposing bridges are identified, and we have
shaded the two operators to clarify which ones are identical. Placing
the two further operators into the dodecagon results in the three
inequivalent bridge configurations~2.1.1, 2.1.2, and 2.1.3
in~\tabref{tab:maxgraphs}. Placing one operator into one of the
bridges in~\eqref{eq:torus2op3br} results in graph~2.2. We do not need
to consider placing both operators into bridges, as the resulting
graph would not have a maximal number of edges (and thus can be
obtained from a maximal graph by deleting edges).

Finally, we have to consider cases where no two operators are
connected by more than two bridges. In this case, it is easy to
convince oneself that all pairs of operators must be connected by
exactly two bridges. We can classify the cases by picking one operator
(1) and enumerating the possible orderings of its bridges to the other
three operators (2,3,4). It turns out that there are only two
distinguishable orderings (up to permutations of the operators):
(2,3,2,4,3,4) and (2,3,4,2,3,4). In each case, there is only one way
to distribute the remaining bridges (such that no two operators are
connected by more than two bridges):
\begin{equation*}
\includegraphicsbox{FigTorusCase31pre}
\;\rightarrow\;
\includegraphicsbox{FigTorusCase31lab}
\;,\qquad
\includegraphicsbox{FigTorusCase32pre}
\;\rightarrow\;
\includegraphicsbox{FigTorusCase32lab}
\;.
\end{equation*}
These are the graphs~3.1 and~3.2 in~\tabref{tab:maxgraphs}.
This completes the classification of maximal graphs.
In \appref{app:bottom-up-graphs},
we discuss an alternative way (an algorithm that can be implemented
for example in \mathematica) of obtaining the complete set of maximal
graphs for any genus and any number of operator insertions.

\paragraph{Non-Maximal Polygonizations.}

In the above classification of maximal graphs,
each edge stands for one or more parallel propagators. In order to account for
all possible ways of contracting four operators on the torus, we also
have to admit cases where some edges carry zero propagators. We
capture those cases by also summing over graphs with fewer edges. All
of these can be obtained from the set of maximal graphs by iteratively removing
edges. When we remove edges from all maximal graphs in all possible
ways, many of the resulting graphs will be identical, so those have to
be identified in order to avoid over-counting.

\paragraph{Hexagonalization.}

\begin{figure}
\centering
\includegraphics[width=\textwidth]{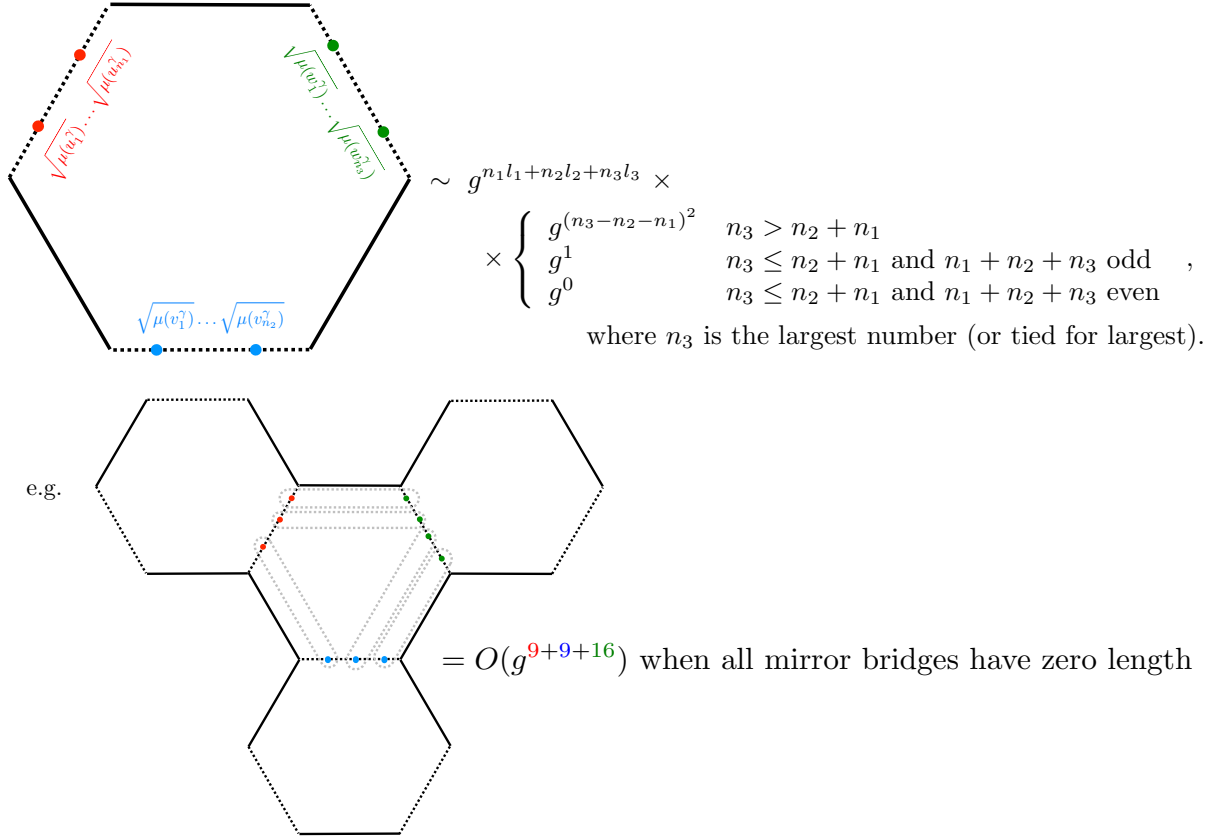}
\caption{To estimate at which loop order a given sprinkling pattern
will start contributing, we can
focus on each hexagon. We absorb in each hexagon one half (\ie the square
root) of the measures and mirror particle
propagation factors of the three adjacent mirror edges. We can then estimate the loop order of a given
populated hexagon by noting that this object has residues where
particles decouple among themselves. For example, the middle hexagon in the
bottom picture must cost no coupling
since it contains residues where
all particles annihilate, leaving an empty hexagon whose expectation
value is just $1$. In other words, in this example, what costs
(a lot!) of loops is to create the particles in the surrounding
hexagons; once they are created, they can freely propagate through the
middle hexagon (\eg following the interior of the dashed regions) and
that costs no coupling at all. The general loop counting is presented
for completeness at the top. It follows
by noticing that after the
decoupling, one is left with mirror particles only on one edge. The integrand for this case
has a trivial matrix part due to the unitarity of the $S$ matrix,
and it is a product ($i<j$) of terms
$h(u_{i}, u_{j})h(u_{j},u_{i}) \sim g^4$,
where $h$ is the hexagon dynamical factor. See the
Appendices~\ref{app:WeakCouplingApendix} and~\ref{app:mirrorparticles}
for the two-particle case.
See also~\cite{Eden:2018vug}.}
\label{fig:counting}
\end{figure}

The next step in our prescription is to tile all graphs of the
polygonization with hexagon form factors, which we refer to as the
\emph{hexagonalization} of the correlator. For many of the maximal
graphs, the hexagonalization is straightforward, as every face has
three edges connecting three operators, giving room to exactly one
hexagon. But some maximal graphs, and in particular graphs with fewer
edges, include higher polygons, which have to be subdivided into
several hexagons. A polygon with $m$ edges (and $m$ cusps)
subdivides into $m-2$ hexagons, which are separated by $m-3$
\emph{zero-length bridges} (ZLBs). In this way, the torus with four
punctures always gets subdivided into eight hexagons.%
\footnote{A surface of genus $g$ with $n$ punctures will be subdivided
into $2n+4g-4$ hexagons. }
Later on, each of these hexagons will be dressed with virtual
particles placed on the \emph{mirror edges} or \emph{bridges}
which will generate the quantum corrections to the correlator under
study, and which we refer to as \emph{sprinkling}.
The general counting of loop order involved in a general sprinkling is illustrated in~\figref{fig:counting}.

\begin{figure}
\centering
\includegraphics[scale=0.85]{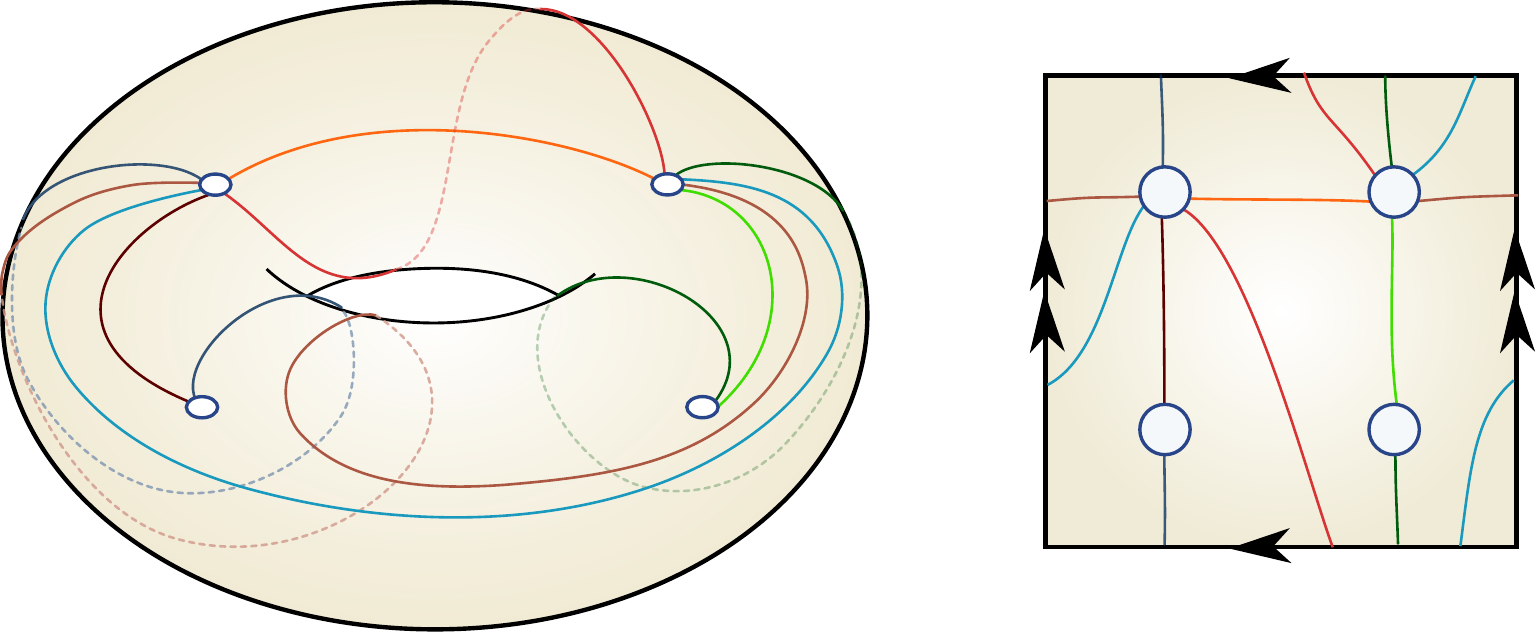}
\caption{Skeleton graph obtained from maximal graph~1.1 by removing
all horizontal lines and some diagonal connections as explained in the main text.}
\label{fig:skeleton}
\end{figure}

\begin{figure}[ht]
\centering
\includegraphics{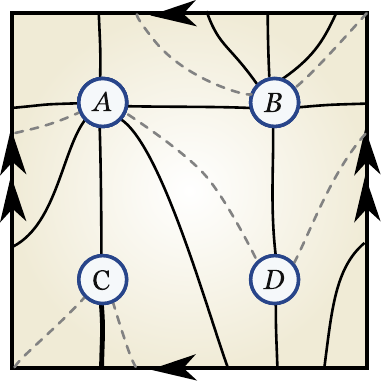}
\caption{The skeleton graph of~\figref{fig:skeleton} can
be completed to a hexagonalization by inserting a zero-length bridge
(ZLB, dashed lines) into each of the four octagons. This
decomposes the four-punctured torus into eight hexagons.}
\label{fig:caseAZLB}
\end{figure}

\begin{figure}
\centering
\includegraphicsbox[width=0.65\textwidth]{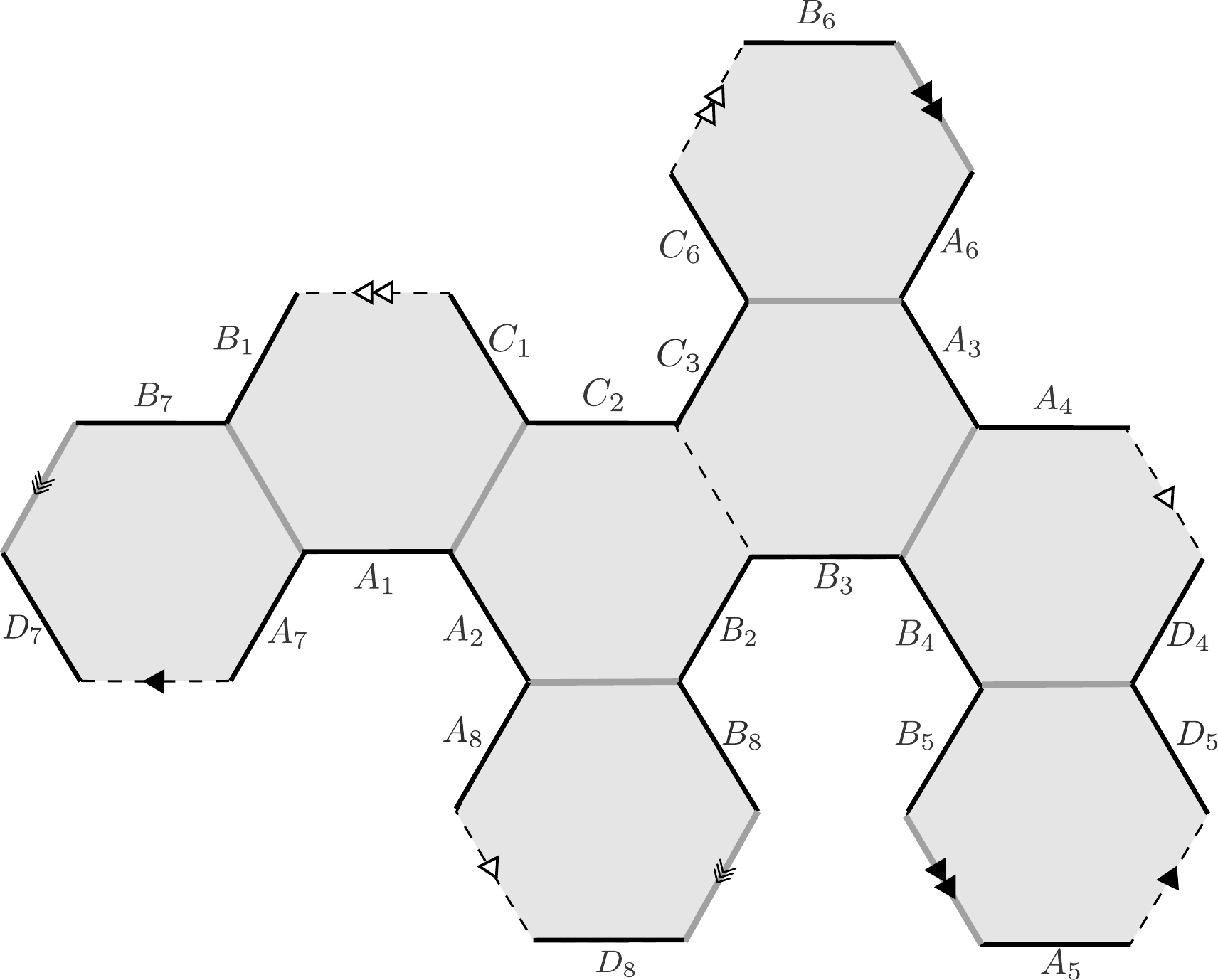}
\caption{The hexagon tiling associated to the hexagonalization
in~\figref{fig:caseAZLB}. The solid black lines attach to the four
operators, the solid gray lines carry one or more propagators, and the
dashed lines are zero-length bridges.}
\label{fig:caseAplan}
\end{figure}

Let us illustrate the hexagonalization with an example. Take the
maximal graph~1.1 of~\tabref{tab:maxgraphs}, and remove the horizontal
lines in the middle, as well as the diagonal lines connecting the
lower operator with the lower two corners. The resulting graph is
depicted in~\figref{fig:skeleton}.
It has eight edges that divide the torus into four octagons. Each
octagon gets subdivided into two hexagons by one zero-length bridge,
as shown in~\figref{fig:caseAZLB}.
In this case, the hexagonalization meant nothing but reinstating the
deleted bridges as ZLBs.
We can now draw the hexagon decomposition in a way that makes the
hexagonal tiles more explicit. This results in the hexagon tiling
shown in~\figref{fig:caseAplan}.

Dressing a skeleton graph such as the one in~\figref{fig:skeleton}
with ZLBs is not unique: Each octagon has two diagonals that we could
choose to become ZLBs. The final answer will be independent of this
choice. This property of the hexagonalization is called flip
invariance~\cite{Fleury:2016ykk}. Hence we can choose any way to cut
bigger polygons into hexagons.

\paragraph{Ribbon Graph Automorphisms and Symmetry Factors.}

When we perform the sum over all graphs and all bridge lengths on the
torus (or higher-genus surface), we need to multiply
some graphs by appropriate symmetry factors.
The graphs we have been classifying are ribbon graphs.
In order to understand the symmetry factors, we will take a closer
look at the formal definition of these ribbon graphs.
A ribbon graph is an ordinary graph together with a cyclic ordering of
the edges at each vertex.%
\footnote{See~\cite{MulasePenkava} for a nice review.}
More formally, ribbon graphs are defined
through \emph{pairing schemes}: Let $\mathcal{V}$ be a collection of
non-empty ordered sets $V_j$,
\begin{equation}
\mathcal{V}=\brc{V_1,\dots,V_v}
\,,\qquad
V_j=\brk{V_{j1},\dots,V_{j\ell_j}}
\,,
\end{equation}
and let
$\tilde{\mathcal{V}}=\brc{V_{11},\dots,V_{1\ell_1},\dots,V_{v1},\dots,V_{v\ell_v}}$
be the union of all $V_j$. A pairing scheme
$P=\brk{\mathcal{V},p}$ is defined by a bijective pairing map
$p:\tilde{\mathcal{V}}\to\tilde{\mathcal{V}}$ with $p^2=1$ and
$p(V)\neq V$ for all $V\in\tilde{\mathcal{V}}$. Each ordered set $V_j$
of $P$ is called a \emph{vertex} of $P$ of degree $\ell_j$.
In our context, each vertex $V_j$ represents one of the operators, and the
$V_{ji}$ label the (half-)bridges attached to operator $j$. The degree
$\ell_j$ is the number of bridges attached to the operator.
$P$~defines a ribbon graph, but also specifies a marked \emph{beginning} of
the ordered sequence of edges (bridges) attached to each vertex. Pairing schemes
are promoted to ribbon graphs by the natural action of the group of
\emph{orientation-preserving isomorphisms}
\begin{equation}
G=\prod_{k=1}^m\grp{S}_{n_k}\rtimes\brk{\Integers/k\Integers}^{n_k}
\,.
\label{eq:ribbongraphgroup}
\end{equation}
Here, $n_k$ is the number of vertices of degree $k$, $m$ is the
maximal degree, $\grp{S}_{n_k}$
permutes vertices of the same degree, and
$\brk{\Integers/k\Integers}^{n_k}$ rotates vertices of degree $k$.
Each orbit $G.P$ of the group action defines a ribbon graph. In other words, a ribbon
graph $\Gamma$ associated with a pairing scheme $P$ is the equivalence
class of $P$ with respect to the action of $G$.

Typically an element of the group~\eqref{eq:ribbongraphgroup} maps a
given pairing scheme $P$ to a different pairing scheme $P'$ (by
permuting vertices and/or
shifting the marked beginnings of the ordered sequences of edges/bridges at
each vertex/operator). However,
some group elements may map a pairing scheme $P$ to itself. If
$\Gamma$ is a ribbon graph associated with a pairing scheme $P$, then
the subgroup of~\eqref{eq:ribbongraphgroup} preserving $P$ is called
the \emph{automorphism group} $\aut(\Gamma)$ of $\Gamma$.%
\footnote{The automorphism group is independent of the
choice of pairing scheme $P$ representing $\Gamma$.}

Assigning a positive real number to each edge of a ribbon graph
promotes it to a \emph{metric ribbon graph}. The number assigned to a
given edge is called the \emph{length} of that edge. Therefore, a
graph with assigned bridge lengths is a metric ribbon graph (with
integer edge lengths). The notion of automorphism group
extends to metric ribbon graphs in an obvious way.

In the sum over graphs and bridge lengths, we need to divide each
graph with assigned bridge lengths (metric graph) by the size of its
automorphism group. These are the symmetry factors mentioned at the
beginning of this paragraph.

Let us illustrate the idea with an example. Consider the following
rather symmetric ribbon graph with eight edges, with all bridge
lengths set to one:
\begin{equation}
\includegraphicsbox{FigTorusCasePmarked1}
\,\xrightarrow{g\in G}\,
\includegraphicsbox{FigTorusCasePmarked2}
\,\xrightarrow{\text{shift}}\,
\includegraphicsbox{FigTorusCasePmarked1}
\label{eq:automorphismexample1}
\end{equation}
In the left picture, the graph is represented by an arbitrarily chosen
pairing scheme, where the beginnings/ends of the edge sequences at
each vertex are indicated by the small blue cuts. The second picture
shows the pairing scheme obtained by applying an isomorphism $g\in G$
that cyclically rotates all vertices by two sites. In the second step,
we shift the cycles along which we cut the torus in order to represent
it in the plane. As a result, we see that the pairing scheme after
applying $g$ is the same as the original pairing scheme on the left.
Thus this graph has to be counted with a symmetry factor of $1/2$
(there is no other non-trivial combination of rotations that leave the
graph invariant, and hence the automorphism group has size $2$).
If we increase the bridge length on two of the edges to two, we find
the following:
\begin{equation}
\includegraphicsbox{FigTorusCasePmarked3}
\,\xrightarrow{g\in G}\,
\includegraphicsbox{FigTorusCasePmarked4}
\end{equation}
As can be seen from the pictures, applying the same group element to
the original pairing scheme results in a different pairing scheme that
cannot be brought back to the original by any trivial operation. In
this case, the automorphism group is trivial, and the graph has to be
counted with trivial factor $1$.

The symmetry factors can also be understood from the point of view of
field contractions: When writing the sum over contractions as a sum
over graphs and bridge lengths, we pull out an overall factor
of $k^4$ that accounts for all possible rotations of the four
single-trace operators. For some graphs and choices of bridge lengths,
non-trivial rotations of the four operators can lead to identical
contractions, which are thus over-counted by the overall factor $k^4$.
This can be seen explicitly in the above
example~\eqref{eq:automorphismexample1}. Dividing by the size of the
automorphism group exactly cancels this over-counting.

\subsection{Stratification}
\label{sec:stratification}

The fact that we are basing the contribution at a given genus $g$ on
the sum over graphs of genus~$g$ is of course natural from the point
of view of perturbative gauge theory: Each graph with assigned bridge
lengths is equivalent to a Feynman graph of the free theory. Summing
over graphs of genus~$g$ and over bridge lengths (weighted by
automorphism factors) is therefore equivalent to summing over all
free-theory Feynman graphs of genus~$g$. All perturbative corrections
associated to a given graph are captured by the product of hexagon
form factors as well as the sums and integrations over mirror states
associated to that graph. It is clear that this prescription cannot be
complete, as it does not include loop corrections that increase the
genus of the underlying free graph. It also omits contributions from disconnected free graphs
that become connected after adding interactions. In other words, it does not
include contributions from handles or connections formed purely by virtual processes.
We can include such contributions by drawing lower-genus and disconnected graphs on a
genus-$g$ surface in all possible ways, and tessellating the genus-$g$
surface into hexagons \emph{including the handles not covered by the
lower-genus graph}.
Weighting such contributions by the same
genus-counting factor $N^{2-2g-n}$ as the honestly genus-$g$ graphs,
we include \emph{all} virtual processes that contribute at this genus.
In other words, the sum over graphs in~\eqref{eq:mainformula} has to be replaced as
\begin{equation}
\sum_{\Gamma\in\mathbold{\Gamma}}=
\sum_{g=0}^\infty
\sum_{\substack{\text{graphs }\Gamma\\\text{of }{\text{genus }g}}}
\qquad
\rightarrow
\qquad
\sum_{g=0}^\infty
\sum_{\Gamma\in\Sigma_g}
\,,
\label{eq:stratification1}
\end{equation}
where $\Sigma_g$ is the set of all graphs, connected or
disconnected, of genus $g$ or smaller. For
graphs whose genus is smaller than $g$, the symbol $\Gamma\in\Sigma_g$ has to
carry not only the information of the graph itself, but also of its
embedding in the genus-$g$ surface. The embedding can for example be
encoded by marking all pairs of faces of the graph to which an extra
handle is attached.

While this prescription solves the problem of
capturing all genus-$g$ contributions, it also spoils the result by
including genuine lower-genus contributions. Namely, the loop
expansion of the hexagon gluing (sum over mirror states) will also
include processes where one or more extra handles (those not covered
by the graph) remain completely void. Such void handles can be pinched.
Pinching a handle reduces the genus, hence such contributions do not
belong to the genus-$g$ answer. However, we can get rid of these
unwanted contributions by subtracting the same lower-genus graphs, but
now drawn on a surface where a handle has been pinched. Pinching a
handle reduces the genus by one, leaving two marked points on the
reduced surface. For an $n$-point function, we hence have to subtract
all $n$-point graphs drawn on a genus $(g-1)$ surface with $2$ marked
points. Such contributions naturally come with the correct
genus-counting factor $N^{2-2(g-1)-(n+2)}=N^{2-2g-n}$. Hence we have
to refine~\eqref{eq:stratification1} to
\begin{equation}
\text{RHS of~\eqref{eq:stratification1}}
\quad
\rightarrow
\quad
\sum_{g=0}^\infty
\biggbrk{
\sum_{\Gamma\in\Sigma_g}
-
\sum_{\Gamma\in\Sigma_{g-1}^2}
}
\,,
\label{eq:stratification2}
\end{equation}
where $\Sigma_{g-1}^2$ is the set of all graphs of genus $(g-1)$ or
smaller embedded in a genus $(g-1)$ surface, with two marked points
inserted into any two faces of the graph (or both marked points
inserted into the same face). This subtraction correctly removes all
excess contributions from the first sum that have exactly one void
handle. In contrast, the excess contributions with two void handles
are contained \emph{twice} in the subtraction sum, once for each
handle that can be pinched. We have to re-add these contributions once
by further refining~\eqref{eq:stratification2} to
\begin{equation}
\text{RHS of~\eqref{eq:stratification2}}
\quad
\rightarrow
\quad
\sum_{g=0}^\infty
\biggbrk{
\sum_{\Gamma\in\Sigma_g}
-
\sum_{\Gamma\in\Sigma_{g-1}^2}
+
\sum_{\Gamma\in\Sigma_{g-2}^4}
}
\,,
\label{eq:stratification3}
\end{equation}
where now $\Sigma_{g-2}^4$ is the set of all graphs of genus $(g-2)$ or
smaller embedded in a genus $(g-2)$ surface, with two pairs of marked points
inserted into any four (or fewer) faces of the graph.
This procedure iterates, leading to the refinement
\begin{equation}
\text{RHS of~\eqref{eq:stratification3}}
\quad
\rightarrow
\quad
\sum_{g=0}^\infty
\sum_{m=0}^g
(-1)^m\sum_{\mathclap{\Gamma\in\Sigma_{g-m}^{2m}}}
\,.
\label{eq:stratificationExtra}
\end{equation}

Under the degenerations discussed thus far, the Riemann surface stays
connected. There are also degenerations that split the Riemann surface
into two components by pinching an intermediate cylinder. Also these
degenerations have to be subtracted in order to cancel unwanted
contributions (that originate from disconnected propagator graphs, or from purely
virtual ``vacuum'' loops). Such degenerations split a Riemann surface of genus $g$
with $n$ punctures into two components with genus $g_1$ and $g_2$ that
contain $n_1$ and $n_2$ punctures, such that $g_1+g_2=g$ and $n_1+n_2=n$. Each
component carries one marked point that remains from pinching.
Such contributions also come with the correct genus-counting factor
\begin{equation}
N^{2-2 g_1-(n_1+1)}N^{2-2 g_2-(n_2+1)}=N^{2-2g-n}
\,.
\end{equation}
Again, the pinching process can iterate, splitting the surface into
more and more components.%
\footnote{Starting with a surface of genus $g$ with $n$ punctures, the maximum
number of iterated degenerations (of both types described above) is
$3g+n-3$, resulting in a surface with $2g+n-2$ components, where each
component is a pair of pants (sphere with three punctures and/or
marked points).
This bound is saturated when we perform the
reduction starting from a maximally disconnected planar graph that
is embedded on the surface in a disk-like region (\ie without any windings). For even $n$, a maximally disconnected planar graph has $n/2$
components, each consisting of two operators connected by a single
bridge. In this case, the maximal degeneration consists of spheres
that contain either one component of the graph and one marked point,
or no part of the graph and three marked points. For odd $n$, a
maximally disconnected planar graph has $(n-1)/2$ components, where one of
the components is a triangular three-point graph (because
every operator has at least one bridge attached). In this case, the
maximal number of degenerations is $3g+n-4$, resulting in $2g+n-3$
surface components.}
We will comment on this type of contributions at the end of
\secref{sec:strat} and in \appref{app:disconnected}.

Summing all possible degenerations with their respective signs, we
arrive at the following final formula, which
is a further refinement of~\eqref{eq:stratificationExtra}:
\begin{equation}
\text{RHS of~\eqref{eq:stratificationExtra}}
\quad
\rightarrow
\quad
\sum_{g=0}^\infty
\sum_{c=1}^{2g+n-2}\mspace{-8mu}
\sum_{\tau\in\mathbold{\tau}_{g,c,n}}\mspace{-3mu}
(-1)^{\sum_im_i/2}\sum_{\Gamma\in\Sigma_\tau}
\equiv
\mathcal{S}\circ
\sum_{\Gamma\in\mathbold{\Gamma}}
\,.
\label{eq:stratification4}
\end{equation}
Here, $c$ counts the number of components of the surface, and the sum over
$\tau$ runs over the set of all genus-$g$ topologies with $c$
components and $n$ punctures:
\begin{equation}
\mathbold{\tau}_{g,c,n}=\bigbrc{
\lrbrc{\brk{g_1,n_1,m_1},\dotsc,\brk{g_c,n_c,m_c}}
\big|
\sum_in_i=n
\,,\;
\sum_i\brk{g_i+m_i/2}-c+1=g
}\,,
\label{eq:topologies}
\end{equation}
where $(g_i,n_i,m_i)$ labels the genus, the number of punctures, and
the number of marked points on component $i$. Finally, we sum over the
set $\Sigma_\tau$ of all graphs $\Gamma$ (connected and disconnected)
that are compatible with the topology $\tau$ and that are embedded in
the surface defined by $\tau$ in all inequivalent possible ways
($\Gamma$ may cover all or only some components of the surface).

In the rightmost expression, we have defined the \emph{stratification
operator} $\mathcal{S}$, which implements the refinement of adding and
subtracting graphs on surfaces of genus ${\leq}g$ with marked points
as just explained. It appears intricate as it stands, but we will see
below that it turns out less complicated than it looks.

We motivated this proposal from gauge theory considerations. We could
have arrived at the very same expression by following string moduli space
considerations as explained in the introduction, by carefully subtracting the
boundary of the discretized moduli space~\cite{Chekov:1992ns,Chekhov:1995cq}.%
\footnote{The map between the moduli space
and metric ribbon graphs induces a cell decomposition on the moduli
space. The highest-dimensional cells are covered by graphs with a
maximal number of edges. Cell boundaries are reached by sending some
bridge length to zero. (The neighboring cell is reached by flipping
the resulting ZLB and making its length positive again.) The moduli
space $\mathcal{M}_{g,n}$ itself also has a boundary, which is reached when a
handle (cylinder) becomes infinitely thin. In terms of ribbon graphs, this
boundary is reached when all bridges traversing a cylinder reduce to
zero size. The minimal number of bridges traversing a cylinder is two,
hence the moduli space boundaries have complex codimension one.
The highest-dimensional
cells (bulk of the space) have complex dimension $3g+n-3$, which
explains the maximal number of iterated degenerations. The alternating
sign in~\eqref{eq:stratification4} is also natural from this point of view.}
%

\paragraph{Example.}

Let us illustrate the above construction with an important example.
Consider the correlator for four equal-length single-trace operators
$\op{Q}_1,\dots,\op{Q}_4$ that are chosen such that the fields in
$\op{Q}_1$ cannot contract with the fields in $\op{Q}_4$, and the
fields in $\op{Q}_2$ cannot contract with the fields in $\op{Q}_3$.
Correlators of this type are studied throughout the rest of this
paper. For such correlators, there is only one planar graph:
\begin{equation}
\includegraphicsbox{FigPlanarGraph}
\,.
\label{eq:planargraph}
\end{equation}
At genus one, stratification requires that we include contributions
from this graph drawn on a torus in all possible ways. An obvious way
of drawing the planar graph on the torus is (the torus is drawn as a
square, opposing sides of the square have to be identified)
\begin{equation}
\includegraphicsbox{FigTorusStrat1lab}
\,.
\label{eq:stratex0}
\end{equation}
Pinching the handle of the torus leads back to the original graph
drawn on the plane, with two marked points remaining where the handle
got pinched:
\begin{equation}
\includegraphicsbox{FigTorusStrat1lab}
\quad
\xrightarrow{\;\text{pinching}\;}
\quad
\includegraphicsbox{FigTorusStrat1sublab}
\,.
\label{eq:stratex0sub}
\end{equation}
According to the stratification prescription, the contribution
from~\eqref{eq:stratex0} has to be added, whereas the contribution
from~\eqref{eq:stratex0sub} (right-hand side) has to be subtracted in
the computation of the genus-one correlator. Of course there are many
more ways to draw the planar graph on a torus. Finding all such ways
amounts to adding an empty handle to the planar graph in all possible
ways. This in turn is equivalent to inserting two marked points into
the planar graph in all possible ways, which mark the insertion points
of the added handle. In other words, we can find all ways of drawing
the planar graph on the torus by drawing graphs of the type shown
on the right-hand side of~\eqref{eq:stratex0sub}. The two marked points can either be put
into faces of the original graph, as in~\eqref{eq:stratex0sub}, but
they can also be put \emph{inside bridges}---a bridge stands for a
collection of parallel propagators, hence it can be split in two by an
extra handle. Going through all possibilities, we find the
seven types of
contributions listed in~\tabref{tab:stratexcyc}.
\begin{table}[tbp]
\centering
\begin{tabular}{|c@{\quad}|@{\quad}c|}
\hline
& \\[-1ex]
    \begin{tabular}{c@{\;\;}c@{\;\;}c}
    \includegraphicsbox{FigTorusStrat1} &
    $\rightarrow$ &
    \includegraphicsbox{FigTorusStrat1sub}
    \\ && \\[-2.5ex]
    $(1)$ && $(1')$
    \\[1.5ex]
    \multicolumn{3}{c}{
    \hfill
    \begin{tabular}{@{}cc@{}}
    \raisebox{2ex}{$\searrow$} &
    \includegraphicsbox{FigTorusStrat1sub2}
    \\ & \\[-2.5ex]
    & $(1'')$
    \end{tabular}
    }
    \end{tabular}
&
    \begin{tabular}{c@{\;\;}c@{\;\;}c}
    \includegraphicsbox{FigTorusStrat2} &
    $\rightarrow$ &
    \includegraphicsbox{FigTorusStrat2sub}
    \\ && \\[-2.5ex]
    $(2)$ && $(2')$
    \end{tabular}
\\[-2ex] & \\
\hline
& \\[-1ex]
    \begin{tabular}{c@{\;\;}c@{\;\;}c}
    \includegraphicsbox{FigTorusStrat3} &
    $\rightarrow$ &
    \includegraphicsbox{FigTorusStrat3sub}
    \\ && \\[-2.5ex]
    $(3)$ && $(3')$
    \end{tabular}
&
    \begin{tabular}{c@{\;\;}c@{\;\;}c}
    \includegraphicsbox{FigTorusStrat4} &
    $\rightarrow$ &
    \includegraphicsbox{FigTorusStrat4sub}
    \\ && \\[-2.5ex]
    $(4)$ && $(4')$
    \end{tabular}
\\[-2ex] & \\
\hline
& \\[-1ex]
    \begin{tabular}{c@{\;\;}c@{\;\;}c}
    \includegraphicsbox{FigTorusStrat5} &
    $\rightarrow$ &
    \includegraphicsbox{FigTorusStrat5sub}
    \\ && \\[-2.5ex]
    $(5)$ && $(5')$
    \end{tabular}
&
    \begin{tabular}{c@{\;\;}c@{\;\;}c}
    \includegraphicsbox{FigTorusStrat6} &
    $\rightarrow$ &
    \includegraphicsbox{FigTorusStrat6sub}
    \\ && \\[-2.5ex]
    $(6)$ && $(6')$
    \end{tabular}
\\[-2ex] & \\
\hline
\multicolumn{2}{|c|}{} \\[-1ex]
\multicolumn{2}{|c|}{
    \begin{tabular}{c@{\;\;}c@{\;\;}ccc}
    \includegraphicsbox{FigTorusStrat7} &
    \;$\rightarrow$\; &
    \includegraphicsbox{FigTorusStrat7sub} &
    , &
    \includegraphicsbox{FigTorusStrat7sub2}
    \\ &&&& \\[-2.5ex]
    $(7)$ && $(7')$ && $(7'')$
    \end{tabular}
}
\\[-2ex] \multicolumn{2}{|c|}{} \\
\hline
\end{tabular}
\caption{List of stratification contributions for a genus-one
four-point correlator $\vev{\op{O}_1\dots\op{O}_4}$ of equal-weight
operators $\op{O}_1,\dotsc,\op{O}_4$, where $\op{O}_1$
cannot contract with $\op{O}_4$, and $\op{O}_2$ cannot contract with
$\op{O}_3$. Each case has to be summed over all inequivalent labelings
of the four operators. Unprimed contributions $(i)$ are planar graphs
drawn on a torus and thus have to be counted with a positive sign.
Primed contributions $(i')$ are obtained from their unprimed
counterparts by pinching a handle and thus have to be counted with a
negative sign. Doubly primed contributions $(i'')$ are obtained by
pinching off the entire torus, they also have to be counted with a
negative sign.}
\label{tab:stratexcyc}
\end{table}

In the table, we have listed unlabeled graphs, which have to be summed
over inequivalent labelings. One may wonder why we have not included a
variant of case~(1) where the two marked points are ``inside'' the
planar graph. In fact, this other case is included in the sum over
labelings of case~(1): Putting the two marked points ``inside'' the
graph is equivalent to turning the graph~(1) ``inside out'', which
amounts to reversing the cyclic labeling of the four operators.
Similarly for case~(3), the case where the exterior marked point sits
inside the central face is included in the sum over labelings.

We will see below that mirror particle contributions may cancel
propagator factors of the underlying free-theory graph. We therefore
have to also sum over graphs containing propagators that are
ultimately not admitted by the external operators. From an operational
point of view, this is equivalent to only restricting the operator
polarizations at the very end of the computation. For operators of
equal weight but generic polarizations, the only planar four-point
graph besides~\eqref{eq:planargraph} is the ``tetragon graph''
\begin{equation}
\raisebox{1ex}{\includegraphicsbox{FigPlanarGraphNonCyc}}
\;.
\label{eq:planargraphnoncyc}
\end{equation}
Putting this graph on a torus in all possible ways, we find eight
inequivalent cases, listed in~\tabref{tab:stratexnoncyc} and labeled
$(8)$--$(15)$.
\begin{table}[tbp]
\centering
\begin{tabular}{|c@{\quad}|@{\quad}c|}
\hline
\multicolumn{2}{|c|}{} \\[-1ex]
\multicolumn{2}{|c|}{
    \begin{tabular}{c@{\;\;}c@{\;\;}ccc}
    \includegraphicsbox{FigTorusStrat8} &
    \;$\rightarrow$\; &
    \includegraphicsbox{FigTorusStrat8sub} &
    , &
    \includegraphicsbox{FigTorusStrat8sub2}
    \\ &&&& \\[-2.5ex]
    $(8)$ && $(8')$ && $(8'')$
    \end{tabular}
}
\\[-2ex] \multicolumn{2}{|c|}{} \\
\hline
& \\[-1ex]
    \begin{tabular}{c@{\;\;}c@{\;\;}c}
    \includegraphicsbox{FigTorusStrat9} &
    $\rightarrow$ &
    \includegraphicsbox{FigTorusStrat9sub}
    \\ && \\[-2.5ex]
    $(9)$ && $(9')$
    \end{tabular}
&
    \begin{tabular}{c@{\;\;}c@{\;\;}c}
    \includegraphicsbox{FigTorusStrat10} &
    $\rightarrow$ &
    \includegraphicsbox{FigTorusStrat10sub}
    \\ && \\[-2.5ex]
    $(10)$ && $(10')$
    \end{tabular}
\\[-2ex] & \\
\hline
\multicolumn{2}{|c|}{} \\[-1ex]
\multicolumn{2}{|c|}{
    \begin{tabular}{c@{\;\;}c@{\;\;}ccc}
    \includegraphicsbox{FigTorusStrat11} &
    \;$\rightarrow$\; &
    \includegraphicsbox{FigTorusStrat11sub} &
    , &
    \includegraphicsbox{FigTorusStrat11sub2}
    \\ &&&& \\[-2.5ex]
    $(11)$ && $(11')$ && $(11'')$
    \end{tabular}
}
\\[-2ex] \multicolumn{2}{|c|}{} \\
\hline
& \\[-1ex]
    \begin{tabular}{c@{\;\;}c@{\;\;}c}
    \includegraphicsbox{FigTorusStrat12} &
    $\rightarrow$ &
    \includegraphicsbox{FigTorusStrat12sub}
    \\ && \\[-2.5ex]
    $(12)$ && $(12')$
    \end{tabular}
&
    \begin{tabular}{c@{\;\;}c@{\;\;}c}
    \includegraphicsbox{FigTorusStrat13} &
    $\rightarrow$ &
    \includegraphicsbox{FigTorusStrat13sub}
    \\ && \\[-2.5ex]
    $(13)$ && $(13')$
    \end{tabular}
\\[-2ex] & \\
\hline
& \\[-1ex]
    \begin{tabular}{c@{\;\;}c@{\;\;}c}
    \includegraphicsbox{FigTorusStrat14} &
    $\rightarrow$ &
    \includegraphicsbox{FigTorusStrat14sub}
    \\ && \\[-2.5ex]
    $(14)$ && $(14')$
    \end{tabular}
&
    \begin{tabular}{c@{\;\;}c@{\;\;}c}
    \includegraphicsbox{FigTorusStrat15} &
    $\rightarrow$ &
    \includegraphicsbox{FigTorusStrat15sub}
    \\ && \\[-2.5ex]
    $(15)$ && $(15')$
    \end{tabular}
\\[-2ex] & \\
\hline
\end{tabular}
\caption{Additional stratification contributions for a genus-one
four-point correlator $\vev{\op{O}_1\dots\op{O}_4}$ of equal-weight
operators $\op{O}_1,\dotsc,\op{O}_4$ of generic polarizations. These
have to be included even if some operators are polarized such that
they ultimately cannot contract, because mirror particle contributions
may cancel propagator factors of the underlying free graph.}
\label{tab:stratexnoncyc}
\end{table}
For the graph~\eqref{eq:planargraphnoncyc}, all faces are
equivalent. Therefore, it is clear that all ways of placing one or two marked
points into the several faces are equivalent (up to operator relabelings).
Therefore, we include only one representative of all these variants.
As for the cases listed in~\tabref{tab:stratexcyc}, the
stratification prescription requires that the unprimed contributions
should be added, while the primed contributions should be subtracted.

Thus far, we have accounted for pinchings where the handle of the
torus becomes infinitely thin. However, for cases~(1),~(7),~(8) and~(11) there is
another way to pinch, where one separates the whole torus from the
graph, leaving an empty torus with one marked point, and the graph on a sphere
with one marked point inside the face that previously contained the
torus. These cases are labeled~($1''$),~($7''$),~($8''$) and~($11''$)
in~\tabref{tab:stratexcyc} and~\tabref{tab:stratexnoncyc}, and have to be subtracted as well.

For connected graphs, these two types of degenerations are all that
can occur at genus one, since
these are the only types of degenerations a torus admits, as illustrated in
\figref{fig:pinching1} and \figref{fig:pinching2}.
Disconnected graphs do not contribute to any computation in
this paper, and hence are not considered here.

To summarize, the effect of stratification at genus one, for
correlators of the type considered here, is that the sum over
genus-one graphs has to be augmented by a sum over the unprimed graphs
(with positive sign) and a sum over the primed graphs (with negative
sign) of~\tabref{tab:stratexcyc} and~\tabref{tab:stratexnoncyc}:
\begin{equation}
\vev{\op{Q}_1\dotsc\op{Q}_4}^{(g=1)}=
\Bigbrk{\begin{tabular}{@{}c@{}}\texttt{genus-one}\\\texttt{graphs}\end{tabular}}
+\underbrace{
    \frac{k^2}{\Nc^4}\sum_{i=1\vphantom{\}}}^{14} S_{(i)}
}_{\substack{\equiv\;\texttt{secretly}\\\texttt{planar}}}
\underbrace{\mbox{}
    -\frac{k^2}{\Nc^4}\sum_{i=1}^{14}S_{(i')}
    -\frac{k^2}{\Nc^4}\sum_{i\in\brc{1,7,8,11}}\mspace{-20mu}S_{(i'')}
}_{\equiv\;\texttt{subtraction}}
\,,
\label{eq:stratificationgenus1}
\end{equation}
where $S_{(i)}$, $S_{(i')}$, and $S_{(i'')}$ stand for the full contributions (sums
over bridge lengths and mirror states) of the respective graphs.
Note that, by construction,
the genus-one stratification
formula~\eqref{eq:stratificationgenus1} is sufficiently general to
hold for half-BPS operators $\op{Q}_i$ of arbitrary polarizations
$\alpha_i$ (but equal weights $k_i$).

\subsection{Subtractions}
\label{sec:subtractions}

We now explain how to compute the contributions from graphs associated
with the degenerate Riemann surfaces, namely $(i')$'s
and $(i'')$'s
in~\tabref{tab:stratexcyc} and~\tabref{tab:stratexnoncyc}.

\begin{figure}[t]
\centering
\includegraphics[scale=0.7]{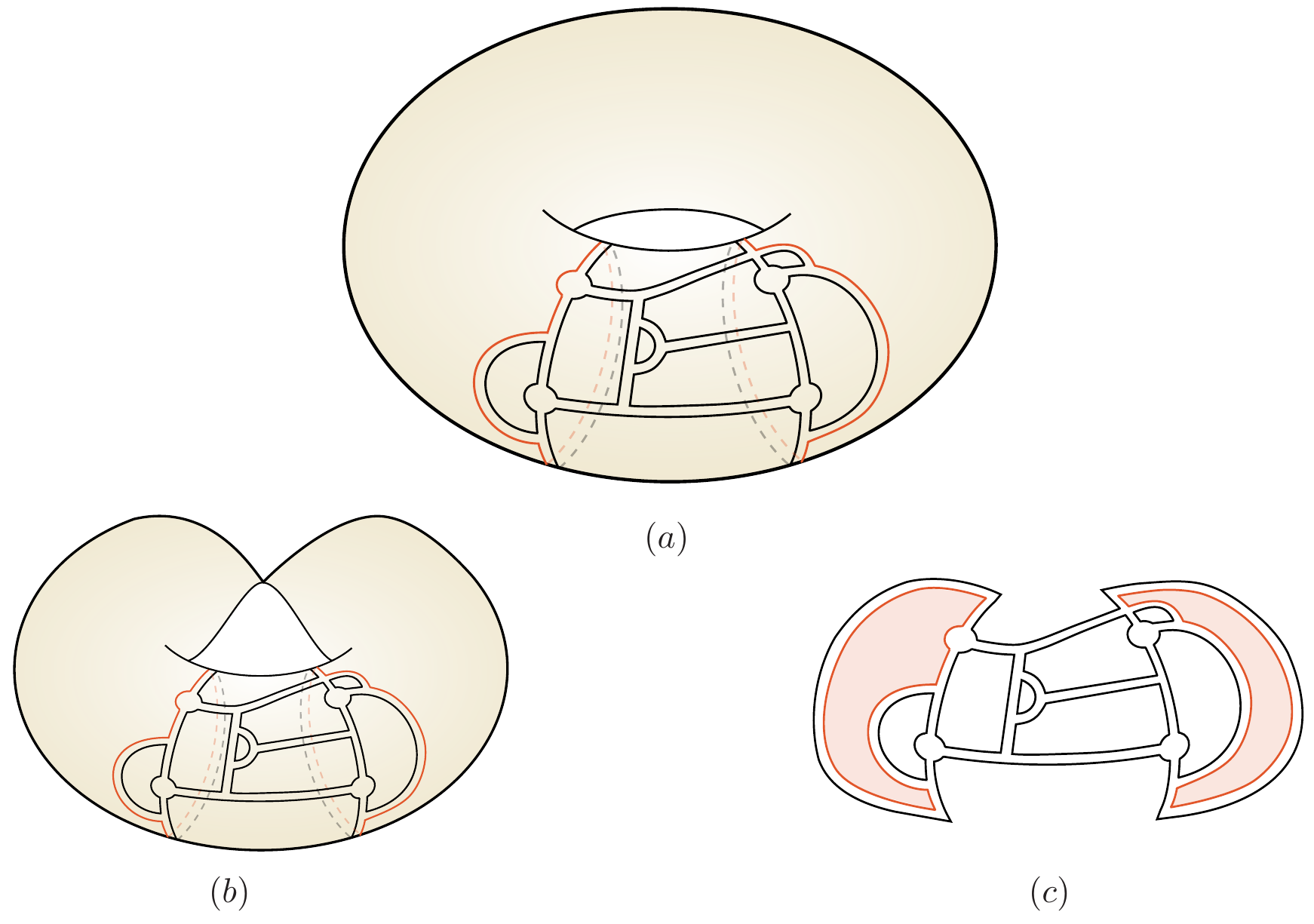}
\caption{Diagrammatic interpretation of marked points. (a) The diagram
corresponding to the degenerate Riemann surface. (b) Pinching
procedure. (c) The diagram after pinching. The red regions in (c)
correspond to the marked points discussed in the previous
subsection.}
\label{fig:pinching1}
\end{figure}

\paragraph{Marked Points as Holes in Planar Diagrams.}

The first step of the computation is to better understand what the marked
points ($\otimes$'s in \tabref{tab:stratexcyc}) represent. For this
purpose, it is useful to look at the corresponding Feynman graphs in
the double-line notation. An example Feynman diagram that
contributes to a stratification subtraction is depicted in
\figref{fig:pinching1}. Although drawn on a torus, it is essentially a
planar diagram, and therefore corresponds to a degenerate Riemann
surface. After the degeneration of the torus (see
\figref{fig:pinching1}(b)), the pinched handle becomes two red regions as
shown in \figref{fig:pinching1}(c), which are the faces of the original
planar diagram. We thus conclude that, at the diagrammatic level,
inserting two marked points on the sphere amounts to specifying two
holes/faces of all planar Feynman graphs.
For a planar graph $\mathcal{G}$ with $F$ faces,
there are $\operatorname{Binomial}(F,2)=F(F-1)/2$ different
ways of specifying two holes in two different faces of the graph.
Thus the contribution of a graph with
two marked points in different faces (denoted by $\mathcal{G}_{2\otimes}$) is given in
terms of the contribution of the original graph
$\mathcal{G}$ as
\begin{align}
\mathcal{G}_{2\otimes}=\frac{F(F-1)}{2}\times\mathcal{G}\,,
\label{eq:G2otimes}
\end{align}
where $F$ is the number of faces in $\mathcal{G}$.
This provides a clear diagrammatic interpretation of the marked points,
but it does not immediately tell us how to compute them using
integrability, since one cannot in general isolate the contributions
of individual Feynman diagrams in the integrability computation. To
perform the computation, we need to relate them to yet another object
that we discuss below.

The key observation is that the same factor $F(F-1)/2$
appears when we shift the rank of the gauge group: Consider the
planar Feynman diagram $\mathcal{G}$ in U$(\Nc)$ $\superN=4$ SYM,
and
change the rank from $\Nc$ to $\Nc +1$. Since each face of the planar
diagram gives a factor of $\Nc$, the shift of $\Nc$ produces the
following change in the final result
\begin{align}
\mathcal{G}
\quad
\xrightarrow{\;\Nc\to\Nc+1\;}
\quad
\lrbrk{\frac{\Nc+1}{\Nc}}^{F}\mathcal{G}
=\lrbrk{1+\frac{F}{\Nc}+\frac{F(F-1)/2}{\Nc^2}+\cdots}\mathcal{G}
\,.
\label{eq:GshiftN}
\end{align}
This offers a reasonably simple way to compute the contribution from
the degenerate Riemann surface: Namely we just need to
\begin{enumerate}
\item Take the planar result and shift the rank of the gauge group from $\Nc$ to $\Nc+1$.
\item Expand it at large $\Nc$ and read off the $1/\Nc^2$ correction.
\end{enumerate}
With this procedure, one can automatically obtain the correct
combinatorial factor without needing to break up the planar results
into individual Feynman diagrams.

Before applying this to our computations, let us add some
clarifications: Firstly, when we shift $\Nc$ to $\Nc+1$, we keep
the Yang--Mills coupling constant $\gym$ fixed, \emph{not} the 't~Hooft
coupling constant $\lambda=\gym^2\Nc$. Put differently, we must shift the value
of $\lambda$ when we perform the shift of $\Nc$. Secondly, the planar
correlators to which we perform the shift must be \emph{unnormalized}:
If we normalize the planar correlators so that the two-point function
is unit-normalized, the shift of $\Nc$ will no longer produce the
correct combinatorial factor dependent on $F$.

It is now straightforward to evaluate the contribution from
degenerate Riemann surfaces explicitly. The planar connected correlator
for BPS operators of weights (lengths) $k_i$ admits the following expansion
\begin{align}
G_{\{k_1,\dotsc,k_n\}}^{(\Nc)}=\Nc^{\mathcal{K}+2-n}\sum_{\ell=0}^{\infty}c_\ell\,\lambda^{\ell}\,,
\end{align}
where $c_\ell$ is a coefficient independent of $\Nc$ and $\lambda$, and
$\mathcal{K}=\sum_i k_i/2$. Shifting $\Nc$ to $\Nc+1$, we obtain
\begin{align}
G_{\{k_1,\dotsc,k_n\}}^{(\Nc+1)}
=\Nc^{\mathcal{K}+2-n}\sum_{\ell=0}^{\infty}c_\ell\,\lambda^{\ell}
\lrsbrk{
1
+\frac{\mathcal{K}+2-n+\ell}{\Nc}
+\frac{1}{\Nc^2}\begin{pmatrix}\mathcal{K}+2-n+\ell\\2\end{pmatrix}
+\dots
}
\end{align}
We thus conclude that the correlator
$G_{\{k_1,\dotsc,k_n\}}^{2\otimes}$ with two extra marked points
inserted into two different faces in all possible ways is given by
\begin{align}
\left.G_{\{k_1,\dotsc,k_n\}}^{2\otimes}\right|_{O(\lambda^{\ell})}
=\begin{pmatrix}\mathcal{K}+2-n+\ell\\2\end{pmatrix}
\times\left.G_{\{k_1,\dotsc,k_n\}}^{(\Nc)}\right|_{O(\lambda^{\ell})}\,.
\end{align}
Once we get this formula, we can then normalize both sides, since the
normalization for BPS operators does not depend on $\lambda$.

\begin{figure}[t]
\centering
\includegraphics[scale=0.8]{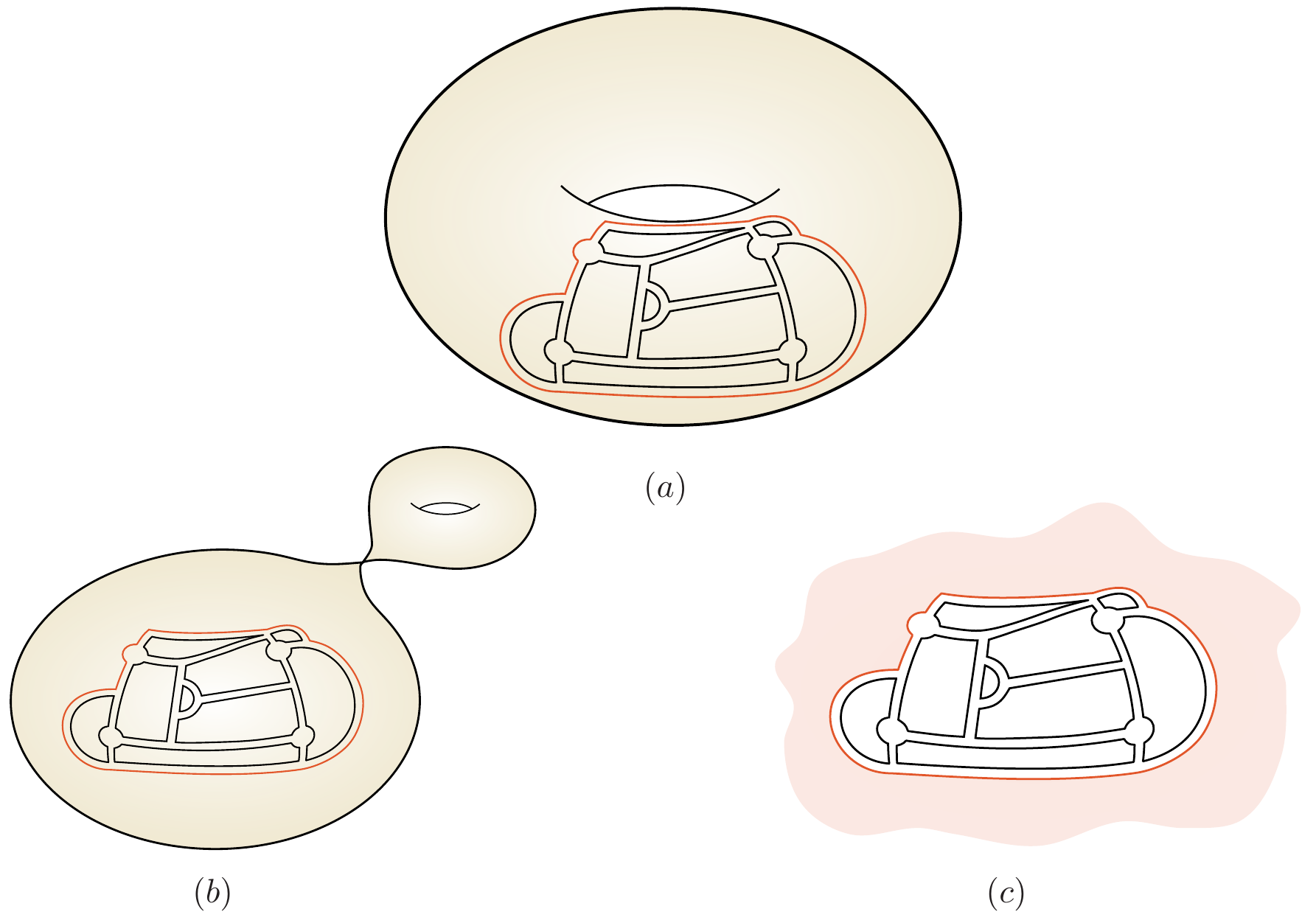}
\caption{Degeneration to a single marked point. In addition to the
degeneration shown in \figref{fig:pinching1}, there is a yet another class
of degenerations which produces a sphere with a single marked point.
They correspond to the diagrams shown in (a) which degenerates into
(c) depicted above. The red region in (c) corresponds to a marked
point.}
\label{fig:pinching2}
\end{figure}

So far, we have been discussing the degeneration in which a handle
degenerates into a pair of marked points.
The other type of degeneration, in which the surface is split in two
by pinching an intermediate cylinder, is exemplified in \figref{fig:pinching2}.
As shown in this figure,
this type of degeneration produces a single marked point on the planar
surface. Therefore, the analogue of \eqref{eq:G2otimes} in those cases
reads
\begin{align}
\mathcal{G}_{\otimes}=F\mathcal{G}\,,
\end{align}
where again $F$ is the number of faces in the Feynman graph $\mathcal{G}$. The
combinatorial factor $F$ in this case can also be related to the shift of
$\Nc$; namely it corresponds to the $O(1/\Nc)$ term in the expansion
\eqref{eq:GshiftN}. We therefore conclude that the correlator with a
single extra marked point is given by
\begin{align}\label{eq:derivativelam}
\left.G_{\{k_1,\dotsc,k_n\}}^{\otimes}\right|_{O(\lambda^{\ell})}=(\mathcal{K}+2-n+\ell)\times \left.G_{\{k_1,\dotsc,k_n\}}^{(\Nc)}\right|_{O(\lambda^{\ell})}\,.
\end{align}
In total, the subtraction for a correlator on the torus at order $O(\lambda^{\ell})$ is given by
\begin{align}
\left.(\texttt{subtraction})\right|_{O(\lambda^{\ell})}
=\begin{pmatrix}\mathcal{K}+3-n+\ell\\2\end{pmatrix}
\times\left.(\texttt{planar})\right|_{O(\lambda^{\ell})}
\end{align}
where (\texttt{subtraction}) denotes the subtraction piece while
(\texttt{planar}) is a planar correlator.

\paragraph{Decomposition into Polygons at One Loop.}

The formula above computes the full $k$-loop subtraction all at once.
However, it is practically more useful to decompose the subtraction into the contributions associated
with individual tree-level diagrams, so we can observe cancellations with other contributions more straightforwardly.

This can be done rather easily by generalizing the argument we just
presented: As shown in \tabref{tab:stratexcyc}, the degeneration of a
Riemann surface with a tree-level graph leads to polygons (\ie faces) with one or
two marked points.\footnote{Polygons and their expectation value at
one loop are discussed in full detail in the next section.} To
evaluate these polygons, we just need to keep in mind that
each polygon admits the expansion
\begin{align}
(\polygon)=\Nc \sum_{\ell}p_\ell\,\lambda^{\ell}\,.
\end{align}
The overall factor $\Nc$ comes from the fact that the edges of the
polygon constitute a closed index-loop. Although we do not normally
associate such a factor with each polygon, here it is crucial to
include that factor\footnote{This is essentially because we need to
consider \emph{unnormalized} correlators, as explained above.} to
count the faces correctly.

The rest of the argument is identical to the one before: Shifting
$\Nc$ to $\Nc+1$ and reading off the $1/\Nc$ and $1/\Nc^2$ terms, we
get
\begin{equation}
\begin{aligned}
(\polygon)_{\otimes}\big|_{O(\lambda^\ell)}&=(1+\ell)\times(\polygon)\big|_{O(\lambda^\ell)}\,,\\
(\polygon)_{2\otimes}\big|_{O(\lambda^\ell)}&=\frac{(1+\ell)\ell}{2}\times(\polygon)\big|_{O(\lambda^\ell)}\,.
\end{aligned}
\end{equation}
Here $(\polygon)_{\otimes}$ and $(\polygon)_{2\otimes}$
denote the contributions from a polygon with one or two marked points
respectively. Using the fact that the $O(\lambda^{0})$ term for each
polygon is just unity\footnote{Here we are dropping the overall $\Nc$
factor as in the rest of this paper.}, one can also write an explicit
weak-coupling expansion as
\begin{equation}
\begin{aligned}
(\polygon)_{\otimes}&=1+2 (\polygon)\big|_{O(\lambda)}+\dots\,,\\
(\polygon)_{2\otimes}&=0+ (\polygon)\big|_{O(\lambda)}+\dots\,.
\end{aligned}
\label{eq:markedpolywc}
\end{equation}
These formulae will be used intensively below.

\paragraph{Worldsheet Interpretation.}

Let us end our discussion on the subtraction by mentioning the
worldsheet interpretation of the marked points. This is more or less
obvious from the way we performed the computation: Shifting the rank
of the gauge group from $\Nc$ to $\Nc+1$ amounts to adding a probe
D3-brane in AdS. It is well-known that the probe brane sitting at some
finite radial position $z$ describes the Coulomb branch of
$\superN=4$ SYM, in which the gauge group is broken from ${\rm
U}(\Nc+1)$ to ${\rm U}(\Nc)\times {\rm U}(1)$. In our case, we are not
breaking any conformal symmetry, and therefore the probe brane must
sit at the horizon of AdS ($z=\infty$ in Poincar\'e coordinates).

This suggests that the marked points that we have been discussing
correspond to boundary states describing the probe brane at the
horizon. Furthermore, our computation \eqref{eq:derivativelam} implies
that the $n$-point tree-level string amplitude with an insertion of a
hole is related to the same amplitude without insertion as%
\footnote{The formula
is reminiscent of the famous soft dilaton theorem~\cite{Ademollo:1975pf},
although it does not seem that there exists any obvious relation
between the two.}
\begin{align}
\lambda^{\mathcal{K}-2+n}A_{\text{sphere+hole}}=\frac{\partial }{\partial \lambda}\left(\lambda^{\mathcal{K}-2+n}A_{\text{sphere}}\right)\,.
\end{align}
It would be interesting to verify this prediction by a direct
worldsheet computation.

Let us finally add that, although the argument above gives a
worldsheet interpretation of the marked points, it does not explain
why such boundary states are relevant for the analysis of the
degenerate worldsheet. It would be desirable to find a worldsheet
explanation for this, which does not rely on the Feynman-diagrammatic
argument presented in this section.

\subsection{Dehn Twists and Modular Group}

\begin{figure}[t]
\centering
\includegraphicsbox[height=2.7cm]{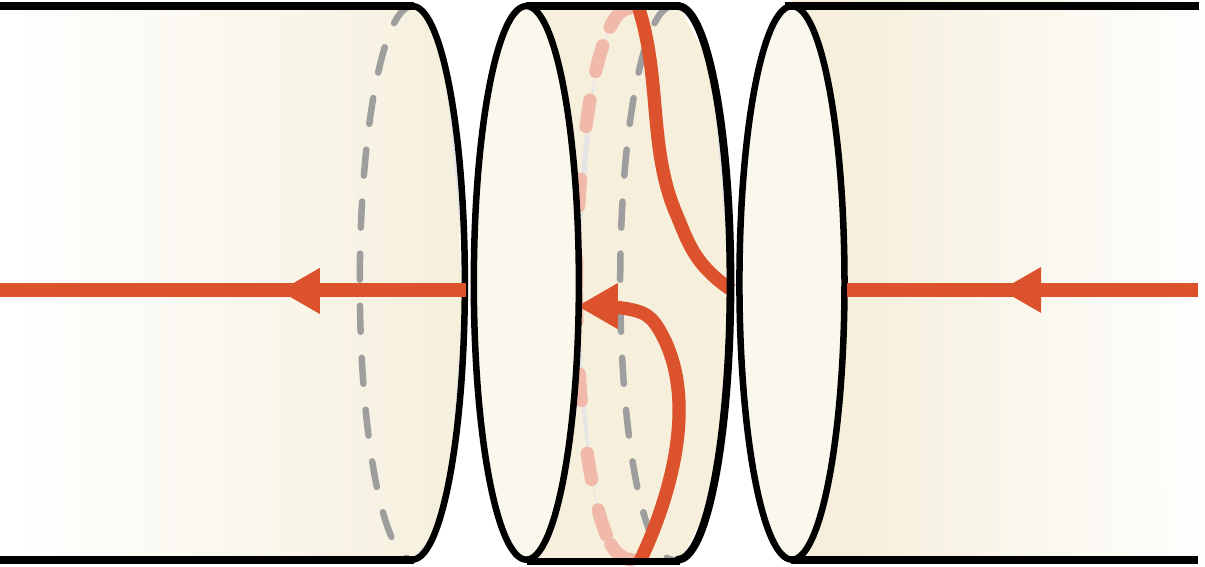}
\qquad\qquad
\includegraphicsbox[height=2.7cm]{FigTorusStrat4Cycle}
\qquad\qquad
\includegraphicsbox[height=2.7cm]{FigTorusStrat2subCycle}
\caption{Dehn twists: \emph{Left:} In red, we represent a path along the worldsheet that
undergoes a complete cycle in the cylindrical piece where a Dehn
twist was performed. \emph{Middle and Right:} Stratification
contributions that get added (graph on a torus, middle figure) and
subtracted (planar graph with marked points, right figure), with
shaded regions that form non-trivial cycles,
on which one can perform
Dehn twists that leave the embedding of the graph invariant.}
\label{fig:dehntwist}
\end{figure}

The backbone of our formula~\eqref{eq:mainformula} is a summation over
(skeleton) graphs. When we construct the complete set of graphs on a surface
of given genus, we implicitly identify graphs that only differ by
``twists'' of a handle. For example, we treat the genus-one graphs
\begin{equation}
\includegraphicsbox{FigTorusCase31}
\quad
\text{and}
\quad
\includegraphicsbox{FigTorusCase31twisted}
\label{eq:dehntwistexample}
\end{equation}
as identical. This makes perfect sense from a weak-coupling
perturbative point of view: Wick contractions only carry information
about the ordering of bridges around each operator, not on the
particular way in which the graph is embedded in a given surface. Hence the
two graphs~\eqref{eq:dehntwistexample} \emph{are} identical as Feynman
graphs. Modding out by such twists is also natural from the
string-worldsheet perspective. The summation over graphs represents
the integration over the moduli space of complex structures of the
string worldsheet. The ``twists'' mentioned above are called
\emph{Dehn twists}. More formally, a Dehn twist is defined as an
operation that cuts a cylindrical piece (the neighborhood of a cycle)
out of a Riemann surface (the worldsheet), performs a $2\pi$ twist on
this piece, and glues it back in, see~\figref{fig:dehntwist}.
Such Dehn twists leave the complex structure of the Riemann surface
invariant, and hence should be modded out by when integrating over the
moduli space. In fact, Dehn twists are isomorphisms that are not
connected to the identity. They form a complete set of generators for
the modular group (mapping class group) for surfaces of any genus and
with any number of operator insertions (boundary components).%
\footnote{At genus one, the modular group is $\grp{PSL}(2,\Integers)$,
and it is generated by Dehn twists along the two independent cycles of
the torus.}
Since all Dehn twists act as identities in the moduli space as well as
on Feynman diagrams, it is natural to mod out by Dehn twists in all
stages of the computation.

While modding out by Dehn twists is natural and straightforward in the
summation over free-theory graphs (as we have been doing implicitly),
it has non-trivial implications for the summation over mirror states,
especially for the stratification contributions. By their nature, all
stratification contributions contain non-trivial cycles that do not
intersect with the graph of propagators: For the terms that get added,
non-trivial cycles can wind the handles
not covered by the graph, and for the terms that get subtracted,
non-trivial cycles can wind around the
isolated marked points
(see \figref{fig:dehntwist} for examples).
Obviously, performing a Dehn twist on a neighborhood of such cycles
neither alters the graph itself, nor its embedding in the surface. But
once we fully tessellate the surface by a choice of zero-length
bridges (and dress them with mirror magnons), such Dehn twists
\emph{will} alter (twist) the embedding of those bridges (ZLBs) on the
surface. For example, the two graphs
\begin{equation}
\includegraphicsbox{FigTorusStrat4tess}
\quad
\text{and}
\quad
\includegraphicsbox{FigTorusStrat4tesstwisted}
\label{eq:zlbtesstwist}
\end{equation}
are related by a Dehn twist on a vertical strip in the middle of the
picture, which only acts on the zero-length bridges (dashed lines).
Since we anyhow do not sum over different ZLB-tessellations, but
rather just pick one choice of ZLBs for each propagator graph, it looks like such
twists need not concern us. However, notice that one can always
transform a Dehn-twisted configuration of ZLBs back to the untwisted
configuration via a sequence of flip moves on the ZLBs. As long as all
participating mirror states are vacuous, these flip moves are trivial
identities. However, as soon as we dress the ZLBs (and other bridges)
with mirror magnons, flip moves will non-trivially map (sets of)
excitation patterns, \ie distributions of mirror magnons, to each
other. Hence we have the situation that a given distribution of mirror
magnons on a fixed choice of ZLB-tessellation might secretly be
related to another distribution (or set of distributions) of magnons
on the same, but now Dehn-twisted ZLB-tessellation. Since part of our
interpretation of the sums over mirror magnons is that they probe the
neighborhood of the discrete point in the moduli space represented by
the underlying propagator graph, it seems natural to identify
distributions of mirror magnons that are related in the way just
described. We are therefore led to add the following element to our prescription:
\begin{equation}
\setlength{\fboxsep}{10pt}
\framebox{\text{\parbox{\textwidth-2.5cm}{%
Among all mirror-magnon contributions that are related to each
other via Dehn twists followed by sequences of bridge flips, take only
one representative into account. In other words, all mirror-magnon
contributions that are related to each other via Dehn
twists and sequences of bridge flips are identified.}}}
\label{eq:dehntwistrule}
\end{equation}
The one-loop evaluation of all relevant stratification
contributions in~\secref{sec:strat} will lend quantitative support to this prescription.

\section{Multi-Particles and Minimal Polygons}

\begin{figure}[t]
\centering
\includegraphics[scale=0.5]{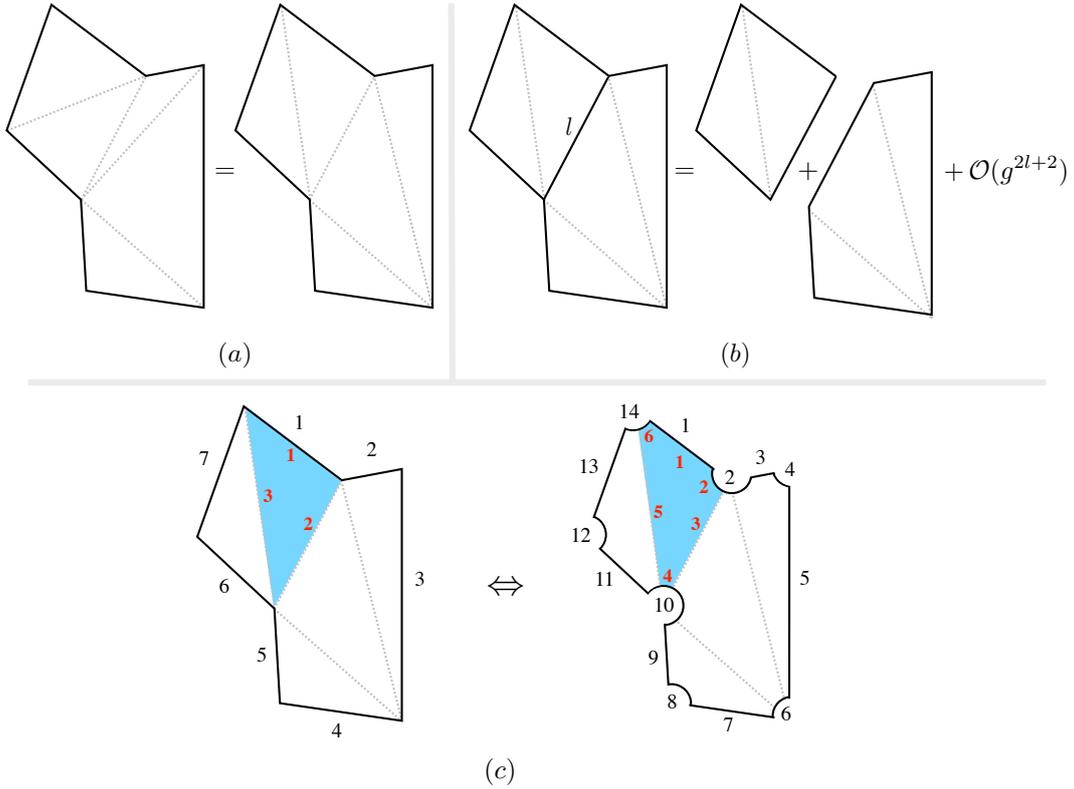}
\caption{(a) An example of a minimal polygon.
A minimal polygon is by definition a polygon that when
triangulated/hexagonalized only contains zero-length bridges. This means
that all internal mirror edges contribute
at one-loop order if one inserts a mirror particle on them.
It can be hexagonalized in several different ways, and
all ways of doing so should give the same integrability
result when summing over mirror particles.
(b)~A general polygon may have zero-length and
non-zero-length bridges, and it can
be divided into minimal
polygons. Inserting mirror particles in non-zero-length
bridges is more costly at weak coupling.
(c)~Two different ways of defining a polygon with
physical operators on its edges. It is possible to shrink the
operators to points or to blow them up to finite size.
In the first case the surface is triangulated (only mirror edges),
and in the second case it is hexagonalized (as many
physical as mirror edges).}
\label{fig:ThreeFigures}
\end{figure}

We think of a polygon as the inside of the face of a larger Feynman
diagram, with the outer edges being propagators in that diagram.
Depending on whether we blow up the physical operators or not, the
same polygon can be either thought of as an $n$-gon (with $n$ mirror edges), or a
$2n$-gon (with $n$ mirror edges and $n$ physical edges), as illustrated
in \figref{fig:ThreeFigures}c. When we do blow up the physical
operators we speak of \emph{hexagonalizing} the polygon, otherwise
we say that we \emph{triangulate} it. In the hexagonalization
picture, every other edge of each hexagon is formed by a segment (in
color space) of a physical operator. In the triangulation picture, the
physical operators sit at the cusps of the triangles. Of course, both
pictures describe
the very same thing, as indicated in \figref{fig:ThreeFigures}c.

\begin{figure}[t]
\centering
\includegraphics[scale=0.5]{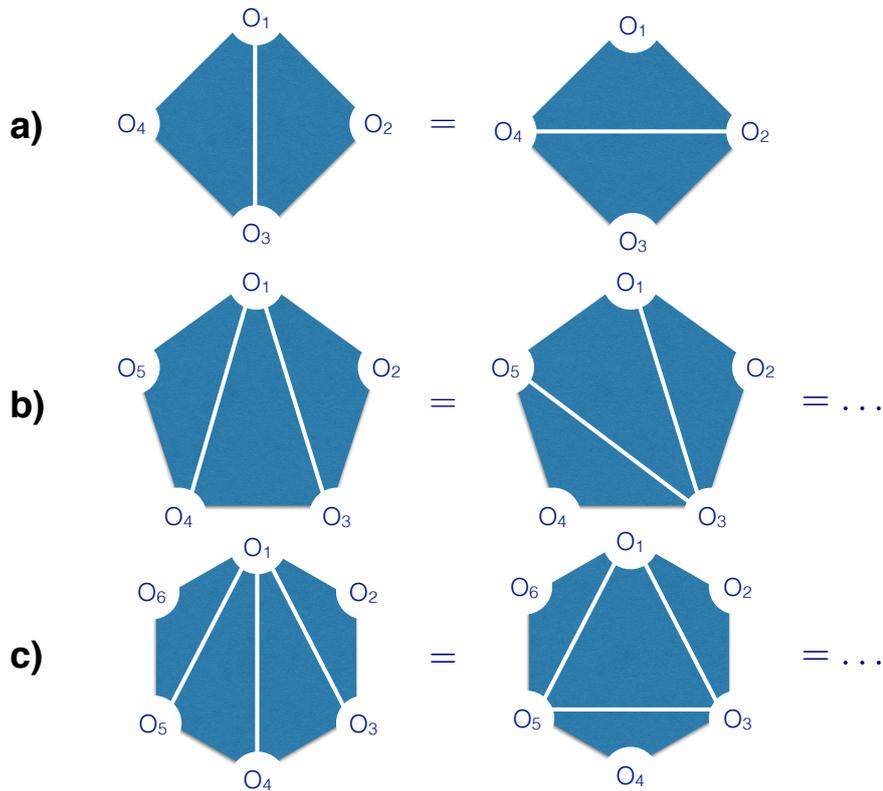}
\caption{(a) An octagon and its two possible tessellations related by
flipping and studied in~\cite{\Hexagonalizationone}. (b)~A decagon and
its various tessellations studied in~\cite{\Hexagonalizationtwo}.
(c)~The dodecagon has a few new features compared to the previous two
examples. Here, different tessellations can be computed by very
different integrability processes, with different numbers of mirror
particles involved, see \figref{fig:FlipExamples2}.}
\label{fig:FlipExamples}
\end{figure}

There can be non-zero-length bridges in the interior of the polygon, as
indicated in \figref{fig:ThreeFigures}b. When computing the
expectation value of a polygon, we triangulate/hexagona\-lize it and
insert mirror particles at all the mirror edges. When these edges are
such non-zero-length bridges, this is more costly at weak coupling, as
indicated in \figref{fig:ThreeFigures}b, so the expectation value of
such polygons breaks down into polygons where all internal
bridges have zero length. We call such polygons \emph{minimal polygons}.
For large bridges, this decomposition holds up to a large number of
loops. In this paper, we focus only on such minimal polygons,
such as the one in \figref{fig:ThreeFigures}a.

A minimal polygon can be hexagonalized in different ways, as illustrated in \figref{fig:ThreeFigures}a,
and an important consistency condition is that all these tessellations ought
to give the same result. Three further examples are illustrated in
\figref{fig:FlipExamples}. The first was considered in~\cite{\Hexagonalizationone},
the second in~\cite{\Hexagonalizationtwo},
and the third will be discussed later in this paper.

\paragraph{Variables.}

Minimal polygons are functions of the labels of the physical operators
at their perimeter, namely of the operator positions $x_i$ and
internal polarizations $\alpha_i$ (for minimal polygons, the operator weights
$k_i$ are irrelevant). Due to conformal symmetry and R-symmetry, minimal polygons
can only be functions of spacetime cross ratios and cross ratios
formed out of the internal polarizations. In this paper, we focus on
four-point functions, and will use the familiar variables
\begin{equation}
z\bar z=\frac{x_{12}^2\,x_{34}^2}{x_{13}^2\,x_{24}^2}
\,,\qquad
(1-z)(1-\bar z)=\frac{x_{14}^2\,x_{23}^2}{x_{13}^2\,x_{24}^2}
\,,\qquad
x_{ij}\equiv\abs{x_i-x_j}
\,.
\label{eq:zdef}
\end{equation}
For cross ratios of the internal polarizations, we similarly choose
\begin{equation}
\alpha\bar\alpha=\frac{(\alpha_1\cdot\alpha_2)(\alpha_3\cdot\alpha_4)}{(\alpha_1\cdot\alpha_3)(\alpha_2\cdot\alpha_4)}
\,,\qquad
(1-\alpha)(1-\bar\alpha)=\frac{(\alpha_1\cdot\alpha_4)(\alpha_2\cdot\alpha_3)}{(\alpha_1\cdot\alpha_3)(\alpha_2\cdot\alpha_4)}
\,.
\label{eq:alphadef}
\end{equation}
In the following, we will consider more general minimal polygons that
depend on $n$ external operators. However, we will restrict all
operators to lie in the same plane, in spacetime as well as in the
internal polarization space, as this is sufficient for our purposes.
For every choice of four operators, we can form spacetime and
polarization cross ratios exactly as in~\eqref{eq:zdef}
and~\eqref{eq:alphadef}, and an $n$-point polygon in these restricted
kinematics depends on $(n-3)$ sets of such cross ratios.%
\footnote{In the plane, distances
factorize as $x^2_{ab}=x_{a,b} \bar x_{a,b}$, and the $R$-charge inner products do the
same, $y_a\cdot y_b = y_{a,b} \bar y_{a,b}$.
As such, when we will deal with functions of cross ratios made out of
four physical and $R$-charge positions they always come in multiples of
four such as $z={x_{a,b}x_{c,d}}/{x_{a,c}x_{b,d}}$, $\bar z={\bar
x_{a,b}\bar x_{c,d}}/{\bar x_{a,c}\bar x_{b,d}}$,
$\alpha={y_{a,b}y_{c,d}}/{y_{a,c}y_{b,d}}$ and $\bar \alpha={\bar
y_{a,b}\bar y_{c,d}}/{\bar y_{a,c}\bar y_{b,d}}$. When dealing with
such quantities we often use the obvious short-hand notation $f(z)$ to
indicate $f(z,\bar z,\alpha,\bar \alpha)$, see for example~\eqref{eq:3ptFinal} below.}

\subsection{One-Loop Polygons and Strings from Tessellation Invariance}
\label{sec:1-loop-polygons}

\begin{figure}[t]
\centering
\includegraphics[scale=0.5]{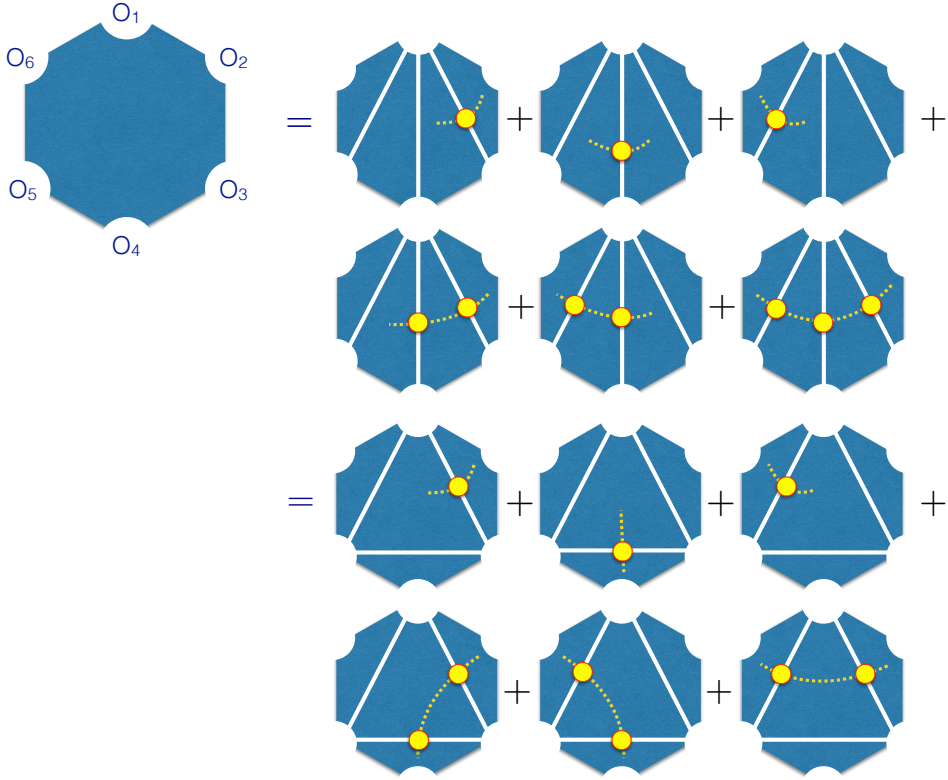}
\caption{A tessellation of the dodecagon can contain paths where a
mirror particle propagates through four different hexagons, as
illustrated in the last graph in the second line. In another
tessellation, a particle can propagate for at most three hexagons, as
illustrated in the second example. Equating both, we can read off the
larger propagation (three-particle) contribution from the smaller ones
(two-particle and one-particle), as shown
in~\eqref{eq:3ptReduce}.}
\label{fig:FlipExamples2}
\end{figure}

To fully compute a $2n$-gon vacuum expectation value, we should insert
any number of mirror particles at all hexagon junctions and
integrate over their rapidities. At one-loop order, things simplify:
According to the loop-counting shown in \figref{fig:counting}, we
only need to sum over multi-particle strings which are associated to
paths that connect one hexagon to another, never passing twice
through the same hexagon. To construct the corresponding
multi-particle string, we insert exactly one mirror particle whenever
the path intersects a mirror edge. In sum, the one-loop $2n$-gon is
obtained by picking a tessellation at one's choice, and summing over all
multi-particle one-loop strings on that tessellation. See
\figref{fig:FlipExamples2} for an example.

Each mirror edge joins two hexagons into an octagon involving four
operators. Hence two cross ratios are associated to each mirror edge
in a natural way. For a mirror line~$i$ connecting operator $\op{O}_a$
with $\op{O}_c$, where the two adjacent hexagons further connect to
operators $\op{O}_b$ and $\op{O}_d$, we define the variable $z_i$
parametrizing the associated cross ratios as
(note the dependence on the orientation of the sequence of operators around the perimeter)
\begin{equation}
\includegraphicsbox{FigCrossRatios}\;\; :
\qquad
z_i\bar z_i=\frac{x_{ab}^2\,x_{cd}^2}{x_{ad}^2\,x_{bc}^2}
\,,\qquad
(1-z_i)(1-\bar z_i)=\frac{x_{ac}^2\,x_{bd}^2}{x_{ad}^2\,x_{bc}^2}
\,.
\label{eq:zidef}
\end{equation}
The corresponding polarization cross ratios are defined accordingly.
With these definitions, we denote the contribution of a multi-particle
one-loop string traversing $n$ mirror edges as
\begin{equation}
\mathcal{M}^{(n)}(z_1,\dotsc,z_n)
\,,
\end{equation}
where the variables $z_i$ parametrize the cross ratios
associated to the $n$ mirror edges as in~\eqref{eq:zidef}, and we are
suppressing the obvious dependencies on $\bar z_i$ and the polarization cross ratios.

By exploiting the above-mentioned invariance under tessellation
choice, one can determine the contribution from any multi-particle
string $\mathcal{M}^{(n)}$ from the knowledge of the one- and two-particle
contributions alone.
As an illustration, consider the dodecagon example in
\figref{fig:FlipExamples2}. In the second tessellation, only two-particle
strings appear, while for the first tessellation, the sum includes a
contribution with three particles. Equating both sums, we can relate the
three-particle contribution to the one- and two-particle strings as
\begin{align}
\mathcal{M}&^{(3)}(z_1,z_2,z_3)=
\nn\\ &
-\lrbrk{
\mathcal{M}^{(1)}(z_1)
+\mathcal{M}^{(1)}(z_2)
+\mathcal{M}^{(1)}(z_3)
+\mathcal{M}^{(2)}(z_1,z_2)
+\mathcal{M}^{(2)}(z_2,z_3)
}
\nn\\ &
+\mathcal{M}^{(1)}\lrbrk{\frac{1}{z_2}}
+\mathcal{M}^{(1)}\bigbrk{z_1(1-z_2)}
+\mathcal{M}^{(1)}\lrbrk{\frac{z_2 z_3}{z_2-1}}
+\mathcal{M}^{(2)}\lrbrk{\frac{1}{z_2},z_1(1-z_2)}
\nn\\ &
+\mathcal{M}^{(2)}\lrbrk{z_1(1-z_2),\frac{z_2 z_3}{z_2-1}}
+\mathcal{M}^{(2)}\lrbrk{\frac{z_2 z_3}{z_2-1},\frac{1}{z_2}}
\,.
\label{eq:3ptReduce}
\end{align}
Here, the variables $z_1$, $z_2$, and $z_3$ parametrize the cross
ratios associated to the three mirror edges of the first tessellation
in \figref{fig:FlipExamples2} (from right to left). Hence,
$\mathcal{M}^{(1)}(z_1)$ equals the first contribution in
\figref{fig:FlipExamples2}, $\mathcal{M}^{(1)}(z_2)$ equals the second
contribution, and so on.%
\footnote{A convenient choice of operator positions to
obtain the arguments of all contributions is
\begin{equation*}
\mathcal{O}_1: 0
\,,\quad
\mathcal{O}_2: z_1
\,,\quad
\mathcal{O}_3: \infty
\,,\quad
\mathcal{O}_4: 1
\,,\quad
\mathcal{O}_5: \frac{1}{1-z_2}
\,,\quad
\mathcal{O}_6: \frac{1}{1-z_2 +z_2 z_3}
\,.
\end{equation*}}
In the above expression, it is implicit that the other, suppressed
variables undergo the same substitutions as the $z_i$ variables, \eg
\begin{equation}
\mathcal{M}^{(1)}\lrbrk{\frac{z_2z_3}{z_2-1}}
\equiv
\mathcal{M}^{(1)}\lrbrk{\frac{z_2z_3}{z_2-1},\frac{\bar z_2\bar
z_3}{\bar z_2-1},\frac{\alpha_2\alpha_3}{\alpha_2-1},\frac{\bar\alpha_2\bar\alpha_3}{\bar\alpha_2-1}}
\,,
\end{equation}
where we have, by slight abuse of notation, used
$(\alpha_i,\bar\alpha_i)$ to parametrize the polarization cross ratios.
Using the explicit known
results for one and two particles~\cite{\Hexagonalizationone,\Hexagonalizationtwo}
\begin{align}
\label{eq:oneandtwoparticle}
\mathcal{M}^{(1)}(z)
&=m(z)+m(z^{-1})
\,,
\\
\mathcal{M}^{(2)}(z_1, z_2)
&=m\lrbrk{\frac{z_1-1}{z_1z_2}}+m\lrbrk{\frac{1-z_1+z_1z_2}{z_2}}
+m\bigbrk{z_1(1-z_2)}-m(z_1)-m(z_2^{-1})
\,,\nn
\end{align}
we find for the three-particle one-loop string:
\begin{multline}
\mathcal{M}^{(3)}(z_1,z_2,z_3)=
m\lrbrk{\frac{1-z_1+z_1z_2}{z_1z_2z_3}}
+m\bigbrk{z_1(1-z_2+z_2z_3)}
\\
+m\lrbrk{\frac{(1-z_2)(-1+z_1-z_1z_2+z_1z_2z_3)}{z_2z_3} }
-m\bigbrk{z_1(1-z_2)}
-m\lrbrk{\frac{z_2-1}{z_2z_3}}
\,.
\label{eq:3ptFinal}
\end{multline}
The cross ratios appearing in the argument of the three-particle
contribution are defined as in~\eqref{eq:zidef}.
Here, the main building block function $m(z)$ is given by
\begin{equation}
m(z)\equiv
g^2\,\frac{(z+\bar{z})-(\alpha+\bar{\alpha})}{2}F^{(1)}(z,\bar z)
\,,
\label{eq:mdefinition}
\end{equation}
with the one-loop conformal box integral
\begin{equation}
F^{(1)}(z,\bar z)
=\frac{1}{z-\bar{z}}\lrbrk{
2\Li_2(z)-2\Li_2(\bar{z})
+\log(z\bar{z})\log\lrbrk{\frac{1-z}{1-\bar{z}}}
}
\,.
\label{eq:F1expr}
\end{equation}
The building block function $m(z)$
satisfies the following important identities:
\begin{equation}
m(0) = m(1) = m(\infty) =0
\,,\qquad
m(z)+m(1-z) = 0
\,.
\label{eq:midentities}
\end{equation}
Note that there is another type of three-particle contribution besides
the one discussed above. It appears in an ``alternating'' tessellation
of the same dodecagon:
\begin{equation}
\includegraphicsbox[trim={0 16.5cm 0 0.5cm}, scale=0.50]{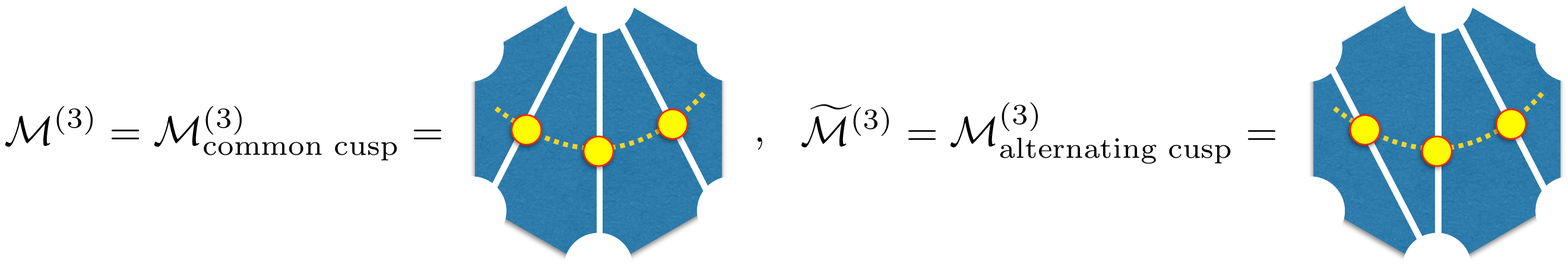} \, .
\label{2ThreeParticleContributions}
\end{equation}
The ``alternating cusp'' three-particle string can be derived in the
same way as the ``common cusp'' string by equating the alternating
tessellation to one of the two tessellations shown in \figref{fig:FlipExamples2}.

\begin{figure}[t]
\centering
\includegraphics[scale=0.35]{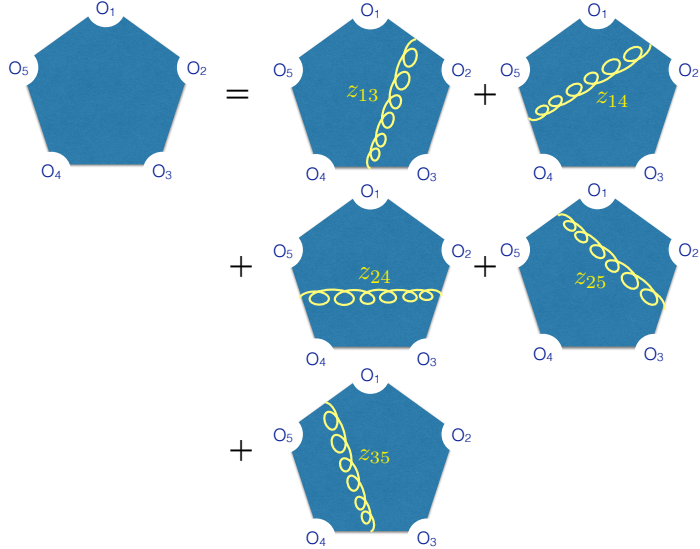}
\caption{A $2n$-gon decomposes into a sum of \emph{gluon
exchange-like} contributions between all non-neighboring edges, with
each exchange given by a function $m(z_{ij})$, as shown in~\eqref{eq:2n-gon}.}
\label{fig:exampleDodecagon}
\end{figure}

\begin{figure}[t]
\centering
\includegraphics[scale=0.84]{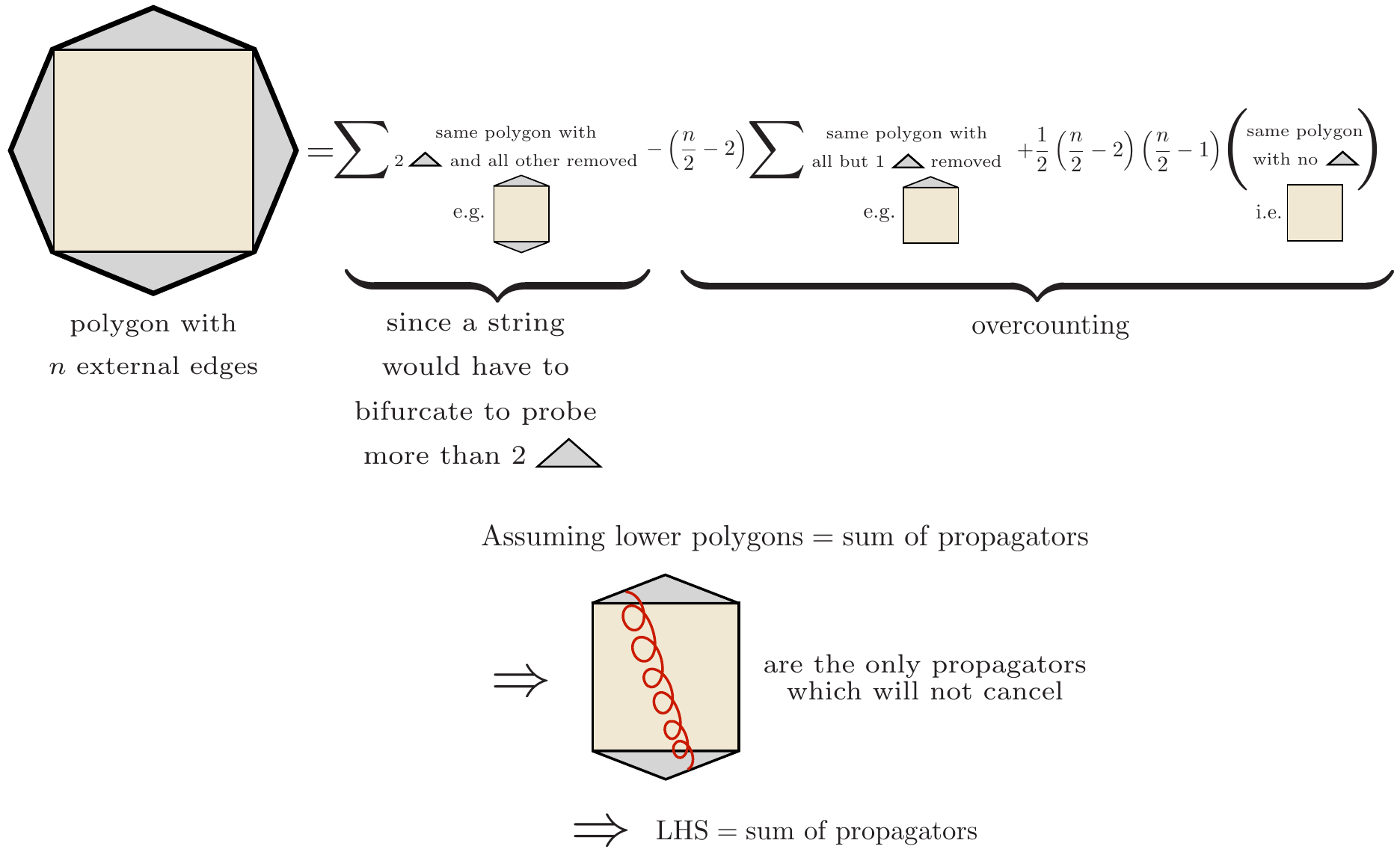}
\caption{Proof of~\eqref{eq:2n-gon} by induction for an even number of
external edges. For an odd number, a proof can be found in a similar
way. The combination in the first line amounts to the statement that
all strings in such symmetric tessellations can probe zero, one, or two
outer triangles. In order to probe more than two triangles, the
string would have to bifurcate. All possible strings are of course
contained in the first sum, but there is an obvious over-counting,
which is removed by the last two terms.}
\label{fig:proof}
\end{figure}

By playing with tessellations of higher $2n$-gons in a similar way, we can
derive, in the fashion described above, all multi-particle one-loop
contributions, and therefore also all higher polygon one-loop expectation values in terms of
contributions involving only one-particle and two-particle strings.
Writing the latter in terms of the building block function $m(z)$ via~\eqref{eq:oneandtwoparticle},
the resulting expression for a general $2n$-gon, for instance, is remarkably simple
and reads
\begin{equation}
\label{eq:2n-gon}
\polygon(1,\dotsc,2n)=\sum_{\substack{[i,i+1],[j,j+1]:\\\text{non-consecutive}}}m\left(z_{i,j}
\equiv\frac{x_{i, j+1}x_{i+1, j}}{x_{i, i+1}x_{j+1,j}} \right)
\,.
\end{equation}
We illustrate the formula in \figref{fig:exampleDodecagon} for the
example of a decagon.
In writing~\eqref{eq:2n-gon}, we cyclically identified the operator
labels, namely $n+1 \equiv 1 \mod n$. The sum runs
over all possible pairs of non-consecutive edges at the
perimeter, $[i,i+1]$ and $[j,j+1]$.%
\footnote{Written more explicitly, we perform the sum over a pair of
indices $(i,j)$ under the condition $i\neq j$, $i+1\neq j$ and $i\neq
j+1$ modulo $n$.}
Roughly speaking, the sum
in~\eqref{eq:2n-gon} corresponds to a summation of all possible
gluon-exchange diagrams that one can draw inside the $n$-point graph.%
\footnote{\label{fn:YMlines}This does \emph{not} mean that each $m(z)$ is given by the corresponding
gluon-exchange diagram, since $m(z)$ should also know about the scalar
contact interaction. What is true is that each $m(z)$
\emph{contains} the corresponding gluon-exchange contribution.
The correspondence between the function $m(z)$
and perturbation theory was made more precise in~\cite{Eden:2017ozn}:
$m$ equals a YM-line exchange in an $\superN=2$ formulation
of $\superN=4$ SYM. We will explore this point further in
\appref{app:NpointPlanar}.}
This general result can actually be proved by induction, as illustrated in \figref{fig:proof}.

\subsection{Tests and Comments}
\label{sec:testsandcomments}

We conclude this section with some further checks and comments.

\subsubsection*{Flip Invariance}

We have assumed tessellation invariance to derive the $2n$-gon
formula~\eqref{eq:2n-gon}. Consistently, the result makes no reference to a particular tessellation,
hence it is manifestly invariant under tessellation choice.

\subsubsection*{Order Invariance}

We can think of each multi-particle string contribution as a mirror-particle
propagation. The direction of propagation ought to be
irrelevant, provided we properly read off the cross ratios for the
associated process as in~\eqref{eq:zidef}. This translates into
\beq
\mathcal{M}^{(2)}(z_1,z_2)= \mathcal{M}^{(2)}(z_2^{-1},z_1^{-1})
\,, \quad
\mathcal{M}^{(3)}(z_1,z_2,z_3)=
\mathcal{M}^{(3)}(z_3^{-1},z_2^{-1},z_1^{-1})
\,, \quad
\dots
\,,
\eeq
which we can indeed verify using the explicit formulas.

\subsubsection*{Reduction to Known $2n$-Gons}

For the octagon $(n=4)$, there are two different pairs of
non-consecutive edges; $[1,2],[3,4]$ and $[4,1],[2,3]$. It is easy to
see that these two contributions lead to $m(z)$ and $m(z^{-1})$
respectively. Therefore, we recover the previous result~\cite{\Hexagonalizationone}.
Similarly, one can check that our formula
reproduces the result for the decagon $(n=5)$. In this case, there are
five different pairs of non-consecutive edges, and they correspond to
the five terms in the decagon~\cite{\Hexagonalizationtwo} represented in
\figref{fig:exampleDodecagon}:
\begin{multline}
\label{decagon}
\mathcal{M}^{(1)}(z_1)+\mathcal{M}^{(1)}(z_2)+\mathcal{M}^{(2)}(z_1,z_2)=
\\
m(z_1^{-1})+m(z_2)+m\left(\frac{z_1-1}{z_1z_2}\right)+m\left(\frac{1-z_1+z_1z_2}{z_2}\right)+m\bigbrk{z_1(1-z_2)}
\,.
\end{multline}
%

\subsubsection*{OPE Limit}

Starting from the dodecagon, one should be able to recover the result
for the decagon by taking the limit $z_3\to 0$. This can be easily
seen by using the properties~\eqref{eq:midentities}. Since the result
is manifestly flip-invariant, any OPE limit is essentially equivalent
and has a good behavior.

\subsubsection*{Extremal and Next-to-Extremal Correlators}

The $n$-point extremal and next-to-extremal correlators have
non-renormalization
properties~\cite{D'Hoker:1999ea,Eden:1999kw,Eden:2000gg}.
Using our conjectural form of the $2n$-gon contribution, one can verify that the
one-loop corrections are zero for those
kinds of correlators, see \appref{app:NpointPlanar} for details of
the planar case.

\subsubsection*{Decoupling Limit}

\begin{figure}[t]
\centering
\includegraphics[trim={0 6.1cm 0 0}, scale=0.5]{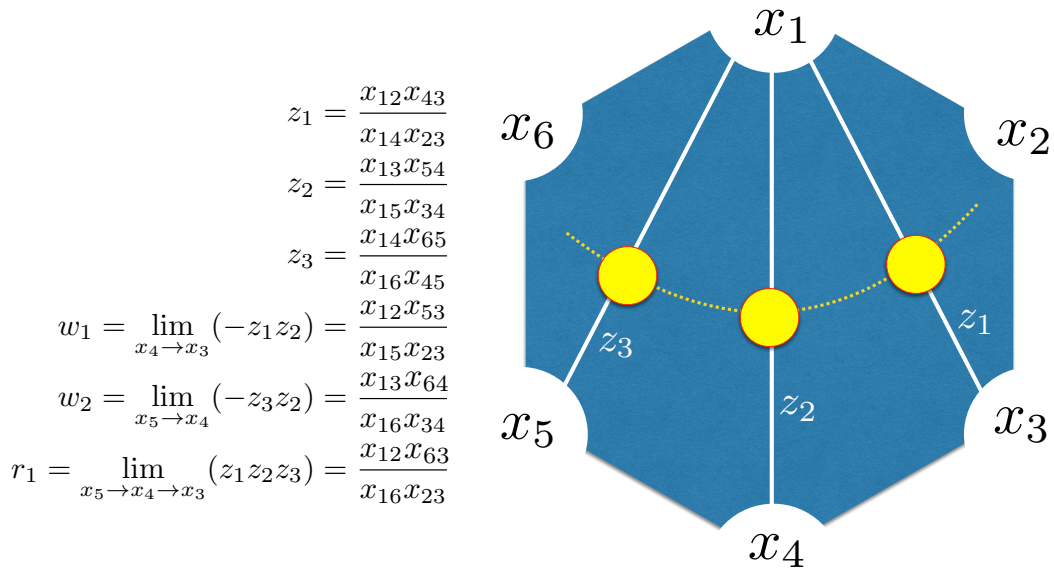}
\caption{A dodecagon and its cross ratios. Collapsing $x_{i+1}\to x_{i}$ eliminates a
slice -- a hexagon -- in the figure. The double limit
$x_{i+2}\to x_{i+1} \to x_{i}$ reduces a $2n$-gon to a $2(n-2)$-gon.
Mirror-state propagations in such polygons are reduced accordingly. From a
form factor point of view, the corresponding sums collapse into
the coinciding rapidity region.}
\label{fig:LimitsFig}
\end{figure}

We can reduce multi-particle strings to strings involving less steps
by collapsing hexagons in the tessellation. For example, if we take
$x_4\to x_3$ in \figref{fig:LimitsFig}, we reduce the dodecagon to a
decagon, and correspondingly the three-particle contribution reduces to
a two-particle contribution. If we further send $x_5\to x_4\to x_3$, we
reduce it further to an octagon, and we end up with a single-particle
contribution. When taking these limits, some cross ratios diverge and
others vanish. For example, $x_4\to x_3$ corresponds to $z_1/z_2 \to 0$
with $z_1 z_2 = -w_1$ fixed. In this limit, we nicely find indeed
\begin{equation}
\mathcal{M}^{(3)}(z_1,z_2,z_3) \to \mathcal{M}^{(2)}(w_1,z_3)
\,, \qquad
\text{as } z_1/z_2 \to 0 \text{ with } z_1 z_2 = -w_1 \text{ fixed} \,,
\end{equation}
in perfect agreement with the above expectations. From the
integrability/form-factor point of view, this limit corresponds to the
so-called decoupling limit, where consecutive rapidities are forced to
become equal, and the corresponding hexagons collapse into measures and
disappear.%
\footnote{From this integrability/form-factor point of view,
one can expect these decoupling relations to hold to all
loops.}
Similarly, we find
\begin{align*}
\mathcal{M}^{(3)}(z_1,z_2,z_3) &\to \mathcal{M}^{(2)}(z_1,w_2) \,, &
\text{as } z_2/z_3 &\to 0 \text{ with } z_2 z_3 = -w_2 \text{ fixed}
\,,\\
\mathcal{M}^{(2)}(z_1,w_2) &\to \mathcal{M}^{(1)}(r_1) \,, &
\text{as } z_1/w_2 &\to 0 \text{ with } z_1 w_2 = -r_1 \text{ fixed}
\,,
\end{align*}
and many other similar relations at higher points.

\subsubsection*{Pinching at One Loop}

\begin{figure}[t]
\centering
\includegraphics[scale=0.5]{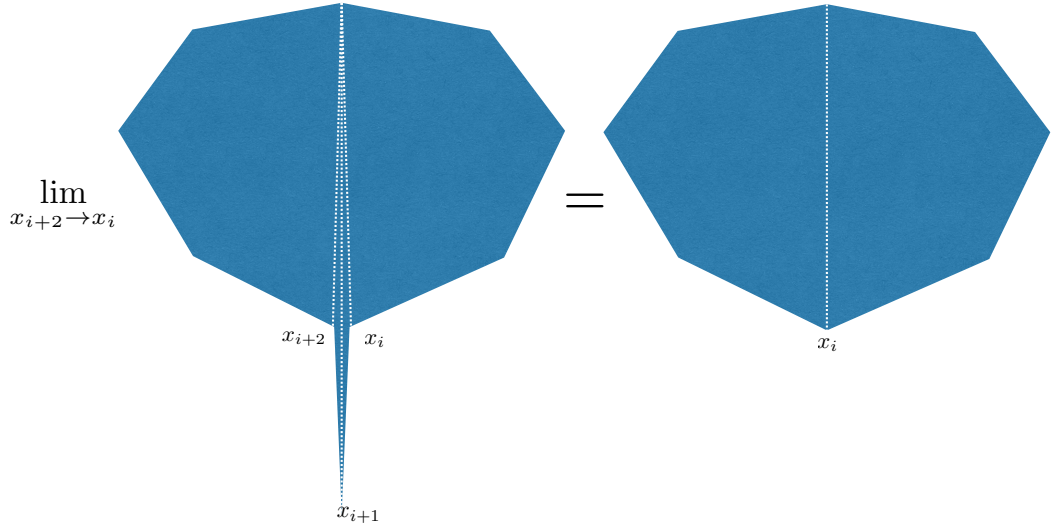}
\vspace{-2.7cm}\caption{Pinching at one loop.
As a consequence of both the form of the 2$n$-gon and
the properties of the function $m(z)$, the limit when the cusps at position
$i$ and $i+2$ have the same position $x_i$
equals a smaller polygon with two fewer cusps.
This limiting polygon does not depend on the
position and $R$-charge of the initial middle cusp $i+1$.}
\label{fig:PolygonPinching}
\end{figure}

Another nice limit of any polygon is the one where cusps $i$ and $i+2$
go to the same position. When doing so, they pinch the edge ending at
cusp $i+1$ and basically remove it, as illustrated in
\figref{fig:PolygonPinching}. This limit removes all traces of the operator
which got sandwiched between cusps $i$ and $i+2$,
\beq
\lim_{x_{i+2}\to x_ i }
\polygon(x_1,\dots,{\color{red}x_i},{\color{Blue}x_{i+1}},{\color{red}x_{i+2}},\dots,x_n)
= \polygon(x_1,\dots,{\color{red}x_i},\dots,x_n)
\,.
\label{eq:pinchrule}
\eeq
This identity is actually quite powerful and very useful for us. For
four-point functions, for instance, all cusps are located at one of the four
possible space-time insertions, so there will naturally be many
repetitions of labels, which can be reduced with this rule. For example:
\beq
\polygon(1,{\color{blue}2},{\color{red}3},{\color{blue}2},4,3,1,{\color{blue}3},{\color{red}2},{\color{blue}3})
\to\polygon({\color{blue}1},{2},4,3,{\color{blue}1},{\color{red}3})
\to\polygon(2,4,3,1)
\,.
\label{examplePinching}
\eeq
For four-point functions, we can use the following simple \mathematica code to
simplify arbitrary one-loop polygons:
\definecolor{mathgreen}{RGB}{0,90,39}
\lstset{
  language=Mathematica,
  basicstyle={\small\ttfamily},
  columns=fullflexible,
  keepspaces=true,
  xleftmargin=1em,
  emph={polygon,m,z,n},
  emphstyle={\color{blue}},
  emph=[2]{L_,L,i,j},
  emphstyle=[2]{\color{mathgreen}}
}
\begin{lstlisting}
polygon[L_] := Block[
  {n=Length[L], x={-1,1/(2z-1),1,0}[[ L[[ Mod[#,n,1] ]] ]]&},
  Sum[
    m[(x[i]-x[j+1])(x[i+1]-x[j])/((x[i]-x[i+1])(x[j+1]-x[j])) // Simplify],
    {i, 1, n}, {j, i+2, n-Boole[i==1]}
  ] /. {
    m[1-z] -> -m[z], m[(-1+z)/z] -> -m[1/z],
    m[z/(-1+z)] -> -m[1/(1-z)], m[0] -> 0, m[1] -> 0
  }]
\end{lstlisting}
It implements~\eqref{eq:2n-gon}, taking into account the
functional identities~\eqref{eq:midentities} of the $m(z)$ building block. Running
\lstinline"polygon[{1,2,3,2,4,3,1,3,2,3}]", for instance, would simply
yield~$m(z)+m(1/z)$, which is the very same as
\lstinline"polygon[{2,4,3,1}]", as expected according to~\eqref{examplePinching}.

\subsubsection*{One-Loop Octagons}

Below, we will need the expressions for one-loop octagons,
hence we will quote them here. The one-loop octagon was computed
in~\cite{Fleury:2016ykk}. Due to the dihedral symmetry of the
one-loop polygons~\eqref{eq:2n-gon}, permutations of the four corners
generate only three independent functions, corresponding to the
orderings 1--2--4--3, 1--2--3--4, and 1--3--2--4 of the four operators around the perimeter
of the octagon. Permutations of the four operators are generated by the
following variable transformations:
\begin{align}
3 &\leftrightarrow 4:&
z &\leftrightarrow \frac{z}{z-1} \,,&
\bar z &\leftrightarrow \frac{\bar z}{\bar z-1} \,,&
\alpha &\leftrightarrow \frac{\alpha}{\alpha-1} \,,&
\bar\alpha &\leftrightarrow \frac{\bar\alpha}{\bar\alpha-1}
\,,\nn\\
2 &\leftrightarrow 4:&
z &\leftrightarrow (1-z) \,,&
\bar z &\leftrightarrow (1-\bar z) \,,&
\alpha &\leftrightarrow (1-\alpha) \,,&
\bar\alpha &\leftrightarrow (1-\bar\alpha)
\,.
\end{align}
Using the identities
\begin{equation}
F^{(1)}\lrbrk{\frac{1}{z},\frac{1}{\bar z}}=z\bar z\,F^{(1)}(z,\bar z)
\,,\qquad
F^{(1)}\lrbrk{1-z,1-\bar z}=F^{(1)}(z,\bar z)
\label{eq:F1ids}
\end{equation}
for the conformal box integral, as well as the
identity~\eqref{eq:midentities} for the building block function
$m(z)$, we find for the three independent functions:
\begin{align}
\polygon(1,2,4,3)
&=m(z)+m\lrbrk{\frac{1}{z}}
\nn\\ &\mspace{-100mu}
=\frac{g^2}{2}\lrbrk{
2(z+\bar z)
-(\alpha+\bar\alpha)\lrbrk{1+\frac{z\bar z}{\alpha\bar\alpha}}
}F^{(1)}(z,\bar z)
\,,\nn\\
\polygon(1,2,3,4)
&=m\lrbrk{\frac{z}{z-1}}+m\lrbrk{\frac{z-1}{z}}
=-m\lrbrk{\frac{1}{1-z}}-m\lrbrk{\frac{1}{z}}
\nn\\ &\mspace{-100mu}
=\frac{g^2}{2}\lrbrk{
-2
+\lrbrk{\alpha+\bar\alpha}\frac{z\bar z}{\alpha\bar\alpha}
-\lrbrk{\alpha+\bar\alpha-2}\frac{(1-z)(1-\bar z)}{(1-\alpha)(1-\bar\alpha)}
}F^{(1)}(z,\bar z)
\,,\nn\\
\polygon(1,3,2,4)
&=m(1-z)+m\lrbrk{\frac{1}{1-z}}
=-m(z)+m\lrbrk{\frac{1}{1-z}}
\nn\\ &\mspace{-100mu}
=\frac{g^2}{2}\lrbrk{
-2(z+\bar z)
+(\alpha+\bar\alpha+2)
+(\alpha+\bar\alpha-2)\frac{(1-z)(1-\bar z)}{(1-\alpha)(1-\bar\alpha)}
}F^{(1)}(z,\bar z)
\,.
\label{eq:oneloopoctagons}
\end{align}
%

\subsubsection*{Integrability}

At this point, we have derived the multi-particle contributions at
one-loop order, starting from the one- and two-particle contributions using flip invariance. An
obvious follow-up question is whether the result agrees with the
integrability computation.
In fact, we compute the three-particle contribution
using integrability in \appref{app:mirrorparticles}, using the weak-coupling
expansions of \appref{app:WeakCouplingApendix}, and it agrees with
the result of this section. This
lends additional support for the
correctness of the $2n$-gon formula~\eqref{eq:2n-gon}.
The multi-particle integrands are huge and
complicated, and we were not able to compute
the multi-particle contributions in general.
It would be interesting to study these integrands systematically.

\subsubsection*{Beyond Polygons}

\begin{figure}
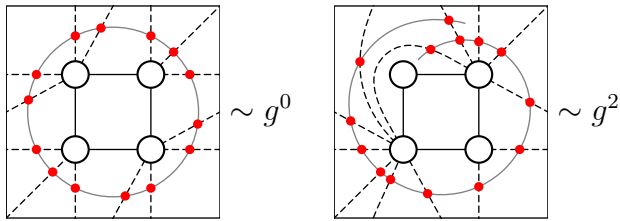

\centering
$\includegraphicsbox{FigLoopExample}\sim g^0$
\quad
$\includegraphicsbox{FigSpiralExample}\sim g^2$
\caption{``Loops'' and ``spirals'' naively start contributing at tree
and one-loop order, by the loop counting of \figref{fig:counting}.
They appear very difficult to evaluate from hexagons.}
\label{fig:loopsandspirals}
\end{figure}

While we can compute any one-loop string that is bounded by a polygon
via the formula~\eqref{eq:2n-gon}, there are further excitation
patterns that, by the loop counting shown in \figref{fig:counting},
could contribute at one-loop order. Namely, all stratification graphs
(\tabref{tab:stratexcyc} and \tabref{tab:stratexnoncyc}) contain
non-trivial cycles that do not intersect the graph. Hexagonalizing the
surface with zero-length bridges, strings of excitations can
wrap the cycle to form ``loops'' or ``spirals'', see
\figref{fig:loopsandspirals}. These types of contributions seem very
difficult to compute from hexagons. At the same time, it appears very
plausible that they are related to simpler configurations by Dehn
twists. Since we are not able to honestly evaluate these
contributions, we will have to resort to a (well-motivated)
prescription to avoid them. We will come back to this point
in~\secref{sec:strat}.

\section{Data}
\label{sec:data}

Let us now introduce the \emph{data} which we will
later use to check our proposal.
Computing correlators in perturbation theory is a hard task
in the planar limit, and an even harder task beyond the planar limit,
hence there is not that much data available. We will use here
results from the nice works of Arutyunov, Penati, Santambrogio and
Sokatchev~\cite{Arutyunov:2003ae,Arutyunov:2003ad}, who studied an interesting
class of four-point correlation functions of single-trace half-BPS
operators~\eqref{eq:BPSop}. The
authors of~\cite{Arutyunov:2003ae,Arutyunov:2003ad} studied the case
where all operators have equal weight $k$. In this case, the
contributions to the correlator can be organized by powers of the
propagator structures
\begin{equation}
X\equiv\frac{\alpha_1\cdot \alpha_2 \,\alpha_3\cdot \alpha_4}{x_{12}^2 x_{34}^2}
\,,\quad
Y\equiv\frac{\alpha_1\cdot \alpha_3 \,\alpha_2\cdot \alpha_4}{x_{13}^2 x_{24}^2}
\,,\quad\text{and}\quad
Z\equiv\frac{\alpha_1\cdot \alpha_4 \,\alpha_2\cdot \alpha_3}{x_{14}^2 x_{23}^2}
\,.
\label{eq:XYZ}
\end{equation}
They further specialized to operator polarizations $\alpha_i$ with
$\alpha_1\cdot\alpha_4=\alpha_2\cdot\alpha_3=0$,%
\footnote{A more invariant statement is
that the R-charge cross-ratio $(\alpha_1\cdot \alpha_4)(\alpha_2\cdot
\alpha_3)/(\alpha_1\cdot \alpha_3)(\alpha_2\cdot \alpha_4)=0$.}
such that the loop correlator $G_k \equiv
\vev{\op{Q}^k_1\op{Q}^k_2\op{Q}^k_3\op{Q}^k_4}
-\vev{\op{Q}^k_1\op{Q}^k_2\op{Q}^k_3\op{Q}^k_4}\suprm{tree}$
takes the form
\begin{equation}
G_k=\sum_{m=0}^{k}\mathcal{F}_{k,m}\,X^mY^{k-m}
\,.
\label{eq:corr}
\end{equation}
The functions $\mathcal{F}_{k,m}$ constitute the quantum corrections
that multiply the respective propagator structures, and they only depend
on the conformally invariant cross ratios~\eqref{eq:zdef}.
Expanding in the coupling,
\begin{equation}
\mathcal{F}_{k,m}=\sum_{\ell=1}^\infty g^{2\ell}\mathcal{F}_{k,m}^{(\ell)}(z,\bar z)
\,,\qquad
g^2=\frac{\lambda}{16\pi^2}
\,,
\label{eq:Fkm}
\end{equation}
we finally isolate the functions $\mathcal{F}_{k,m}^{(\ell)}$ against
which we will check our integrability computations in later sections.
The one-loop and two-loop contributions
$\mathcal{F}_{k,m}^{(1)}(z,\bar z)$ and
$\mathcal{F}_{k,m}^{(2)}(z,\bar z)$ have been computed
in~\cite{Arutyunov:2003ae,Arutyunov:2003ad} at the full non-planar
level. Two key ingredients
appear in their result. The first one are the conformal box and
double-box functions
\begin{align}
F^{(1)}(z,\bar z)
&=
\frac{x_{13}^2x_{24}^2}{\pi^2}
\int\frac{\dd^4x_5}{x_{15}^2x_{25}^2x_{35}^2x_{45}^2}
\, = \, \includegraphicsbox{FigBoxInt}
\,,
\label{eq:box}
\\
F^{(2)}(z,\bar z)
&=
\frac{x_{13}^2x_{24}^2x_{14}^2}{(\pi^2)^2}
\int\frac{\dd^4x_5\,\dd^4x_6}{x_{15}^2x_{25}^2x_{45}^2x_{56}^2x_{16}^2x_{36}^2x_{46}^2}
\, = \, \includegraphicsbox{FigDoubleBoxInt}
\,,
\label{eq:doublebox}
\end{align}
whose expressions in terms of polylogarithms are quoted in~\eqref{eq:F1expr}
and~\eqref{eq:FintermsofPolylogs}.

The second main ingredient are the so-called color factors, which
consist of color contractions of four symmetrized traces from the four
operators,
dressed with insertions of gauge group structure constants $f_{ab}{}^c$.
For instance, we have%
\footnote{Here, $\tr\brk{\brk{a_1\dots a_k}}\equiv\tr\brk{T^{(a_1}\dots
T^{a_k)}}$ denotes a totally symmetrized trace of adjoint gauge group
generators $T^a$.}
\begin{multline}
C\suprm{c}_{k,m}=
\frac{f_{abe}f_{cd}{}^ef_{pqt}f_{rs}{}^t}{2m!^2(k-m-2)!^2}
\tr\brk{\brk{d_1\dots d_{k-m-2}a_1\dots a_mbd}}
\tr\brk{\brk{a_1\dots a_mb_1\dots b_{k-m-2}ar}}
\\\times
\tr\brk{\brk{d_1\dots d_{k-m-2}c_1\dots c_mcp}}
\tr\brk{\brk{c_1\dots c_mb_1\dots b_{k-m-2}qs}}
\,,
\label{eq:Ccdef}
\end{multline}
which we can represent pictorially as
\begin{equation}
C\suprm{c}_{k,m}=
\includegraphicsbox{FigCc}
\,.
\end{equation}
At two loops, $C\suprm{c}$ as well as three other color factors $C\suprm{a}$,
$C\suprm{b}$, and $C\suprm{d}$ appear. The one-loop correlator is
expressed in terms of a single color factor $C^1$. The various color
factors differ from~\eqref{eq:Ccdef} only in the distribution of
structure constants $f_{ab}{}^c$ on the four single-trace operators.
Due to supersymmetry, the loop correction functions can be written as%
\footnote{This structure is due to the fact that $G_k$ contains a
universal prefactor $R$, see~\cite{Arutyunov:2002fh}
and~\appref{app:data-details}.}
\begin{equation}
\mathcal{F}_{k,m}^{(\ell)}
=
\mathcal{\widetilde{F}}_{k,m}^{(\ell)}
+(t-s-1)\mathcal{\widetilde{F}}_{k,m-1}^{(\ell)}
+s\mathcal{\widetilde{F}}_{k,m-2}^{(\ell)}
\,.
\label{eq:FFtilde}
\end{equation}
In terms of color factors and box integrals, the
functions $\mathcal{\widetilde{F}}_{k,m}$ read~\cite{Arutyunov:2003ae,Arutyunov:2003ad}
\begin{align}
\mathcal{\widetilde{F}}_{k,m}^{(1)}(z,\bar z)
&=\frac{C^1_{k,m}}{k^2\Nc^{2k+1}}F^{(1)}(z,\bar z)
\label{eq:Ftilde1}
\,,\\
\mathcal{\widetilde{F}}_{k,m}^{(2)}(z,\bar z)
&=\frac{4}{k^2\Nc^{2k+2}}\biggsbrk{
  \sfrac{1}{4}\bigbrk{2C^{\mathrm{b}'}-C\suprm{d}+\brk{2C\suprm{b}-C\suprm{d}}s+C\suprm{d}t}
  \bigbrk{F^{(1)}(z,\bar z)}^2
  \nn\\&\mspace{-20mu}
  +(C\suprm{c}-C\suprm{d})F^{(2)}(z,\bar z)
  +(C\suprm{d}-C^{\mathrm{a}'})F^{(2)}_{1-z}(z,\bar z)
  +(C\suprm{d}-C\suprm{a})F^{(2)}_{z/(z-1)}(z,\bar z)
},
\label{eq:Ftilde2}
\end{align}
where all color factors $C^i$ depend on $k$ and $m$.
We have used the shorthand notation
$C^{i'}_{k,m}=C^i_{k,k-m-2}$, and
\begin{equation}
F^{(2)}_{1-z}(z,\bar z)
\equiv F^{(2)}\lrbrk{1-z,1-\bar{z}}
\,,\quad
F^{(2)}_{z/(z-1)}(z,\bar z)
\equiv\frac{1}{\abs{1-z}^{2}}F^{(2)}\lrbrk{\frac{z}{z-1},\frac{\bar z}{\bar z-1}}
\,.
\label{eq:F2variants}
\end{equation}
In order to compare
with our integrability predictions, we need to explicitly evaluate
the color factors. This turns out to be a fun yet
involved calculation, which we did in two steps. First, we have
explicitly performed the contractions with \mathematica for different
values of $k$ and $m$; for some
coefficients up to $k=8$, for others up to $k=9$.
Expanding the color factors to subleading order in $1/\Nc$,
\begin{align}
C^1_{k,m}&=\Nc^{2k-1}k^4\lrbrk{\Csphere^1_{k,m}+\Ctorus^1_{k,m}\Nc^{-2}+\order{\Nc^{-4}}}
\,,\nn\\
C^i_{k,m}&=\Nc^{2k}k^4\lrbrk{\Csphere^i_{k,m}+\Ctorus^i_{k,m}\Nc^{-2}+\order{\Nc^{-4}}}
\,,\quad
i\in\brc{\mathrm{a,b,c,d}}
\,,
\label{eq:colorfactorexpansion}
\end{align}
the results for
the subleading color coefficients are displayed in~\tabref{tab:CCsub}.
Depending on the algorithm, the computation can take very long (up to
${\sim}1$ day on $16$ cores for a single coefficient at fixed $k$ and $m$) and
becomes memory intensive (up to ${\sim}100\,$GB) at intermediate
stages.%
\footnote{Very likely, the performance can be greatly improved by using
more specialized and better-scaling tools such as \form.}
The leading coefficients
\begin{equation}
\Csphere^1_{k,m}=-2k^4
\,,\qquad
\Csphere^{\mathrm{a},\mathrm{d}}_{k,m}=\half\Csphere\suprm{c}_{k,m}=k^4
\,,\qquad
2\Csphere\suprm{b}_{k,m}=(1+\delta_{m,0})k^4
\,,
\label{eq:colorfactorleading}
\end{equation}
are straightforwardly
computed~\cite{Arutyunov:2003ae,Arutyunov:2003ad}.

\begin{table}
\centering
\begin{tabular}{rrr@{\;\;}rr@{\;\;}r@{\;\;}r@{\;\;}rr@{\;\;}r@{\;\;}r@{\;\;}r}
\toprule
$k$ & $m$ & $\half\Ctorus^{1,\grp{U}}_{k,m}$
                   & $\half\Ctorus^{1,\grp{SU}}_{k,m}$
                            & $\Ctorus^{\mathrm{a},\grp{U}}_{k,m}$
                                    & $2\Ctorus^{\mathrm{b},\grp{U}}_{k,m}$
                                            & $\half\Ctorus^{\mathrm{c},\grp{U}}_{k,m}$
                                                    & $\Ctorus^{\mathrm{d},\grp{U}}_{k,m}$
                                                            & $\Ctorus^{\mathrm{a},\grp{SU}}_{k,m}$
                                                                    & $2\Ctorus^{\mathrm{b},\grp{SU}}_{k,m}$
                                                                            & $\half\Ctorus^{\mathrm{c},\grp{SU}}_{k,m}$
                                                                                    & $\Ctorus^{\mathrm{d},\grp{SU}}_{k,m}$ \\
\midrule
$2$ & $0$ & $1$                  & $1$                 & $0$                 & $-2$                & $-1$                & $-1$                & $0$    & $-2$   & $-1$   & $-1$ \\
\midrule
$3$ & $0$ & $1$                  & $9$                 & $-5$                & $-2$                & $-1$                & $-1$                & $-9$   & $-18$  & $-9$   & $-9$ \\
$3$ & $1$ & $1$                  & $9$                 & $0$                 & $3$                 & $-1$                & $-1$                & $0$    & $-5$   & $-9$   & $-9$ \\
\midrule
$4$ & $0$ & $-5$                 & $13$                & $-7$                & $10$                & $5$                 & $5$                 & $-25$  & $-26$  & $-13$  & $-13$ \\
$4$ & $1$ & $-12$                & $24$                & $4$                 & $15$                & $13$                & $14$                & $-23$  & $-21$  & $-23$  & $-22$ \\
$4$ & $2$ & $-5$                 & $13$                & $0$                 & $21$                & $5$                 & $5$                 & $0$    & $3$    & $-13$  & $-13$ \\
\midrule
$5$ & $0$ & $-23$                & $9$                 & $-1$                & $46$                & $23$                & $23$                & $-33$  & $-18$  & $-9$   & $-9$ \\
$5$ & $1$ & $-51$                & $13$                & $31$                & $47$                & $55$                & $59$                & $-33$  & $-17$  & $-9$   & $-5$ \\
$5$ & $2$ & $-51$                & $13$                & $39$                & $76$                & $55$                & $59$                & $-9$   & $12$   & $-9$   & $-5$ \\
$5$ & $3$ & $-23$                & $9$                 & $0$                 & $63$                & $23$                & $23$                & $0$    & $31$   & $-9$   & $-9$ \\
\midrule
$6$ & $0$ & $-61$                & $-11$               & $20$                & $122$               & $61$                & $61$                & $-30$  & $22$   & $11$   & $11$ \\
$6$ & $1$ & $-126$               & $-26$               & $92$                & $107$               & $135$               & $144$               & $-8$   & $7$    & $35$   & $44$ \\
$6$ & $2$ & $-159$               & $-59$               & $139$               & $187$               & $175$               & $191$               & $39$   & $87$   & $75$   & $91$ \\
$6$ & $3$ & $-126$               & $-26$               & $110$               & $201$               & $135$               & $144$               & $35$   & $101$  & $35$   & $44$ \\
$6$ & $4$ & $-61$                & $-11$               & $0$                 & $139$               & $61$                & $61$                & $0$    & $89$   & $11$   & $11$ \\
\midrule
$7$ & $0$ & $-129$               & $-57$               & $65$                & $258$               & $129$               & $129$               & $-7$   & $114$  & $57$   & $57$ \\
$7$ & $1$ & $-249$               & $-105$              & $198$               & $205$               & $265$               & $281$               & $54$   & $61$   & $121$  & $137$ \\
$7$ & $2$ & $-343$               & $-199$              & $323$               & $366$               & $379$               & $415$               & $179$  & $222$  & $235$  & $271$ \\
$7$ & $3$ & $-343$               & $-199$              & $331$               & $455$               & $379$               & $415$               & $187$  & $311$  & $235$  & $271$ \\
$7$ & $4$ & $-249$               & $-105$              & $229$               & $404$               & $265$               & $281$               & $121$  & $260$  & $121$  & $137$ \\
$7$ & $5$ & $-129$               & $-57$               & $0$                 & $261$               & $129$               & $129$               & $0$    & $189$  & $57$   & $57$ \\
\midrule
$8$ & $0$ & $-239$               & $-141$              & $145$               & $478$               & $239$               & $239$               & $ 47$  & $282$  & $141$  & $141$ \\
$8$ & $1$ & $-434$               & $-238$              & $362$               & $353$               & $459$               & $484$               & $166$  & $157$  & $263$  & $288$ \\
$8$ & $2$ & ${\color{red}-619}$  & ${\color{red}-423}$ & $606$               & $627$               & $683$               & $747$               & $410$  & $431$  & $487$  & $551$ \\
$8$ & $3$ & ${\color{red}-692}$  & ${\color{red}-496}$ & $710$               & $841$               & $773$               & $854$               & $514$  & $645$  & $577$  & $658$ \\
$8$ & $4$ & ${\color{red}-619}$  & ${\color{red}-423}$ & $623$               & $869$               & $683$               & $747$               & $427$  & $673$  & $487$  & $551$ \\
$8$ & $5$ & $-434$               & $-238$              & $410$               & $701$               & $459$               & $484$               & $263$  & $505$  & $263$  & $288$ \\
$8$ & $6$ & $-239$               & $-141$              & $  0$               & $443$               & $239$               & $239$               & $  0$  & $345$  & $141$  & $141$ \\
\midrule
$9$ & $0$ & ${\color{red}-405} $ & ${\color{red}-277}$ & $273              $ & ${\color{red}810} $ & ${\color{red}405} $ & ${\color{red}405} $ & $ 145$ & $ 554$ & $ 277$ & $ 277$ \\
$9$ & $1$ & ${\color{red}-697} $ & ${\color{red}-441}$ & ${\color{red}599} $ & ${\color{red}565} $ & ${\color{red}733} $ & ${\color{red}769} $ & $ 343$ & $ 309$ & $ 477$ & $ 513$ \\
$9$ & $2$ & ${\color{red}-1005}$ & ${\color{red}-749}$ & ${\color{red}1005}$ & ${\color{red}986} $ & ${\color{red}1105}$ & ${\color{red}1205}$ & $ 749$ & $ 730$ & $ 849$ & $ 949$ \\
$9$ & $3$ & ${\color{red}-1193}$ & ${\color{red}-937}$ & ${\color{red}1266}$ & ${\color{red}1377}$ & ${\color{red}1337}$ & ${\color{red}1481}$ & $1010$ & $1121$ & $1081$ & $1225$ \\
$9$ & $4$ & ${\color{red}-1193}$ & ${\color{red}-937}$ & ${\color{red}1273}$ & ${\color{red}1554}$ & ${\color{red}1337}$ & ${\color{red}1481}$ & $1017$ & $1298$ & $1081$ & $1225$ \\
$9$ & $5$ & ${\color{red}-1005}$ & ${\color{red}-749}$ & ${\color{red}1033}$ & ${\color{red}1449}$ & ${\color{red}1105}$ & ${\color{red}1205}$ & $ 777$ & $1193$ & $ 849$ & $ 949$ \\
$9$ & $6$ & ${\color{red}-697} $ & ${\color{red}-441}$ & ${\color{red}669} $ & ${\color{red}1110}$ & ${\color{red}733} $ & ${\color{red}769} $ & $ 477$ & $ 854$ & $ 477$ & $ 513$ \\
$9$ & $7$ & ${\color{red}-405} $ & ${\color{red}-277}$ & $0                $ & ${\color{red}701} $ & ${\color{red}405} $ & ${\color{red}405} $ & $   0$ & $ 573$ & $ 277$ & $ 277$ \\
\bottomrule
\end{tabular}
\caption{Subleading coefficients of color factors from explicit
(laborious) contractions are presented in black. By fitting
appropriate polynomials in $k$ and $m$, we can obtain the general
expressions for the various color factors, which then allow us to
complete the table with the new values in red. The result depends on
the choice of gauge group indicated as a superscript
U for $U(\Nc)$ and SU for $SU(\Nc)$.}
\label{tab:CCsub}
\end{table}

Secondly, we used the fact that by their combinatorial nature, it is
clear that the various color factors should be polynomials in $k$ and
$m$ (up to boundary cases at extremal values
of $k$ or $m$). By looking at all ways in which the propagators among
the four operators can be distributed on the torus, one finds that the
polynomial can be at most quartic.%
\footnote{This fact is best understood by looking
at~\tabref{tab:maxcycgraphs} and~\eqref{eq:lengthsum} below.}
Any closed formula for these color factors
therefore has to be a quartic polynomial in $k$ and $m$.
A general polynomial of this type has $15$ coefficients.
Matching those against the (overcomplete) data points
in~\tabref{tab:CCsub} yields the desired formulas for the color
factors. The color factor~\eqref{eq:Ccdef}, for instance, takes the relatively involved form
\begin{multline}
C_{k,m}^c= \Nc^{2k} k^4
\Bigbrk{
2k^4
+\frac{1}{6\Nc^2}\Bigsbrk{
k^4 + 2 k^3 (-1 + 2 m) + k^2 (-1 +  6 m + 42 m^2)
\\
- 2 k (11 + 49 m +  99 m^2 + 46 m^3)
+ 2 (18 +  70 m + 127 m^2 +  92 m^3 + 23 m^4)
\\
+4(k-1)^2(-2+\delta_{m,0}+\delta_{m,k-2})}
+O(\Nc^{-4})
} \, ,
\end{multline}
for an $\grp{SU}(\Nc)$ gauge group, while the last line would be absent for
the $\grp{U}(\Nc)$ theory. Further details and explicit expressions for all
relevant color factors are presented in~\appref{app:data-details}.
Putting all these ingredients together, we finally obtain the desired
one-loop and two-loop expressions shown in~\tabref{tab:FkmUdata}.
\begin{table}
\centering
\fbox{%
\begin{minipage}{-0.5cm+\textwidth}
\begin{align*}
&\mathcal{F}_{k,m}^{(1),\mathrm{U}}(z,\bar z)=
\\&\mspace{10mu}
-\frac{2 k^2}{\Nc^{2}}\Biggbrc{
    {\color{mypurple}t}+\frac{1}{\Nc^2}\biggsbrk{
        \bigbrk{\bigsbrk{17r^2-\sfrac{7}{4}}k^2+\sfrac{9}{2}k+3}s_+
        -r\bigbrk{\bigsbrk{\sfrac{34r^2}{3}-\sfrac{7}{2}}k^3+9k^2+\sfrac{35}{3}k}s_-
        \nn\\&\mspace{130mu}
        +\bigbrk{
            \bigsbrk{\sfrac{17r^4}{6}-\sfrac{7r^2}{4}+\sfrac{11}{32}}k^4
            +\bigsbrk{\sfrac{9r^2}{2}-\sfrac{13}{8}}k^3
            +\bigsbrk{\sfrac{r^2}{6}+\sfrac{15}{8}}k^2
            -\sfrac{1}{2}k
        }t
    }
\nn\\&\mspace{40mu}
       {\color{blue}\mbox{}
       -
       \biggbrk{{\color{mypurple}1}+\frac{k\brk{k^3-6k^2+23k-6}}{12\Nc^2}}
        \bigbrk{(t-1)\delta_m^0+s\delta_m^1+\delta_m^{k-1}+(t-s)\delta_m^k}
       }
        \nn\\&\mspace{40mu}
       {\color{blue}\mbox{}
       +\frac{\brk{k+1}\brk{k^2-22k-9}}{3\Nc^2}
        \bigbrk{s\delta_m^0+\delta_m^k}}
}{\color{red}F^{(1)}}
+\order{\Nc^{-6}}
\,,
\\[1ex]
&\mathcal{F}_{k,m}^{(2),\mathrm{U}}(z,\bar z)=
\\&\mspace{20mu}
\frac{4 k^2}{\Nc^{2}}\Biggsbrk{
\biggbrc{
    {\color{mypurple}t}+\frac{1}{\Nc^2}\biggsbrk{
        \bigbrk{\bigsbrk{17r^2-\sfrac{7}{4}}k^2+\sfrac{9}{2}k+3}s_+
        -r\bigbrk{\bigsbrk{\sfrac{34r^2}{3}-\sfrac{7}{2}}k^3+9k^2+\sfrac{35}{3}k}s_-
        \nn\\&\mspace{40mu}
        +\bigbrk{
            \bigsbrk{\sfrac{17r^4}{6}-\sfrac{7r^2}{4}+\sfrac{11}{32}}k^4
            +\bigsbrk{\sfrac{9r^2}{2}-\sfrac{13}{8}}k^3
            +\bigsbrk{\sfrac{r^2}{6}+\sfrac{15}{8}}k^2
            -\sfrac{1}{2}k
        }t
    }
}{\color{red}F^{(2)}}
\nn\\&\mspace{20mu}
+\biggbrc{
    {\color[rgb]{0.6,0,0.8}\frac{t^2}{4}}+\frac{1}{\Nc^2}\biggsbrk{
        \bigbrk{\bigsbrk{\sfrac{17r^2}{2}-\sfrac{7}{8}}k^2+3k+\sfrac{7}{4}}s_-^2
        -\sfrac{31}{2}rks_+s_-
        +\sfrac{7}{2}s_+^2
        \nn\\&\mspace{40mu}
        +\sfrac{1}{4}\bigbrk{
            \bigsbrk{\sfrac{29r^4}{6}-\sfrac{11r^2}{4}+\sfrac{15}{32}}k^4
            +\bigsbrk{\sfrac{17r^2}{2}-\sfrac{21}{8}}k^3
            -\bigsbrk{\sfrac{23r^2}{6}-\sfrac{39}{8}}k^2
            -\sfrac{9}{2}k+2
        }t^2
        \nn\\&\mspace{40mu}
        -r\bigbrk{\bigsbrk{\sfrac{23r^2}{3}-\sfrac{9r}{4}}k^3+\sfrac{29}{4}k^2+\sfrac{11}{6}k}ts_-
        +\bigbrk{\bigsbrk{\sfrac{43r^2}{4}-\sfrac{13}{16}}k^2+\sfrac{11}{4}k}ts_+
    }
}{\color{red}\bigbrk{F^{(1)}}^2}
\nn\\&\mspace{20mu}
+\frac{1}{\Nc^2}\biggsbrk{
    \sfrac{r}{2}\brk{5k^2-1k}{\color{red}F^{(2)}_{\mathrm{A,-}}}
    +\bigbrk{\bigsbrk{\sfrac{7r^2}{2}-\sfrac{1}{8}}k^2+\sfrac{1}{4}k+3}{\color{red}F^{(2)}_{\mathrm{A,+}}}
    \nn\\&\mspace{40mu}
    +\sfrac{r}{2}\brk{5k^2+13k}{\color{red}F^{(2)}_{\mathrm{B,-}}}
    -\bigbrk{\bigsbrk{\sfrac{7r^2}{2}-\sfrac{1}{8}}k^2+\sfrac{11}{4}k+6}{\color{red}F^{(2)}_{\mathrm{B,+}}}
    \nn\\&\mspace{40mu}
    -r\bigbrk{\bigsbrk{\sfrac{7r^2}{6}-\sfrac{1}{8}}k^3+\sfrac{3}{2}k^2+\sfrac{10}{3}k}{\color{red}F^{(2)}_{\mathrm{C,-}}}
    -\bigbrk{\bigsbrk{\sfrac{5r^2}{4}-\sfrac{19}{48}}k^3+\bigsbrk{\sfrac{3r^2}{2}+\sfrac{7}{8}}k^2+\sfrac{1}{3}k}{\color{red}F^{(2)}_{\mathrm{C,+}}}
}
}
\nn\\&\mspace{20mu}
{\color{blue}\mbox{}+\mathcal{F}_{k,m}^{(2),\mathrm{U,bdry}}(z,\bar z)}
+\order{\Nc^{-6}}
\,.
\end{align*}
\vspace{0mm}
\end{minipage}}
\caption{Perturbative one-loop and two-loop \emph{data} taken
from~\cite{Arutyunov:2003ae,Arutyunov:2003ad}, explicitly expanded to
include the first non-planar correction, which can be directly matched
against our integrability computation. Leading terms of order
$\Nc^{-2}$ form the planar contribution, whereas terms of order
$\Nc^{-4}$ constitute the first non-planar correction. All dependence on $k$
and $m$ is explicitly shown, via $r=m/k-1/2$.
The variables $s$, $t$, and $s_\pm$, as well as the various
combinations of double-box functions $F^{(2)}$ are defined
in~\eqref{eq:rst}, \eqref{eq:F2variants}, and~\eqref{eq:F2combs}. We
show the result for gauge group $\grp{U}(\Nc)$, since this is what we
will match with our integrability computation. We have highlighted the
box integrals ({\color{red}red}), the planar terms ({\color{mypurple}purple}) as well as terms that only
contribute at extremal values of $m$ ({\color{blue}blue}). The
expression for such boundary terms for $\mathcal{F}^{(2)}$ is deferred
to~\tabref{tab:Fkm2Udatabdry}.}
\label{tab:FkmUdata}
\end{table}
\begin{table}[t]
\centering
\fbox{%
\begin{minipage}{-0.5cm+\textwidth}
\begin{align*}
{\color{blue}\mathcal{F}_{k,m}^{(2),\mathrm{U,bdry}}(z,\bar z)}&=
\frac{4 k^2}{\Nc^{2}}\Biggsbrk{
\biggbrc{
-\biggbrk{{\color{mypurple}1}+\frac{k\brk{k^3-6k^2+23k-6}}{12\Nc^2}}{\color{blue}\bigbrk{t_-\delta_m^0+s\delta_m^1}}
\nn\\&\mspace{200mu}
+\frac{\brk{k+1}\brk{k^2-22k-9}}{3\Nc^2}s{\color{blue}\delta_m^0}
}{\color{red}F^{(2)}}
\nn\\&\mspace{-100mu}
+
\biggbrc{
\biggbrk{{\color{mypurple}1}+\frac{(k-1)(k^3+3k^2-46k+36)}{12\Nc^2}}{\color{blue}\bigbrk{\brk{s+t-t^2}\delta_m^0-s(s+1)\delta_m^1+s^2\delta_m^2}}
\nn\\&\mspace{-70mu}
+\frac{1}{\Nc^2}\biggsbrk{
\Bigbrc{
\bigbrk{5k^2+15k-7}-\bigbrk{5k^2+43k+21}s
+\sfrac{1}{3}(k+1)\bigbrk{k^2-40k-3}t
}s{\color{blue}\delta_m^0}
\nn\\&\mspace{-46mu}
-\frac{1}{3}\Bigbrc{
\bigbrk{k^3-9k^2+14k-3}
-\bigbrk{k^3-27k^2-k+3}s
\nn\\&\mspace{+20mu}
-\bigbrk{2k^3-24k^2+34k-15}t
}{\color{blue}\bigbrk{(t-1)\delta_m^0+s\delta_m^1}}
}
}\frac{1}{4}{\color{red}\bigbrk{F^{(1)}}^2}
\nn\\&\mspace{-100mu}
+
\biggbrc{
\biggbrk{{\color{mypurple}1}+\frac{(k-2)_4}{12\Nc^2}}{\color{blue}\bigbrk{\delta_m^0+\brk{t_--s}\delta_m^1+s\delta_m^2}}
\nn\\&\mspace{-70mu}
+\frac{1}{\Nc^2}\biggsbrk{
2\bigbrk{k^2+3k+3}s{\color{blue}\delta_m^0}
+2k\brk{k+1}{\color{blue}\bigbrk{t_-\delta_m^0+s\delta_m^1}}
+\sfrac{1}{2}\brk{k-3}\brk{k+2}{\color{blue}\delta_m^k}
\nn\\&\mspace{-20mu}
-\sfrac{1}{6}k\bigbrk{k^2-3k+8}{\color{blue}\bigbrk{\delta_m^{k-1}+(t-s)\delta_m^k}}
}
}{\color{red}F^{(2)}_{1-z}}
}
+\text{(crossing)}
\end{align*}
\vspace{0mm}
\end{minipage}}
\caption{Terms that contribute to $\mathcal{F}_{k,m}^{(2)}$ at
extremal values of $m$, see~\tabref{tab:FkmUdata}. Here, $t_-=(t-1)$.
The term ``(crossing)'' stands for a repetition of the complete
preceding expression, with the replacements~\eqref{eq:strcrossing}
and~\eqref{eq:F12crossing} as well as $m\to(k-m)$. Again, planar terms
are marked {\color{mypurple}purple}.}
\label{tab:Fkm2Udatabdry}
\end{table}
We show the result for gauge group $\grp{U}(\Nc)$, since this is what
we will compare to with our integrability computation. Corresponding
expressions for gauge group $\grp{SU}(\Nc)$ as well as further details
are given in~\appref{app:data-details}.
The expressions in~\tabref{tab:FkmUdata} are written in terms of the
variables $z$, $\bar z$, and $k$, as well as the combinations
\begin{equation}
r=\frac{m}{k}-1/2
\,,\quad
s=\abs{z}^2
\,,\quad
s_\pm=s\pm 1
\,,\quad
t=\abs{1-z}^2
\,.
\label{eq:rst}
\end{equation}
Besides the box integrals~\eqref{eq:box}, \eqref{eq:doublebox},
and~\eqref{eq:F2variants}, the following combinations of double-box
integrals occur:
%
\begin{equation}
F^{(2)}_{\mathrm{A},\pm}
=\abs{z}^2F^{(2)}_{z/(z-1)}\pm F^{(2)}_{1-z}
\,,\quad
F^{(2)}_{\mathrm{B},\pm}
=\abs{z}^2F^{(2)}_{1-z}\pm F^{(2)}_{z/(z-1)}
\,,\quad
F^{(2)}_{\mathrm{C},\pm}
=\abs{1-z}^2\bigbrk{F^{(2)}_{z/(z-1)}\pm F^{(2)}_{1-z}}
\, .
\label{eq:F2combs}
\end{equation}
We have suppressed the arguments $(z,\bar z)$ of all box functions for brevity.

The formulas are written such that crossing invariance is manifest:
The crossing transformation $x_1 \leftrightarrow x_4$ implies
\begin{equation}
X \leftrightarrow Y
\,,\quad
z \rightarrow 1/z
\,,\quad
\bar z \rightarrow 1/\bar z
\,,
\end{equation}
and hence crossing invariance of $G_k$~\eqref{eq:corr} is equivalent to
\begin{equation}
\mathcal{F}^{(\ell)}_{k,m}(z,\bar z)
=
\mathcal{F}^{(\ell)}_{k,k-m}(1/z,1/\bar z)
\,.
\end{equation}
Because of the transformations
\begin{equation}
s \rightarrow 1/s
\,,\quad
t \rightarrow t/s
\,,\quad
s_\pm \rightarrow \pm s_\pm/s
\,,\quad
r \rightarrow -r
\,,
\label{eq:strcrossing}
\end{equation}
and
\begin{equation}
F^{(1)} \rightarrow s F^{(1)}
\,,\quad
F^{(2)} \rightarrow s F^{(2)}
\,,\quad
F^{(2)}_{1-z} \rightarrow s F^{(2)}_{z/(z-1)}
\,,\quad
F^{(2)}_{z/(z-1)} \rightarrow s F^{(2)}_{1-z}
\,,
\label{eq:F12crossing}
\end{equation}
as well as the fact that all functions~\eqref{eq:F2combs} with
$+/-$ subscript are even/odd under crossing $x_1 \leftrightarrow
x_4$, it is clear that the expressions in~\tabref{tab:FkmUdata} are
indeed crossing invariant.

\paragraph{Remark.}

One immediate observation is that (up to an overall numerical
prefactor) the coefficient of the double-box
integral $F^{(2)}(z,\bar z)$ in the two-loop function
$\mathcal{F}^{(2)}_{k,m}$ equals the coefficient of the single-box
integral $F^{(1)}(z,\bar z)$ in the one-loop function
$\mathcal{F}^{(1)}_{k,m}$. As we shall see below, this
fact has a straightforward explanation from the perspective of the
integrability computation. In short, the one-loop function is a sum of
terms where only a single polygon (surrounded by non-zero-length
bridges) is excited. At two loops, the term proportional to
$F^{(2)}(z,\bar z)$ stems from the same sum of terms, where now the
single polygon is excited to two loops.
This pattern likely extends to higher loops.

\section{Contribution from Stratification}
\label{sec:strat}

Here, we want to evaluate the stratification contributions at genus one listed
in~\tabref{tab:stratexcyc} and~\tabref{tab:stratexnoncyc} at one-loop order. That is, we want to
evaluate the contributions $S_{(i)}$, $S_{(i')}$, and $S_{(i'')}$
in~\eqref{eq:stratificationgenus1}. As we have seen
in~\secref{sec:1-loop-polygons}, the one-loop expression for any
hexagonalization is given by the sum over all ``one-loop strings'',
where every one-loop string is a path that starts inside any hexagon,
ends in any other (or possibly the same) hexagon, and that crosses any
number of zero-length bridges, but no non-zero-length bridge. Every
crossing of any bridge by the path creates one excitation on that
bridge. For every closed, simply connected polygon, the number of such
one-loop strings is finite. For the graphs in~\tabref{tab:stratexcyc},
it is clear that a one-loop string can wind a cycle of the torus (or a
marked point) any number of times, and hence there is an infinite
number of one-loop strings. For example, the following magnon-patterns
all start contributing at one-loop order (for the loop-counting, see \figref{fig:counting}):
\begin{equation}
\includegraphicsbox{FigTorusStrat41}
\,,\quad
\includegraphicsbox{FigTorusStrat42}
\,,\quad
\includegraphicsbox{FigTorusStrat43}
\,,\quad
\includegraphicsbox{FigTorusStrat44}
\,,\quad\dotsc
\label{eq:twistedgraphs}
\end{equation}
Here, each of the red dots stands for a mirror magnon, and we
have also indicated (in gray) a path that connects them.

At present, we do not have the technology
to compute one-loop strings that form closed cycles, or that cross any
edge more than once (we call such strings ``spirals''). However, it is reasonable to assume that almost
all one-loop string contributions will either be projected out by our
Dehn-twist prescription~\eqref{eq:dehntwistrule}, or cancel between the torus
contributions $(i)$ and their pinched degenerations $(i')$ and $(i'')$ shown
in~\tabref{tab:stratexcyc} and~\tabref{tab:stratexnoncyc}.
Our working assumption is that all one-loop strings that either form
closed loops, or cross any bridge more than once, will either be projected
out by Dehn twists, or cancel with the stratification subtractions (or
sum to zero). We will therefore not take such contributions into
account.

Another limitation that we are facing is the
mapping among magnon configurations under flipping zero-length
bridges.
Even after dropping one-loop strings that cross bridges more than
once, there remain configurations that look related through Dehn
twists and bridge flips (for example all contributions in~\eqref{eq:twistedgraphs}).
Flipping any number of zero-length bridges should leave the
total contribution of the graph invariant, but it will non-trivially
map magnon configurations to each other. This map is technically quite
involved, and we have not evaluated it
except in the simplest cases (a single magnon on a single bridge)~\cite{Fleury:2016ykk}.
What we will assume is the following identification: Consider a one-loop string
of excitations traversing an otherwise empty handle across a number of
zero-length bridges. Imagining the
string of excitations as a continuous path, performing a Dehn twist on
such a handle adds a cycle to the path (string of excitations), as
well as to all zero-length bridges that also traverse the handle.
Subsequently performing flip moves of these zero-length bridges, we can restore the graph of zero-length
bridges to what it was before the Dehn twist. Effectively, this
operation adds a cycle to the path (string of excitations), and otherwise leaves
the graph invariant. Among all one-loop excitation strings
related by such operations, we only take one representative into
account. For example, all one-loop strings shown
in~\eqref{eq:twistedgraphs} are related by this operation, and hence
we would take only one of them into account. Even though we cannot
prove that all one-loop strings related under this operation indeed map
to each other one-to-one under Dehn twists and flip moves, we will see
in all examples below that one-loop strings related in this way indeed contribute
identical terms.

To summarize, we will evaluate the
stratification contributions at one loop using the following
prescription:
\begin{equation}
\mspace{-50mu}
\parbox{\textwidth-1.5cm}{%
\vspace{-1.2ex}
\raggedright
\begin{itemize}
\item Add up all one-loop strings that do not form closed loops and
that do not cross any bridge more than once (in the same direction).\footnotemark
\item Among all remaining excitation patterns, identify those that are
related to each other via Dehn twists that act on the path that
constitutes the one-loop string, but leaves the configuration of
zero-length bridges invariant.
\end{itemize}
\vspace{-1.2ex}
}
\label{eq:oneloopprescription}
\end{equation}
\footnotetext{The restriction ``in the same direction'' is relevant only
for the stratification contribution (1), and will be explained below.}%
We cannot rigorously show that our prescription is correct, but
we will see below that it produces the right answer. Given the
limitations in our present computational ability, it is the best we
can do.

In the following, we will consider the unprimed contributions (1)--(14)
of~\tabref{tab:stratexcyc} and~\tabref{tab:stratexnoncyc}. The primed contributions $(i')$ and $(i'')$ that have
to be subtracted were evaluated in~\secref{sec:subtractions}. In order
to evaluate the cancellations among primed and unprimed contributions,
we will use the identities given in~\eqref{eq:markedpolywc} that we reproduce
here:
\begin{equation}
\begin{aligned}
(\polygon)_{\otimes}&=1+2 (\polygon)\big|_{O(\lambda)}+\cdots\,,\\
(\polygon)_{2\otimes}&=0+ (\polygon)\big|_{O(\lambda)}+\cdots\,.
\end{aligned}
\label{eq:markedpolyrepeat}
\end{equation}
They
immediately imply that at tree level the contributions $(i)$ and $(i')$ (and $(i'')$
for $i=1,7,8,11$) of~\tabref{tab:stratexcyc}
and~\tabref{tab:stratexnoncyc} perfectly cancel each other
separately for each $i=1,\dotsc,14$. The first non-trivial effect of
stratification therefore occurs at one loop, and we will evaluate the
various contributions in the following, starting with the
simplest case.

\paragraph{Contribution (5).}

For case~$(5)$, the only non-vanishing contributions can come from
excitations of the two octagon faces that involve all four operators.
But these faces are exactly replicated in case~$(5')$, and hence the contributions $S_{(5)}$
and $S_{(5')}$ perfectly cancel each other. This cancellation relies
on the fact that polygons with one marked point at tree level equal the same polygons without insertions
as shown in~\eqref{eq:markedpolyrepeat}.

\paragraph{Contribution (6).}

This contribution works the same as contribution~(5): The only
non-vanishing one-loop contributions come from excitations in one of
the two faces that involve all four operators, which are exactly
replicated in contribution~$(6')$, and therefore perfectly cancel.

\paragraph{Contribution (7).}

Due to the identity~\eqref{eq:markedpolyrepeat} for a polygon with
two marked points, and the fact that a polygon with only two different
operators receives no loop corrections, contribution~$(7')$ vanishes.
By the same arguments as for cases $(5)$ and $(6)$, the
contributions~$S_{(7)}$ and~$S_{(7'')}$ perfectly cancel each other at
one-loop order.

\paragraph{Contributions (8)-(12).}

For the cases~(8) to~(12), all faces involve at most three out of
four operators. Therefore, we do not expect corrections at one-loop and the result
is simply the tree level one. This in turn will be canceled by the subtractions.

\paragraph{Contribution (4).}

Next, we will consider case~(4) of~\tabref{tab:stratexcyc}.
Picking an operator labeling, and shifting the
fundamental domain of the torus on which the graph is drawn, we can
depict this contribution as
\begin{equation}
\includegraphicsbox{FigTorusStrat4tess}
\label{eq:stratex4tess}
\end{equation}
Here, we have also
indicated a choice of zero-length bridges across the
handle not covered by the graph. Similar to case~$(5)$, we do not have
to consider one-loop excitations of the other faces, as these are
replicated in the pinched graph~$(4')$, and thus manifestly cancel.
Inside the face that wraps the torus, any non-vanishing one-loop excitation
string will have to involve hexagons that touch all four operators.
We have picked a tessellation that isolates operators $\op{O}_3$ and $\op{O}_4$ as
much as possible, such that any potentially non-zero string will have
to connect the hexagon that involves operator $\op{O}_3$ with the hexagon
that involves operator $\op{O}_4$. The only potentially non-zero
excitation strings that do not cross any bridge more than once are
exactly the two leftmost contributions of~\eqref{eq:twistedgraphs}:
\begin{equation}
\includegraphicsbox{FigTorusStrat41}
\quad
\includegraphicsbox{FigTorusStrat42}
\,,
\label{eq:stratex4}
\end{equation}
Here, each of the red dots stands for mirror
particles, and we
have also indicated (in gray) the path that connects them.
The left excitation pattern is equal to the one-loop (clockwise)
$\polygon(1,2,4,2,1,3)$, which vanishes by pinching (all other
one-loop excitation patterns in this polygon vanish, since they
involve at most three out of the four operators):
\begin{equation}
\includegraphicsbox{FigTorusStrat4PolyE}
=\includegraphicsbox{FigTorusStrat4Poly}
=\includegraphicsbox{FigTorusStrat4PolyRed}
=0
\,.
\end{equation}
The excitation pattern shown on the right of~\eqref{eq:stratex4} is
related to the one on the left by a Dehn twist according to our
working prescription~\eqref{eq:oneloopprescription}, hence we should not
take it into account. We can still evaluate this contribution in order
to check the consistency of our prescription. And indeed, the right one-loop string again
equals the (Dehn-twisted) one-loop
$\polygon(1,2,4,2,1,3)$ and thus vanishes by pinching.
Stratification requires that we subtract the contribution of
graph~$(4')$ in~\tabref{tab:stratexcyc}, which is obtained from $(4)$ by
pinching the handle not covered by the genus-zero graph.
In fact, because two-operator polygons receive no loop
corrections, the two-operator polygons with insertions of a single
marked point also receive no loop corrections,
and hence we trivially find that $S_{(4)}-S_{(4')}=0$.

\paragraph{Contribution (13).}

The case~$(13)$ will produce a vanishing contribution exactly by the same
argument as in the previous case~$(4)$.

\paragraph{Contribution (14).}

Let us consider the case~$(14)$ of~\tabref{tab:stratexnoncyc}. We again pick a tessellation of the
empty handle that isolates two operators as much as possible (in this
case~$\op{O}_2$ and~$\op{O}_3$):
\begin{equation}
\begin{tabular}{c@{}cc@{}c}
\includegraphicsbox{FigTorusStrat141}
& \ , &
\includegraphicsbox{FigTorusStrat142}
& \ .
\\
(a) & & (b) &
\end{tabular}
\label{eq:contrib14}
\end{equation}
Since a one-loop string can only be non-vanishing when it involves
hexagons that together touch all four operators, the two string
configurations above are the only potentially non-zero contributions.
The other faces involve three operators and hence contribute at tree
level only. They in turn will be canceled by the
subtraction~$S_{(14')}$. In addition to the excitation patterns shown above, we
could have also considered other string configurations that could
potentially contribute at one loop. But it is easy to see that these would
unavoidably involve placing two excitations in the same bridge, forming
a path that crosses that bridge twice in the same direction. By our
prescription, we do not take these cases into account.

The contributions (a) and (b) above are related by Dehn twists
according to our prescription~\eqref{eq:oneloopprescription}.
Consistently, it is simple to see that they produce identical results.
Namely, both cases evaluate to
\begin{equation}
\includegraphicsbox{FigTorusStrat14PolyE}
=\includegraphicsbox{FigTorusStrat14Poly}
=\includegraphicsbox{FigTorusStrat14PolyRed}
=\polygon(1,2,4,3)
\,.
\label{eq:contrib14poly}
\end{equation}
The subtraction~$S_{(14')}$ does not produce any contribution at
one-loop, as all of its polygons involve only three operators.
As a final step, we need to perform a sum over all non-equivalent
labels of the vertices. As the graph is drawn on a torus, there are
twelve inequivalent labelings (the same graph on a sphere has only two
inequivalent labelings):
\begin{equation}
\begin{tabular}{@{}r@{\;\;}c@{\;\;}c@{\;\;}c@{\;\;}c@{\;\;}c@{\;\;}c@{\;\;}c@{\;\;}c@{\;\;}c@{\;\;}c@{\;\;}c@{\;\;}c@{}}
\toprule
Labeling: &
$1243$ & $2134$ & $1342$ & $2431$ &
$1234$ & $2143$ & $1432$ & $2341$ &
$1324$ & $3142$ & $1423$ & $3241$ \\
Propag.: &
\multicolumn{2}{c}{$\displaystyle X^pY^qZ^r$} &
\multicolumn{2}{c}{$\displaystyle Y^pX^qZ^r$} &
\multicolumn{2}{c}{$\displaystyle X^pZ^qY^r$} &
\multicolumn{2}{c}{$\displaystyle Z^pX^qY^r$} &
\multicolumn{2}{c}{$\displaystyle Y^pZ^qX^r$} &
\multicolumn{2}{c}{$\displaystyle Z^pY^qX^r$} \\
One loop: &
\multicolumn{4}{c}{$\polygon(1,2,4,3)$} &
\multicolumn{4}{c}{$\polygon(1,2,3,4)$} &
\multicolumn{4}{c}{$\polygon(1,3,2,4)$} \\
\bottomrule
\end{tabular}
\end{equation}
The first line shows the labelings, reading clockwise starting with
the upper left operator in~\eqref{eq:contrib14}. The various labelings
come with different propagator structures, shown in the second line,
where $p$, $q$, and $r$ are the bridge lengths of the graph. The last
line shows the one-loop polygon the respective labeling evaluates to
(the $\polygon$ function obeys a dihedral symmetry).
In the sum over bridge lengths $p$, $q$, and $r$, all terms cancel due
to the identity~\eqref{eq:midentities}
\begin{align}
&\polygon(1,2,4,3)+\polygon(1,2,3,4)+\polygon(1,3,2,4)
\nn\\
&\qquad =m\lrbrk{z}+m\lrbrk{\frac{1}{z}}
+m\lrbrk{1-z}+m\lrbrk{\frac{1}{1-z}}
+m\lrbrk{\frac{z}{z-1}}+m\lrbrk{\frac{z-1}{z}}
\nn\\
&\qquad =m\lrbrk{z}+m\lrbrk{\frac{1}{z}}
-m\lrbrk{z}+m\lrbrk{\frac{1}{1-z}}
-m\lrbrk{\frac{1}{1-z}}-m\lrbrk{\frac{1}{z}}
\nn\\
&\qquad =0
\,.
\end{align}
We therefore find that $S_{(14)}=0$, and hence trivially
$S_{(14)}-S_{(14')}=0-0=0$. This case is different from all previous
(and subsequent) cases in that the cancellation occurs among graphs with different
labelings and bridge lengths.

\paragraph{Contribution (15).}

Picking a tessellation for graph $(15)$ of~\tabref{tab:stratexnoncyc},
we find, similar to the previous cases, only two potentially non-zero
one-loop contributions compatible with the first rule
of~\eqref{eq:oneloopprescription}:
\begin{equation}
\begin{tabular}{cc}
\includegraphicsbox{FigTorusStrat151}\,, & \includegraphicsbox{FigTorusStrat152}\,. \\
(a) & (b)
\end{tabular}
\end{equation}
By isolating it in a one-loop polygon, we find that the one-loop
string (a) evaluates to
\begin{equation}
\includegraphicsbox{FigTorusStrat151PolyE}
=\includegraphicsbox{FigTorusStrat151Poly}
=\includegraphicsbox{FigTorusStrat151PolyRed}
=0
\,.
\end{equation}
Case (b) is related to (a) by a Dehn twist according to the
prescription in~\eqref{eq:oneloopprescription}, hence for consistency
it should also evaluate to zero. And indeed one finds:
\begin{equation}
\includegraphicsbox{FigTorusStrat152PolyE}
=\includegraphicsbox{FigTorusStrat152Poly}
=\includegraphicsbox{FigTorusStrat152PolyRed}
=0
\,.
\end{equation}
The subtraction $(15')$ trivially evaluates to zero at one loop
by~\eqref{eq:markedpolyrepeat}, since the marked points are inserted
into polygons that involve only two and three out of the four
operators. We thus again find $S_{(15)}-S_{(15')}=0-0=0$.

\paragraph{Contribution (3).}

Now consider case~$(3)$. Again, we only need to consider excitations
of the face that wraps the torus, as all other excitations manifestly
cancel against the corresponding excitations in the pinched graph~$(3')$.
Picking a labeling and a tessellation
that isolates operators $\op{O}_2$ and $\op{O}_3$ as much as possible,
We find the following potentially non-zero one-loop excitation
patterns (we have slightly distorted the graph, and have shifted the
fundamental domain of the torus):
\begin{equation}
\begin{tabular}{cccc}
\includegraphicsbox{FigTorusStrat3a} &
\includegraphicsbox{FigTorusStrat3b} &
\includegraphicsbox{FigTorusStrat3c} &
\includegraphicsbox{FigTorusStrat3d}
\\
(a) &
(b) &
(c) &
(d)
\end{tabular}
\,.
\end{equation}
Again we are dropping the string configurations involving two
excitations placed on the same bridge according to our
prescription~\eqref{eq:oneloopprescription}.
For contribution (a) we find:
\begin{equation}
\includegraphicsbox{FigTorusStrat3aPolyE}
=\includegraphicsbox{FigTorusStrat3aPoly}
=\includegraphicsbox{FigTorusStrat3aPolyRed}
=\polygon(1,3,4,2)
\,.
\label{eq:stratex3a}
\end{equation}
Similarly, contribution (d) gives:
\begin{equation}
\includegraphicsbox{FigTorusStrat3dPolyE}
=\includegraphicsbox{FigTorusStrat3dPoly}
=\includegraphicsbox{FigTorusStrat3dPolyRed}
=\polygon(1,3,4,2)
\,.
\label{eq:stratex3d}
\end{equation}
Contributions (b) and (c) are related by a Dehn twist according to our
prescription~\eqref{eq:oneloopprescription}. For
consistency, both should give identical answers. Indeed we find for
contribution (b):
\begin{equation}
\includegraphicsbox{FigTorusStrat3bPolyE}
=\includegraphicsbox{FigTorusStrat3bPoly}
=\includegraphicsbox{FigTorusStrat3bPolyRed1}
=\includegraphicsbox{FigTorusStrat3bPolyRed2}
=0
\,,
\end{equation}
and for contribution (c):
\begin{equation}
\includegraphicsbox{FigTorusStrat3cPolyE}
=\includegraphicsbox{FigTorusStrat3cPoly}
=\includegraphicsbox{FigTorusStrat3cPolyRed1}
=\includegraphicsbox{FigTorusStrat3cPolyRed2}
=0
\,.
\label{eq:stratex3c}
\end{equation}
The contributions~\eqref{eq:stratex3a} and~\eqref{eq:stratex3d} each
equal the one-loop octagon $\polygon(1,3,4,2)$.
Since the one-loop octagon with one
marked point in contribution~$(3')$ (see~\tabref{tab:stratexcyc})
equals twice the same one-loop octagon with no
insertion by~\eqref{eq:markedpolyrepeat}, it is clear that $(3)$ and $(3')$ perfectly cancel
each other: $S_{(3)}-S_{(3')}=0$.

\paragraph{Contribution (2).}

Let us now list the possible one-loop excitation patterns for the
stratification graph~$(2)$. Picking an operator labeling as
well as a tessellation, we find:
\begin{equation}
\begin{tabular}{cccccc}
\includegraphicsbox{FigTorusStrat2a} &
\includegraphicsbox{FigTorusStrat2b} &
\includegraphicsbox{FigTorusStrat2c} &
\includegraphicsbox{FigTorusStrat2d} &
\includegraphicsbox{FigTorusStrat2e} &
\includegraphicsbox{FigTorusStrat2f}
\\
{\small (a)} &
{\small (b)} &
{\small (c)} &
{\small (d)} &
{\small (e)} &
{\small (f)}
\end{tabular}
\,.
\label{eq:stratex2e}
\end{equation}
These are all potentially non-zero one-loop excitation
patterns according to the prescription~\eqref{eq:oneloopprescription}:
We have picked a tessellation that isolates operators $\op{O}_2$ and
$\op{O}_3$ as much as possible.
Every non-zero excitation pattern has to have an excitation next to
operator $2$ (two choices), and an excitation next to operator $3$
(two choices). Otherwise no cross-ratio can be formed. The six
excitation patterns shown are all possible completions of the
$2\times2$ cases that involve at most one particle on any edge, up to
Dehn twists.
Contribution~\eqref{eq:stratex2e}(a) equals the one-loop polygon
\begin{equation}
(\text{a}):\quad
\includegraphicsbox{FigTorusStrat2aPolyE}
=
\includegraphicsbox{FigTorusStrat2aPoly}
=-\polygon(1,2,4,3)
\,.
\label{eq:stratex2aPoly}
\end{equation}
Here, we have used the code below~\eqref{examplePinching}
which implements the 2$n$-gon
formula~\eqref{eq:2n-gon}.
Similarly, for contribution~\eqref{eq:stratex2e}(b), we find
\begin{equation}
(\text{b}):\quad
\includegraphicsbox{FigTorusStrat2bPolyE}
=
\includegraphicsbox{FigTorusStrat2bPoly}
=-\polygon(1,2,4,3)
\,.
\label{eq:stratex2bPoly}
\end{equation}
Also the
contributions~\eqref{eq:stratex2e}(c)--(f) can be isolated as
individual one-loop polygons, which in turn can be evaluated using
pinching and decoupling identities. Explicitly, the result is:
\begin{align}
(\text{c}):\quad
\includegraphicsbox{FigTorusStrat2cPolyE}
&=
\includegraphicsbox{FigTorusStrat2cPoly}
=\polygon(1,3,4,2)
=\polygon(1,2,4,3)
\,,
\\
(\text{d}):\quad
\includegraphicsbox{FigTorusStrat2dPolyE}
&=
\includegraphicsbox{FigTorusStrat2dPoly}
=\polygon(1,2,4,3)
\,,
\\
(\text{e}):\quad
\includegraphicsbox{FigTorusStrat2ePolyE}
&=
\includegraphicsbox{FigTorusStrat2ePoly}
=\polygon(1,3,4,2)
=\polygon(1,2,4,3)
\,.
\\
(\text{f}):\quad
\includegraphicsbox{FigTorusStrat2fPolyE}
&=
\includegraphicsbox{FigTorusStrat2fPoly}
=\polygon(1,2,4,3)
\,.
\end{align}
Each of the contributions (c)--(f) gives
the same answer $\polygon(1,2,4,3)$.
Since each of the two octagons with one marked point in
contribution~$(2')$ of~\tabref{tab:stratexcyc} at one loop evaluates
to two times the same octagon without insertions,
they cancel the terms (c)--(f), leaving only the sum of
contributions~\eqref{eq:stratex2e}(a)--(b).

In order to further test the consistency of our
prescription~\eqref{eq:oneloopprescription}, we can compute
Dehn-twisted versions of the one-loop strings (a) and (b):
\begin{equation}
\begin{tabular}{cc}
\includegraphicsbox{FigTorusStrat2aTwisted} &
\includegraphicsbox{FigTorusStrat2bTwisted}
\\
{\small (a$^{\mathrm{t}}$)} &
{\small (b$^{\mathrm{t}}$)}
\end{tabular}
\,,
\end{equation}
and check that their contributions equal those of their untwisted
counterparts. Again, both one-loop strings can be isolated as
individual one-loop polygons. For contribution (a$^{\mathrm{t}}$), we
find
\begin{equation}
(\text{a}^{\mathrm{t}}):\quad
\includegraphicsbox{FigTorusStrat2aTwistedPolyE}
=\includegraphicsbox{FigTorusStrat2aTwistedPoly}
=\includegraphicsbox{FigTorusStrat2aTwistedPolyRed}
=-\polygon(1,2,4,3)
\,,
\end{equation}
where, in the last equation, we have again used the general one-loop
polygon formula~\eqref{eq:2n-gon}. Similarly, we find for contribution
(b$^{\mathrm{t}}$):
\begin{equation}
(\text{b}^{\mathrm{t}}):\quad
\includegraphicsbox{FigTorusStrat2bTwistedPolyE}
=\includegraphicsbox{FigTorusStrat2bTwistedPoly}
=\includegraphicsbox{FigTorusStrat2bTwistedPolyRed}
=-\polygon(1,2,4,3)
\,.
\end{equation}
Indeed, these Dehn-twisted contributions equal their untwisted
versions~\eqref{eq:stratex2aPoly} and~\eqref{eq:stratex2bPoly}.

\paragraph{Contribution (1).}

Let us finally turn to case~$(1)$. Picking a particular tessellation, we
find the following potentially non-zero excitation patterns:
\begin{equation}
\begin{tabular}{ccc}
\includegraphicsbox{FigTorusStrat1a} &
\includegraphicsbox{FigTorusStrat1b} &
\includegraphicsbox{FigTorusStrat1c}
\\
(a) &
(b) &
(c)
\end{tabular}
\,.
\label{eq:stratex1all}
\end{equation}
Pattern (a) readily evaluates to
$\polygon(1,1,3,4,4,2)$, which equals the one-loop octagon
$\polygon(1,3,4,2)$, which in turn equals half the contribution
of the planar graph~\eqref{eq:planargraph}. The contributions (b) and
(c) require some comments: These contributions include two
excitations on a single zero-length bridge. Even though we have thus far
discarded excitation patterns with more than one excitation on any bridge, we want to
argue that we should still include these contributions. All patterns
with multiple excitations on a single bridge that we have excluded thus
far had the form of a string of excitations that crossed a single
bridge twice \emph{in the same direction}. For the cases (b) and (c)
in~\eqref{eq:stratex1all}, the string of excitations crosses a bridge
twice, but in \emph{opposite directions}. As indicated at the
beginning of this section, we postulate that such
excitation patterns should be included. Next comes the question of
\emph{computing} these contributions. Because the excitation pattern
spans such a large part of the graph, it cannot be localized inside a
compact polygon. For case (b), the best we can do is to cut out the
inside of the square formed by the propagator bridges, and to cut
along the horizontal zero-length bridge that connects $\op{O}_4$ to
itself:
\begin{equation}\label{eq:complicated}
\includegraphicsbox{FigTorusStrat1b}
\quad \rightarrow \quad
\includegraphicsbox{FigTorusStrat1bStrip}
\end{equation}
We have no rigorous way of computing this contribution. It would be
easily computable if the horizontal, doubly excited zero-length bridge
that connects $\op{O}_1$ to itself was flipped: The flipped bridge would
connect $\op{O}_4$ with itself, and would no longer be crossed by the
string of excitations. After flipping, the three strings of
excitations become:
\begin{equation}
\begin{tabular}{ccc}
\includegraphicsbox{FigTorusStrat1aFlipped} &
\includegraphicsbox{FigTorusStrat1bFlipped} &
\includegraphicsbox{FigTorusStrat1cFlipped}
\\ && \\[-2.2ex]
(a$^\mathrm{f}$) &
(b$^\mathrm{f}$) &
(c$^\mathrm{f}$)
\end{tabular}
\,.
\label{eq:stratex1allflipped}
\end{equation}
The full one-loop answer for contribution (1) should be invariant
under flipping any bridge. But a priori, it is not clear that the individual
excitation patterns map to each other one-to-one. However, one
immediately finds that the pattern (a$^\mathrm{f}$) equals $\polygon(1,3,4,4,2,1,4)$, which after pinching
equals $\polygon(1,3,4,2)$, and thus indeed equals the
pattern (a) one-to-one. Since flipping the bridge does not alter the
string of excitations (a), we will assume that (b) and (c) are also individually
invariant under this flip. The contribution (b) then becomes
$\polygon(1,3,4,4,4,4,4,2)$, which equals
$\polygon(1,3,4,2)$, which equals half the
contribution of the planar graph, just as (a)=(a$^\mathrm{f}$) did.
Applying the same analysis to excitation pattern (c), but now flipping
the horizontal bridge that connects $\op{O}_4$ to itself, we find that
also the contribution (c) equals half the contribution of the
planar graph. In total, under the above flip-invariance assumption, we
thus find that the non-trivial part (without considering the internal polygon) of the stratification
contribution $(1)$ equals $3/2$ times the contribution of the planar
graph, or, equivalently, $3$ times the contribution of the one-loop
octagon. By the identities~\eqref{eq:markedpolyrepeat}, we find
that the non-trivial part of contribution~$(1')$
evaluates to the one-loop octagon, and the non-trivial part
of contribution~$(1'')$ gives two times the planar octagon. Hence in the
sum, we find that $S_{(1)}-S_{(1')}-S_{(1'')}=0$.

\paragraph{Summary and Result.}

We have demonstrated in the preceding paragraphs that almost all
stratification contributions $S_{(i)}$, $S_{(i')}$, and $S_{(i'')}$
are either zero, or directly cancel each other.
We should stress that all cancellations among primed and unprimed contributions hold at
the level of individual graphs \emph{with assigned bridge lengths and
operator labelings}:
There is a one-to-one map between the bridges of graphs $S_{(i)}$,
$S_{(i')}$, and $S_{(i'')}$ for fixed $i$. Therefore, for all
graphs $(i)$ and for any labeling of its operators as well as
any distribution of propagators on the bridges of that graph (\ie
any choice of bridge lengths), there is a corresponding operator
labeling and distribution
of propagators on the bridges of the associated pinched graph $(i')$
(and $(i'')$).
Hence the cancellations trivially extend to the full sum over all
operator labelings and bridge lengths, for any value of the weight~$k$.

The only remaining non-zero contributions from stratification at
one-loop order are the terms~\eqref{eq:stratex2aPoly}
and~\eqref{eq:stratex2bPoly}, which both evaluate to
$(-\polygon(1,2,4,3))$. We immediately note that their sum equals minus
the one-loop contribution of the simple planar
graph~\eqref{eq:planargraph}
\begin{equation}
\includegraphicsbox{FigPlanarGraph}
\label{eq:planargraphrepeat}
\end{equation}
on the sphere, which evaluates to
$2\times\polygon(1,2,4,3)$. Also, because the stratification contribution
stems from graph ($2$) in \tabref{tab:stratexcyc}, it is clear that
the sum over operator labelings and bridge lengths produces the same
answer for the stratification as for the planar graph.
We therefore conclude that the genus-one stratification
contribution~\eqref{eq:stratificationgenus1} at one-loop order equals minus the planar correlator,
\begingroup
\setlength{\fboxsep}{10pt}
\begin{empheq}[box=\fbox]{equation}
G_{1,1}^{\text{stratification}}
=\biggbrk{\sum_{i=1}^{14}S_{(i)}-\sum_{i=1}^{14}S_{(i')}-\sum_{\mathclap{i\in\brc{1,7,8,11}}}S_{(i'')}}
=-G_{0,1}
\label{eq:stratificationresult}
\end{empheq}
\endgroup
where we have decomposed the correlator as
\begin{equation}
\vev{\op{Q}_1\dotsc\op{Q}_4}
=\frac{k^2}{\Nc^2}\sum_{g,\ell}\frac{\lambda^\ell}{\Nc^{2g}}\,G_{g,\ell}
\,,
\end{equation}
and the
prefactor $k^2$ comes from the overall normalization
of~\eqref{eq:mainformula}.
We note that the result~\eqref{eq:stratificationresult} even holds for
generic internal polarizations $\alpha_i$ (but equal weights $k_i$),
since the graph~\eqref{eq:planargraph} is the only graph contributing
to the general-polarization one-loop correlator at genus zero:
The only other planar equal-weight graph~\eqref{eq:planargraphnoncyc}
\begin{equation*}
\raisebox{1ex}{\includegraphicsbox{FigPlanarGraphNonCyc}}
\end{equation*}
gives no contribution at one loop, because all of its faces are
hexagons framed by non-zero-length bridges.

In order to evaluate the stratification result or, equivalently, the
planar one-loop correlator~\eqref{eq:stratificationresult}, we have
to sum over inequivalent operator labelings and bridge lengths.
In this case, there are only
three distinct labelings. Using the operator lineup
in~\eqref{eq:planargraphrepeat} and
going clockwise (or equivalently going upwards
in~\eqref{eq:stratex2e}), we have the possible orderings 1--2--4--3
(used above in the derivation of~\eqref{eq:stratex2aPoly} and~\eqref{eq:stratex2bPoly}), 1--4--2--3, and
1--2--3--4. Making use of the dihedral symmetry of the $\polygon$
function~\eqref{eq:2n-gon}, summing over bridge lengths, and inserting
the respective propagator factors, we thus find
\begin{multline}
G_{1,1}^{\text{stratification}}
=-2\biggbrk{
\sum_{p=1}^{k-1}X^{p} Y^{k-p} \polygon(1,2,4,3)
+\sum_{p=1}^{k-1}Z^{p} Y^{k-p} \polygon(1,3,2,4)
\\
+\sum_{p=1}^{k-1}X^{p} Z^{k-p} \polygon(1,2,3,4)
}
\,,
\label{eq:strat2nonzero}
\end{multline}
where the sums run over $p=1, \ldots, k-1$, because all bridges in the
graph must be occupied by at least one propagator.
Writing the internal polarization cross ratios $\alpha$, $\bar\alpha$~\eqref{eq:alphadef}
in terms of the propagator structures $X$, $Y$, and $Z$~\eqref{eq:XYZ} via
\begin{equation}
\alpha\bar\alpha=\frac{X}{Y}\,z\bar z
\,,\qquad
(1-\alpha)(1-\bar\alpha)=\frac{Z}{Y}\,(1-z)(1-\bar z)
\,,
\label{eq:alphaXYZ}
\end{equation}
the octagon functions~\eqref{eq:oneloopoctagons} become
\begin{align}
\polygon(1,2,4,3)
&=\frac{g^2}{2}\biggsbrk{
1-\frac{Y}{X}+z \bar z \lrbrk{1-\frac{X}{Y}}+(1-z)(1-\bar z)\lrbrk{\frac{Z}{X}+\frac{Z}{Y}-2}
}F^{(1)}(z)
\,,\nn\\
\polygon(1,2,3,4)
&=\frac{g^2}{2}\biggsbrk{
\frac{Y}{X}+\frac{Y}{Z}-2+z\bar z \lrbrk{1-\frac{X}{Z}}+(1-z)(1-\bar z) \lrbrk{1-\frac{Z}{X}}
}F^{(1)}(z)
\,,\nn\\
\polygon(1,3,2,4)
&=\frac{g^2}{2}\biggsbrk{
1-\frac{Y}{Z}+z\bar z\lrbrk{\frac{X}{Y}+\frac{X}{Z}-2}+(1-z)(1-\bar z)\lrbrk{1-\frac{Z}{Y}}
}F^{(1)}(z)
\,.
\label{eq:octagonsXYZ}
\end{align}
Plugging these expressions into~\eqref{eq:strat2nonzero}, we recover
the result for the planar one-loop correlator
\begingroup
\setlength{\fboxsep}{10pt}
\begin{empheq}[box=\fbox]{equation}
G_{1,1}^{\text{stratification}}
=-G_{0,1}
=2 R \;\sum_{\mathclap{\substack{p,q,r\geq0\\p+q+r=k-2}}}\; X^p Y^q Z^r F^{(1)}(z)
\label{eq:stratfinalanswer}
\end{empheq}
\endgroup
with the universal polynomial factor $R$ due to
supersymmetry~\cite{Arutyunov:2002fh}
\begin{equation}
R\equiv\bigbrk{Y-Z+z\brk{Z-X}}\bigbrk{Y-Z+\bar{z}\brk{Z-X}}\,.
\label{eq:preR}
\end{equation}
We have computed the stratification contribution for arbitrary
polarizations $\alpha_i$. In order to compare to the data presented in
\secref{sec:data}, we might take the $Z=0$ limit of the result.

This computation shows the importance of summing over \emph{all} tree
level graphs, even those containing $Z$ propagator structures,
and only at the end take the particular limit $Z\rightarrow 0$
for comparison with the available data.
The reason is that, as we dress such graphs with mirror
particles, the overall dependence on the propagator structures can be
different from what it was at tree level. This comes about due to the
fact that the one-loop correction to the polygon carries itself a
dependence on the $R$-charge cross ratios, see the expression~\eqref{eq:mdefinition}
of the building block for the one-loop
polygons. As a consequence, the dependence on $Z$ of the tree-level
configurations might get canceled at one-loop order, resulting in
a contribution which is relevant to match the $Z=0$ data. Let us consider
one further example for illustration.
Take the following graph:
\begin{equation}
\mathcal{G}=\includegraphicsbox{FigTorusExampleZCancel}
\;,
\end{equation}
where we have explicitly drawn the propagators, assigned labels to the
vertices, and indicated the two faces in two different shades of gray.
This graph amounts to the following one-loop contribution
\begin{equation}
\mathcal{G}^{(1)}=\frac{Y^3 Z}{\Nc^4}\bigbrk{\polygon(4,1,3,2)+\polygon(4,2,3,1,3,1,3,1)}\,.
\end{equation}
After replacing the explicit expression for the corresponding
polygon~\eqref{eq:2n-gon}, we arrive at the result
\begin{equation}
\mathcal{G}^{(1)}=-\frac{g^2}{\Nc^4}Y^2
\lrsbrk{
Y \bigbrk{Z (z \bar{z}+z+\bar{z}-2)-X z \bar{z}}
+Z \bigbrk{(z-1) Z (\bar{z}-1)-X z \bar{z}}+Y^2
}
F^{(1)}
\,,
\end{equation}
which, after setting $Z=0$, results in a non-zero contribution.

\paragraph{Comparison with Perturbation Theory.}

We have seen above that the only non-trivial stratification
contribution to the correlator stems from graph~$(2)$. More
specifically, its origin are the contributions~(a) and~(b)
in~\eqref{eq:stratex2e}.%
\footnote{Even though we have no precise way of telling which of the
contributions in~\eqref{eq:stratex2e} are canceled by the pinched
graph~($2'$), it is reasonable to assume that the pinched graph
cancels the contributions (c)--(f).}
We will see that this matches beautifully with the expectation from
gauge theory. Stratification is supposed to reproduce perturbative
contributions to the genus-one correlator that stem from planar graphs
in the free theory. At fixed $k$ and $m$, that is at fixed propagator
structure $X^mY^{k-m}$, there is only one planar graph:
\begin{equation}
\includegraphicsbox{FigPlanarGraphProps}
\end{equation}
We are looking for one-loop decorations of this graph that contribute
to subleading order in $1/\Nc^2$ (\ie at genus one). All one-loop
processes are $\superN=2$ YM (super-gluon) lines%
\footnote{We are considering the $\superN=2$
description of $\superN=4$ SYM
with only external hypermultiplet fields,
see~\cite{Eden:2017ozn} and \appref{app:NpointPlanar}.}
between either two vertical or two horizontal propagators:
\begin{equation}
\includegraphicsbox{FigPlanarGraphProps1}
=\includegraphicsbox{FigPlanarGraphPropsGluon1}
+\includegraphicsbox{FigPlanarGraphPropsGluon2}
+\includegraphicsbox{FigPlanarGraphPropsGluon3}
+\includegraphicsbox{FigPlanarGraphPropsGluon4}
+\text{(vertical)}
\,.
\label{eq:pertstrat}
\end{equation}
Here, $\text{(vertical)}$ stands for similar contributions of a
vertical YM line connecting two horizontal propagators. Kinematically,
all ways of attaching the horizontal YM line to two vertical propagators
are identical. The only differences are powers of $1/\Nc^2$ (depending
on the genus of the one-loop graph) as well as relative signs: Each
end of the YM line can attach to a given propagator from either
side, at the cost of a relative sign, due to the antisymmetry of the
gauge structure constants $f_{abc}$:
\begin{equation}
\includegraphicsbox{FigGluonFlip1}
=-\includegraphicsbox{FigGluonFlip2}
\,.
\end{equation}
Of course there are many more
ways to connect the YM lines to two vertical propagators, but one can
easily see that all contributions except the ones shown cancel each
other, due to these relative signs. The third and fourth figure
in~\eqref{eq:pertstrat} have genus one, hence they are suppressed by
one factor of $1/\Nc^2$ compared to the first two figures (which
are planar). Also, the third and fourth figure carry a relative
sign, since one structure constant is flipped compared to the first
two figures. Hence we find
\begin{align}
\includegraphicsbox{FigPlanarGraphProps1}
&=\lrbrk{1-\frac{1}{\Nc^2}}\lrbrk{
\includegraphicsbox{FigPlanarGraphPropsGluon1}
+\includegraphicsbox{FigPlanarGraphPropsGluon2}
+\text{(vertical)}
}
\nn\\
&=\lrbrk{1-\frac{1}{\Nc^2}}\includegraphicsbox{FigPlanarGraphProps1Planar}
\,,
\end{align}
which exactly matches the result~\eqref{eq:stratificationresult}.
Moreover, the non-planar (third and fourth) terms
in~\eqref{eq:pertstrat} can be drawn on the torus as
\begin{equation}
\includegraphicsbox{FigPlanarGraphPropsGluon3}
=
\includegraphicsbox{FigTorusStrat2Gluon3}
\,,\qquad
\includegraphicsbox{FigPlanarGraphPropsGluon4}
=
\includegraphicsbox{FigTorusStrat2Gluon4}
\,,
\end{equation}
and hence can be associated to graphs of the type $(2)$
in~\tabref{tab:stratexcyc}.

\paragraph{Disconnected Graphs.}

Before ending this section, let us finally comment on a small subtlety:
In addition to the graphs considered so far, one can in principle
consider disconnected graphs drawn on a torus. Here, either both
components can be planar, or one of them may have genus one.
Clearly, by $1/\Nc$ power counting, without interactions, neither case contributes to the
same order as non-planar connected four-point graphs.
However, much like the secretly planar graphs, we
cannot simply discard them, since they can become
of the same order in $1/\Nc$
at high enough loop order, once they are dressed by a sufficient
number of gluon propagators. Therefore, when performing the
stratification procedure, we do need to include them in principle.

\begin{figure}
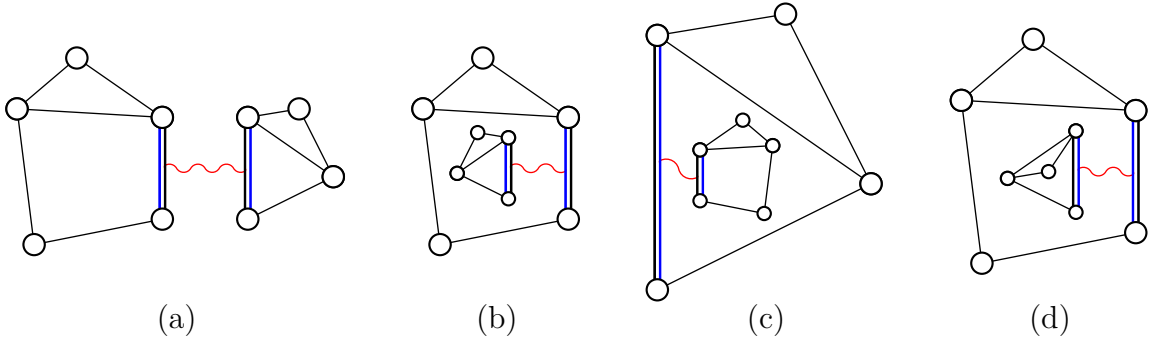

\centering
\begin{tabular}{c@{\qquad}c@{\qquad}c@{\qquad}c}
\includegraphicsbox{FigDisconnected1} &
\includegraphicsbox{FigDisconnected2} &
\includegraphicsbox{FigDisconnected3} &
\includegraphicsbox{FigDisconnected4}
\\
(a) & (b) & (c) & (d)
\end{tabular}
\caption{One-loop contributions from a generic disconnected graph at
the planar level. A gluon line may connect to any bridge on either
side, hence it can connect to two bridges in four different ways.
Kinematically, all four cases are equivalent, the only difference lies
in the color structure. The figures show that all four cases
contribute to the same order in $1/\Nc$. Since the four different
cases come with different signs ((b) and (c) come with a relative
minus sign compared to (a) and (d)), they cancel each other. We have
highlighted the double-line structure of the propagators involved in
the exchange interaction.}
\label{fig:disconnected1loop}
\end{figure}

Unfortunately, at the time of writing this article, we have not
succeeded in evaluating the contributions from these graphs if both
components are planar,%
\footnote{The contribution of graphs where one (two-point) component
is non-planar can be shown to vanish by similar arguments as for the
stratification contributions computed above.}
owing to
the existence of so many zero-length bridges. We thus \emph{assumed}
that their contributions at one loop vanish, once the subtraction and
the Dehn twist are taken into account. We should nevertheless stress
that this is a reasonable assumption: Firstly, in perturbation theory,
it is clear that such graphs cannot give rise to non-planar contributions
at one loop. This implies that the contribution from such disconnected
graphs will be canceled by the subtractions, as was the case for (some
of) the secretly planar graphs that we discussed in this section.
(From a perturbation-theory point of view, one can actually argue that
even the planar contribution from such graphs is zero. See the
discussion in \figref{fig:disconnected1loop}.) Secondly, although we
could not compute the contribution from disconnected graphs on a
torus, we could show, using the stratification and the Dehn twist,
that the contributions from disconnected graphs on a sphere vanish
at one loop. This will be demonstrated in \appref{app:disconnected}.
Let us also emphasize that, although the computation is sometimes
hard, the proposal we made is quite concrete and can be tested if one
has infinite computational ability. It would be an important future
task to complete the computation and prove or disprove the
cancellation that we assumed.

\paragraph{Stratification Summary and Discussion.}

We carefully analyzed fourteen contributions listed in
\tabref{tab:stratexcyc} and \tabref{tab:stratexnoncyc}, adding
all the secretly planar graphs and subtracting all pinched surfaces.
At the end of a laborious analysis, the punch line is amazingly
simple: These terms almost cancel each other completely. (Only
contribution (2) in \tabref{tab:stratexcyc} ends up not
canceling!) In the end, the result is simply minus one times the
planar result.

In the light of such a simple result, one might wonder if all this
stratification business, with all these involved considerations on
boundaries of moduli space subtleties are a huge overkill. Could it be
that, even at higher loops, the stratification ends up boiling down to
some simple terms proportional to lower-genus contributions?

Definitely \emph{not}!

On the contrary, at sufficiently high loops, the stratification is in
fact \emph{the most important} contribution, since, for any given
size of the external operators, the tree-level skeleton graphs only
exist up to some fixed genus order. So higher-genus contributions are
actually given uniquely by the stratification procedure. Therefore, if
we consider the full $1/N_c$ expansion, the stratification contributes
to all corrections and is the sole contributor starting at some genus
order. As an example, for $k=2$, we can only draw planar skeleton
graphs, hence all higher-genus corrections to this correlator --
starting already with the torus -- will come uniquely from the
stratification procedure!

Given the simplicity of the final one-loop
result~\eqref{eq:stratfinalanswer}, and the importance of the
stratification
at higher loops and higher genus, it is absolutely critical to
streamline its analysis. For that, we will likely need to better
understand the nature of the various exotic contributions, such as the
spirals and loops discussed above.

\section{Checks and Predictions}

\subsection{Finite \texorpdfstring{$k$}{k} Checks}
\label{sec:k2345}

We now proceed to test the integrability predictions against the data
described in \secref{sec:data}, starting with a few examples for
finite $k$.
At finite $k$, the relevant graphs are typically far from the maximal
ones. As described earlier, they can be obtained by successively
removing edges from the maximal graphs until each operator is connected by at most $k$
bridges, discarding the duplicate ones on the way. On top of this, we
should sum over all inequivalent labelings of the vertices and sum
over all bridge length assignments such that each operator is
connected by exactly $k$ propagators.
The \emph{statistics} of the polygonization procedure for the five
lowest $k$ cases is summarized in~\tabref{tab:numberofgraphsperk}. It
is apparent that the number of graphs grows very quickly both with
$k$ and with the genus, and therefore we have resorted to a \mathematica code
to generate them.
\begin{table}
\centering
\begin{tabular}{r@{\,:\quad}rrrr}
  \toprule
  $k$ &
  $2$ &
  $3$ &
  $4$ &
  $5$ \\
  \midrule
  $g=0$ &
  $3$ &
  $8$ &
  $15$ &
  $24$ \\
  $g=1$ &
  $0$ &
  $32$ &
  $441$ &
  $2760$ \\
  \bottomrule
\end{tabular}
\caption{Number of connected labeled graphs with specified bridge
lengths at genus $g=0$ and $g=1$ for
various values of $k$.}
\label{tab:numberofgraphsperk}
\end{table}
%

\subsubsection{\texorpdfstring{$k=2,3$}{k=2,3}}

\begin{figure}[t]
\centering
\includegraphics[scale=1.1]{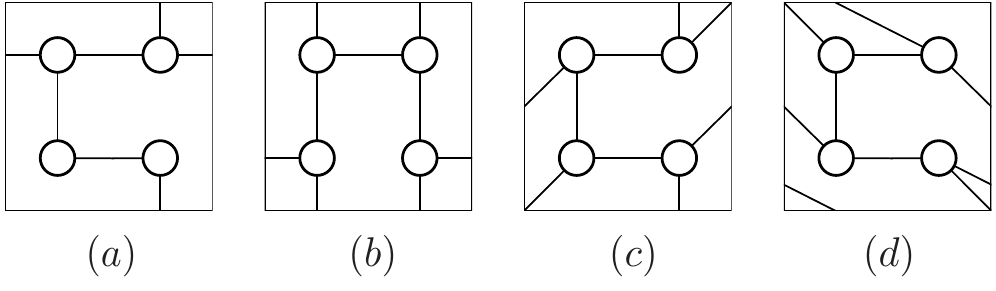}
\caption{Non-planar graphs for $k=3$.}
\label{fig:k3nonplanar}
\end{figure}

In the simplest $k=2$ example, it turns out that one cannot draw any
graph with the topology of a torus, since each operator will be
connected by at most two bridges. The single connected graph with this
constraint is depicted in~\eqref{eq:planargraph}. Therefore, the whole contribution
should come from the stratification
result~\eqref{eq:stratfinalanswer}, which in this case simply reads
\beq
\vev{\op{Q}^2_1\op{Q}^2_2\op{Q}^2_3\op{ Q}^2_4}^{1-\text{loop}}_{(g=1)}
=\frac{8g^2}{\Nc^4} \, R\, F^{(1)}(z)
\,.
\eeq
For the case of $k=3$, we already encounter non-planar graphs, as depicted in~\figref{fig:k3nonplanar}.
After assigning labels to the vertices and lengths to the bridges compatible with the operators' $R$-charges, one
generates $32$ distinct configurations as indicated in the
corresponding entry of~\tabref{tab:numberofgraphsperk}. Regardless of
the assignments, the graphs $(a),\,(b)$ and $(d)$
of~\figref{fig:k3nonplanar} produce a vanishing contribution.
The vanishing of the cases $(a)$ and $(b)$ can be anticipated only by
successive use of the pinching limit of the polygon as illustrated in the
expression~\eqref{eq:pinchrule}. For example, consider the case $(a)$ and
label the vertices from 1 to 4 in a clockwise order starting from the top left operator. There
is a single face corresponding to an icosagon ($20$-gon) bounded by
the bridges. Taking into account the order of the vertices along this
boundary, we have the one-loop contribution given by
$\polygon(1,2,1,3,4,2,1,2,4,3)$.
We now apply
several pinching limits to reduce that sequence down to $\polygon(1,3)$ which
would correspond to a two-point function, and that is zero by
supersymmetry.

The graph $(d)$ is decomposed into a hexagon and an octadecagon, and
both vanish once we use the corresponding one-loop expression as given
in~\eqref{eq:2n-gon}.

The only non-trivial graph is $(c)$, which produces a non-zero result.
However, after summing over all labelings, those contributions simply cancel
out. Therefore, the non-planar graphs do not contribute, and once again
we expect to obtain the final result simply from the stratification
contribution~\eqref{eq:stratfinalanswer}, which reads in this case
\beq
\vev{\op{ Q}^3_1\op{Q}^3_2\op{Q}^3_3\op{Q}^3_4}^{1-\text{loop}}_{(g=1)}
=\frac{18 g^2 }{\Nc^4} (X+Y+Z)\,R\,F^{(1)}(z)\,.
\eeq
For comparison with perturbative data, we now consider the case $Z=0$.
We find that for the two cases considered here:
\beq
\vev{\op{Q}^k_1\op{Q}^k_2\op{Q}^k_3\op{Q}^k_4}^{1-\text{loop}}_{(g=1)}
\xrightarrow{\;Z=0\;}
-\frac{1}{\Nc^2}G_{k,(g=0)}^\text{1-loop}
\quad
\text{for}
\quad
k=2,3
\,.
\eeq
and this perfectly matches with the data shown in~\tabref{tab:FkmUdata}.

\subsubsection{\texorpdfstring{$k=4$}{k=4}}

The case $k=4$ is significantly more involved than the previous ones.
The number of non-planar graphs is $57$, and they give $441$ distinct
physical configurations when operator labelings and bridge
lengths are chosen.

Let us consider one example in detail. Among the 441 graphs with
assigned labels and bridge lengths, we have the following example
\begin{equation}
\includegraphicsbox{FigTorusExamplek4}
\;,
\label{examplek4}
\end{equation}
where each solid line now corresponds to a propagator. This graph is
decomposed into two polygons: An octagon (dark gray) and a
hexadecagon (light gray). Accounting for the corresponding
propagators, we have that this contribution is given by
\beq
\frac{X^2 Y^2}{\Nc^4} \bigbrk{ \polygon(1,2,4,3) + \polygon(1, 2, 4, 3, 1, 3, 4, 3) }\,.
\eeq
We can now simply use the expression for the corresponding polygon
using~\eqref{eq:2n-gon} to get the final result. Alternatively, we
observe that using the pinching limit, the hexadecagon degenerates
into an octagon as follows,
\beq
\polygon(1, 2, 4, {\color{red}3}, 1, {\color{red}3}, 4, 3) \rightarrow \polygon(1,2,4,{\color{blue}3},4,{\color{blue}3}) \rightarrow \polygon(1,2,4,3)\,.
\eeq
Now plugging in the corresponding expression for the one-loop octagon
from~\eqref{eq:octagonsXYZ}, we get that this graph produces
\beq
-\frac{g^2}{\Nc^4} X Y \left(X^2 z \bar{z}+Y^2+X Y (z\bar{z}-2z-2\bar{z}+1)+Z(X+Y)(z+\bar{z}-z\bar{z}-1)\right) F^{(1)}(z)\,.
\eeq
All other graphs are equally straightforward to compute as this
example. Upon summing over the 441 graphs and adding the
stratification contribution~\eqref{eq:stratfinalanswer}, we
recover
the prefactor $R$ and the final result is given by
\beq
\vev{\op{Q}^4_1\op{Q}^4_2\op{Q}^4_3\op{Q}^4_4}^{1-\text{loop}}_{(g=1)}
=-\frac{32 g^2}{\Nc^4} \left(5\, (X^2+ Y^2+ Z^2)+12\, (X Y+X Z+Y Z)\right) \, R \,F^{(1)}\,.
\eeq
After setting $Z=0$ and comparing with the data of
\tabref{tab:FkmUdata} for $k=4$, we find again a perfect agreement.

\subsubsection{\texorpdfstring{$k=5$}{k=5}}

We have extended our analysis to the case $k=5$, which involves $2760$
distinct graphs. The procedure is no different from the previous cases,
and we simply display here the result from the summation over all
those genus-one graphs, together with the stratification contribution.
Once again, we recover the universal prefactor $R$~\eqref{eq:preR}
and the outcome reads
\beq
\begin{aligned}
\vev{\op{Q}^5_1\op{Q}^5_2\op{Q}^5_3\op{Q}^5_4}^{1-\text{loop}}_{(g=1)}
=&-\frac{50 g^2}{\Nc^4} \Bigl(108\, X Y Z+ 23 \left(X^3+Y^3+Z^3\right)\\
&+51 \left(X^2 Y+X^2 Z+X Y^2+X Z^2+Y^2 Z+Y Z^2\right) \Bigr) \, R \,F^{(1)}\,.
\end{aligned}
\eeq
When $Z=0$ we again recover the perturbative result of \tabref{tab:FkmUdata}.

To summarize the findings of this section: By summing over genus-one
graphs and adding the stratification contribution determined in
\secref{sec:strat}, we computed the four-point correlator for a generic
polarization of the external BPS operators. We compared these results
with data for the particular polarization studied in literature, namely
when $Z=0$, and found a perfect match in all cases, which strongly
corroborates our proposal. The $Z\neq 0$ results are simple
predictions of the hexagonalization procedure, which would be nice to
check against a direct perturbative computation.

\subsection{Checks at Large \texorpdfstring{$k$}{k}}

\subsubsection{\texorpdfstring{$k\gg 1$}{k>>1}: Leading Order}

Another interesting case that we will focus on in the following are
contributions $\mathcal{F}_{k,m}$ where both $m$ and $(k-m)$ are
large, that is we look at the limit $k\gg1$ with $0<m/k<1$.
In this regime, the four operators are connected by a parametrically
large number $\order{k}$ of propagators. This implies that graphs where the propagators
connecting any two operators are distributed on as many bridges as
possible outweigh all other graphs by combinatorial factors. In other words, graphs where any
bridge is only filled with a few (or zero) propagators are suppressed
by powers of $1/k$. Namely, the sum over distributions of $n$
propagators on $j$ bridges at large $n$ expands to
\begin{equation}
\sum_{\substack{n_0\leq n_1,\dots,n_j\leq n\\\sum_in_i=n}}1
=\frac{n^{j-1}}{(j-1)!}
+\frac{j(1-2n_0)n^{j-2}}{2(j-2)!}
+\order{n^{j-3}}\,,
\quad
j\in\Integers_{>0}
\,.
\label{eq:lengthsum}
\end{equation}
This combinatorial dominance greatly reduces the
number of contributing graphs:
For a correlator with generic polarizations $\alpha_i$, only the
maximal graphs~1.1, 2.1.1--2.1.3, 3.1, and 3.2 of
\tabref{tab:maxgraphs} contribute to the
leading order in $1/k$, since all other graphs have fewer bridges. For
these graphs, every face has room for exactly one hexagon, and thus
all mirror magnons live on bridges with a large number $\order{k}$ of
propagators, which means that all quantum corrections are delayed.
However, in this work, we consider operator polarizations with
$(\alpha_1\cdot\alpha_4)=(\alpha_2\cdot\alpha_3)=0$, which do not
admit propagator structures of the type
$Z\equiv(\alpha_1\cdot\alpha_4)(\alpha_2\cdot\alpha_3)/x_{14}^2x_{23}^2$,
see~\eqref{eq:corr}. In other words, there are no contractions between operators $1$ and $4$, and
no contractions between operators $2$ and $3$. Hence, even at large $k$, the dominant graphs
will leave room for zero-length bridges and thus admit quantum
corrections already at one-loop order.

Before diving into the computation,
let us quote the leading and first subleading terms in $1/k$ of
our data from~\tabref{tab:FkmUdata} for reference (subleading terms are shown in gray):
\begin{align}
  \mathcal{F}_{k,m}^{(1),\mathrm{U}}
  (z,\bar z)
  \big|_{\text{torus}}
  &=
\nn\\&\mspace{-110mu}
-\frac{2 k^2}{\Nc^{4}}\biggbrc{
        \Bigbrk{
            \bigsbrk{\sfrac{17r^4}{6}-\sfrac{7r^2}{4}+\sfrac{11}{32}}k^4
            {\color{gray}\mbox{}+\bigsbrk{\sfrac{9r^2}{2}-\sfrac{13}{8}}k^3}
        }t
        {\color{gray}\mbox{}-r\bigsbrk{\sfrac{34r^2}{3}-\sfrac{7}{2}}k^3s_-+\order{k^{2}}}
}{\color{red}F^{(1)}}
\,,
\label{eq:Fkm1Ulargek}
\\[1ex]
  \mathcal{F}_{k,m}^{(2),\mathrm{U}}
  (z,\bar z)
  \big|_{\text{torus}}
  &=
\nn\\&\mspace{-90mu}
\frac{4 k^2}{\Nc^{4}}\Biggsbrk{
\biggbrc{
        \Bigbrk{
            \bigsbrk{\sfrac{17r^4}{6}-\sfrac{7r^2}{4}+\sfrac{11}{32}}k^4
            {\color{gray}\mbox{}+\bigsbrk{\sfrac{9r^2}{2}-\sfrac{13}{8}}k^3}
        }t
        {\color{gray}\mbox{}-r\bigsbrk{\sfrac{34r^2}{3}-\sfrac{7}{2}}k^3s_-}
}{\color{red}F^{(2)}}
\nn\\&\mspace{-72mu}
+\biggbrc{
        \Bigbrk{
            \bigsbrk{\sfrac{29r^4}{24}-\sfrac{11r^2}{16}+\sfrac{15}{128}}k^4
            {\color{gray}\mbox{}+\bigsbrk{\sfrac{17r^2}{8}-\sfrac{21}{32}}k^3}
        }t^2
        {\color{gray}\mbox{}-r\bigsbrk{\sfrac{23r^2}{3}-\sfrac{9r}{4}}k^3ts_-}
}{\color{red}\bigbrk{F^{(1)}}^2}
\nn\\&\mspace{-72mu}
{\color{gray}\mbox{}-\bigsbrk{\sfrac{5r^2}{4}-\sfrac{19}{48}}k^3{\color{red}F^{(2)}_{\mathrm{C,+}}}+\order{k^{2}}}
}
\,.
\label{eq:Fkm2Ulargek}
\end{align}

\paragraph{Polygonization: Maximal Cyclic Graphs.}

Since there are no contractions between operators $1$ and $4$ and
between operators $2$ and $3$, we need to
consider graphs where the four operators are cyclically connected, as
in \mbox{$1$---$2$---$4$---$3$---$1$}
(later we will see that non-cyclic graphs are also important).
We can obtain all possible
graphs of this type by deleting bridges from the maximal graphs listed in~\tabref{tab:maxgraphs}.
Among all cyclically connected graphs, we
only consider graphs where as many bridges as possible are filled. We
will call those ``maximal cyclic graphs''. These will be the only
graphs that contribute at leading order in $1/k$. All further
graphs only contribute to subleading orders in $1/k$, and can be
obtained by setting further bridge lengths to zero.

Starting from any of the $16$ cases listed in~\tabref{tab:maxgraphs},
we can obtain cyclic graphs by grouping the four operators into two
pairs and deleting all bridges that connect the members of either
pair. Doing this in all possible ways for all the $16$ graphs,
and discarding non-maximal%
\footnote{Here, by non-maximal graphs we mean graphs with fewer than 8
bridges. All such graphs can be obtained from the maximal cyclic
graphs listed in~\tabref{tab:maxcycgraphs} by deleting further edges.}
as well as duplicate graphs, we end up with
the complete set of maximal cyclic graphs A through Q displayed
in~\tabref{tab:maxcycgraphs}.
\begin{table}[t]
\centering
\begin{tabular}{cccc}
\includegraphicsbox{FigTorusCaseA} &
\includegraphicsbox{FigTorusCaseB} &
\includegraphicsbox{FigTorusCaseC} &
\includegraphicsbox{FigTorusCaseD}
\\
\rule{0pt}{2.5ex}%
A &
B &
C &
D
\\[1ex]
\includegraphicsbox{FigTorusCaseE} &
\includegraphicsbox{FigTorusCaseF} &
\includegraphicsbox{FigTorusCaseG} &
\includegraphicsbox{FigTorusCaseH}
\\
\rule{0pt}{2.5ex}%
E &
F &
G &
H
\\[1ex]
\includegraphicsbox{FigTorusCaseI} &
\includegraphicsbox{FigTorusCaseJ} &
\includegraphicsbox{FigTorusCaseK} &
\includegraphicsbox{FigTorusCaseL}
\\
\rule{0pt}{2.5ex}%
I &
J &
K &
L
\\[1ex]
\includegraphicsbox{FigTorusCaseM} &
\includegraphicsbox{FigTorusCaseN} &
\includegraphicsbox{FigTorusCaseP} &
\includegraphicsbox{FigTorusCaseQ}
\\
\rule{0pt}{2.5ex}%
M &
N &
P &
Q
\end{tabular}
\caption{Inequivalent maximal cyclic graphs on the torus.}
\label{tab:maxcycgraphs}
\end{table}
For example, consider the maximal graph~1.1. We can delete either all
vertical or all horizontal lines; these two cases are equivalent and
give
\begin{equation}
\includegraphicsbox{FigTorusCase11}
\;\rightarrow\;
\includegraphicsbox{FigTorusCase11A}
\;,
\end{equation}
which is easily recognized as case~A. Alternatively, we could delete
all diagonal lines, which gives case~P. In fact, for all cases~1.1--1.5.3 we
do not need to consider deleting the diagonal lines, as the resulting
configurations will (by construction) always be covered by the
cases~2.1.1--3.2. Moving on to cases~1.2.1 and~1.2.2, up to
operator relabelings, all ways of deleting bridges (keeping the
diagonal ones) lead to equivalent configurations:
\begin{equation}
\includegraphicsbox{FigTorusCase121}
\;,\;
\includegraphicsbox{FigTorusCase122}
\quad\rightarrow\quad
\includegraphicsbox{FigTorusCase121B}
\;,
\end{equation}
which we recognize as case~B. The derivation of the further cases C
through Q from the maximal graphs in~\tabref{tab:maxgraphs} is listed
in~\appref{app:cyclic-graphs}.

The large-weight limit brings about another simplification:
In~\secref{sec:k2345} above, we saw that magnons carrying non-trivial
R-charges may cancel $Z$ propagator structures~\eqref{eq:XYZ} such
that the final result is free of $Z$'s. Such cancellations cannot
occur here, since all graphs
of~\tabref{tab:maxcycgraphs} dissect the torus into four octagons
separated by large bridges. Such octagons do not leave enough room for
$Z$ propagator cancellations.%
\footnote{Such cancellations would require a $1/Z$ type excitation on
a non-zero-length bridge of type $Z$, but excitations on
non-zero-length bridges are delayed.}
Hence we do not have to include graphs containing $Z$ propagators.

Looking at the cases A through Q, we find that the bridge
configurations of the cases A, C, D, E, F, H, I, J, N, and K imply a constraint on $m$:
Either $m=0$, or $m=k$.
Hence, even though no further bridges can be added to these graphs (under the
cyclicity constraint), these cases are suppressed at large $m$ and $(k-m)$, and only
the cases~B, G, L, M, P, and~Q remain (these were called B, A, C, D,
E, and~F in our previous publication~\cite{Bargheer:2017nne}).
For these graphs, we now have to consider all possible operator
labelings, taking care that some seemingly different labelings in fact
produce identical bridge configurations. In addition, each labeled
graph comes with a combinatorial factor from the distribution of
propagators on the various bridges according to~\eqref{eq:lengthsum}.
We list all inequivalent labelings for the relevant graphs as well as
their combinatorial factors in~\tabref{tab:labeltab}.
\begin{table}
\centering
\begin{tabular}{ccc}
\toprule
Case & Inequivalent Labelings & Combinatorial Factor\\
\midrule
B & $(1,2,4,3), (2,1,3,4), (3,4,2,1), (4,3,1,2)$ & $m^3(k-m)/6$ \\
B & $(1,3,4,2), (3,1,2,4), (2,4,3,1), (4,2,1,3)$ & $m(k-m)^3/6$ \\
G & $(1,2,4,3), (3,4,2,1)$ & $m^4/24$ \\
G & $(1,3,4,2), (2,4,3,1)$ & $(k-m)^4/24$ \\
L & $(1,2,4,3), (3,4,2,1), (2,1,3,4), (4,3,1,2)$ & $m^2/2\cdot(k-m)^2/2$ \\
M & $(1,2,4,3), (2,1,3,4), (1,3,4,2), (3,1,2,4)$ & $m^2(k-m)^2/2$ \\
P & $(1,2,4,3)$ & $m^2(k-m)^2/2$ \\
Q & $(1,2,4,3)$ & $m^2(k-m)^2$ \\
\bottomrule
\end{tabular}
\caption{All inequivalent operator labelings for the graphs that
contribute to leading order in $1/k$, together with their
combinatorial factors according to~\eqref{eq:lengthsum}. The order of
the labels runs clockwise, starting at the top left operator in the graphs
of~\tabref{tab:maxcycgraphs}.}
\label{tab:labeltab}
\end{table}
For case P, all operator labelings are equivalent. Beyond that, it has
an extra symmetry: Every pair of operators is connected by a pair of
bridges. Exchanging the members of all pairs simultaneously amounts to
a cyclic rotation of the four operators and thus leaves the
configuration invariant. This operation is an example of a graph
automorphism, see the last part of \secref{sec:polygonization}, in
particular~\eqref{eq:automorphismexample1}. The naive sum over bridge
lengths gives a combinatorial factor
$m^2(k-m)^2$, which thus has to be corrected by a factor of $1/2$.

\paragraph{Sprinkling: One and Two Loop Check.}

The previous maximal cyclic graphs polygonalize the torus into four
octagons each, generating some toroidal polyhedra. We represent their
corresponding nets in~\tabref{tab:hexagons} for easier visualization.
The one-loop and two-loop computations can then be performed
straightforwardly from a single particle sitting in the single ZLB of
each octagon. Such contributions can be easily computed to any desired
loop order using the ingredients of \appref{app:mirrorparticles}
(at one loop we can simply use the $\polygon$ function of
\secref{sec:1-loop-polygons}).
At one loop this is the only particle configuration contributing. At
two loops, we have to consider in addition two virtual particles in
different octagons, which essentially amounts to the one-loop octagon
squared. The contribution of two virtual particles inserted in the
same octagon turns out to be delayed to four loops as shown in
\appref{app:mirrorparticles}. The final step is then to sum over
the labelings of the vertices, weighted by the combinatorial factors
arising from the different ways of distributing the propagators among
the bridges. \tabref{tab:labeltab} contains the details of these
combinatorics. We have performed this calculation
in~\cite{Bargheer:2017nne} and found a perfect agreement with the
large $k$ data~\eqref{eq:Fkm1Ulargek} and~\eqref{eq:Fkm2Ulargek}.

\begin{table}[tbp]
\centering
\begin{tabular}{|@{\hspace{0.1cm}}c@{\hspace{0.1cm}}|c||@{\hspace{0.1cm}}c@{\hspace{0.1cm}}|c|}
\hline
\footnotesize{Case} & \footnotesize{Hexagonalization} & \footnotesize{Case} & \footnotesize{Hexagonalization}
\\ \hline &&& \\[-1ex]
\footnotesize{B} & \includegraphicsbox[scale=0.37]{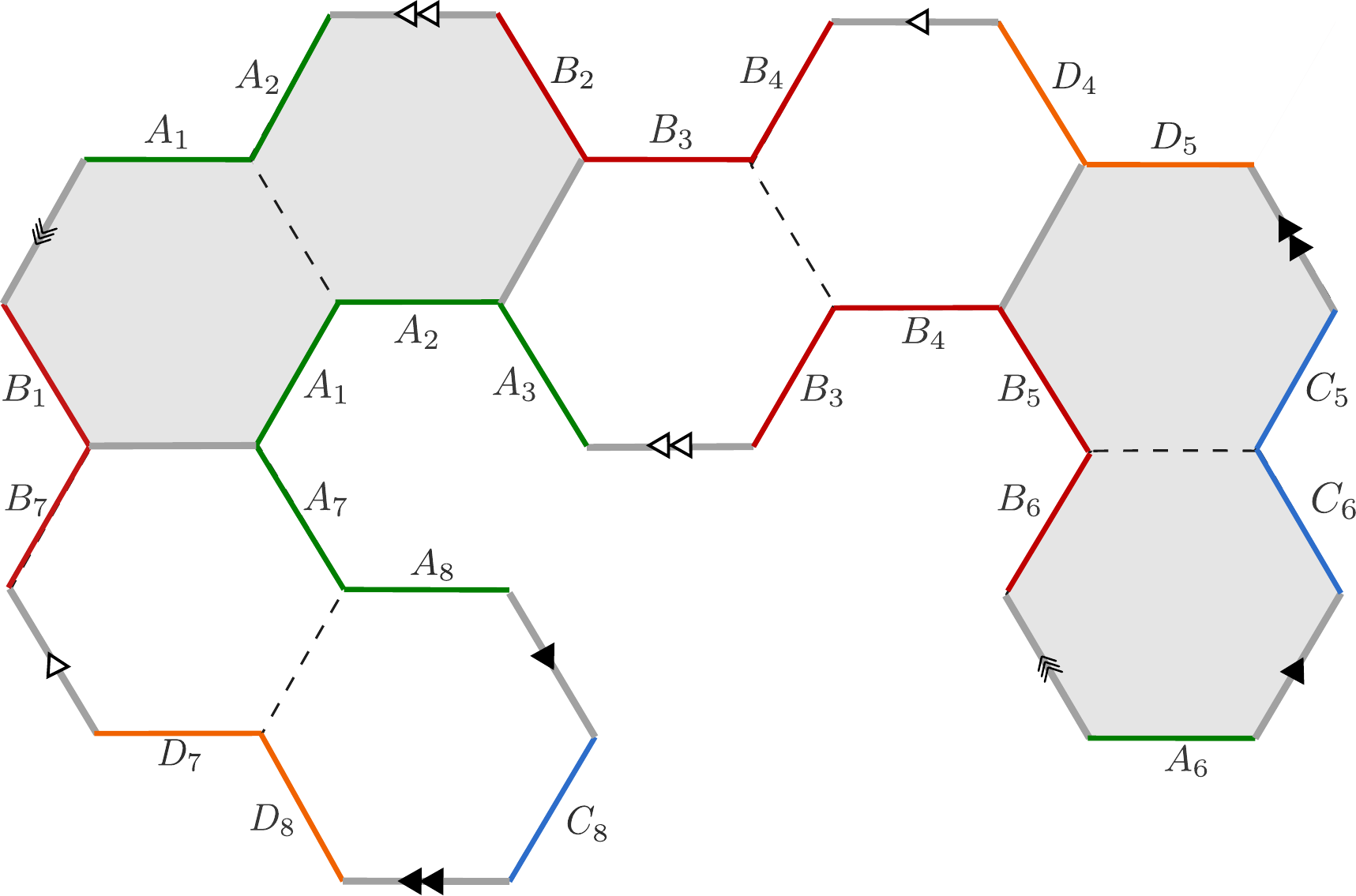}
&
\footnotesize{G} & \includegraphicsbox[scale=0.37]{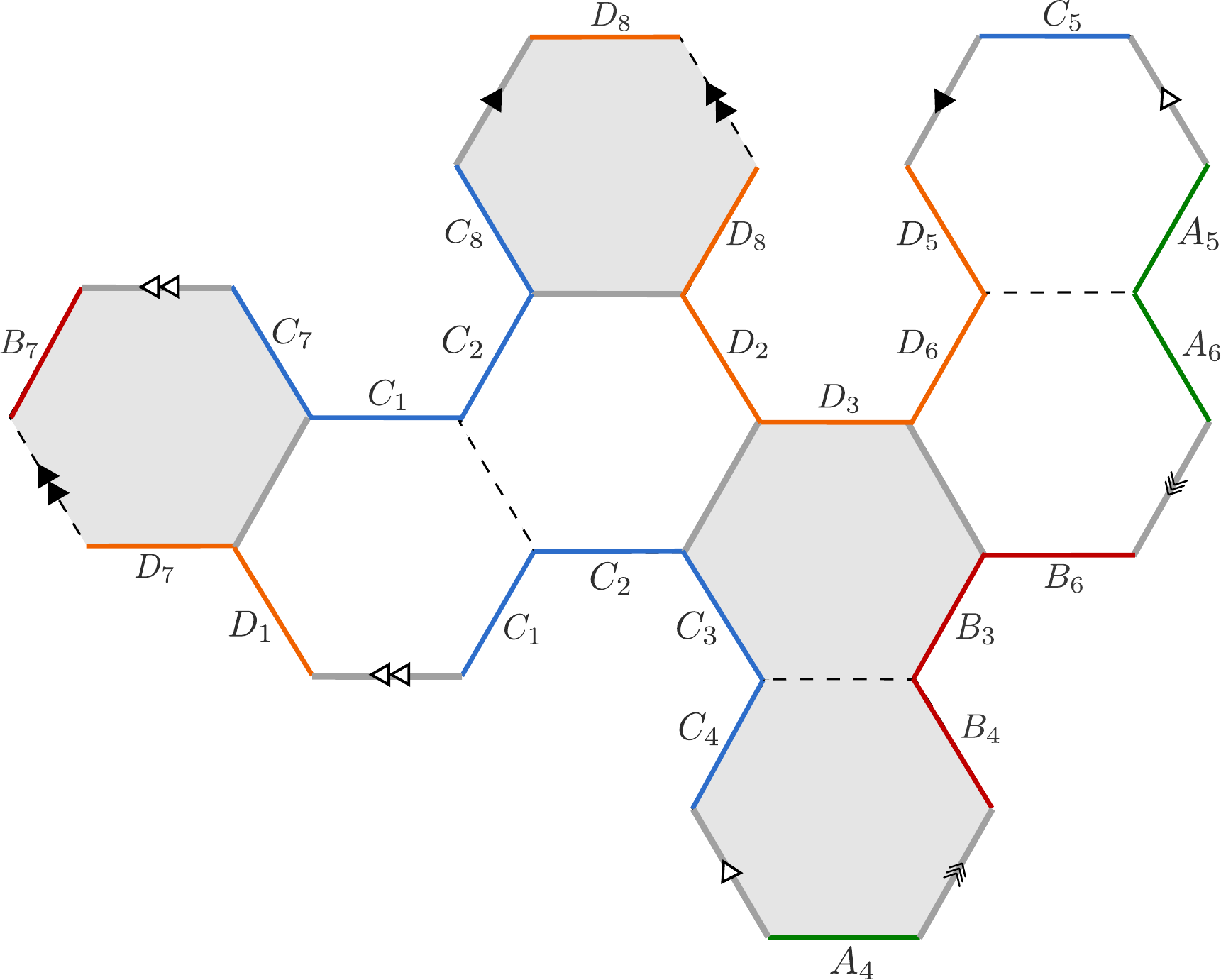}
\\ &&& \\[-1ex] \hline &&& \\[-1ex]
\footnotesize{L} & \includegraphicsbox[scale=0.37]{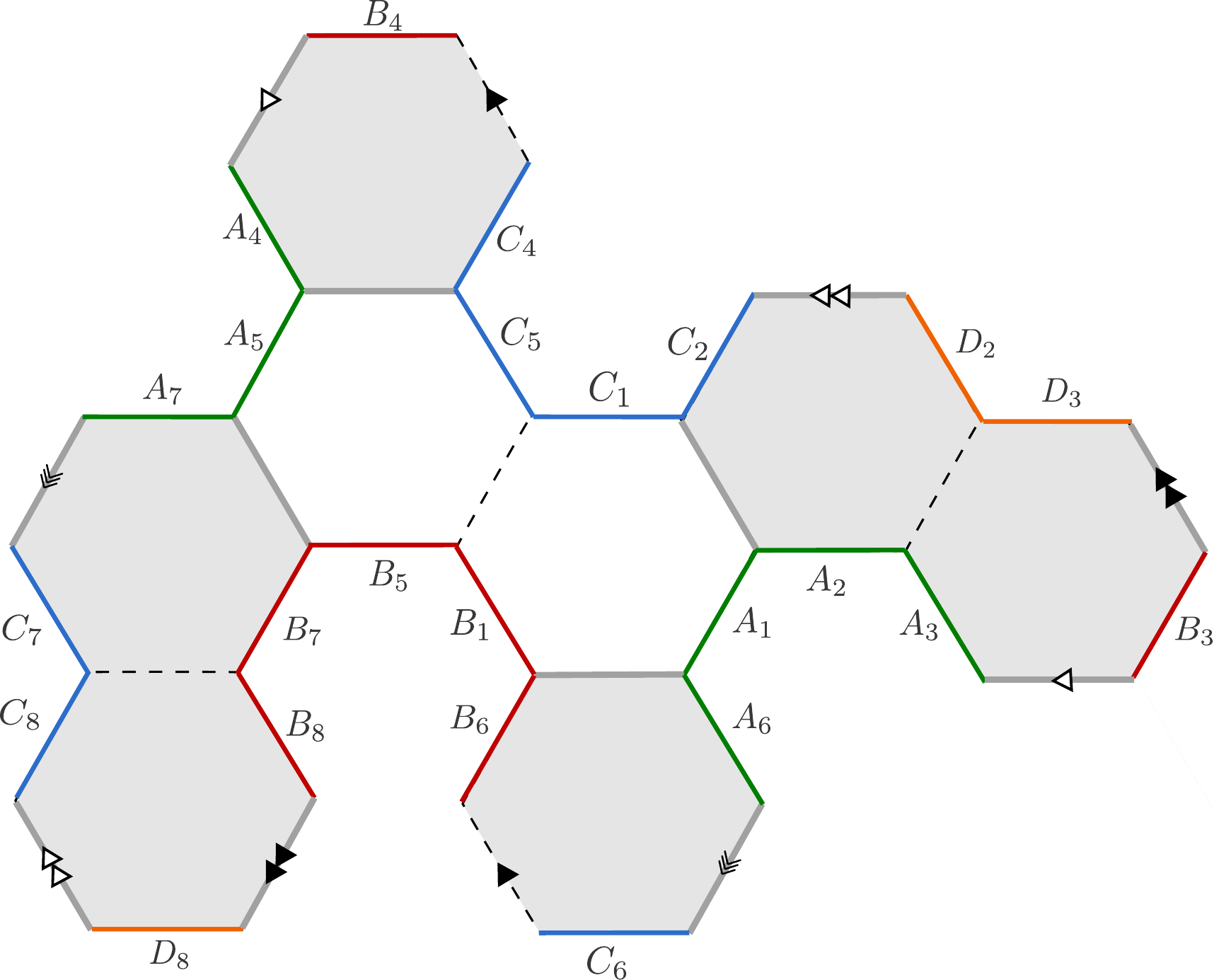}
&
\footnotesize{M} & \includegraphicsbox[scale=0.37]{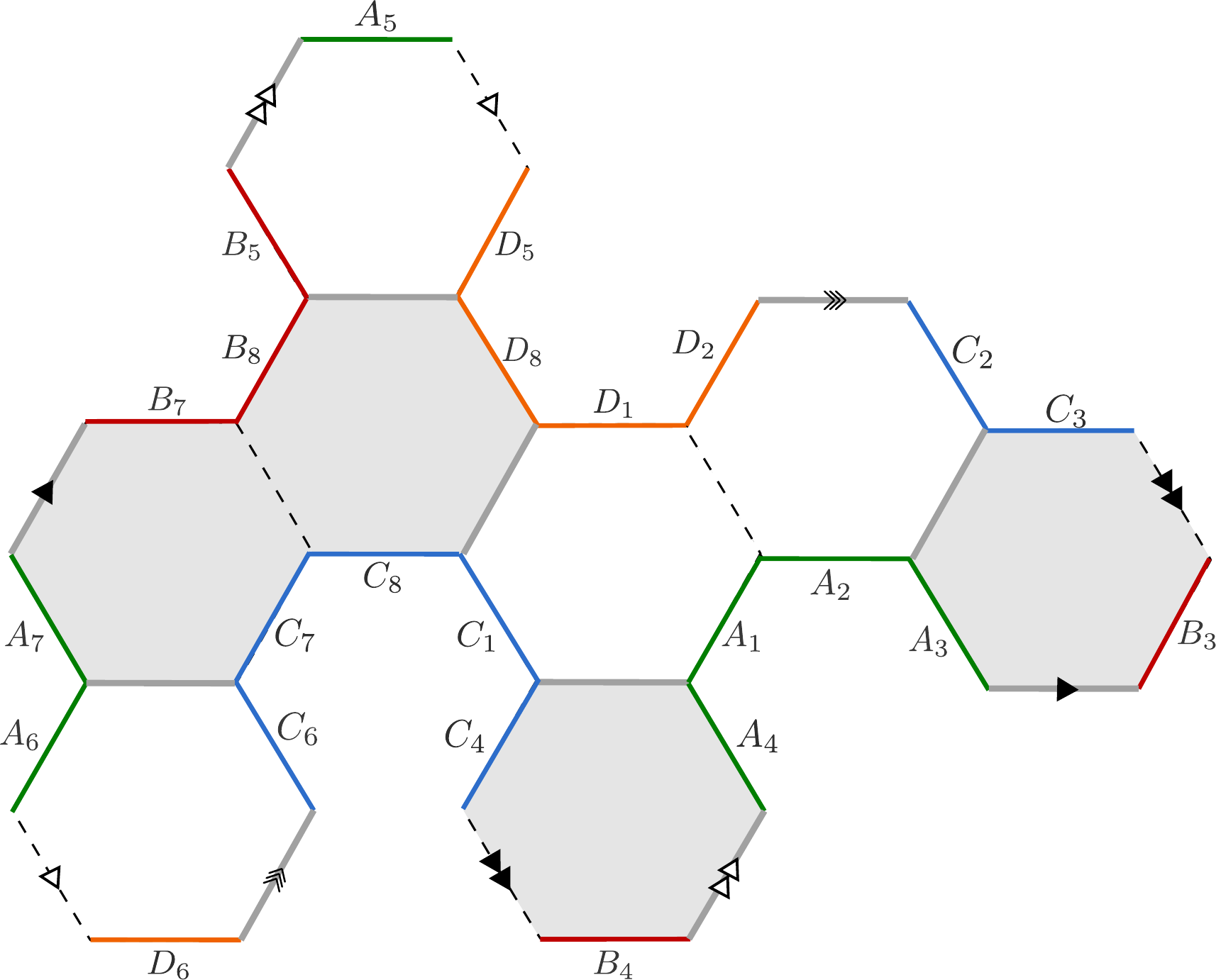}
\\ &&& \\[-1ex] \hline &&& \\[-1ex]
\footnotesize{P} & \includegraphicsbox[scale=0.37]{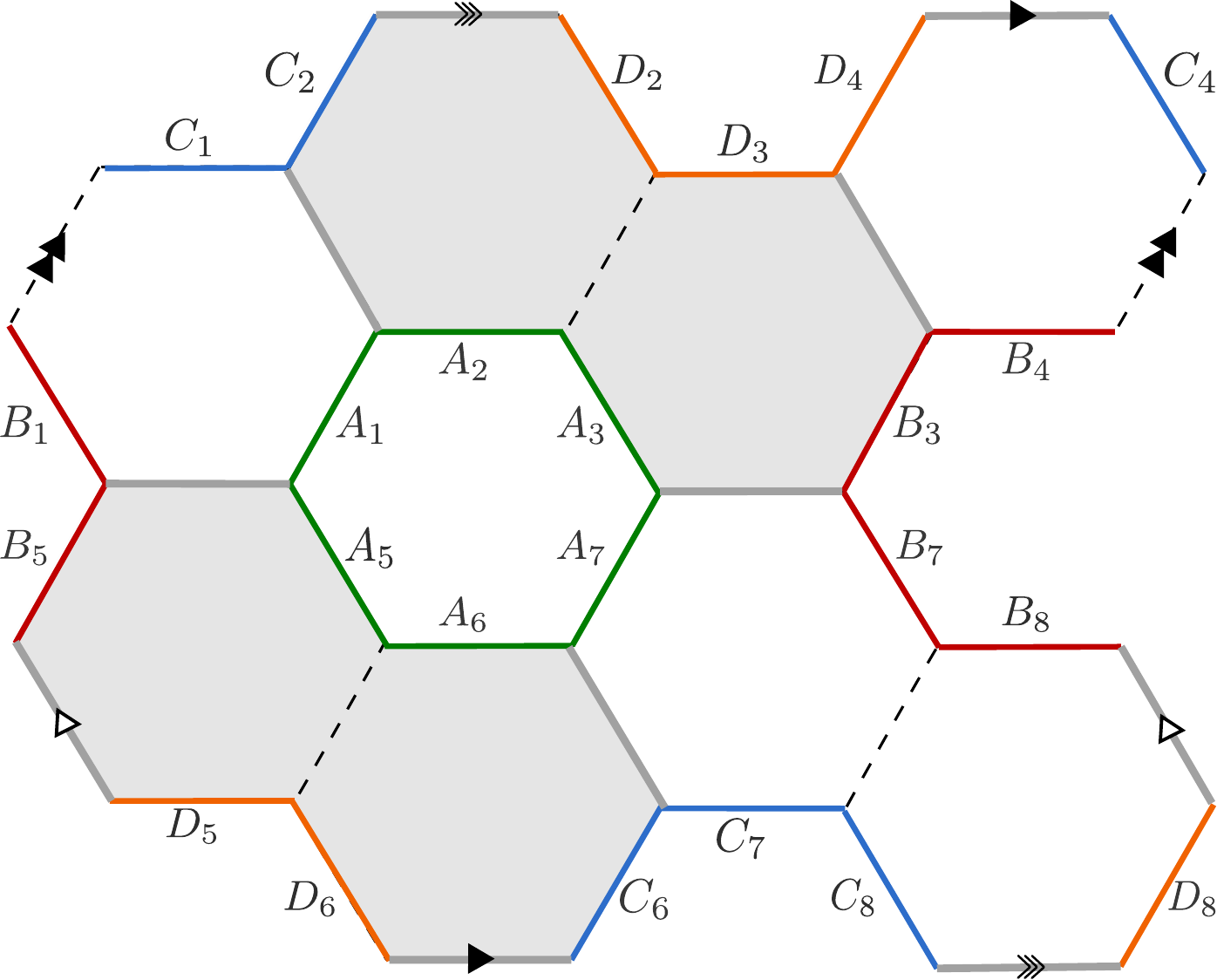}
&
\footnotesize{Q} & \includegraphicsbox[scale=0.37]{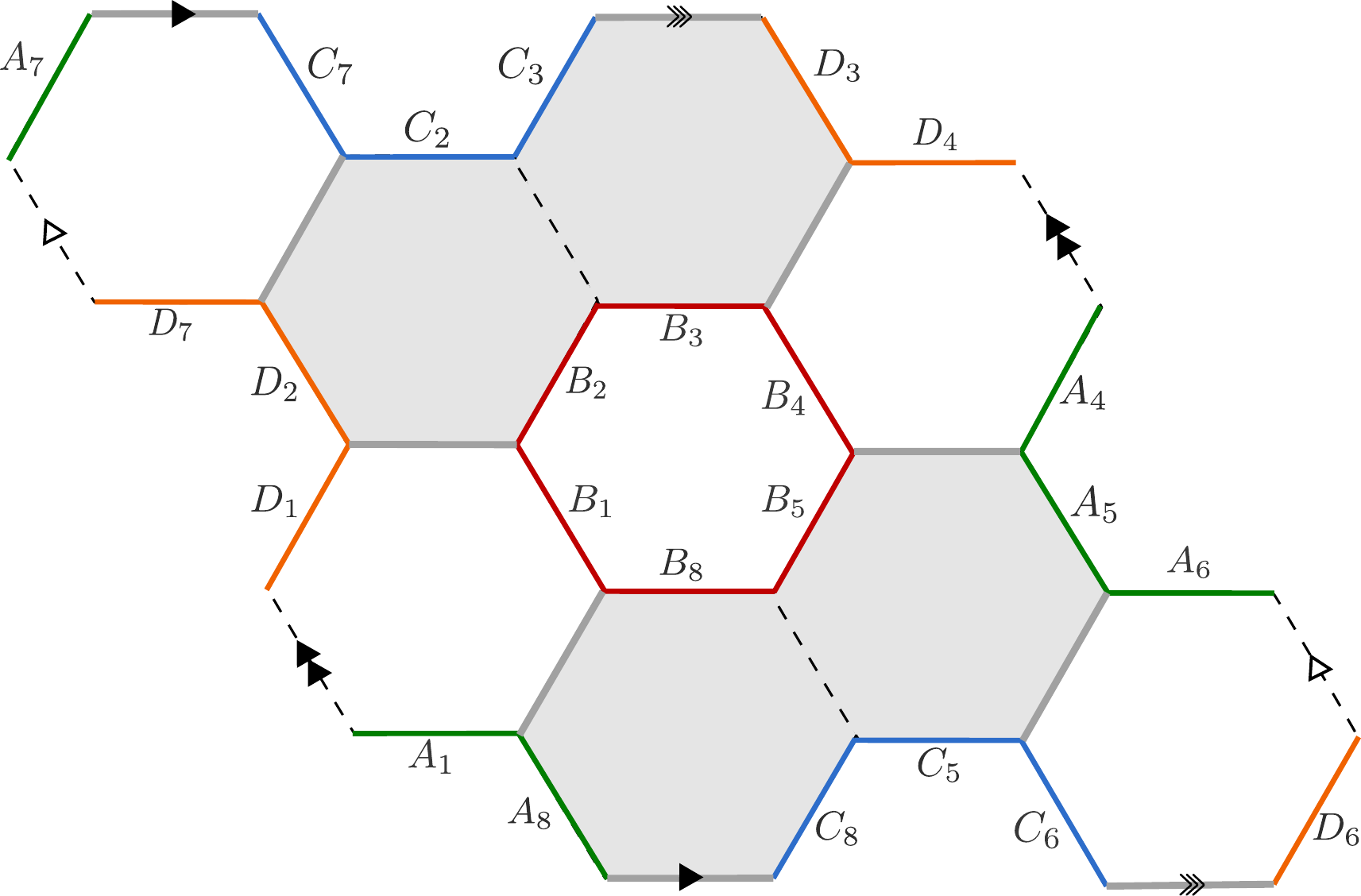}
\\[-1ex] &&& \\ \hline
\end{tabular}
\caption{After completing the relevant skeleton graphs B, G, L, M, P,
and Q with the missing ZLBs,
we obtain complete hexagonalizations of the four-punctured torus.
The outcome is that each configuration is decomposable into~$8$
hexagons, or~$4$ minimal octagons, using the terminology of
\secref{sec:1-loop-polygons}. The distinct octagons are colored in white
and gray.
The colored edges correspond to the physical ones, with the operators
being labeled as A, B, C, and D. The subscript in each edge label indicates to
which of the eight hexagons the respective edge belongs. Later on, we will specify the
labels
A, B, C, and D of the operators. For each hexagonalization, the dashed
lines correspond to the ZLBs, while the solid gray bridges have non-zero
lengths.}
\label{tab:hexagons}
\end{table}

\paragraph{Sprinkling: Three Loop Prediction.}

As far as we know,
there is no available non-planar perturbative data at three loops.
The planar case, however, was computed in~\cite{Drummond:2013nda}.
Here, we are going to make a prediction for the three-loop
result at leading order in large $k$ using integrability.
In principle, one can keep going and make predictions
for arbitrary order in~$g^2$, and it would be very interesting
to try to re-sum the series. At three loops, one has
the following possible contributions:
\begin{enumerate}[topsep=1.5ex]
\item Three-loop correction of the one-particle octagon.
\item Two mirror particles inserted at different octagons.
\item Three mirror particles inserted at three different octagons.
\item Multiple mirror particles inserted in the same octagon.
\end{enumerate}
One can show that
contribution 3 is only present for case P,
because all other cases only have
two octagons involving four operators, and
an octagon involving only three (or two) operators
vanishes as
the relevant cross-ratio for gluing is either $0$, $1$, or $\infty$.
The contribution 4 kicks in only at four loops -- the
case of two mirror particle in the same edge
is computed in the \appref{app:mirrorparticles} -- and thus is not
relevant here.

The new
ingredient for the three-loop computation,
when compared to the two-loop calculation
performed in~\cite{Bargheer:2017nne}, is
the one-particle mirror contribution
in a ZLB expanded to three loops,
given by (see~\appref{app:mirrorparticles} for details):
\begin{multline}
\mathcal{M}^{(1)}(z)=
\lrbrk{z+\bar{z}-\frac{\alpha+\bar\alpha}{2}\lrbrk{1+\frac{z\bar z}{\alpha\bar\alpha}}}
\times\\\times
\bigbrk{
g^2F^{(1)}(z,\bar{z})
-2g^4F^{(2)}(z,\bar{z})
+6g^6F^{(3)}(z,\bar{z})
+\order{g^8}
} \, ,
\label{eq:OneParticleThreeLoop}
\end{multline}
where $F^{(3)}$ is the three-loop ladder integral defined as
\begin{equation}
F^{(3)}(z, \bar{z}) = \frac{x_{13}^2 x^2_{24} x^4_{14}}{\pi^6}
\int \frac{d^4 x_5 d^4 x_6 d^4 x_7}{x_{15}^2 x_{45}^2 x^2_{35}
x^2_{56} x^2_{16} x^2_{46} x^2_{67} x^2_{17} x^2_{47} x^2_{27} }
\, = \, \includegraphicsbox{FigTripleBoxInt}
\,.
\label{eq:Ladder3}
\end{equation}
The one-loop and two-loop ladder integrals $F^{(1)}$ and $F^{(2)}$ are
defined in~\eqref{eq:box} and~\eqref{eq:doublebox} (see also the
expression~\eqref{eq:FintermsofPolylogs} in terms of polylogarithms), and the cross
ratios $z$, $\bar z$ are defined in~\eqref{eq:zidef}.

Using the hexagonalized graphs of
\tabref{tab:hexagons}, the combinatorial factors of
\tabref{tab:labeltab}, and adding all mirror
particle corrections, one arrives at the three-loop prediction
\begin{align}
\mathcal{F}_{k,m}^{(3),\mathrm{U}}(z,\bar z)\big|_{\text{torus}}=
&-\frac{12 k^2}{\Nc^{4}}\Biggsbrk{
\biggbrc{
    \Bigbrk{\bigsbrk{\sfrac{17r^4}{6}-\sfrac{7r^2}{4}+\sfrac{11}{32}}k^4}t
    {\color{gray}\mbox{}+\mathcal{O}(k^3)}
}{\color{red}F^{(3)}}
\nn\\
&+\biggbrc{
    \Bigbrk{\bigsbrk{\sfrac{29r^4}{18}-\sfrac{11r^2}{12}+\sfrac{5}{32}}k^4}t^2
    {\color{gray}\mbox{}+\mathcal{O}(k^3)}
}{\color{red}F^{(2)}F^{(1)}}
\nn\\
&+\biggbrc{
    \Bigbrk{\bigsbrk{\sfrac{(1-4 r^2)^2}{96}}k^4}t^3
    {\color{gray}\mbox{}+\mathcal{O}(k^3)}
}{\color{red}\bigbrk{F^{(1)}}^3}
}
\,.
\label{eq:Fkm3Ulargek}
\end{align}
%

\subsubsection{\texorpdfstring{$k\gg 1$}{k>>1}: Subleading Order}
\label{sec:subleadingSection}

Next, we are going to compute
the subleading contribution in the large-$k$ expansion, \ie the terms of order $\mathcal{O}(k^3)$,
at one-loop order using integrability.
We find an agreement with the perturbative data (gray)
given in~\eqref{eq:Fkm1Ulargek}.
As described in detail below, the subleading computation receives contributions from three
different sources: The graphs used in the leading-order computation, the graphs
obtained from the leading-order graphs by deleting one
bridge, and the ``deformed'' graphs which are graphs
having one pair of propagators of type $Z$.

\paragraph{Leading Cyclic Graphs.}

The graphs B, G, L, M, P, and Q used
in the leading-order
computation also contribute
at subleading order in large $k$.
The integrability contribution is computed
exactly as in the leading-order case, in particular
one uses the same set of hexagons of
\tabref{tab:hexagons},
however one considers the subleading
contribution to the combinatorial factors
given in~\eqref{eq:lengthsum}, with $n_0=1$.
Recall that to obtain a final term
with the propagator structure $X^m Y^{k-m}$
at one-loop order, it is necessary
to consider also the neighboring tree-level
graphs with propagators $X^{m-1} Y^{k-m+1}$ and
$X^{m+1} Y^{k-m-1}$. This follows because
the mirror particles carry $R$-charge and they can change
the propagator structure of a tree-level
graph~\cite{Bargheer:2017nne}, which is seen explicitly by the ratios
of $X$, $Y$, and $Z$ propagator factors in the prefactor
of~\eqref{eq:OneParticleThreeLoop} rewritten via~\eqref{eq:alphaXYZ}:
\begin{multline}
\lrbrk{z+\bar{z}-\frac{\alpha+\bar\alpha}{2}\lrbrk{1+\frac{z\bar z}{\alpha\bar\alpha}}}
\\
=\frac{1}{2}\biggsbrk{
1-\frac{Y}{X}+z \bar z \lrbrk{1-\frac{X}{Y}}+(1-z)(1-\bar z)\lrbrk{\frac{Z}{X}+\frac{Z}{Y}-2}
}
\,.
\label{eq:octagonprefactor}
\end{multline}
One important remark is that,
differently from the leading case,
where the combinatorial factor of each graph is universal,
in the subleading case the combinatorial factor
changes when considering the neighboring graphs.
As an example, \tabref{tab:CombinatorialSubleadingLeading}
shows the combinatorial factors relevant for case~B.

\begin{table}
\centering
\begin{tabular}{ccc}
\toprule
\begin{tabular}{@{}c@{}}
Tree-Level \\
\vspace{-0.5mm}
Propagators
\end{tabular}
 & Labelings:
\begin{tabular}{@{}l@{\,}l@{}}
(1,2,4,3), & (2,1,3,4), \\
\vspace{-0.5mm}%
(3,4,2,1), & (4,3,1,2).
\end{tabular}
& Labelings:
\begin{tabular}{@{}l@{\,}l@{}}
(1,3,4,2), & (3,1,2,4), \\
\vspace{-0.5mm}%
(2,4,3,1), & (4,2,1,3).
\end{tabular} \\
\midrule
$ X^{m} Y^{k-m}$      & $-(k-m) m^2 - \frac{m^3}{6}$               & $- \frac{1}{6} (k-m)^3 - (k-m)^2 m $ \\
$ X^{m-1} Y^{k-m+1} $ & $ - \frac{3}{2} (k-m) m^2 $                & $- \frac{3}{2} (k-m)^2 m $ \\
$ X^{m+1} Y^{k-m-1}$  & $ - \frac{1}{2} (k-m) m^2 - \frac{m^3}{3}$ & $- \frac{1}{3} (k-m)^3 - \frac{1}{2} (k-m)^2 m $ \\
\bottomrule
\end{tabular}
\caption{The combinatorial factors used for the subleading
computation of the graph~B. The
case~B has eight inequivalent
labelings, see \tabref{tab:labeltab}.}
\label{tab:CombinatorialSubleadingLeading}
\end{table}

\paragraph{Subleading Cyclic Graphs.}

In addition to the leading-order graphs,
there will be contributions from cyclic graphs
that are obtained
from the cases B, G, L, M, P, and Q by removing one of their bridges. Deleting a
bridge in all possible ways, and identifying identical graphs, we find
seven inequivalent subleading cyclic graphs,
see~\tabref{tab:sublcycgraphs}. The number of inequivalent
labelings is indicated in the parentheses below each graph.
The hexagonalization of the subleading cyclic
graphs can be obtained from the hexagonalization
of the leading cyclic graphs given previously
by replacing the corresponding line that was deleted
in the process by a zero-length
bridge. The final step is to add the mirror particles.
In this case, we have one-, two- and three-particle
contributions, because there are four hexagons sharing
bridges of zero length in a sequence. Thus at one-loop order
for the integrability computation,
one uses the expressions for both
the octagon and the dodecagon of~\eqref{eq:2n-gon}.
In addition, the relevant combinatorial factors can be read
from the leading term of formula~\eqref{eq:lengthsum}.

\begin{table}
\centering
\begin{tabular}{cccc}
\includegraphicsbox{FigTorusCase121B1}
& \includegraphicsbox{FigTorusCase151G1}
& \includegraphicsbox{FigTorusCase211L1}
& \includegraphicsbox{FigTorusCase211L2} \\
B.1 (8) & G.1 (4) & L.1 (8) & L.2 (8) \\
\includegraphicsbox{FigTorusCase211L3}
& \includegraphicsbox{FigTorusCaseP1}
& \includegraphicsbox{FigTorusCaseQ1} & \\
L.3 (8) & P.1 (4) & Q.1 (4) &
\end{tabular}
\caption{The seven inequivalent graphs that are obtained by deleting
one bridge from graphs B, G, L, M, P, or Q
of~\tabref{tab:maxcycgraphs}. These graphs contribute
at subleading order in $k$. The parentheses show the number of
inequivalent labelings that each graph has.}
\label{tab:sublcycgraphs}
\end{table}
%

\paragraph{Deformed Graphs.}

At subleading order in large $k$, there is room
for so-called \emph{deformed graphs}.
The mirror particles carry $R$-charge,
in other words they depend on the $R$-charge
cross ratios $\alpha$ and $\bar{\alpha}$, as seen for example in~\eqref{eq:octagonprefactor}.
Hence the final $R$-charge structure of
a graph depends not only on the tree-level propagators, but also
on the mirror particles.
For example, graphs that include a propagator of the type
$Z\equiv(\alpha_1\cdot\alpha_4)(\alpha_2\cdot\alpha_3)/x_{14}^2x_{23}^2$
can give a final term free of $Z$'s
after the
inclusion of the mirror corrections, which is thus compatible with
our chosen polarizations~\eqref{eq:corr}, and gives a
non-zero contribution in the limit $Z \rightarrow 0$. We have already encountered
the same phenomenon when we performed checks for finite $k$.
In the sum over graphs, we
hence must include graphs with $Z$ propagators.

Graphs containing one or more propagators of type $Z$ (and
otherwise only large bridges filled with many propagators) will be called \emph{deformed
graphs}. At one loop, the relevant deformed graphs include
only one pair of $Z$
propagators connecting two disjoint pairs of operators. We can
classify all such graphs by starting with the set of maximal
graphs listed in~\tabref{tab:maxgraphs}, declaring two of the
bridges to become $Z$ propagators, and deleting other bridges such
that the graph becomes subleading in $k$. In the limit of large $k$
that we consider, extremal graphs with $m=0$ or $(k-m)=0$
will not contribute. Taking into account that
one of the bridges attaching to each operator
in~\tabref{tab:maxgraphs} will become a $Z$ propagator, this means
that we only need to consider the graphs 1.2.1, 1.2.2, 1.5.3, 2.1.1,
2.1.2, 2.1.3, 3.1, and 3.2. Starting with these, and deleting
bridges\,/\,replacing bridges by $Z$ propagators, we arrive at the
set of inequivalent deformed graphs shown
in~\tabref{tab:deformedgraphs}.
\begin{table}[t]
\centering
\begin{tabular}{ccccc}
\includegraphicsbox{FigTorusCase121defA} &
\includegraphicsbox{FigTorusCase121defB} &
\includegraphicsbox{FigTorusCase121defC} &
\includegraphicsbox{FigTorusCase122defA} &
\includegraphicsbox{FigTorusCase122defB} \\
1.2.1A (8) & 1.2.1B (8) & 1.2.1C (8) & 1.2.2A (8) & 1.2.2B (8) \\
\includegraphicsbox{FigTorusCase153defA} &
\includegraphicsbox{FigTorusCase153defB} &
\includegraphicsbox{FigTorusCase211defA} &
\includegraphicsbox{FigTorusCase211defB} &
\includegraphicsbox{FigTorusCase211defC} \\
1.5.3A (4) & 1.5.3B (4) & 2.1.1A ({\color{blue}$6\cdot8$}) & 2.1.1B ({\color{blue}$6\cdot8$}) & 2.1.1C ({\color{blue}$6\cdot8$}) \\
\includegraphicsbox{FigTorusCase212defA} &
\includegraphicsbox{FigTorusCase212defC} &
\includegraphicsbox{FigTorusCase213defA} &
\includegraphicsbox{FigTorusCase213defB} & \\
2.1.2A (8) &
2.1.2C (8) & 2.1.3A (8) & 2.1.3B (8) & \\
\includegraphicsbox{FigTorusCase31defA} &
\includegraphicsbox{FigTorusCase31defB} &
\includegraphicsbox{FigTorusCase31defC} &
\includegraphicsbox{FigTorusCase31defD} &
\\
3.1A (${\color[rgb]{0,0.6,0}2{\cdot}4},{\color{blue}8}$) &
3.1B (${\color{blue}2{\cdot}8}$) &
3.1C (${\color[rgb]{0,0.6,0}2{\cdot}4},{\color{blue}8}$) &
3.1D (${\color[rgb]{0,0.6,0}3{\cdot}4},{\color{blue}2{\cdot}8}$) &
\end{tabular}
\caption{All inequivalent deformed graphs that contribute to the first
subleading order in $1/k$. Wiggly lines stand for $Z$ propagators.
Graphs with crossed bridges stand for classes of graphs where any one
of the crossed bridges is deleted. The parentheses show the number of
inequivalent labelings. Recall that for the hexagonal graphs, opposite
edges of the outer hexagon are identified.}
\label{tab:deformedgraphs}
\end{table}
Alternatively, the graphs in~\tabref{tab:deformedgraphs} can be
obtained by starting with the graphs B, G, L, M, P, and Q
of~\tabref{tab:maxcycgraphs}, and inserting $Z$ propagators as well as
deleting one bridge in all possible ways.

After having determined all deformed graphs, the next
step is the hexagonalization. This is done by adding
bridges of zero length to the graphs, and dividing them into
eight hexagons. Due to the flip invariance
of the mirror particle corrections, any different set
of zero-length bridges will give the same final
result. In the case of the deformed graphs,
the multi-particle contribution will show up, and
we use the expression for the octagon, decagon, and dodecagon of~\eqref{eq:2n-gon}.
In order to perform the integrability computation
for the deformed graphs, one
uses that $\alpha$ and $\bar{\alpha}$
are determined by the equations~\eqref{eq:alphaXYZ}.
The limit $Z \rightarrow 0$ is only taken after adding
the mirror-particle corrections to a graph.
Similar to the case of the leading cyclic graphs, to get a final term proportional to
$ X^{m} Y^{k-m} $ at one-loop, one has to consider the
set of graphs corresponding to the
tree-level terms $X^{m-1} Y^{k-m} Z $ and $X^{m} Y^{k-m-1} Z$.
Most of the graphs of~\tabref{tab:deformedgraphs} give a vanishing contribution.
One example of a non-vanishing
graph is
\begin{align}
\text{Case} \; 1.2.1\mathrm{C}
& = \sum_{s=0}^{1}
\frac{4}{6}\lrbrk{
m^3\mathcal{M}^{(1)}(1-z)
+(k-m)^3\mathcal{M}^{(1)}\lrbrk{\frac{z}{z-1}}
} X^{m-s}\,Y^{k-m-1+s}\,Z
\nn\\
& \xrightarrow{\;Z\to0\;} \,
\frac{1}{3}\lrbrk{k^3 - 3 k^2 m + 3 k m^2 - 2 m^3 }
( z \bar{z} -1 ) F^{(1)} (z, \bar{z})\, X^m Y^{k-m} \, .
\end{align}

\paragraph{Summary.}

The subleading integrability result is obtained
by summing three different kinds of contributions
which were described
above. The final result agrees with the
perturbative data. It is possible to use the same steps
to compute the predictions for the remaining orders in~$k$.

\section{Conclusions}

We performed detailed tests of our proposal on the application of the
hexagon formalism to non-planar correlators at weak coupling. The basic strategy is the same as in the planar case; we
first draw all possible tree-level diagrams on a given Riemann
surface, dissect them into hexagonal patches, and glue those patches back
together by summing over complete sets of intermediate (mirror) states. The key
new idea that is essential in the non-planar case is the
procedure called stratification: We first computed all
contributions coming from tree-level graphs drawn on a torus, including
the graphs that are actually planar. After doing so, we subtracted the
contributions from degenerate Riemann surfaces, which in turn can
be computed by taking the planar results and shifting the rank of the
gauge group. The procedure was tested against available perturbative data,
and the results agree perfectly.

What we developed in this paper may be viewed as a bottom-up approach
to construct a new way of performing string perturbation theory, based
on the triangulation of the worldsheet. The central object in our
formalism is the hexagon, which is a branch-point twist operator on
the worldsheet. The idea of using the twist operator for constructing
higher-genus surfaces is not new; it was one of the
motivations for Knizhnik to conduct detailed studies of twist
operators~\cite{Knizhnik:1987xp}. It also showed up in other important
contexts such as the low-energy description of matrix string
theory~\cite{Motl:1997th,Dijkgraaf:1997vv}. In this sense, the hexagon
formalism is yet another instance of ``old wine in new bottles'',
which we have been encountering multiple times in recent
years.\footnote{Other instances are the conformal bootstrap and the
S-matrix bootstrap.}

There are several obvious next steps. It would be important to extend the computation to
higher loops, both in $\lambda$ and in $1/\Nc$. Also desirable would be
to tie up several loose ends in our arguments: For instance, in the
discussion in \secref{sec:strat}, we estimated the contribution
from certain magnon configurations~\eqref{eq:complicated} by claiming
that they are related to simpler configurations via Dehn twists and
flip transformations. It would be nice to perform a direct
computation of such configurations and show the flip invariance
explicitly.

One practical obstacle for doing such computations is the complexity
of the multi-particle integrands. Even for the two- and three-magnon
contributions at one loop which were studied in this paper, the
integrands are horrendously complicated. Given the simplicity of the
final answer, it would be worth trying to find a better way to
organize the integrand. This will eventually be crucial if we were to
perform more complicated and physically interesting computations,
such as taking the strong-coupling limit and reproducing the
supergravity answers. Another strategy is to avoid dealing with the
complicated integrand for now, and look for simplifying limits. In flat
space, it was shown by Gross and Mende that the high-energy string
scattering takes a remarkably simple and universal
form~\cite{Gross:1987kza,Gross:1987ar}. The results were later used by
Mende and Ooguri, who succeeded in Borel-resumming the higher-genus
contributions in the same limit~\cite{Mende:1989wt}. In our context,
the analogue of the high-energy limit would be played by large
operator lengths (charges). As already observed in this paper, taking the
large-charge limit simplifies the computation drastically. It is
therefore interesting to analyze the limit in more detail, and possibly
try to re-sum the $1/\Nc$ corrections~\cite{TillPedroFrankToAppear,TillPedroFrankVascoWorkInProgress}. It
would be even more exciting if we could make a quantitative prediction
for the non-perturbative corrections by analyzing the large-order
behavior of the $1/\Nc$ expansion~\cite{Shenker:1990uf}, which one could test
against the direct instanton
computation~\cite{Alday:2016tll,Alday:2016jeo,Alday:2016bkq,Korchemsky:2017ttd}.%
\footnote{The conclusion of the large-$\Nc$ results to appear
in~\cite{TillPedroFrankToAppear} is bitter-sweet in this respect:
while one does observe the famous $g!^2$ behavior typical for
instanton-like large-genus behavior, this is further multiplied by
$1/g!^4$ arising from the kinematics of large operators, hence this
effect is not yet seen as sharply as one would like.}

The relation between the summation over graphs and the integration
over the moduli space of Riemann surfaces deserves further study.
As mentioned in the introduction, one big puzzle in this regard is the
fact that the summation over graphs is discrete, while the moduli
space is continuous. In the study of simple matrix models, such
a discretization of the moduli space was attributed to the topological
nature of the dual worldsheet theory. We should however note that the
discretization could take place even in non-topological worldsheet
theories, namely in the light-cone quantization of the DLCQ background~\cite{Motl:1997th,Dijkgraaf:1997vv,Grignani:2000zm}.
This is in fact
closer to our context since, in the generalized light-cone gauge, the
lengths of the string becomes proportional to the angular momentum in
$\grp{S}^5$, which takes discrete values. To make more progress on these
points, it would perhaps be helpful to study the recently proposed
worldsheet action for the DLCQ background~\cite{Bergshoeff:2018yvt},
which is suited for quantization in the conformal gauge, and
clarify how the conformal-gauge computation reproduces
the light-cone gauge expectation
that the moduli space gets discretized.

As a final remark, let us emphasize that the results in this paper are
just the first steps in the application of integrability to
non-planar observables: Firstly, it would also be interesting to
understand other non-planar quantities, such as non-planar
anomalous dimensions of single-trace operators, and anomalous
dimensions of double-trace operators. See~\cite{Eden:2017ozn} for
an important initial attempt.%
\footnote{See also~\cite{Ben-Israel:2018ckc} for an extension of the
amplitude\,/\,Wilson loop duality to the first non-planar order.}
 Secondly, although it is remarkable that
integrability can reproduce non-planar quantities, the computation
performed in this paper is almost as complicated as the direct
perturbative computation, and as we include more and more mirror
particles, we face the integrand challenges alluded to above. Is there
something better we can do? Can we reformulate this formalism, for
instance, by combining it with the quantum spectral curve~\cite{Gromov:2013pga}?
In fact, there are already two data points which
indicate that the quantum spectral curve could be useful for analyzing
correlation functions~\cite{Cavaglia:2018lxi, Giombi:2018qox}. Whatever the upgraded formalism will be, we expect that the
results in this paper will be useful in finding it.

\pdfbookmark[1]{Acknowledgments}{acknowledgments}
\subsection*{Acknowledgments}

We would like to thank Benjamin Basso, Nathan Berkovits, Niklas
Beisert, Vasco Gon\c{c}alves, Frank Coronado, Vladimir Kazakov,
Gregory Korchemsky, Juan Maldacena, Tristan McLoughlin, and Soo-Jong Rey for discussions.
We also thank Vasco Gon\c{c}alves for helping us
with the expansion of the mirror bound-state $S$-matrix elements.
TB would like to thank DESY Hamburg for its support and hospitality
during all stages of this work, Perimeter Institute for two very
fruitful visits, and KITP Santa Barbara for an extended stay while this
project was carried out.
TF would like to thank the warm
hospitality of the Laboratoire
de Physique Th\'eorique de l'Ecole Normale
Sup\'erieure and Nordita, where part of this work was done.
TF would like to thank also CAPES/INCTMAT process
88887.143256/2017-00
for financial support.
The work of JC and TF was supported by the
People Programme (Marie Curie Actions) of the European
Union's Seventh Framework Programme FP7/2007-2013/
under REA Grant Agreement No 317089 (GATIS),
by the European Research Council (Programme ``Ideas''
ERC-2012-AdG 320769 AdS-CFT-solvable), from the
ANR grant StrongInt (BLANC-SIMI-4-2011). The work of SK was supported by DOE grant number DE-SC0009988.
This research was supported in part by the National Science Foundation
under Grant No.\ NSF PHY17-48958.
This research was supported in part by Perimeter Institute for
Theoretical Physics. Research at Perimeter Institute is supported by
the Government of Canada through the Department of Innovation, Science
and Economic Development and by the Province of Ontario through the
Ministry of Research and Innovation.

\appendix

\section{Details on Non-Planar Data}
\label{app:data-details}

In~\eqref{eq:FFtilde}--\eqref{eq:Ftilde2}, we represented the quantum corrections
$\mathcal{F}_{k,m}$~\eqref{eq:Fkm} to the four-point correlator
$G_k$~\eqref{eq:corr}
in terms of the
conformal box~\eqref{eq:box} and
double-box functions~\eqref{eq:doublebox},
as well as color factors $C^1_{k,m}$ and $C^i_{k,m}$,
$i\in\brc{\mathrm{a,b,c,d}}$.
In the following, we will explain the color factors and their
evaluation in more detail. We will also give further expressions for
$\mathcal{F}_{k,m}$ as well as $\mathcal{\widetilde{F}}_{k,m}$. The
expressions depend on the choice of gauge group, and we will present
results for both $\grp{U}(\Nc)$ and $\grp{SU}(\Nc)$.

\paragraph{Color Factors.}

The color factors $C^i_{k,m}$
consist of color contractions of four symmetrized traces from the four
operators,
dressed with insertions of gauge group structure constants $f_{ab}{}^c$.
The one-loop color factor reads~\cite{Arutyunov:2003ae}:
\begin{multline}
C^1_{k,m}=
\frac{f_{pqe}f_{rs}{}^e}{(m+1)!^2(k-m-2)!^2}
\tr\brk{\brk{a_1\dots a_{k-1}p}}
\tr\brk{\brk{a_1\dots a_{m+1} c_{m+2}\dots c_{k-1}s}}
\\
\times
\tr\brk{\brk{a_{m+2}\dots a_{k-1}c_1\dots c_{m+1} q}}
\tr\brk{\brk{c_1\dots c_{k-1}r}}
\,,
\label{eq:C1}
\end{multline}
and the two-loop color factors are~\cite{Arutyunov:2003ad}:
\begin{align}
C\suprm{a}_{k,m}=&\frac{f_{abe}f_{cd}{}^ef_{pqt}f_{rs}{}^t}{2m!^2(k-m-1)!(k-m-3)!}
\nn\\&
\times
\tr\brk{\brk{d_1\dots d_{k-m-1}a_1\dots a_ma}}
\tr\brk{\brk{a_1\dots a_mb_1\dots b_{k-m-3}bdp}}
\nn\\&
\times
\tr\brk{\brk{d_1\dots d_{k-m-1}c_1\dots c_mr}}
\tr\brk{\brk{c_1\dots c_mb_1\dots b_{k-m-3}cqs}}
\,,
\nn\\
C\suprm{b}_{k,m}=&\frac{f_{abe}f_{cd}{}^ef_{pqt}f_{rs}{}^t}{4m!^2(k-m-2)!^2}
\nn\\&
\times
\tr\brk{\brk{d_1\dots d_{k-m-2}a_1\dots a_mbp}}
\tr\brk{\brk{a_1\dots a_mb_1\dots b_{k-m-2}cs}}
\nn\\&
\times
\tr\brk{\brk{d_1\dots d_{k-m-2}c_1\dots c_maq}}
\tr\brk{\brk{c_1\dots c_mb_1\dots b_{k-m-2}dr}}
\,,
\nn\\
C\suprm{c}_{k,m}=&\frac{f_{abe}f_{cd}{}^ef_{pqt}f_{rs}{}^t}{2m!^2(k-m-2)!^2}
\nn\\&
\times
\tr\brk{\brk{d_1\dots d_{k-m-2}a_1\dots a_mbd}}
\tr\brk{\brk{a_1\dots a_mb_1\dots b_{k-m-2}ar}}
\nn\\&
\times
\tr\brk{\brk{d_1\dots d_{k-m-2}c_1\dots c_mcp}}
\tr\brk{\brk{c_1\dots c_mb_1\dots b_{k-m-2}qs}}
\,,
\nn\\
C\suprm{d}_{k,m}=&\frac{f_{abe}f_{cd}{}^ef_{pqt}f_{rs}{}^t}{2m!^2(k-m-2)!^2}
\nn\\&
\times
\tr\brk{\brk{d_1\dots d_{k-m-2}a_1\dots a_mbp}}
\tr\brk{\brk{a_1\dots a_mb_1\dots b_{k-m-2}as}}
\nn\\&
\times
\tr\brk{\brk{d_1\dots d_{k-m-2}c_1\dots c_mcq}}
\tr\brk{\brk{c_1\dots c_mb_1\dots b_{k-m-2}dr}}
\,.
\label{eq:Cabcd}
\end{align}
Here, $\tr\brk{\brk{a_1\dots a_k}}\equiv\tr\brk{T^{(a_1}\dots
T^{a_k)}}$ denotes a totally symmetrized trace of adjoint gauge group
generators $T^a$, without $1/n!$
prefactor.
In the above formulas, $0\leq m\leq k-2$ for
$C^{1,\mathrm{b},\mathrm{c},\mathrm{d}}_{k,m}$, and $0\leq m\leq k-3$ for
$C\suprm{a}_{k,m}$, whereas $C\suprm{a}_{k-2}\equiv0$.
Pictorially, we can represent the color factors~as
\begin{alignat}{3}
C^1_{k,m}&=
\includegraphicsbox{FigC1aligned}
\,,&\quad
C\suprm{a}_{k,m}&=
\includegraphicsbox{FigCaaligned}
\,,&\quad
C\suprm{b}_{k,m}&=
\includegraphicsbox{FigCbaligned}
\,,
\nn\\
&&
C\suprm{c}_{k,m}&=
\includegraphicsbox{FigCcaligned}
\,,&
\quad
C\suprm{d}_{k,m}&=
\includegraphicsbox{FigCdaligned}
\,,
\label{eq:colorfactorpics}
\end{alignat}
where the big circles are the operator traces, the dots are structure
constants, the thin lines are single color contractions, and the thick lines
are multiple color contractions. For $C^1$, the horizontal thick lines
stand for $(m+1)$ propagators, while the vertical thick lines stand
for $(k-m-1)$ propagators. For the two-loop color factors
$C\suprm{a}$, $C\suprm{b}$, $C\suprm{c}$, and $C\suprm{d}$, the
horizontal lines stand for $m$ propagators and the vertical lines
stand for $(k-m-2)$ propagators.

Expanding the color factors to subleading order in $1/\Nc$~\eqref{eq:colorfactorexpansion},
the leading coefficients~\eqref{eq:colorfactorleading} are straightforwardly
computed~\cite{Arutyunov:2003ae,Arutyunov:2003ad}.
The subleading coefficients $\Ctorus$ are much harder to obtain. Their
computation is outlined in the following.

\paragraph{Color Algebra.}

We will evaluate the color contractions using the fission and fusion rules
\begin{gather}
\tr\brk{T^aBT^aC}=\gamma\lrbrk{\tr(B)\tr(C)-\frac{n}{\Nc}\tr(BC)}
\,,\\
\tr\brk{T^aB}\tr\brk{T^aC}=\gamma\lrbrk{\tr(BC)-\frac{n}{\Nc}\tr(B)\tr(C)}
\,,
\end{gather}
with $n=0$ for gauge group $\grp{U}(\Nc)$, and $n=1$ for gauge group
$\grp{SU}(\Nc)$. The gauge group generators $T^a$ are normalized via
\begin{equation}
\tr\brk{T^aT^b}=\gamma\,\delta^{ab}\,.
\end{equation}
The fusion and fission rules follow from the completeness relation
\begin{equation}
(T^a)^i{}_j(T^a)^k{}_l
=
\gamma\lrbrk{\delta^i_l\delta^k_j-\frac{n}{\Nc}\delta^i_j\delta^k_l}
\,.
\end{equation}
We set $\gamma=1$ to match the normalization
of~\cite{Arutyunov:2003ae,Arutyunov:2003ad}. The structure constants
are normalized~to
\begin{equation}
\comm{T^a}{T^b}=if^{ab}{}_c\,T^c
\,,
\end{equation}
such that
\begin{align}
f_{abc}&=-i\tr\brk{\comm{T_a}{T_b}T_c}=-i\tr\brk{T_aT_bT_c}+i\tr\brk{T_cT_bT_a}
\,,\\
f_{abe}f_{cd}{}^e&=-\tr\brk{\comm{T_a}{T_b}\comm{T_c}{T_d}}
\nn\\
&=-\tr\brk{T_aT_bT_dT_e}+\tr\brk{T_aT_bT_eT_d}+\tr\brk{T_aT_dT_eT_b}-\tr\brk{T_aT_eT_dT_b}
\,.
\label{eq:f2}
\end{align}
%

\paragraph{Results of Contractions.}

We have explicitly performed the contractions in~\eqref{eq:C1}
and~\eqref{eq:Cabcd} with \mathematica for various different values
of $k$ and $m$, for
some coefficients up to $k=8$, for others up to $k=9$. The results for
the subleading color coefficients are displayed in~\tabref{tab:CCsub} (page~\pageref{tab:CCsub}).
Depending on the algorithm, the computation can take very long (up to
${\sim}1$ day on $16$ cores for a single coefficient at fixed $k$ and $m$) and
becomes memory intensive (up to ${\sim}100\,$GB) at intermediate
stages.

As indicated in the main text, the
subleading color coefficients $\Ctorus^1_{k,m}$, $\Ctorus^i_{k,m}$ have to
be polynomials in $k$ and $m$ (up to boundary cases at extremal values
of $k$ or $m$). This is best understood by noting that collecting all
contractions contributing to the first subleading order in
$1/\Nc^2$ amounts to summing over all ways in which the propagators
in~\eqref{eq:colorfactorpics} can be distributed on the torus. This in
turn is equivalent to summing over all ``cyclic'' graphs on the torus (graphs with
no edge (bridge) connecting diagonally opposite operators), as
well as over all ways in which the fat propagators
in~\eqref{eq:colorfactorpics} can be distributed on the edges of the
graphs, and over all ways in which the structure constants
$f_{ab}{}^ef_{cde}$ can be inserted. At large values of $k$ (with
finite and fixed $m/k$), graphs with a maximal number of edges (bridges) will be
combinatorially dominating. At leading order in $1/k$,
these are exactly the graphs shown in~\tabref{tab:maxcycgraphs} (page~\pageref{tab:maxcycgraphs}). Due
to~\eqref{eq:lengthsum}, two operators connected by $n$ propagators
distributed on $j$ bridges will contribute a factor $n^{j-1}$. Looking
at the graphs in~\tabref{tab:maxcycgraphs}, one finds that the maximal
power of $k$ is four. Hence the polynomials representing the color
factors will be quartic.
Any closed formula for $\Ctorus^i_{k,m}$
therefore has to be a quartic polynomial in $k$ and $m$.
A general polynomial of this type has $15$ coefficients.
Matching those against the $\grp{U}(\Nc)$ data points in~\tabref{tab:CCsub} yields
the following solutions:
\begin{align}
\Ctorus_{k,m}^{1,\grp{U}}
&=
-\frac{1}{6}\Bigbrk{
k^4 + 2 k^3 (-1 + 2 m) + k^2 ( -1 + 6 m + 30 m^2)
\nn\\&\mspace{20mu}
- 2 k ( 11 + 49 m + 75 m^2 + 34 m^3)
+ 2 (1 + m)^2 (18 + 34 m + 17 m^2)
}\,,
\label{eq:C1subU}
\\
\label{eq:CasubU}
\Ctorus_{k,m}^{\mathrm{a},\grp{U}}
&=
\frac{1}{12}\Bigbrk{
k^4 + 2 k^3 (-2 + 2 m) + k^2 (-1 +        54 m^2)
\\&\mspace{20mu}
- 2 k (22 + 55 m + 126 m^2 + 58 m^3)
+ 2 (54 + 129 m + 181 m^2 + 123 m^3 + 29 m^4)
}\,,
\nn\\
\label{eq:CbsubU}
\Ctorus_{k,m}^{\mathrm{b},\grp{U}}
&=
\frac{1}{24}\Bigbrk{
k^4 + 2 k^3 (-3 + 2 m) + k^2 (47 - 54 m + 54 m^2)
\\&\mspace{20mu}
- 2 k (63 +    m +  21 m^2 + 58 m^3)
+ 2 (54 +  60 m +  31 m^2 +  48 m^3 + 29 m^4)
}(1+\delta_{m,0})
\nn\\&\mspace{360mu}
+ 1/3 (-18 + 26 k - 12 k^2 + k^3) \delta_{m,0}
\,,
\nn\\
\label{eq:CcsubU}
\Ctorus_{k,m}^{\mathrm{c},\grp{U}}
&=
\frac{1}{6}\Bigbrk{
k^4 + 2 k^3 (-1 + 2 m) + k^2 (-1 +  6 m + 42 m^2)
\\&\mspace{20mu}
- 2 k (11 + 49 m +  99 m^2 + 46 m^3)
+ 2 (18 +  70 m + 127 m^2 +  92 m^3 + 23 m^4)
}\,,
\nn\\
\label{eq:CdsubU}
\Ctorus_{k,m}^{\mathrm{d},\grp{U}}
&=
\frac{1}{12}\Bigbrk{
k^4 + 2 k^3 (-1 + 2 m) + k^2 (-1 +  6 m + 54 m^2)
\\&\mspace{20mu}
- 2 k (11 + 49 m + 123 m^2 + 58 m^3)
+ 2 (18 +  70 m + 151 m^2 + 116 m^3 + 29 m^4)
}\,.
\nn
\end{align}
For gauge group $\grp{SU}(\Nc)$, we find
\begin{align}
\Ctorus_{k,m}^{1,\grp{SU}}
&=
\Ctorus_{k,m}^{1,\grp{U}}
- 4 (k-1)^2 (-2 + \delta_{m,0} + \delta_{m,k-2})
\,,
\label{eq:C1subSU}
\\
\Ctorus_{k,m}^{\mathrm{a},\grp{SU}}
&=
\Ctorus_{k,m}^{\mathrm{a},\grp{U}}
+   (k-1)^2 (-4 + 2 \delta_{m,0} + \delta_{m,k-3})
\,,
\label{eq:CasubSU}
\\
\Ctorus_{k,m}^{\mathrm{b},\grp{SU}}
&=
\Ctorus_{k,m}^{\mathrm{b},\grp{U}}
+   (k-1)^2 (-2 +                  \delta_{m,k-2}) + \delta_{k,2}
\,,
\label{eq:CbsubSU}
\\
\Ctorus_{k,m}^{\mathrm{c},\grp{SU}}
&=
\Ctorus_{k,m}^{\mathrm{c},\grp{U}}
+ 4 (k-1)^2 (-2 +   \delta_{m,0} + \delta_{m,k-2})
\,,
\label{eq:CcsubSU}
\\
\Ctorus_{k,m}^{\mathrm{d},\grp{SU}}
&=
\Ctorus_{k,m}^{\mathrm{d},\grp{U}}
+ 2 (k-1)^2 (-2 +   \delta_{m,0} + \delta_{m,k-2})
\,.
\label{eq:CdsubSU}
\end{align}
In all cases, there are more data points than degrees of freedom in
the quartic polynomial. Moreover, one can convince oneself that the
difference between $\grp{U}(\Nc)$ and $\grp{SU}(\Nc)$ gauge groups
should not depend on $m$, and should be at most quadratic in $k$. We
can thus be fairly confident that the results are correct for general
$k$ and $m$.

\paragraph{Analytic Check.}

We can perform an analytic check of the
expressions~\eqref{eq:C1subU}--\eqref{eq:CdsubU} by studying the limit
of large $k$ with $0<m/k<1$ fixed and finite. As outlined in the
previous paragraph, we can organize the contractions in the color
factors~\eqref{eq:C1} and~\eqref{eq:Cabcd} at the first subleading
order in $1/\Nc^2$ as a sum over graphs on the torus. At leading order
in large $k$, only graphs with a maximal number of bridges will
contribute, all other graphs will be combinatorially suppressed due
to~\eqref{eq:lengthsum}. The contributing graphs are exactly the ones
listed in~\tabref{tab:maxcycgraphs}. For each of those graphs, we have
to sum over all inequivalent labelings of the four operators, over all
possible combinations of non-zero bridge lengths on the edges of the
graph, and over all possible insertions of $f_{ab}{}^ef_{cde}$ terms
(expanded as in~\eqref{eq:f2}). For each fixed configuration of bridge
lengths, the sum over all planar contractions compatible with those
bridge lengths (from the total trace symmetrizations) gives a factor
$k^4$ from cyclic rotations of the four operators, times a factor
$(m+1)!^2(k-m-2)!^2$ (for $C^1_m$), $m!^2(k-m-1)!(k-m-3)!$ (for
$C\suprm{a}_m$), or $m!^2(k-m-2)!^2$ (for $C\suprm{b,c,d}_m$), which
cancel the combinatorial denominators in~\eqref{eq:C1}
and~\eqref{eq:Cabcd}. We will now go through the graphs
of~\tabref{tab:maxcycgraphs} and find the number of inequivalent
labelings as well as the combinatorial factors from the summation over bridge lengths. The insertions of
$f_{ab}{}^ef_{cde}$ terms will be considered below.
\begin{description}[leftmargin=2\parindent,labelindent=\parindent]
\item[Cases A, C, D, E, F, H, J, N, K:]
The bridge configurations of these cases imply a constraint on $m$:
Either $m=0$, or $m=k-2$ ($m=k-3$ for $C\suprm{a}_m$, $m=k-1$ for
$C^1$), which would set the lengths of either the vertical or the horizontal bridges
in~\eqref{eq:colorfactorpics} to zero.
Hence, these cases are suppressed at large $m$ and $k$.
\item[Case B:]
This case has $4\cdot2=8$ inequivalent operator labelings. Depending
on the labeling, the sum over bridge lengths~\eqref{eq:lengthsum} for
large $m$ and $k$
gives a factor of either $m^3(k-m)/6$ (call this Case B.1), or
$m(k-m)^3/6$ (call this Case B.2). Each subcase has $4$ inequivalent
operator labelings.
\item[Case G:]
This bridge configuration is symmetric under a horizontal flip of the
graph in~\tabref{tab:maxcycgraphs}. Hence there are four
inequivalent operator labelings. For one half of the operator
labelings, the sum over bridge lengths for large $m$ and $k$ gives
$m^4/24$, for the other half it gives $(k-m)^4/24$.
\item[Case L:]
This bridge configuration is symmetric under exchange of the top right
and the bottom left operators. Hence there are four inequivalent
operator labelings for this case. For all operator labelings, the sum
over bridge lengths for large $m$ and $k$ gives $m^2/2\cdot(k-m)^2/2$.
\item[Case M:]
This bridge configuration is symmetric under simultaneous exchange of
the top left with the bottom left operator and the top right with the
bottom right operator. Hence there are four inequivalent operator
labelings. For large $m$ and $k$, the sum over bridge lengths gives
$m^2(k-m)^2/2$.
\item[Case P:]
For this bridge configuration, all operator labelings are equivalent.
There is one more symmetry: Every pair of operators is connected by
two bridges. Exchanging all such bridge pairs simultaneously leaves
the configuration invariant (the operation is equivalent to a specific
rotation of each operator, see also~\eqref{eq:automorphismexample1}).
The resulting over-counting in the naive sum over bridge lengths needs
to be compensated by a factor of $1/2$. For large $m$ and $k$, the
(naive) sum over bridge lengths gives $m^2(k-m)^2$.
\item[Case Q.]
As for Case P, all operator labelings are equivalent. This graph has
no additional symmetry though. The sum over bridge lengths gives a
factor $m^2(k-m)^2$.
\end{description}
Now we come to the insertion of $f_{ab}{}^ef_{cde}$ factors (called
``$f^2$'' in the following). The $f^2$ factors either attach to three
of the four operators (for $C\suprm{a}$ and $C\suprm{c}$), or to all
four operators (for $C^1$, $C\suprm{b}$ and $C\suprm{d}$). The bridge
configurations A through Q all decompose the torus into four octagons.
One octagon of case B and two octagons of case G involve only two of
the four operators, hence they cannot accommodate an $f^2$ factor. All
other octagons involve either three or all four operators. For all
cases and all operator labelings, inserting an $f^2$ term into an
octagon that involves only three operators produces a zero, since
either none of the four trace terms in~\eqref{eq:f2} contributes, or
all of them contribute and sum to zero. Thus all non-trivial
contributions have both $f^2$ factors inserted into octagons that
involve all four operators. In all such insertions, only one of the
four trace terms of~\eqref{eq:f2} contributes, and the signs of those
terms of the two $f^2$ factors always multiply to $+1$. The
combinatorial factors from inequivalent $f^2$ insertions for the
relevant cases are:
\begin{description}[leftmargin=2\parindent,labelindent=\parindent]
\item[Cases B, G, L, M:] In these three cases, there are two
four-operator octagons. For $C\suprm{a,b,d}$, the two $f^2$ cannot be
inserted into the same octagon, hence there are only two inequivalent
ways to distribute the $f^2$ factors. For $C\suprm{c}$, the two $f^2$
factors can also be inserted into the same octagon, hence there are
four ways to distribute the $f^2$ terms.
\item[Case P:] In this case, each of the four octagons involves all
four operators. Again, the two $f^2$ can be inserted into the same
octagon for $C\suprm{c}$, but not for $C\suprm{a,b,d}$. Hence, there
are $16$ ways to distribute the $f^2$ terms for $C\suprm{c}$, but only
$12$ ways to do so for $C\suprm{a,b,d}$.
\item[Case Q:] In this case, each of the four octagons involves only three
of the four operators, and hence there are no non-trivial $f^2$ insertions.
\end{description}
Summarizing the above,
at large $k$ with $\tilde{m}=m/k$ fixed and $0<\tilde{m}<1$,
we find the combinatorial structure displayed in \tabref{tab:colorfactorslargek}.
\begin{table}
\centering
\begin{tabular}{clcc@{\;}rrrrr}
\toprule
Case & $\sum_{\text{bridges}}$                    & Labelings & $f^2:$ & $C^1$ & $C\suprm{a}$ & $C\suprm{b}$ & $C\suprm{c}$ & $C\suprm{d}$ \\
\midrule
B.1  & $k^4\tilde{m}^3(1-\tilde{m})/6$            & $4$       &        & $ -2$ & $2$          & $2$          & $4$          & $2$  \\
B.2  & $k^4\tilde{m}(1-\tilde{m})^3/6$            & $4$       &        & $ -2$ & $2$          & $2$          & $4$          & $2$  \\
G.1  & $k^4\tilde{m}^4/24$                        & $2$       &        & $ -2$ & $2$          & $2$          & $4$          & $2$  \\
G.2  & $k^4(1-\tilde{m})^4/24$                    & $2$       &        & $ -2$ & $2$          & $2$          & $4$          & $2$  \\
L    & $k^4\tilde{m}^2/2\cdot(1-\tilde{m})^2/2$   & $4$       &        & $ -2$ & $2$          & $2$          & $4$          & $2$  \\
M    & $k^4\tilde{m}^2(1-\tilde{m})^2/2$          & $4$       &        & $ -2$ & $2$          & $2$          & $4$          & $2$  \\
P    & $k^4\tilde{m}^2(1-\tilde{m})^2$            & $1/2$     &        & $ -4$ & $12$         & $12$         & $16$         & $12$ \\
Q    & $k^4\tilde{m}^2(1-\tilde{m})^2$            & $1$       &        & 0&0&0&0&0 \\
all others    & $\order{k^3}$                              &   \dots     &        & \multicolumn{5}{c}{\dots} \\
\bottomrule
\end{tabular}
\caption{List of graphs that contribute to the color factors at large
$k$ with $\tilde{m}=m/k$ fixed and $0<\tilde{m}<1$. Listed are the
combinatorial factors from summing over bridge lengths, the numbers of
inequivalent labelings, and the combinatorial factors from inserting
the (pairs of) structure constants $f_{ab}{}^c$ into the various polygons.}
\label{tab:colorfactorslargek}
\end{table}
Multiplying all factors and summing all cases, we find:
\begin{align}
C^1_m
&=
-\Nc^{2k-1}k^4\biggbrk{2+\frac{k^4}{6\Nc^2}\biggsbrk{
1+4\tilde{m}+30\tilde{m}^2-68\tilde{m}^3+34\tilde{m}^4
}
+\order{k^3}
}
\label{eq:C1sublargek}
\,,\\
C\suprm{a}_m
&=2C\suprm{b}_m=C\suprm{d}_m=
\nn\\
&=
\Nc^{2k}k^4\biggbrk{1+\frac{k^4}{12\Nc^2}\biggsbrk{
1+4\tilde{m}+54\tilde{m}^2-116\tilde{m}^3+58\tilde{m}^4
}
+\order{k^3}
}
\label{eq:Cabdsublargek}
\,,\\
C\suprm{c}_m&=
\Nc^{2k}k^4\biggbrk{2+\frac{k^4}{6\Nc^2}\biggsbrk{
1+4\tilde{m}+42\tilde{m}^2-92\tilde{m}^3+46\tilde{m}^4
}
+\order{k^3}
}
\,.
\label{eq:Ccsublargek}
\end{align}
One can indeed see that the above formulas reproduce the leading terms
of~\eqref{eq:C1subU}--\eqref{eq:CdsubU}. This match is an important
cross-check both of the results~\eqref{eq:C1subU}--\eqref{eq:CdsubU}, and
of the classification of torus contractions in~\tabref{tab:maxcycgraphs}.

\paragraph{The Quantum Corrections.}

Inserting the above expressions~\eqref{eq:C1subU}--\eqref{eq:CdsubU}
for the color factors into the
formulas~\eqref{eq:FFtilde}--\eqref{eq:Ftilde2} yields the one-loop
and two-loop $\grp{U}(\Nc)$ data shown in~\tabref{tab:FkmUdata}
(page~\pageref{tab:FkmUdata}, with the definitions~\eqref{eq:rst}
and~\eqref{eq:F2combs}). For gauge group $\grp{SU}(\Nc)$, we find
\begin{align}
\mathcal{F}^{(1),\mathrm{SU}}_{k,m}(z,\bar z)
&=
\mathcal{F}^{(1),\mathrm{U}}_{k,m}(z,\bar z)
    -\frac{4k^2(k-1)^2}{\Nc^4}\biggsbrk{
        -t
        {\color{blue}\mbox{}+\delta_m^0\bigbrk{(2t-1)+(1+s-t)\delta_m^1-s\delta_m^2}}
    \nn\\&\mspace{40mu}
        {\color{blue}\mbox{}+\delta_m^1\bigbrk{(t+s-1)-s\delta_{m,2}}
        +s\delta_{m,2}+(s-t)\delta_m^0\delta_m^{k-1}-\sfrac{1}{2}(s+1)\delta_m^0\delta_m^k}
    \nn\\&\mspace{40mu}
           +\text{(crossing)}
           }
           {\color{red}F^{(1)}}
\,,\\
\mathcal{F}^{(2),\mathrm{SU}}_{k,m}(z,\bar z)
&=
\mathcal{F}^{(2),\mathrm{U}}_{k,m}(z,\bar z)
    +\frac{4k^2(k-1)^2}{\Nc^4}\Biggsbrk{
    -4t {\color{red}F^{(2)}}-t^2 {\color{red}\bigbrk{F^{(1)}}^2}
    \nn\\&\mspace{20mu}
           {\color{blue}
           \mbox{}+\frac{1}{2}
           \biggbrc{(2t-1)\delta_{m,0}+(t+s-1)\delta_{m,1}+s\delta_{m,2}}
           \Bigbrk{4{\color{red}F^{(2)}}+(t-s){\color{red}\bigbrk{F^{(1)}}^2}}
           }
    \nn\\&\mspace{20mu}
           {\color{blue}
           \mbox{}+\frac{1}{2}
           \biggbrc{\delta_{m,k-2}+(t+s-1)\delta_{m,k-1}+(2t-s)\delta_{m,k}}
           \Bigbrk{4{\color{red}F^{(2)}}+(t-1){\color{red}\bigbrk{F^{(1)}}^2}}
           }
    \nn\\&\mspace{20mu}
           {\color{blue}
           \mbox{}-\biggbrc{s(t-1)\delta_{m,0}+s^2\delta_{m,1}+\delta_{m,k-1}+(t-s)\delta_{m,k}}
           {\color{red}\bigbrk{F^{(1)}}^2}
           }
    \nn\\&\mspace{20mu}
           {\color{blue}
           \mbox{}-\biggbrc{2\delta_{m,0}+(2t-2s-1)\delta_{m,1}+(t+s-1)\delta_{m,2}+s\delta_{m,3}}
           {\color{red}F^{(2)}_{1-z}}
           }
    \nn\\&\mspace{20mu}
           {\color{blue}
           \mbox{}-\biggbrc{\delta_{m,k-3}+(t-s+1)\delta_{m,k-2}+(2t-s-2)\delta_{m,k-1}+2s\delta_{m,k}}
           {\color{red}F^{(2)}_{z/(z-1)}}
           }
    \nn\\&\mspace{20mu}
           {\color{mypurple}\mbox{}
           +\delta_{k,2}
           \biggbrkopen
           \frac{s+1}{2}\Bigbrc{t-(t-1)\delta_m^0-(s+1)\delta_m^1-(t-s)\delta_m^2}{\color{red}\bigbrk{F^{(1)}}^2}
           }
    \nn\\&\mspace{60mu}
           {\color{mypurple}\mbox{}
           +\Bigbrc{2t-2(t-1)\delta_m^0-(2s+1)\delta_m^1-(t-s+1)\delta_m^2}{\color{red}F^{(2)}_{1-z}}
           }
    \nn\\&\mspace{60mu}
           {\color{mypurple}\mbox{}
           +\Bigbrc{2t-(t-s-1)\delta_m^0-(s+2)\delta_m^1-2(t-s)\delta_m^2}{\color{red}F^{(2)}_{z/(z-1)}}
           \biggbrkclose
           }
    }
\,,
\end{align}
where we have suppressed the arguments $(z,\bar z)$ of all box and
double-box functions, and where $\text{(crossing)}$ stands for $s$
times the whole preceding expression with the replacements
\begin{equation}
t \rightarrow t/s
\,,\quad
s \rightarrow 1/s
\,,\quad
m \rightarrow k-m
\,.
\end{equation}
Using the transformations~\eqref{eq:strcrossing}
and~\eqref{eq:F12crossing}, it is easy to verify that the expressions
above are invariant under crossing $x_1 \leftrightarrow x_4$.

Due to supersymmetry, the quantum corrections to the correlator $\vev{\op{Q}_1\dotsc\op{Q}_4}$
contain a universal prefactor $R$~\eqref{eq:preR} that
is usually pulled out,
\begin{equation}
\vev{\op{Q}_1\dotsc\op{Q}_4}^\text{quantum}
=R\sum_{m=0}^{k-2}\sum_{\ell=1}^{\infty}g^{2\ell}\mathcal{\widetilde{F}}_{k,m}^{(\ell)}X^mY^{k-m-2}
\,,\quad
R=z\bar z X^2-(z+\bar z)XY+Y^2
\,.
\label{eq:Ftildesum}
\end{equation}
In the bulk of this work, we have rather used the expansion~\eqref{eq:corr} without $R$ factored
out, because it is better suited for comparison
with our integrability-based computation. The relation between the
different expansion coefficients
$\mathcal{F}_{k,m}$ and $\mathcal{\widetilde{F}}_{k,m}$ is shown
in~\eqref{eq:FFtilde}. For completeness, we also state the
perturbative results for $\mathcal{\widetilde{F}}_{k,m}$. For gauge
group $\grp{U}(\Nc)$, the expressions are:
\begin{align}
\mathcal{\widetilde F}^{(1),\mathrm{U}}_{k,m}(z,\bar z)
&=
  \label{eq:Ftilde1kmU}
  \\&\mspace{-60mu}
  -\frac{2k^2}{\Nc^2}\biggbrc{
  1
         +\frac{1}{\Nc^2}\biggsbrk{
         \bigsbrk{\sfrac{17}{6}\tilde{r}^4-\sfrac{7}{4}\tilde{r}^2+\sfrac{11}{32}}k^4
         +\bigsbrk{\sfrac{9}{2}\tilde{r}^2-\sfrac{13}{8}}k^3
         +\bigsbrk{\sfrac{1}{6}\tilde{r}^2+\sfrac{15}{8}}k^2
         -\sfrac{1}{2}k
  }
  }
  {\color{red}F^{(1)}}
\,,\nn\\[1ex]
\mathcal{\widetilde F}^{(2),\mathrm{U}}_{k,m}(z,\bar z)
&=
  \nn\\&\mspace{-60mu}
  \frac{4k^2}{\Nc^2}
  \Biggsbrk{
  \biggbrc{
  1
         +\frac{1}{\Nc^2}\biggsbrk{
         \bigsbrk{\sfrac{17}{6}\tilde{r}^4-\sfrac{7}{4}\tilde{r}^2+\sfrac{11}{32}}k^4
         +\bigsbrk{\sfrac{9}{2}\tilde{r}^2-\sfrac{13}{8}}k^3
         +\bigsbrk{\sfrac{1}{6}\tilde{r}^2+\sfrac{15}{8}}k^2
         -\sfrac{1}{2}k
         }
  }{\color{red}F^{(2)}}
  \nn\\&\mspace{-40mu}
  +\biggbrc{
  \frac{t}{4}
  +\frac{1}{\Nc^2}\biggsbrk{
  \Bigbrk{
  \bigsbrk{\sfrac{7}{2}\tilde{r}^2-\sfrac{1}{8}}k^2
  +\sfrac{5}{8}k
  -\sfrac{1}{4}
  }s_+
  -\tilde{r}\Bigbrk{
  \bigsbrk{\sfrac{17}{6}\tilde{r}^2-\sfrac{7}{8}}k^3
  +3k^2
  -\sfrac{13}{12}k
  }s_-
  \nn\\&\mspace{-20mu}
  +\Bigbrk{
  \bigsbrk{\sfrac{29}{24}\tilde{r}^4-\sfrac{11}{16}\tilde{r}^2+\sfrac{15}{128}}k^4
  +\bigsbrk{\sfrac{17}{8}\tilde{r}^2-\sfrac{21}{32}}k^3
  -\bigsbrk{\sfrac{23}{24}\tilde{r}^2-\sfrac{39}{32}}k^2
  -\sfrac{9}{8}k
  +\sfrac{1}{2}
  }t
  }
  }{\color{red}\bigbrk{F^{(1)}}^2}
  \nn\\&\mspace{-40mu}
         -\frac{1}{\Nc^2}
         \biggsbrk{
         \tilde{r}\Bigbrc{
         \bigsbrk{\sfrac{7}{6}\tilde{r}^2-\sfrac{1}{8}}k^3
         +\sfrac{3}{2}k^2
         +\sfrac{10}{3}k
         }{\color{red}F^{(2)}_{\mathrm{C},-}}
         \nn\\&\mspace{4mu}
         +\Bigbrc{
         \bigsbrk{\sfrac{5}{4}\tilde{r}^2-\sfrac{19}{48}}k^3
         +\bigsbrk{\sfrac{3}{2}\tilde{r}^2+\sfrac{7}{8}}k^2
         +\sfrac{1}{3}k
         }{\color{red}F^{(2)}_{\mathrm{C},+}}
         }
  \nn\\&\mspace{-40mu}
         {\color{blue}
         \mbox{}+
         \frac{1}{4}
         \biggbrc{1+\frac{(k-1)(k^3+3k^2-46k+36)}{12\Nc^2}}
         \bigbrk{s\delta_{m,0}+\delta_{m,k-2}}
         {\color{red}\bigbrk{F^{(1)}}^2}
         }
  \nn\\&\mspace{-40mu}
         {\color{blue}
         \mbox{}+
         \biggbrc{1+\frac{(k-2)_4}{12\Nc^2}}
         \bigbrk{\delta_{m,0}{\color{red}F^{(2)}_{1-z}}+\delta_{m,k-2}{\color{red}F^{(2)}_{z/(z-1)}}}
         }
  }
         \,,
  \label{eq:Ftilde2kmU}
\end{align}
whereas for gauge group $\grp{SU}(\Nc)$:
\begin{align}
\mathcal{\widetilde F}^{(1),\mathrm{SU}}_{k,m}(z,\bar z)
&=
\mathcal{\widetilde F}^{(1),\mathrm{U}}_{k,m}(z,\bar z)
-\frac{2k^2(k-1)^2}{\Nc^4}\bigbrk{-4{\color{blue}\mbox{}+2\delta_{m,0}+2\delta_{m,k-2}}}{\color{red}F^{(1)}}
\,,\\
\mathcal{\widetilde F}^{(2),\mathrm{SU}}_{k,m}(z,\bar z)
&=
\mathcal{\widetilde F}^{(2),\mathrm{U}}_{k,m}(z,\bar z)
  +\frac{4k^2(k-1)^2}{\Nc^4}\biggsbrk{
  -4{\color{red}F^{(2)}}-t{\color{red}\bigbrk{F^{(1)}}^2}
  \nn\\&\mspace{40mu}
         {\color{blue}\mbox{}
         +2\bigbrc{\delta_{m,0}+\delta_{m,k-2}}{\color{red}F^{(2)}}
         +\frac{1}{2}\bigbrc{(t-s)\delta_{m,0}+(t-1)\delta_{m,k-2}}{\color{red}\bigbrk{F^{(1)}}^2}
         }
  \nn\\&\mspace{40mu}
         {\color{blue}\mbox{}
         -\bigbrc{2\delta_{m,0}+\delta_{m,1}}{\color{red}F^{(2)}_{1-z}}
         -\bigbrc{\delta_{m,k-3}+2\delta_{m,k-2}}{\color{red}F^{(2)}_{z/(z-1)}}
         }
  \nn\\&\mspace{40mu}
         {\color[rgb]{0.6,0,1}\mbox{}
         +\frac{1}{2}\delta_{k,2}\Bigbrk{(s+1){\color{red}\bigbrk{F^{(1)}}^2}+4{\color{red}F^{(2)}_{1-z}}+4{\color{red}F^{(2)}_{z/(z-1)}}}
         }
         }
         \,.
\end{align}
Here,
\begin{equation}
\tilde{r}=\frac{m+1}{k}-\frac{1}{2}
\,,
\end{equation}
and note the definitions~\eqref{eq:rst} and~\eqref{eq:F2combs}.
It is easy to see that the above formulas obey crossing symmetry: Under
the crossing transformation $x_1 \leftrightarrow x_4$,
\begin{equation}
X \leftrightarrow Y
\,,\quad
z \rightarrow 1/z
\,,\quad
\bar z \rightarrow 1/\bar z
\,,\quad
R \rightarrow R/s
\,,
\end{equation}
and hence crossing invariance of~\eqref{eq:Ftildesum} is equivalent to
\begin{equation}
\mathcal{\widetilde F}^{(\ell)}_{k,m}(z,\bar z)
=
s\mathcal{\widetilde F}^{(\ell)}_{k,k-2-m}(1/z,1/\bar z)
\,,
\end{equation}
which is easily verified using~\eqref{eq:strcrossing}
and~\eqref{eq:F12crossing} as well as
$\tilde{r}\rightarrow -\tilde{r}$
under $m \rightarrow k-2-m$.

\paragraph{Remark.}

From the above expressions, we note that
\begin{equation}
\Ctorus\suprm{c}_{k,m}-\Ctorus\suprm{d}_{k,m}=-\frac{1}{2}\Ctorus^1_{k,m}\,,
\end{equation}
which is equivalent to the equality of coefficients in front of
$F^{(1)}$ and $F^{(2)}$ in~\tabref{tab:FkmUdata} as well as
in~\eqref{eq:Ftilde1kmU} and~\eqref{eq:Ftilde2kmU}.
In fact, we have computed
the full $\Nc$ dependence of the color factors
$C^{1,\mathrm{a},\mathrm{b},\mathrm{c},\mathrm{d}}$ for all values
$0\leq m\leq (k-2)\leq 5$, and the statement
\begin{equation}
C\suprm{c}_{k,m}-C\suprm{d}_{k,m}=-\frac{\Nc}{2}C^1_{k,m}
\end{equation}
remains true for any $\Nc$.

This equality of the coefficients (up to overall numerical factors) of
the ladder integrals $F^{(\ell)}$ at any $\ell$-loop order can be understood from integrability: This
term stems from the
one-particle contribution
which at $\ell$ loops is proportional to $F^{(\ell)}$. The prefactor
of the single-particle excitation is given purely by graph
combinatorics, which is independent of the loop order.

\section{Graph Constructions}

\subsection{Bottom-Up Construction of All Graphs}
\label{app:bottom-up-graphs}

In~\secref{sec:polygonization}, we have manually classified all
maximal graphs on the torus (displayed in~\tabref{tab:maxgraphs} on
page~\pageref{tab:maxgraphs}). All other graphs can be obtained by
deleting bridges from these maximal graphs. Here we want to outline an
algorithm that produces all graphs, maximal and non-maximal.
The algorithm can be used
for any genus and for any number
of operators, but it can become very time
consuming.

The main step of the algorithm takes a list of graphs, and
adds to it all graphs obtained by inserting another bridge (that is
homotopically inequivalent to all previous bridges) into any of the graphs
already in the list. The new bridge may attach to an operator in
between two existing bridges, or it may split an existing bridge in two.
Graphs related by rotations or relabelings of the operators or bridges
are identified. Duplicate graphs as well as graphs exceeding the wanted genus are
discarded. This step is iterated, starting with the ``empty'' graph with
$n$ vertices (operators) and no bridges. The algorithm stops once the
iteration step generates no new graphs, and will produce \emph{all}
inequivalent graphs with $n$ vertices whose genus is equal or lower than the
wanted genus.
The maximal graphs are the ones
that exceed the wanted genus when any possible
bridge is added.

\subsection{Cyclic Graphs from Maximal Graphs}
\label{app:cyclic-graphs}

As explained in~\secref{sec:polygonization}, all cyclic graphs are
obtained from the set of maximal graphs in~\tabref{tab:maxgraphs} by
grouping the four operators into pairs and deleting edges that connect
the members of each pair. In the following, we list the descendance of
the cases C through Q from the maximal graphs 1.3 through 3.1. We have
only kept inequivalent cyclic graphs, and have discarded cases that
have a non-maximal number of bridges (the latter can all be obtained
by deleting further bridges from the following maximal graphs):
\newlength{\rightboxsize}
\setlength{\rightboxsize}{4cm}
\newcommand{\onecase}[3]{%
\left.
\begin{tabular}{l@{}}
$\displaystyle
\includegraphicsbox{#1}
\;=\;
\text{\makebox[\rightboxsize][l]{\includegraphicsbox{#2}\quad(case #3)\,.}}$%
\end{tabular}\right.}
\newcommand{\twocases}[6]{%
\left\{
\begin{tabular}{l@{}}
$\displaystyle
\includegraphicsbox{#1}
\;=\;
\text{\makebox[\rightboxsize][l]{\includegraphicsbox{#2}\quad(case #3)\,.}}$\\[6ex]
$\displaystyle
\includegraphicsbox{#4}
\;=\;
\text{\makebox[\rightboxsize][l]{\includegraphicsbox{#5}\quad(case #6)\,.}}$%
\end{tabular}\right.}
\begin{alignat}{2}
\text{Case 1.3:}
\quad
\includegraphicsbox{FigTorusCase13}
\;\longrightarrow\;&&
\twocases{FigTorusCase13C}{FigTorusCaseC.mps}{C}{FigTorusCase13D.mps}{FigTorusCaseD.mps}{D}&
\\[1ex]
\text{Case 1.4.1:}
\quad
\includegraphicsbox{FigTorusCase141}
\;\longrightarrow\;&&
\onecase{FigTorusCase141E}{FigTorusCaseE.mps}{E}&
\\[1ex]
\text{Case 1.4.2:}
\quad
\includegraphicsbox{FigTorusCase142}
\;\longrightarrow\;&&
\onecase{FigTorusCase142F}{FigTorusCaseF.mps}{F}&
\\[1ex]
\text{Case 1.5.1:}
\quad
\includegraphicsbox{FigTorusCase151}
\;\longrightarrow\;&&
\twocases{FigTorusCase151G}{FigTorusCaseG.mps}{G}{FigTorusCase151H.mps}{FigTorusCaseH.mps}{H}&
\\[1ex]
\text{Case 1.5.2:}
\quad
\includegraphicsbox{FigTorusCase152}
\;\longrightarrow\;&&
\twocases{FigTorusCase152I}{FigTorusCaseI.mps}{I}{FigTorusCase152J.mps}{FigTorusCaseJ.mps}{J}&
\\[1ex]
\text{Case 1.6:}
\quad
\includegraphicsbox{FigTorusCase16}
\;\longrightarrow\;&&
\onecase{FigTorusCase16K}{FigTorusCaseK.mps}{K}&
\\[1ex]
\text{Case 2.1.1:}
\quad
\includegraphicsbox{FigTorusCase211}
\;\longrightarrow\;&&
\onecase{FigTorusCase211L}{FigTorusCaseL.mps}{L}&
\\[1ex]
\text{Case 2.1.2:}
\quad
\includegraphicsbox{FigTorusCase212}
\;\longrightarrow\;&&
\onecase{FigTorusCase212M}{FigTorusCaseM.mps}{M}&
\\[1ex]
\text{Case 2.2:}
\quad
\includegraphicsbox{FigTorusCase22}
\;\longrightarrow\;&&
\onecase{FigTorusCase22N}{FigTorusCaseN.mps}{N}&
\\[1ex]
\text{Case 3.1:}
\quad
\includegraphicsbox{FigTorusCase31}
\;\longrightarrow\;&&
\twocases{FigTorusCaseP}{FigTorusCaseP.mps}{P}{FigTorusCase31Q.mps}{FigTorusCaseQ.mps}{Q}&
\end{alignat}
%

\section{Weak Coupling Expansions}
\label{app:WeakCouplingApendix}

We give here the weak-coupling expansion and
the definition of
useful quantities for the integrability computation
of the mirror particle corrections.
Since we give a three-loop prediction in
this work, we have to evaluate
some expansions up to order $g^6$. We have for the
measure and momenta of the mirror particles
\begin{equation}
\begin{aligned}
& \mu_a(u^{\gamma}) = g^2 \, \frac{ a }{(u^2 + \frac{a^2}{4})^2}
- g^4 \, \frac{ a (a^2 - 8 u^2)}{(u^2 + \frac{a^2}{4})^4} +
g^6 \, \frac{a (a^4 - 24
a^2 u^2+ 48 u^4)}{(u^2 + \frac{a^2}{4})^6}
+ \mathcal{O}(g^8) \, ,
\end{aligned}
\label{eq:weakmeasure}
\end{equation}
and
\begin{equation}
\begin{aligned}
& \tilde{p}_a(u^{\gamma}) = u - g^2 \frac{2 u}{(u^2+ \frac{a^2}{4})} +
g^4 \frac{ 2 u (\frac{3 a^2}{4} - u^2)}{(u^2+\frac{a^2}{4})^3} + \mathcal{O}(g^6) \, .
\end{aligned}
\label{eq:weakmomentum}
\end{equation}
In addition, the fused dynamical factor is defined as
\begin{equation}
h_{ab}(u,v) = \sum_{k = - \frac{a-1}{2}}^{\frac{a-1}{2}}
\sum_{l = - \frac{b-1}{2}}^{\frac{b-1}{2}}
h(u^{[2 k]},v^{[2 l]}) \,,
\end{equation}
where we use the usual notation $u^{[n]}\equiv u+in/2$ for imaginary shifts.
In what follows, we will need the mirror rotated fused dynamical
factor $h_{ab}(u^{\gamma}, v^{\gamma})$ and products of it.
It was computed in~\cite{\Hexagonalizationtwo}, and we reproduce it here
for completeness:
\begin{multline}
h_{ab}(u^{\gamma}, v^{\gamma}) = \frac{g^2}{\sigma_{ab}(u^{\gamma},v^{\gamma})}
\frac{\left[ \frac{(a+b)^2}{4}+(u-v)^2 \right]}{\left( \frac{a^2}{4}+u^2 \right)
\left( \frac{b^2}{4}+v^2 \right)}
\times\\\times
\frac{
\mathrm{\Gamma}[-\frac{a}{2}-iu]\,
\mathrm{\Gamma}[\frac{a+b}{2}-i(u-v)]\,
\mathrm{\Gamma}[-\frac{a+b}{2}+i(u-v)]\,
\mathrm{\Gamma}[\frac{b}{2}-iv]
}{
\mathrm{\Gamma}[\frac{a}{2}-iu]\,
\mathrm{\Gamma}[\frac{b-a}{2}-i(u-v)]\,
\mathrm{\Gamma}[\frac{b-a}{2}+i(u-v)]\,
\mathrm{\Gamma}[-\frac{b}{2}-iv]
}\,,
\end{multline}
with
\begin{equation}
\sigma_{ab}(u^{\gamma},v^{\gamma}) =
\frac{
\mathrm{\Gamma}[1-\frac{a}{2}+iu]\,
\mathrm{\Gamma}[1+\frac{a-b}{2}-i(u-v)]\,
\mathrm{\Gamma}[1+\frac{b}{2}-iv]
}{
\mathrm{\Gamma}[1+\frac{a}{2}-iu]\,
\mathrm{\Gamma}[1+\frac{b-a}{2}+i(u-v)]\,
\mathrm{\Gamma}[1-\frac{b}{2}+iv]
}\,.
\end{equation}
Using the expressions above, one can easily deduce that
\begin{equation}
h_{ab}(u^{\gamma},v^{\gamma})\,
h_{ba}(v^{\gamma},u^{\gamma}) =
g^4 \frac{
\lrsbrk{(u-v)^2+\frac{(a+b)^2}{4}}
\lrsbrk{(u-v)^2+\frac{(a-b)^2}{4}}
}{
\lrbrk{u^2+\frac{a^2}{4}}^2
\lrbrk{v^2+\frac{b^2}{4}}^2
}\,.
\end{equation}

\section[Mirror Particle Contributions: Integrability Calculation]{%
Mirror Particle Contributions:
\texorpdfstring{\newline}{}
Integrability Calculation}
\label{app:mirrorparticles}

In this Appendix, we provide details of the
integrability calculation
of the mirror particle contributions used in the main text.
We start considering the one-particle contribution in a zero-length edge,
then we consider the case of two particles in the same edge,
and finally the three-particle contribution involving three hexagons
at one-loop is considered.

\subsection{One-Particle Contribution with \texorpdfstring{$l=0$}{l=0}}

Consider the hexagons $\mathcal{H}_1$ formed by the operators
$\mathcal{O}_{1,4,3}$ and $\mathcal{H}_2$ formed by the
operators $\mathcal{O}_{1,2,4}$, as on the left in
\figref{fig:TheTwoParticleSameEdge}.
The integrand for the one-particle
mirror contribution for gluing the edge 1--4 with $l_{14}=0$ was
given in~\cite{\Hexagonalizationone}. It reads
\begin{equation}
{\rm{int}}^a_{14}(v) = \frac{2 ({\rm{cos}}\phi - {\rm{cosh}}( i \varphi) \, {\rm{cos}} \theta)
\, {\rm{sin}} \, a \phi}{{\rm{sin}} \phi} \mu_a(v^{\gamma})e^{- 2 i \tilde{p}_a (v) {\rm{log}}|z|} \, ,
\end{equation}
where (the cross-ratio $z$ and the $R$-charge cross-ratio $\alpha$ are
defined in~\eqref{eq:zdef} and~\eqref{eq:alphadef})
\begin{equation}
    \phi = -\frac{i}{2} \log\lrbrk{\frac{z}{\bar{z}}} \, , \quad
  \theta = -\frac{i}{2} \log\lrbrk{\frac{\alpha}{\bar{\alpha}}} \, , \quad
i\varphi =  \frac{1}{2} \log\lrbrk{\frac{\alpha\bar{\alpha} }{z\bar{z}}}
\,.
\label{eq:anglesdefinition}
\end{equation}
Using the weak coupling expansions given in~\eqref{eq:weakmeasure}
and~\eqref{eq:weakmomentum}, one can find the integrand up to order $g^6$.
The integral is done by residues and
one gets the one- and two-loop results used in our first paper
\cite{Bargheer:2017nne} and
the three-loop result given in
(\ref{eq:OneParticleThreeLoop}).

\subsection{Two Particles in the Same \texorpdfstring{$l=0$}{l=0} Mirror Edge}

\begin{figure}[t]
\centering
\includegraphics[scale=0.6]{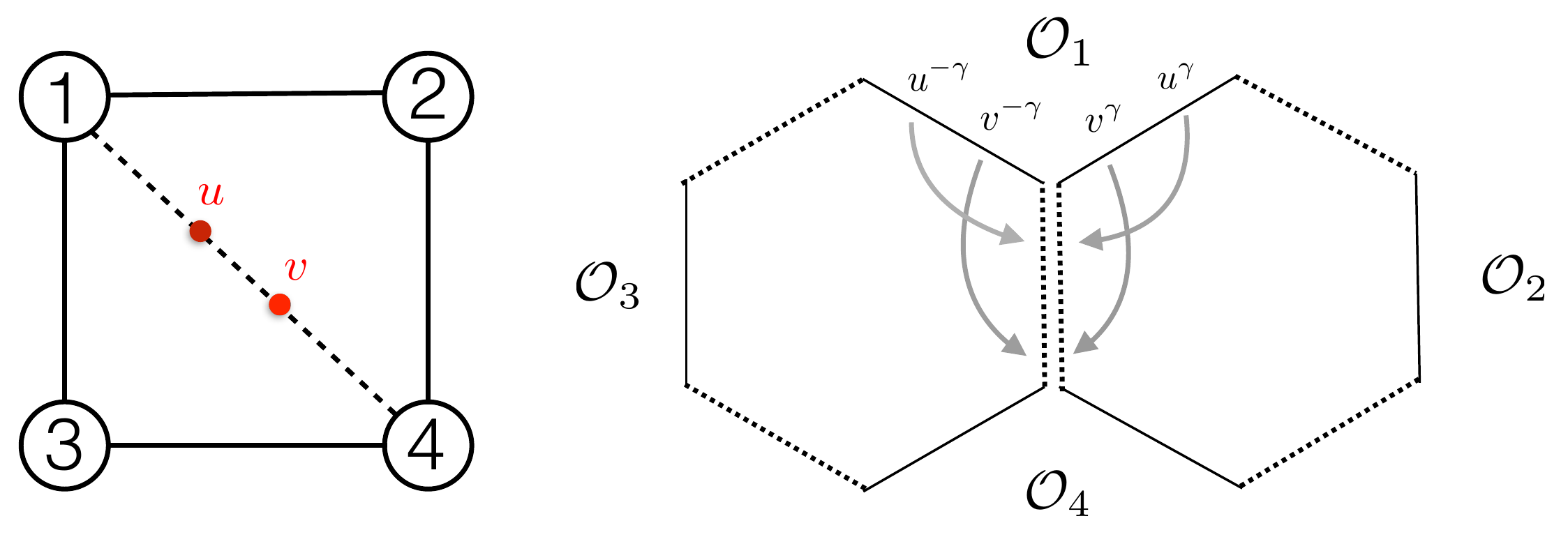}
\caption{The contribution of two particles in the same $l=0$
mirror edge.
There are two hexagons involved, and
we call them left and right hexagons.
By an explicit calculation,
one verifies that it only contributes at four loops and beyond.}
\label{fig:TheTwoParticleSameEdge}
\end{figure}

This subsection is devoted to the computation
of the two-particle contribution in the same
mirror edge shown in \figref{fig:TheTwoParticleSameEdge}.
It will be shown in particular that it contributes
only at four loops.
Recall that the $a$-th mirror bound state $\mathcal{X}_a$
is composed from the
tensor product of two
factors belonging to the $a$-th antisymmetric
representation of
$\mathfrak{su}(2|2)$. A basis for this
representation is $(\alpha_i = 3,4)$
\begin{equation}
\begin{aligned}
& | \psi_{\alpha_1} \ldots \psi_{\alpha_a} \rangle +
\ldots \, , \quad \quad
\; \, \, \quad \; \, \,
 | \phi_1 \psi_{\alpha_1} \ldots
 \psi_{\alpha_{a-1}} \rangle + \ldots \, , \\
& | \phi_2 \psi_{\alpha_1} \ldots \psi_{\alpha_{a-1}} \rangle + \dots \, ,
\quad
\quad \; \, | \phi_1 \phi_2 \psi_{\alpha_1}
\ldots \psi_{\alpha_{a-2}} \rangle
+ \ldots
\,,
\quad
\end{aligned}
\label{eq:boundstatebasis}
\end{equation}
where $(\phi_1,\phi_2,\psi_3,\psi_4)$ form an $\alg{su}(2|2)$
fundamental multiplet, $a$ is called the bound state index, and the
dots stand for permutations.
As discussed in~\cite{\Hexagonalizationone,\Hexagonalizationtwo},
the basis above has to be modified for the
hexagonalization procedure to reproduce the
perturbative data.
It is necessary to add so-called $Z$-markers to some
of the basis states, and the
prescription used here follows from
the one given
in the appendix A of~\cite{\Hexagonalizationtwo}.
The addition of $Z$-markers has two consequences:
They give a contribution
to the weight factors,
and when one
moves and removes them using the rules given in~\cite{\HexagonPaper},
one can get factors of momenta.
Note that a rigorous explanation for the $Z$-marker
prescription is still lacking.
The dressing of the basis states is as follows
(the bar denotes antiparticles)
\begin{equation}
\underbrace{
| Z^{-t^I_u} \bar{\mathcal{X}}^I_a(u^{-\gamma})
Z^{-t_v^J} \bar{\mathcal{X}}^J_b(v^{-\gamma})
\rangle
}_{\text{left hexagon}}
\otimes
\underbrace{
|\mathcal{X}^J_b(v^{\gamma})
Z^{t^J_v}
\mathcal{X}^I_a(u^{\gamma}) Z^{t_u^I}
\rangle
}_{\text{right hexagon}}
\,,
\hspace{-0.4mm}
\label{eq:TheDressTwoParticleSameEdge}
\end{equation}
where $\mathcal{X}_a^I(u)$ is a mirror magnon with bound
state index $a$ and rapidity $u$,
with $I$ being a (flavor) index for the $a$-th bound state representation,
and $\gamma$ denotes the mirror transform that
transports excitations from one edge of the hexagon to the next.
The values of $t^I_u$ and $t^J_v$ depend
on the field content of the bound-state basis elements,
and whether one is considering the ``$+$'' or ``$-$''
dressing. The rules to find the
values of the $t_i$ are:
\begin{align}
\text{``$+$'' } &\text{dressing}: &
\psi_{\alpha}       &\rightarrow \psi_{\alpha} \,, &
\phi_1              &\rightarrow Z^{\frac{1}{2}} \phi_1 \,, &
\phi_2              &\rightarrow Z^{-\frac{1}{2}} \phi_2 \,,
\nn\\ &&
\psi_{\dot{\alpha}} &\rightarrow \psi_{\dot{\alpha}} \,, &
\phi_{\dot{1}}      &\rightarrow Z^{-\frac{1}{2}} \phi_{\dot{1}} \,, &
\phi_{\dot{2}}      &\rightarrow Z^{\frac{1}{2}} \phi_{\dot{2}} \,,
\nn\\[1.5ex]
\text{``$-$'' } &\text{dressing}: &
\psi_{\alpha}       &\rightarrow \psi_{\alpha} \,, &
\phi_1              &\rightarrow Z^{-\frac{1}{2}} \phi_1 \,, &
\phi_2              &\rightarrow Z^{\frac{1}{2}} \phi_2 \,,
\nn\\ &&
\psi_{\dot{\alpha}} &\rightarrow \psi_{\dot{\alpha}} \,, &
\phi_{\dot{1}}      &\rightarrow Z^{\frac{1}{2}} \phi_{\dot{1}} \,, &
\phi_{\dot{2}}      &\rightarrow Z^{-\frac{1}{2}} \phi_{\dot{2}} \,,
\label{eq:ZmarkersPrescription}
\end{align}
where undotted/dotted labels are left/right $\alg{su}(2|2)$
fundamental indices,
and the prescription is to average over
the two different dressings at the end of the
calculation. Within a hexagon form factor $\mathfrak{h}$, one can move
all $Z$-markers to the left, and then remove
them via the rules (see Appendices~C and~F in~\cite{Basso:2015zoa})
\begin{equation}
\chi Z\simeq e^{ip}Z\chi
\,,\qquad
\vev{\mathfrak{h}|Z^n\Psi}=z^n \vev{\mathfrak{h}|\Psi}
\,,\qquad
z^2=e^{-ip}
\,,
\end{equation}
where $\chi$ is a fundamental magnon, and $\Psi$ is a generic
spin-chain state. When removing all $Z$-markers in this way,
it is possible to show that for any value of
$t^I_u$ and $t^J_v$, all momentum factors $e^{ip}$
cancel each other.

The hexagon form factors are matrices in flavor space.
In what follows, we are going to work in the string
frame, where the non-vanishing
components of the one-particle hexagon form factors
are
\begin{equation}
\mathfrak{h}^{1 \dot{2}} = - \mathfrak{h}^{2 \dot{1}}
=1 \, , \quad \quad
\mathfrak{h}^{3 \dot{4}} = - \mathfrak{h}^{4 \dot{3}}
=-i \, .
\label{eq:OneParticleHexagon}
\end{equation}
The contribution from two particles in the same
zero-length bridge is the result
of the following integral
\begin{multline}
\mathcal{M}^{(2)}_{\rm{same \, edge}} (z, \alpha) =
\int \frac{d u}{2 \pi} \frac{d v}{2 \pi}
\sum_{a=1}^{\infty} \sum_{b=1}^{\infty}
\sum_{I,J}
\mu_a(u^{\gamma}) \, \mu_b(v^{\gamma})
\times\\\times
\mathfrak{h} \left[ \bar{\mathcal{X}}^I_a(u^{-\gamma}) \bar{\mathcal{X}}^J_b(v^{-\gamma}) \right]
\mathcal{W} \left[ \mathcal{X}^J_b(v^{\gamma}) \right] \mathcal{W} \left[ \mathcal{X}^I_a(u^{\gamma}) \right]
\mathfrak{h} \left[ \mathcal{X}^J_b(v^{\gamma})
\mathcal{X}^I_a(u^{\gamma}) \right]\, ,
\end{multline}
where the $\mu$'s are the measure factors, and the $\mathcal{W}$'s are weight
factors associated to the particles whose origin is a
$\grp{PSU}(2,2|4)$ transformation that aligns the frames of the two hexagons~\cite{Fleury:2016ykk}.
In order to simplify the calculation of the matrix part (flavor sums),
it is convenient to use the following identity
to have both hexagon form factors
with the same crossed arguments:
\begin{equation}
\mathfrak{h} \left[ \bar{\mathcal{X}}^I_a(u^{-\gamma}) \bar{\mathcal{X}}^J_b(v^{-\gamma}) \right]
=
(-1)^{\bar{I}}
(-1)^{\bar{J}}
\mathfrak{h} \left[
\bar{\mathcal{X}}^{I \mathrm{c}}_a(u^{\gamma})
\bar{\mathcal{X}}^{J \mathrm{c}}_b(v^{\gamma})
\right]
\, ,
\end{equation}
where the superscript $\mathrm{c}$
indicates that the indices $A$ and $\dot{A}$ of the excitations are swapped.
The precise values for the signs can be deduced
from the crossing rules~\cite{\HexagonPaper,Caetano:2016keh}
\begin{equation}
\chi^{ a \dot{b}} \xrightarrow{\;2\gamma\;}
\chi^{b \dot{a}} \, , \quad
\chi^{ \alpha \dot{\beta}} \xrightarrow{\;2\gamma\;} -
\chi^{\beta \dot{\alpha}} \, , \quad
\chi^{ \alpha \dot{a}} \xrightarrow{\;2\gamma\;}
\chi^{a \dot{\alpha}} \, , \quad
\chi^{ a \dot{\alpha}} \xrightarrow{\;2\gamma\;} -
\chi^{\alpha \dot{a}}
\end{equation}
for fundamental magnons $\chi$.
In particular, one has
\begin{equation}
(-1)^{\bar{I}} = (-1)^{\sharp \, {\rm{scalars}}_{\bar{I}} \, + \,
\dot{f}_{\bar{I}}} \, ,
\end{equation}
with $\dot{f}_{\bar{I}}$ the number of fermionic
dotted indices.
The weight factor $\mathcal{W}$ was computed in~\cite{\Hexagonalizationone}, and it
was rewritten in~\cite{\Hexagonalizationtwo} taking both the
$Z$-markers prescription and the ``$+$'' and ``$-$'' dressings
into account as
\begin{equation}
\mathcal{W}^{\pm} [ \mathcal{X}_a^I (u^{\gamma})]
= e^{- 2 i \tilde{p}_a(u^{\gamma}) {\rm{log}} |z|}
\, e^{i L^I \phi} \, e^{i R^I (\theta \pm \varphi)} \, ,
\label{eq:WeightFactor}
\end{equation}
where the angles were defined in~\eqref{eq:anglesdefinition}, and the
eigenvalues $L^I$ and $R^I$ of the generators $L$ and $R$
can be deduced from the
action of these generators
on the fundamental excitations given by (the dotted
indices have opposite eigenvalues)
\begin{equation}
e^{ i L } \psi^1 = e^{\frac{i}{2}} \psi^1 \, , \quad
e^{ i L } \psi^2 = e^{-\frac{i}{2}} \psi^2 \, , \quad
e^{i R} \phi^1 = e^{\frac{i}{2}} \phi^1 \, , \quad
e^{i R} \phi^2= e^{- \frac{i}{2}} \phi^2 \, .
\label{eq:ChargesParticles}
\end{equation}
As a consequence of the formulas above, one
has
\begin{multline}
\mathcal{M}^{(2)}_{\text{same edge}}(z,\alpha)=
\int\frac{d u}{2\pi}\frac{d v}{2\pi}
\sum_{a=1}^{\infty} \sum_{b=1}^{\infty}
\mu_a(u^{\gamma})
\,\mu_b(v^{\gamma})
\,h_{ab}(u^{\gamma},v^{\gamma})
\,h_{ba}(v^{\gamma},u^{\gamma})
\times\\\times
\,e^{- 2 i \tilde{p}_a (u^{\gamma}) \log|z| }
\,e^{- 2 i \tilde{p}_b (v^{\gamma}) \log|z| }
\,\mathcal{F}_{a b}
\,.
\label{eq:M2inter}
\end{multline}
Here, $\mathcal{F}_{ab}$
contains the matrix part and
the flavor-dependent part of the weight factor. It is given by
\begin{multline}
\mathcal{F}_{ab} =
\sum_{I , J} \,
(-1)^{\bar{I}} (-1)^{\bar{J}} \,
\mathcal{W}^{\pm}_{\text{flavor}}
\left[ \mathcal{X}^J_b(v^{\gamma}) \right] \mathcal{W}^{\pm}_{{\rm{flavor}}} \left[ \mathcal{X}^I_a(u^{\gamma}) \right]
\times\\\times
\langle \chi_b^{\bar{J}_a} (v^{\gamma})
\chi_a^{\bar{I}_a} (u^{\gamma})
| S | \chi_a^{\bar{I}_b}(u^{\gamma}) \chi_b^{\bar{J}_b} (v^{\gamma}) \rangle \,
\langle \chi_a^{I_b} (u^{\gamma})
\chi_b^{J_b} (v^{\gamma})
| S | \chi_b^{J_a}(v^{\gamma}) \chi_a^{I_a} (u^{\gamma}) \rangle
\,,
\end{multline}
with $S$ being the mirror bound state
$S$-matrix~\cite{Fleury:2017eph}.
In principle, one can use
the unitarity of the $S$-matrix to simplify
the expression above, however one has to
check that the weight factors do not spoil this
simplification. Indeed, the $S$-matrix
has a block-diagonal
decomposition~\cite{Arutyunov:2008zt,Arutyunov:2009mi,\Hexagonalizationtwo},
and fixing the indices $J_a$ and $I_a$, one can
show that the resulting states, after the action
of the $S$-matrix, have a non-vanishing inner product
only with definite weight-factor eigenstates,
so indeed unitarity can be used.
As an example, let us
select a particular value of $J_a$ and $I_a$ to have the case
Ia of~\cite{\Hexagonalizationtwo}, \ie we have for some $k$ and $l$ that
\begin{equation}
| \chi_b^{J_a}(v^{\gamma}) \chi_a^{I_a} (u^{\gamma})
\rangle^{\rm{Ia}} = | k, l \rangle^{\rm{Ia}} \, , \quad
{\rm{with}} \quad
| k, l \rangle^{\rm{Ia}} =
| \phi_1 \psi_1^{b-k-1} \psi_2^k \rangle
\otimes | \phi_1 \psi_1^{a-l-1} \psi_2^l \rangle
\, .
\end{equation}
The $S$-matrix, when acting on a state of type Ia,
produces a linear combination
of states of type Ia of the form ($N=k+l$)
\begin{equation}
S \cdot |k, l \rangle^{{\rm{Ia}}}
= \sum_{n=0}^{N} H^{k,l}_n \, |N-n,n \rangle^{{\rm{Ia}}} \, .
\end{equation}
As a consequence of the equation above, all final states
have precisely two $\phi_1$'s and the same total number
of $\psi_1$'s and $\psi_2$'s. Thus they have
non-zero inner products only with
definitive weight-factor eigenstates, and this
selects only a particular non-trivial set of values for
$J_b$ and $I_b$.
Using the unitarity of the $S$-matrix, we have
\begin{equation}
\mathcal{F}_{ab} = \frac{1}{2} \, (-1)^a (-1)^b \, (T^{+}_a T^{+}_b +
T^{-}_a T^{-}_b) \, ,
\end{equation}
where the factor of $1/2$ is present because we are
averaging between the ``$+$'' and ``$-$'' dressings, and the $T^{\pm}$ are twisted transfer matrices given by
\begin{equation}
T^{\pm}_a = 2 (-1)^a \bigbrk{ {\rm{cos}} \, \phi - {\rm{cos}} (\theta \pm \varphi)} \frac{{\rm{sin}} \, a \phi}{{\rm{sin}} \phi} \, .
\end{equation}
Notice that the twisted transfer matrices are defined by
\begin{multline}
T^{\pm}_a = {\rm{Tr}}_a \left[ (-1)^F e^{i L \phi} e^{i R (\theta \pm \varphi)} \right]=
\\
(-1)^a \left( e^{i a \phi} \sum_{n=0}^a e^{- 2 i n \phi} -
2 \, {\rm{cos}} (\theta \pm \varphi)
e^{i (a-1) \phi} \sum_{n=0}^{a-1} e^{- 2 i n \phi}
+ e^{i (a-2) \phi} \sum_{n=0}^{a-2} e^{-2 in \phi} \right)
\, .
\end{multline}
Substituting the expression for $\mathcal{F}_{ab}$ in~\eqref{eq:M2inter},
it only remains to evaluate the integral. Using
the weak-coupling expansions given in \appref{app:WeakCouplingApendix},
it is easy to see that this contribution contributes only at
four loops.
The integral is easily evaluated by residues and
at order $g^8$ it gives
\begin{multline}
\mathcal{M}^{(2)}_{\rm{same \, edge}} (z, \alpha) \Big|_{g^8}
= - 2 \left[ \frac{(z-\alpha)^2 (\bar{z}-\alpha)^2}{\alpha^2}
+ \frac{(z-\bar\alpha)^2 (\bar{z}-\bar\alpha)^2}{\bar\alpha^2}
\right]
\times\\\times
\left(
\frac{1}{6} \lrbrk{F^{(2)}(z, \bar{z})}^2 - \frac{1}{2} F^{(1)}(z, \bar{z})\, F^{(3)}(z, \bar{z}) \right) \, ,
\end{multline}
where $F^{(1)}$, $F^{(2)}$ and $F^{(3)}$ were given
in~\eqref{eq:box}, \eqref{eq:doublebox}, and~\eqref{eq:Ladder3}. Another
representation for $F^{(L)}$ is
\begin{equation}
F^{(L)}(z, \bar{z}) = \frac{1}{z - \bar{z}}
\left[ \sum_{m=0}^L \frac{ (-1)^m (2L-m)!}{L! (L-m)! m!} \, {\rm{log}}^m (z \bar{z}) ({\rm{Li}}_{2 L -m} (z) - {\rm{Li}}_{2 L -m} (\bar{z})) \right] \, .
\label{eq:FintermsofPolylogs}
\end{equation}

\subsection{The Three-Particle Contribution}

\begin{figure}
\begin{center}
\includegraphics[clip,height=10cm]{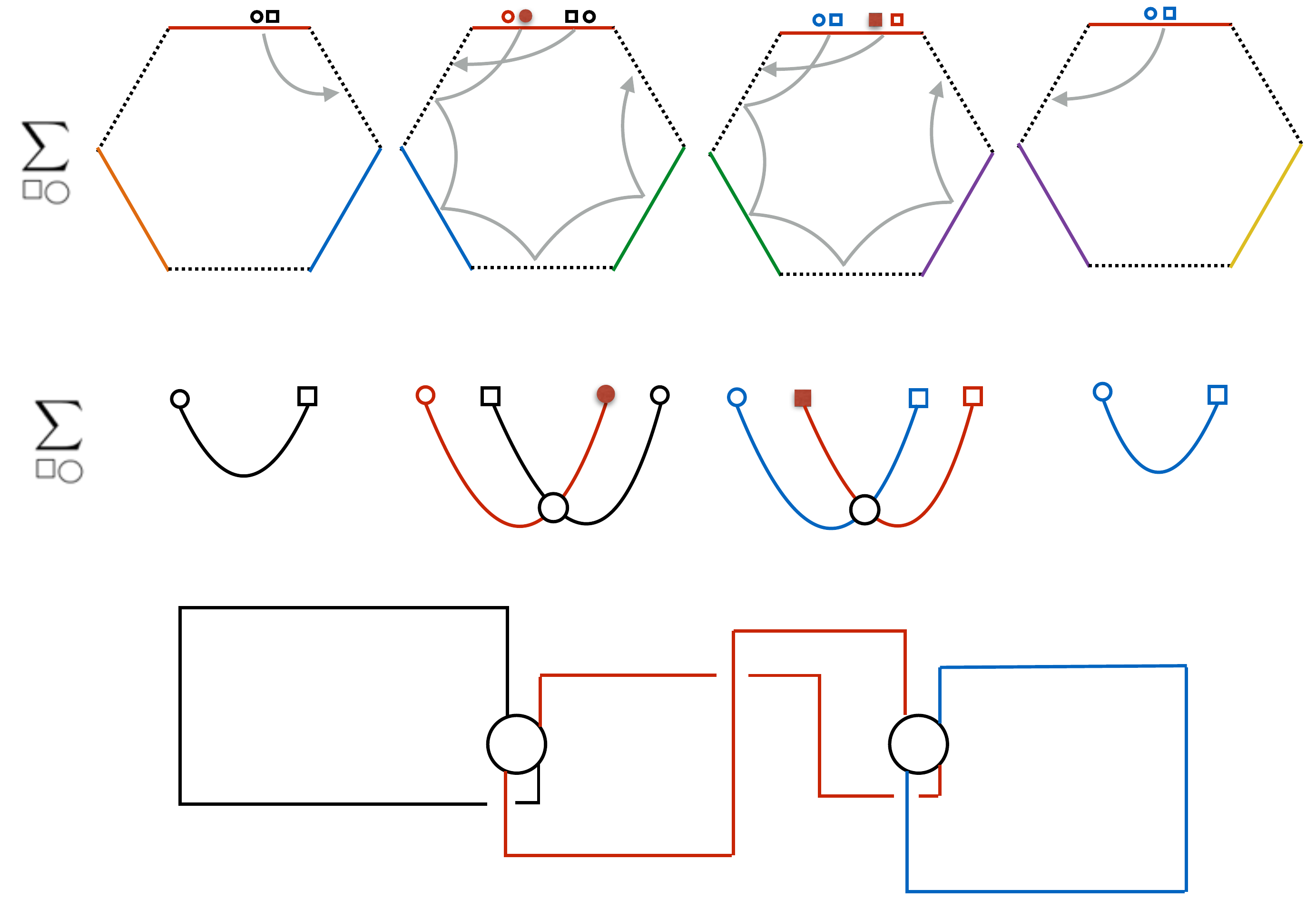}
\end{center}
\vspace{-0.7cm}
\caption{The figure at the top
represents the four hexagons involved in
the three-particle computation.
The circles and the corresponding squares represent
complementary sets of fundamental indices, \ie if one of them equals $1$, then
the other equals $2$, if one equals $3$, the other equals $4$, and
vice versa, see~\cite{Basso:2015zoa}.
The hexagons are glued together using three weight factors $\mathcal{W}$ (not shown in the figure).
We have chosen to rotate some of the
particles of the second and third hexagons by sequences of mirror
transformations.
The figure in the middle represents the contractions
of the flavor indices, and the white circle with four lines denotes a
mirror bound state $S$-matrix. The last figure
schematically shows the sum over the indices denoted
by circles and squares. The sum is not a straight trace, but rather is
weighted by the three weight factors $\mathcal{W}$. We restrict
ourselves to operators that lie in a common plane; in this case the
weight factors are diagonal in the mirror state space.
The result of the last figure is proportional
to the three-particle matrix part. Note
that it involves two mirror bound state $S$-matrices, and, unlike
in the two-particle contribution, the sum represented by the red lines
includes elements that are non-diagonal in the $\alg{su}(2|2)$
preserved by the hexagon.}
\label{fig:TheThreeParticle2}
\end{figure}

Next, we compute the three-particle
contribution,
shown in both \figref{fig:LimitsFig} and \figref{fig:TheThreeParticle2},
using
integrability. Note that this is a particular kind of
three-particle contribution,
as one can flip the line connecting the operators at
positions $x_1$
and $x_5$, such that it connects the operators at position $x_6$
and $x_4$ instead.
These two kinds of three-particle contributions are
related by flipping
invariance, and it is possible to deduce one from
the other.

To compute the three-particle contribution, it is
necessary to evaluate
four hexagon form factors $\mathfrak{h}$, to use
three weight factors $\mathcal{W}$
for gluing the hexagons
together,
and to sum over three mirror bound-state
basis elements $\mathcal{X}^I$, whose
bound state indices are going to be
denoted by $a$, $b$, and $c$.
The three-particle contribution is given by the integral
\begin{multline}
\label{eq:TheExpressionM3}
\mathcal{M}^{(3)}(z_1,z_2,z_3,\alpha_1,\alpha_2,\alpha_3)=
\\
\sum_{a=1}^{\infty}
\sum_{b=1}^{\infty}
\sum_{c=1}^{\infty}
\int\frac{d u_1}{2\pi}\frac{d u_2}{2\pi}
\frac{d u_3}{2 \pi}
\,\mu_a(u_1^{\gamma})
\,\mu_b(u_2^{\gamma})
\,\mu_c(u_3^{\gamma})
\sum_{I,J,K}
\mathfrak{h}[\bar{\mathcal{X}}_c^I(u_3^{-\gamma})]
\,\mathcal{W}[\mathcal{X}_c^I(u_3^{\gamma})]
\times\mspace{60mu}\\\times
\mathfrak{h}[\mathcal{X}_c^I(u_3^{\gamma})\bar{\mathcal{X}}_b^J(u_2^{-\gamma})]
\,\mathcal{W}[\mathcal{X}^J_b(u_2^{\gamma})]
\,\mathfrak{h}[\mathcal{X}_b^J(u_2^{\gamma})\bar{\mathcal{X}}_a^K(u_1^{-\gamma})]
\,\mathcal{W}[\mathcal{X}_a^K(u_1^{\gamma})]
\,\mathfrak{h}[\mathcal{X}_a^K(u_1^{\gamma})]
\,.
\end{multline}
A naive basis, \ie without the $Z$-markers,
for the $a$-th mirror bound state $\mathcal{X}$
is given in~\eqref{eq:boundstatebasis}. The dressing of the states by
$Z$-markers with exponents $t_i$ appearing below
are found using the rules of~\eqref{eq:ZmarkersPrescription} .
Notice the values of $t^I_1$, $t^J_2$ and $t^K_3$ depend
on the field content of the bound state basis elements,
and on whether one is considering the ``$+$'' or the ``$-$''
dressing. We have
\begin{equation}
\underbrace{| Z^{- t^I_3} \bar{\mathcal{X}}
_c^I
(u_3^{- \gamma}) \rangle}_ {\rm{left \, hexagon}}
\hspace{-0.4mm} \otimes \hspace{-0.4mm} \underbrace{|\mathcal{X}^I_c(u_3^{\gamma}) Z^{t^I_3}
Z^{-t_2^J} \bar{\mathcal{X}}^J_b(u_2^{-\gamma})
\rangle}_{\rm{second \, hexagon}}
\hspace{-0.4mm} \otimes \hspace{-0.4mm} \underbrace{|\mathcal{X}^J_b(u_2^{\gamma})
Z^{t^J_2}
Z^{-t_1^K}
\bar{\mathcal{X}}_a^K(u_1^{-\gamma}) \rangle}_{\rm{third
\, hexagon}}
\hspace{-0.4mm} \otimes
\hspace{-0.4mm} \underbrace{|\mathcal{X}_a^K(u_1^{\gamma})
Z^{t_1^K} \rangle}_{\rm{right \, hexagon}} \, .
\end{equation}
Moving all $Z$-markers to the left
and removing them, one gets some
non-trivial factors of momenta that will
contribute to the integrand. The expression
above is equal to
\begin{multline}
e^{i t^I_3 p(u_2^{\gamma})/2}
e^{- i t^J_2 p(u_3^{\gamma})/2}
e^{ i t^J_2 p(u_1^{\gamma})/2}
e^{- i t^K_1 p(u_2^{\gamma})/2}
\,\times\\\times
\underbrace{| \bar{\mathcal{X}}^I_c (u_3^{- \gamma}) \rangle}_{\rm{left \, hexagon}}
\otimes \underbrace{|\mathcal{X}^I_c(u_3^{\gamma}) \bar{\mathcal{X}}^J_b(u_2^{-\gamma}) \rangle}_{\rm{second \, hexagon}}
\otimes \underbrace{|\mathcal{X}^J_b(u_2^{\gamma}) \bar{\mathcal{X}}^K_a(u_1^{-\gamma}) \rangle}_{\rm{third \, hexagon}}
\otimes \underbrace{|\mathcal{X}^K_a(u_1^{\gamma}) \rangle}_{\rm{right \, hexagon}} \, .
\end{multline}
A mirror particle-antiparticle pair
is always created on a mirror
edge shared by two
hexagons. The particle is absorbed by
one of the hexagons, the antiparticle
is absorbed by the other hexagon.
The weight factor originates in the
symmetry transformation needed to bring both hexagons to the same
frame---this transformation acts non-trivially on the mirror
particles as one moves them from one hexagon to the other.
The expression for the weight factor
was given in~\eqref{eq:WeightFactor}, here we give its
expression for the case with more cross ratios
\begin{equation}
\mathcal{W}^{\pm}
[ \mathcal{X}^{I}_{a_i}(u_i^{\gamma})]=
e^{- 2 i \tilde{p}_{a_i}(u_i) {\rm{log}}|z_i|}
e^{i L^I \phi_i} e^{i R^I (\theta_i \pm \varphi_i)}
\, ,
\end{equation}
with
\begin{equation}
e^{i \phi_i} = \sqrt{\frac{z_i}{\bar{z}_i}} \, ,
\quad \quad e^{i \theta_i } = \sqrt{\frac{\alpha_i}{
\bar{\alpha}_i}} \, , \quad \quad
e^{i \varphi_i} = \sqrt{\frac{\alpha_i \bar{\alpha}_i}
{z_i \bar{z}_i}} \, ,
\end{equation}
and the charges of the fundamental excitations
under the generators $L$ and $R$ are given
in~\eqref{eq:ChargesParticles}.

In the expression~\eqref{eq:TheExpressionM3},
the hexagon form factors
corresponding to the left and right
hexagons only have one excitation.
These hexagons have a trivial dynamical part and
they contribute only with a possible sign that
can be computed using a combination of
the one-particle hexagon form factors given in~\eqref{eq:OneParticleHexagon}.
In addition, they imply that
the excitations with rapidities $u_1$ and $u_3$
are both composed of transverse excitations
only, and that their states
are not changed by the scattering with the
particles with rapidity $u_2$.
As a matter of choice,
we are going to mirror-rotate the two
middle hexagon form factors before evaluating
them. One has for the non-zero cases
\begin{equation}
\mathfrak{h} [
\mathcal{X}_c^I (u_3^{\gamma} )
\bar{\mathcal{X}}_b^{J} (u_2^{-\gamma})] =
(-1)^{\sharp \, {\rm{scalars}}_{\bar{J}} \, + \,
f_{\bar{J}}}
\, \mathfrak{h}[\bar{\mathcal{X}}_b^{Jc} (u_2^{ 5 \gamma})
\mathcal{X}_c^I (u_3^{\gamma} )] \, ,
\end{equation}
with $f_{\bar{J}}$ the number of undotted fermionic indices
in the set $\bar{J}$, and
\begin{equation}
\mathfrak{h}[
\mathcal{X}_b^J (u_2^{\gamma})
\bar{\mathcal{X}}_a^K(u_1^{-\gamma})] =
(-1)^a \, \mathfrak{h}[
\bar{\mathcal{X}}_a^{Kc}(u_1^{5\gamma}) \mathcal{X}_b^J (u_2^{\gamma})] \, .
\end{equation}
Notice that an important property of the dynamical
factor of the hexagons that will be used below is
\begin{equation}
h_{ab}(u^{5 \gamma}, v^{\gamma})=
\frac{1}{ h_{ba}( v^{\gamma}, u^{\gamma} ) } \, .
\end{equation}
Collecting the expressions above, we have
\begin{multline}
\mathcal{M}^{(3)}(z_1,z_2,z_3,\alpha_1,\alpha_2,
\alpha_3) =
\sum_{a=1}^{\infty}
\sum_{b=1}^{\infty}
\sum_{c=1}^{\infty}
\int
\frac{d u_1}{2\pi}
\frac{d u_2}{2\pi}
\frac{d u_3}{2\pi}
\frac{\mu_a(u_1^{\gamma})
\,\mu_b(u_2^{\gamma})
\,\mu_c(u_3^{\gamma})
}{
  h_{cb}(u_3^{\gamma},u_2^{\gamma})
\,h_{ba}(u_2^{\gamma},u_1^{\gamma})
}
\times\\\times
  e^{- 2 i \tilde{p}_a (u_1^{\gamma}) {\log|z_1|} }
\,e^{- 2 i \tilde{p}_b (u_2^{\gamma}) {\log|z_2|} }
\,e^{- 2 i \tilde{p}_c (u_3^{\gamma}) {\log|z_3|} }
\, \frac{1}{2} \, \mathcal{F}_{abc}
\,,
\label{eq:ThreeParticleIntegral}
\end{multline}
with%
\footnote{We have changed all the matrix
part arguments: $5\gamma \rightarrow \gamma$.
It seems at least at one-loop that possible additional signs
from a nontrivial monodromy of the $S$-matrix are not important.
Notice that every index breaks in two, for example $I \rightarrow
\{I_a, I_b \}$, however these two indices are related for the first
and the last hexagons to give a nonzero result, as they are fused
one-particle hexagon form factors.}
\begin{multline}
\mathcal{F}_{abc} = \sum_{I,J,K}
(-1)^{\sharp \, {\rm{scalars}}_{\bar{J}} \, + \, f_{\bar{J}}}
\,(-1)^{a}
\,e^{i t^I_3 p(u_2^{\gamma})/2}
\,e^{-i t^J_2 p(u_3^{\gamma})/2}
\,e^{ i t^J_2 p(u_1^{\gamma})/2}
\,e^{-i t^K_1 p(u_2^{\gamma})/2}
\,\times\\\times
  \mathfrak{h}[\bar{\mathcal{X}}_c^I(u_3^{-\gamma})]
\,\mathcal{W}^{\pm}[\mathcal{X}_c^I(u_3^{\gamma})]
\,\mathcal{W}^{\pm}[\mathcal{X}^J_b(u_2^{\gamma})]
\,\mathcal{W}^{\pm}[\mathcal{X}_a^K(u_1^{\gamma})]
\,\mathfrak{h}[     \mathcal{X}_a^K(u_1^{\gamma})]
\times\\\times
\langle
\chi_c^{\bar{I}_a} (u_3^{\gamma}) \chi_b^{\bar{J}_a} (u_2^{\gamma})
| S | \chi_b^{\bar{J}_b}(u_2^{\gamma}) \chi_c^{I_a} (u_3^{\gamma})
\rangle
\,\langle
\chi_b^{J_b} (u_2^{\gamma}) \chi_a^{\bar{K}_a} (u_1^{\gamma})
| S | \chi_a^{K_a}(u_1^{\gamma}) \chi_b^{J_a} (u_2^{\gamma})
\rangle
\,.
\end{multline}
The mirror bound-state $S$-matrix
using the ``hybrid'' convention%
\footnote{In the hybrid
convention, the supercharges schematically act
on the fundamental particles as $Q | \chi \rangle = | Z^{\frac{1}{2}}
\chi \rangle$ and $S | \chi \rangle = | Z^{-\frac{1}{2}} \chi
\rangle$. Notice that the powers of the $Z$-markers differ from both
the spin frame and string frame.}
was derived in~\cite{\Hexagonalizationtwo}
by adapting
the derivation of the physical bound-state $S$-matrix of~\cite{Arutyunov:2009mi}.
The $S$-matrix has a block-diagonal form, and
it can be organized into three cases, depending
on the values $(2,1,0,-1,-2)$ of the following charge (the superscripts
1 and 2 denotes the first and the second bound state being scattered)
\begin{equation}
C_1 = \sharp \phi_1^1 + \sharp \phi_1^2
- \sharp \phi_2^1 - \sharp \phi_2^2 \, .
\end{equation}
A basis
for each case can be found in~\cite{\Hexagonalizationtwo}, and they
are functions of two parameters $k$ and $l$ that
are related with the
number of fields $\psi_2$ within the bound states.
The $S$-matrices are denoted by $H$, $Y$, and $Z$
for the cases I, II, and III respectively.
Notice that the sum in $\mathcal{F}_{abc}$ has
many terms, and each term involves a product of two
$S$-matrix elements that, because of the sum
in $J$ corresponding to the $u_2$ rapidity,
can be diagonal or non-diagonal, see
\figref{fig:TheThreeParticle2}.
Some of the terms do not contribute at one-loop
order, and to select the ones that do contribute,
one has to analyze the dependence
of the $S$-matrix elements on $g^2$.
Using the results of~\cite{\Hexagonalizationtwo}, we have
in a particular basis
\begin{equation}
H^{k,l}_n(u^{\gamma},v^{\gamma}) \; \sim \; \mathcal{O}(1) \, , \quad \quad
Y^{k,l}_{n} (u^{\gamma},v^{\gamma}) \; \sim \; \left(
\begin{array}{cccc}
\order{1}           & \order{g}           & \order{g} & 0 \\
\order{g}           & \order{1}           & 0         & \order{g} \\
\order{\frac{1}{g}} & 0                   & \order{1} & \order{g} \\
0                   & \order{\frac{1}{g}} & \order{g} & \order{1}
\end{array} \right) \, ,
\end{equation}
and
\begin{equation}
Z^{k,l}_{n}(u^{\gamma},v^{\gamma})
\; \sim \;
\left(
\begin{array}{cccccc}
\order{1}           & \order{g}           & \order{g}           & \order{g^2} & \order{g^2} & \order{g^2} \\
\order{\frac{1}{g}} & \order{1}           & \order{1}           & \order{g}   & \order{g}   & \order{g} \\
\order{\frac{1}{g}} & \order{1}           & \order{1}           & \order{g}   & \order{g}   & \order{g} \\
\order{1}           & \order{\frac{1}{g}} & \order{\frac{1}{g}} & \order{1}   & \order{1}   & \order{1}  \\
\order{1}           & \order{g}           & \order{g}           & \order{g^2} & \order{1}   & \order{g^2} \\
\order{1}           & \order{g}           & \order{g}           & \order{g^2} & \order{g^2} & \order{1}
\end{array}
\right)
\,.
\end{equation}
As an example, let us evaluate
one of the contributing terms of $\mathcal{F}_{abc}$,
namely the term
proportional to $\alpha_1 \alpha_2 \alpha_3$
(where $\alpha_i$ are the internal cross ratios, defined as
in~\eqref{eq:zidef}).
One
can show that this term is obtained using the
$+$ dressing and it only involves diagonal
$S$-matrices elements.
We have at one-loop order
\begin{multline}
\mathcal{F}_{abc} \big|^{g^2}_{\alpha_1 \alpha_2 \alpha_3} =
- \sum_{m=0}^{a-1} \sum_{l=0}^{b-1} \sum_{k=0}^{c-1}
(-1)^{a+c} (-1)^{ b(a+c)} \sqrt{\frac{u_3^2 + \frac{c^2}{4}}{u_1^2 +
\frac{a^2}{4}}}
\times\\\times
\frac{1}{z_1 z_2 z_3} \left( \frac{z_1}{\bar{z}_1} \right)^{\frac{a}{2} - m}
 \left( \frac{z_2}{\bar{z}_2} \right)^{\frac{b}{2} - l}
 \left( \frac{z_3}{\bar{z}_3} \right)^{\frac{c}{2} - k}
H^{m,l}_m (u_1^{\gamma},u_2^{\gamma}) \,
H^{l,k}_l (u_2^{\gamma},u_3^{\gamma})
\,.
\end{multline}
After computing $\mathcal{F}_{abc}$ at one loop,
one has to perform the triple integration in~\eqref{eq:ThreeParticleIntegral}.
The integrand is singular for $a=b$ and/or $b=c$, because
it has a pole lying on the integration line. We regularize
the integral using the same $i \epsilon$ prescription used for the
two-particle calculation of~\cite{\Hexagonalizationtwo}, described there
in Appendix~E.
Basically, we close the contours of integration of $u_3$ and $u_2$
from below, and the contour of $u_1$ from above. With this
choice of contours, we do not get the
contributions from the poles $u_1=u_2$ and $u_2=u_3$,
due to the $i \epsilon$ prescription.
The integration can be done for chosen values of
$a$, $b$, and $c$, and this generates a power
series in $z_1$, $z_2$, and $z_3$. Taylor-expanding
the result for the three-particle contribution given in~\eqref{eq:3ptFinal}, one
can show that both series agree.

This concludes the
integrability derivation of the three-particle result.
As mentioned in the main text,
the same result can be derived assuming that
flip invariance holds. It would be very interesting
to integrate~\eqref{eq:ThreeParticleIntegral}
directly and get the full three-particle result instead
of its Taylor expansion. We leave this for future work.

\section{The Planar \texorpdfstring{$n$}{n}-Point Functions of BPS Operators and Non-1EI Graphs}
\label{app:NpointPlanar}

In this Appendix, we further test
the integrability formula for the $2n$-gon
given in~\eqref{eq:2n-gon}.
We start by comparing the integrability
result for the planar $n$-point functions
of length-two BPS operators in a plane
derived using this formula with the
perturbative data at one loop computed in~\cite{Drukker:2008pi}.
It will be shown that both results agree.
In principle, one can compare the results
for BPS operators of any length, however
the combinatorics for the general case
are complicated due to non-trivial cancellations among
the different contributions.
The argument for the general case
will be based on the relation between
the integrability results and the $\superN=2$
formulation of $\superN=4$ SYM that was also used
in~\cite{Eden:2017ozn}.
We end the Appendix with a discussion about non-one-edge-irreducible
graphs at one loop. These graphs were expected
to cancel among themselves for some cases, and they were excluded
from the calculations of four- and five-point
functions in~\cite{\Hexagonalizationone,\Hexagonalizationtwo}. In this paper,
the multi-particle mirror contributions were determined,
and it is now possible to compute the contributions from all graphs
without making any assumption.
It will be shown that this will imply a refinement of the prescription
of the sum over graphs of~\cite{\Hexagonalizationone}.

\subsection{The Correlation Functions of \texorpdfstring{$n$}{n}
BPS Operators}

\subsubsection{The Case of
\texorpdfstring{$n$}{n}
\texorpdfstring{$20^{\prime}$}{20'} Operators}

The connected planar one-loop correlation function of
$n$ BPS operators of lengths $k_i$ was computed
perturbatively in~\cite{Drukker:2008pi}. The result is
\begin{multline}
\langle \mathcal{O}_{k_1}(x_1)
\mathcal{O}_{k_2}(x_2) \ldots \mathcal{O}_{k_n}
(x_n) \rangle \Big|_{\rm{connected}}^{g^2}=
\\
\sum_{i,j,l,p} k_i k_j k_l k_p \;
D_{i j l p} \Disk\Bigsbrk{
\mathcal{O}_{k_i -1}
\mathcal{O}_{k_j -1}
\mathcal{O}_{k_l -1}
\mathcal{O}_{k_p -1};
\;\prod_{\mathclap{m \neq \{i,j,l,p\}}}^n \mathcal{O}_{k_m}
}
\,,
\label{eq:DrukkerGeneral}
\end{multline}
where the summation over $i,j,l,p$ is to be understood
as follows. For every set of four different
indices $\{i,j,l,p\}$, one has only three different
terms in the sum, precisely
$ijlp$, $iljp$ and $ijpl$. In addition,
$\Disk$ means the tree-level correlation function with
all the Wick contraction lines contained inside a
disk, and with the operators listed in the first argument
being inserted in the boundary of the disk, respecting
their cyclic order, and the operators
in the second argument inserted inside
the disk. In evaluating the function $\Disk$, one
also does not consider disconnected graphs.
Notice that the four operators at the boundary of the disk
are already connected to each other by interaction lines lying
outside of the disk, and this has to be taken into account
when classifying the disconnected graphs.
As an example,
the graph where all the operators at the boundary of the disk
are not contracted with the ones inside the disk
is disconnected.
Finally, using the definition of $m$ given in~\eqref{eq:mdefinition},
\begin{equation}
D_{i j l p} = 2 m(z_{ijpl}) d_{ip} d_{jl}
+ 2 m(z_{ipjl}) d_{ij} d_{lp} \, ,
\label{eq:TheDfunction}
\end{equation}
with the cross ratios $z_{ijlk}$ being defined as
\begin{equation}
z_{ijkl} \bar{z}_{ijkl} = \frac{x_{ij}^2 x_{kl}^2}{x_{ik}^2 x_{jl}^2} \, , \quad \quad
(1- z_{ijkl})(1-\bar{z}_{ijkl})
= \frac{x_{il}^2 x_{jk}^2}{x_{ik}^2 x_{jl}^2} \, .
\label{eq:TheCrosszijkl}
\end{equation}
Notice that the function $D$ is invariant
under both a reflection and a cyclic rotation
of its indices due to the properties of the function $m$ given
in~\eqref{eq:midentities}.
This is consistent with the fact
that there are only three terms in the
summation~\eqref{eq:DrukkerGeneral}
for every set of four indices.

Here, we are going to consider the restriction
of the general formula~\eqref{eq:DrukkerGeneral}
to $k_i=2$ for all $i=1,\ldots,n$.
Moreover, in order to compare the perturbative result
with the integrability result, it is enough to
consider the contribution to the sum
in~\eqref{eq:DrukkerGeneral} coming from
a definite set of four indices,
say $\{1,2,3,4\}$ for definiteness.
For this set of indices,
we have that the sum on the right-hand side gives
\begin{multline}
\text{RHS of~\eqref{eq:DrukkerGeneral}} \big|_{\{1,2,3,4\}} =
2^4 \, D_{1 2 3 4} \Disk \Bigsbrk{
\mathcal{O}^1_{1}
\mathcal{O}^2_{1}
\mathcal{O}^3_{1}
\mathcal{O}^4_{1};
\;\prod^n_{\mathclap{m \neq \{1,2,3,4\}}} \mathcal{O}^m_{2}
} +
\\
+2^4 \, D_{1 3 2 4} \Disk \Bigsbrk{
\mathcal{O}^1_{1}
\mathcal{O}^3_{1}
\mathcal{O}^2_{1}
\mathcal{O}^4_{1};
\;\prod^n_{\mathclap{m \neq \{1,2,3,4\}}} \mathcal{O}^m_{2}
}
+2^4 \, D_{1 2 4 3} \Disk \Bigsbrk{
\mathcal{O}^1_{1}
\mathcal{O}^2_{1}
\mathcal{O}^4_{1}
\mathcal{O}^3_{1};
\;\prod^n_{\mathclap{m \neq \{1,2,3,4\}}} \mathcal{O}^m_{2}
} \, .
\label{eq:Drukker1234first}
\end{multline}
The Disk correlation functions appearing above can be computed in a closed form. By the definition
of the Disk function, one has to consider only connected
planar correlators, and there are two distinct cases
that
have to be considered. Firstly, two neighboring boundary
operators can contract, and the remaining operators
form a string starting and ending on the remaining boundary operators. Secondly, it is possible
to have two separate strings starting and ending on two
neighboring boundary operators. We have, for example
\begin{multline}
\Disk \Bigsbrk{
\mathcal{O}^1_{1}
\mathcal{O}^2_{1}
\mathcal{O}^3_{1}
\mathcal{O}^4_{1};
\;\prod^n_{\mathclap{m \neq \{1,2,3,4\}}} \mathcal{O}^m_{2}
} =
\; 2^{(n-4)}
\mspace{-50mu}
\sum_{\substack{i,j,k,l\,\in\,\brc{1,2,3,4}\\(i,j)\text{ and }(k,l)\text{ neighbors}}}
\;\sum_{\substack{\text{permutations}\\\sigma\text{ of }\brc{5,\dotsc,n}}}
\biggsbrk{
d_{ij} d_{k \sigma(5)} d_{\sigma(5)\sigma(6)} \dotsc d_{\sigma(n) l}
\\
+ \sum_{p=5}^{n-1}
d_{i \sigma(5)} d_{\sigma(5) \sigma(6)} \ldots
d_{\sigma(p) j}
d_{k \sigma(p+1)} d_{\sigma(p+1) \sigma(p+2)} \ldots
d_{\sigma(n) l}
}
\,.
\label{eq:closedDisk}
\end{multline}
As an example application of the
formula above, let us
consider the following five-operator case
\begin{equation}
\begin{aligned}
\Disk\left[ \mathcal{O}^1_{1}
\mathcal{O}^2_{1} \mathcal{O}^3_{1}
 \mathcal{O}^4_{1};
\mathcal{O}^5_{2}
\right] = 2 \, ( d_{12} d_{35} d_{54} + d_{34} d_{15} d_{52}
+ d_{14} d_{25} d_{53} + d_{23} d_{15} d_{54} ) \, .
\end{aligned}
\end{equation}
Using the formula~\eqref{eq:closedDisk}, one can compute~\eqref{eq:Drukker1234first}. Recall
that the $D$ function given in~\eqref{eq:TheDfunction} is a sum of two terms and each of them
contains a function $m$ with some argument.
In order to compare with the integrability computation, we can focus on the terms with $m(z_{1423})$ and $m(z_{1324})$,
as the argument for the remaining terms is similar.
Firstly, notice that
\begin{equation}
m(z_{1423})+m(z_{1324})=0 \, ,
\end{equation}
and that both of these functions appear multiplied by $d_{12} d_{34}$ in $D_{1234}$
and $D_{1243}$ respectively.
This implies that the contributions to~\eqref{eq:closedDisk} proportional
to $m(z_{1423})$ and $m(z_{1324})$,
consisting
of strings of operators starting at the operators
$\mathcal{O}_1$ and $\mathcal{O}_3$ and ending
at the operators $\mathcal{O}_2$ and $\mathcal{O}_4$
respectively, cancel among themselves.
For $m(z_{1423})$, one has the final result
\begin{multline}
\eqref{eq:Drukker1234first} \big|_{m(z_{1423})} =
2^{n+1} m(z_{1423}) \, d_{12} d_{34}
\times\\\times
\mspace{-20mu}
\sum_{\substack{\text{permutations}\\\sigma\text{ of }\brc{5,\dotsc,n}}}
\biggsbrk{
d_{14} d_{2\sigma(5)} d_{\sigma(5)\sigma(6)} \dotsc d_{\sigma(n)3}
+ \, d_{23} d_{1 \sigma(5)} d_{\sigma(5)\sigma(6)}
\ldots d_{\sigma(n) 4}
\\[-2ex]
+\sum_{p=5}^{n-1}
d_{1 \sigma(5)} d_{\sigma(5) \sigma(6)} \ldots
d_{\sigma(p) 4}
d_{2 \sigma(p+1)} d_{\sigma(p+1) \sigma(p+2)} \ldots
d_{\sigma(n) 3}
}
\,.
\label{eq:PerturbativeFinalResult}
\end{multline}
The next step is to compare the above result with
the integrability calculation.
The integrability result can be obtained by using the
formula for the 2$n$-gon given in~\eqref{eq:2n-gon}.
The argument of the function $m$ appearing in that
formula is given by the cross ratios
\begin{equation}
z_{i,j} \bar{z}_{i,j} = \frac{x_{i,j+1}^2 x^2_{i+1,j}}{x^2_{i,i+1},x^2_{j,j+1}} \, , \quad
(1-z_{i,j})(1- \bar{z}_{i,j}) = \frac{x_{i,j}^2 x^2_{i+1,j+1}}{x^2_{i,i+1} x^2_{j,j+1}} \, .
\end{equation}
where $i$ and $j$ labels the operators in the polygon
with $i \neq j$, $i+1 \neq j$ and $i \neq j+1$ modulo $n$.
The cross ratios $z_{ijkl}$ were defined in~\eqref{eq:TheCrosszijkl}
and they are related with
$z_{i,j}$ by
\begin{equation}
z_{i,j} = z_{i,j+1,i+1,j} \, .
\label{eq:RelationsBetweenz}
\end{equation}
The connected tree-level graphs of $n$
length-two BPS operators consist of
polygons with $n$ vertices. The integrability computation,
assuming that disconnected
tree-level graphs give zero contribution,
consist in using the 2$n$-gon formula of~\eqref{eq:2n-gon}
for the tree-level connected polygons. Note
that the internal and the external polygons give
the same result, hence one gets a factor of two.
The terms proportional to $m(z_{1423})$
are generated by polygons where
$(i,i+1)=(1,2)$ and $(j,j+1)=(3,4)$ for some $i$ and $j$,
as consequence of the relation~\eqref{eq:RelationsBetweenz}.
Summing over all possible polygons, it is not difficult to
see that the integrability result agrees with
the perturbative result of~\eqref{eq:PerturbativeFinalResult}.
The argument is similar for the
other terms $m(z_{ijkl})$, and this proves the
equality of both computation methods.

\subsubsection{The Case of
\texorpdfstring{$n$}{n} Arbitrary
BPS Operators}

We have shown above that the integrability
result for the $n$-point function
of length-two BPS operators agrees with the perturbative
answer. The perturbative result was computed
using the general result for correlation function of
BPS operators described
by Drukker and Plefka~\cite{Drukker:2008pi}.
In principle, one can use a similar procedure as above for proving the
equality for general BPS operators. However,
the combinatorics for the general case are more complicated,
and one has to take into account nontrivial cancellations between terms with different $D(z_{ijkl})$.
We are going to argue that the integrability result agrees with
the perturbative result by using the $\superN=2$ off-shell
superfield formulation of $\superN=4$ SYM
as discussed in~\cite{Howe:1999hz,Eden:2017ozn}.

The $\superN=4$ supermultiplet decomposes into
a $\superN=2$ supermultiplet and a hypermultiplet.
When computing a correlation function of BPS operators,
it is possible to restrict the polarization vectors to a certain
subspace, and treat the external operators
as containing only hypermultiplets. The polarization
vectors $Y$ (we called these $\alpha_i$ for most of this work) were parametrized as a function of
a complex parameter $\beta_i$ in~\cite{Eden:2017ozn} as follows
\begin{equation}
Y_{\beta_i} = \left( \frac{1+ \beta_i \bar{\beta}_i}{2},
i \frac{1- \beta_i \bar{\beta}_i}{2}, i \, {\rm{Im}} \, \beta_i ,
i \, {\rm{Re}} \, \beta_i, 0 ,0 \right) \, .
\end{equation}
Notice that the polarizations above give,
for a generic value of
the parameters $\beta_i$, a non-zero inner product
between two arbitrary polarizations vectors.
This property of the polarizations is important
for the integrability computation, since
when some of the inner products are zero, it is necessary
to consider deformed graphs, as for example in the
subleading computation of \secref{sec:subleadingSection}.
By a direct computation one has
\begin{equation}
Y_{\beta_i} \cdot Y_{\beta_j} = y_{ij}^2 = (\beta_i -\beta_j)
(\bar{\beta}_i - \bar{\beta}_j) \, .
\end{equation}
The one-loop correlation functions in the $\superN=2$
superfield formalism
are computed by inserting $\superN=2$ YM lines in all possible ways
in all tree-level graphs, see~\cite{Eden:2017ozn} for details.
Take two edges of a tree-level graph,
one connecting
the operators $\mathcal{O}_i$ and
$\mathcal{O}_j$ and the other connecting
the operators $\mathcal{O}_k$ and
$\mathcal{O}_l$.
Deleting two propagators and respecting the cyclic
order, one inserts the following function for computing
the one-loop correction:
\begin{equation}
F_{ij ; kl} = \frac{g^2}{2} T_{ij; kl} \, d_{ij} d_{kl} \, g_{ijkl} \, \tilde{F}_{ij;kl}
\,,
\label{eq:FEden}
\end{equation}
where $T_{ij;kl}$ is a color factor, $d_{ij} = y^2_{ij}/x_{ij}^2$, and
\begin{equation}
\tilde{F}_{ij ; kl} = x_{il}^2 x^2_{jk} -
x_{ik}^2 x_{jl}^2 - x_{ij}^2 x_{kl}^2 \left( \frac{y^2_{il} y^2_{jk}}{y^2_{ij} y^2_{kl}} - \frac{y^2_{ik} y^2_{jl}}{y^2_{ij} y^2_{kl}} \right) \, , \quad g_{ijkl} = \frac{1}{\pi^2}
\int \frac{d^4 x_a}{x^2_{ai} x^2_{aj} x^2_{ak} x^2_{al}}
\,.
\end{equation}
Defining the following cross ratios (similar definitions apply
to the $R$-charge cross ratios $\alpha_{ijkl}$)
\begin{equation}
\tilde{z}_{ijkl} \tilde{\bar{z}}_{ijkl}
= \frac{x^2_{ik} x^2_{jl}}{x^2_{ij} x^2_{kl}} \, , \quad
(1- \tilde{z}_{ijkl})(1- \tilde{\bar{z}}_{ijkl})
= \frac{x^2_{il} x^2_{jk}}{x^2_{ij} x_{kl}^2}
\,,
\end{equation}
it is possible to rewrite~\eqref{eq:FEden} as
\begin{equation}
F_{ij ; kl} = - T_{ij; kl} \, d_{ij} d_{kl} \,
m(z_{ijkl}) \, ,
\label{eq:FEdenM}
\end{equation}
where the function $m(z)$ is defined in~\eqref{eq:mdefinition},
and it is the same function appearing in the formula of the 2$n$-gon.
It is possible to get rid of the minus sign above
by using the last of the $m$ function identities given
in~\eqref{eq:midentities}, and by changing variables $z_{ijkl} = (1-z_{i,k})$.

In the integrability calculation, one hexagonalizes
the tree-level graphs and corrects the tree-level result by
adding the mirror-particle contributions.
It follows from using~\eqref{eq:FEdenM} that disconnected
tree-level graphs give perturbatively zero contribution at one-loop
order, and the only tree-level graphs that one has to
consider are connected graphs that decompose
the sphere into a set of polygonal faces.
It is hard to prove using integrability that disconnected
graphs give a zero contribution at one loop, as the
calculation involves loops and spirals. See
\appref{app:disconnected} for details. However,
using our prescription, it is possible to argue
that they vanish, and all the
integrability contributions from any planar graph can be calculated
using the $2n$-gon formula.
Due to the fact that the same function $m$ appears
in the $2n$-gon integrability formula and in the perturbative
building block $F_{ij;kl}$ defined above, it is easy to see
that the integrability result agrees with the perturbative result
for general $n$-point functions of BPS operators.
In particular, this implies the non-renormalization
of the extremal and next-to-extremal correlation functions
by integrability, as mentioned in
\secref{sec:testsandcomments}.

\subsection{On Non-1EI Graphs}

\begin{figure}
\centering
\includegraphics[scale=0.45]{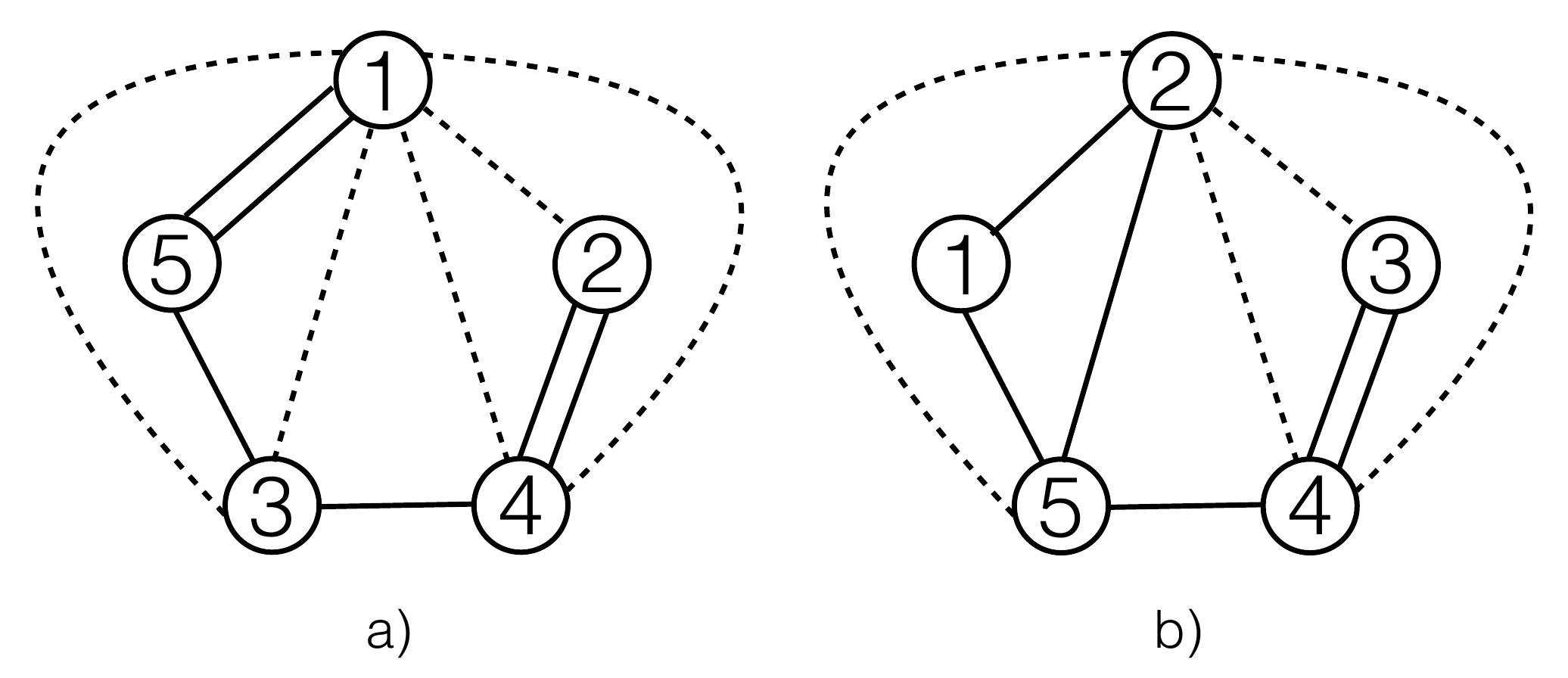}
\caption{The two kinds of non-1EI graphs
that appear in the computation of a five-point
function of three length-two and two length-three
BPS operators. The perturbative result was reproduced
by an integrability calculation in~\cite{\Hexagonalizationtwo} without considering these graphs, in other words
their contributions must vanish.
In this paper we have computed the $n$-particle
contribution, and show that they indeed give a zero contribution.}
\label{fig:FiveNon1E}
\end{figure}

The connected graphs were classified into two types
in~\cite{Fleury:2016ykk}:
The one-edge irreducible (1EI) graphs and the
non-1EI graphs. By definition, 1EI graphs are
graphs that do not become disconnected
when a set of lines connecting any two operators are cut.
Typically, non-1EI graphs have more zero-length bridges,
hence their calculations using integrability are harder
because they involve more multi-particle contributions.
In this work, we have computed these integrability
contributions, and we are in a position to evaluate
all non-1EI planar graphs without making any assumption
about them.

We start by showing that the non-1EI graphs not considered
in the analysis of the five-point function of
three length-two and two length-three BPS operators
done in~\cite{\Hexagonalizationtwo}
do indeed vanish. The graphs are shown in
\figref{fig:FiveNon1E}. Considering that the five-point
function lies in a plane, there are four spacetime cross ratios
characterizing it (similarly for the $R$-charge cross ratios).
They are given by
\begin{equation}
z \bar{z} = \frac{x_{12}^2 x_{34}^2}{x_{13}^2 x_{24}^2}
\, , \quad (1-z)(1-\bar{z}) = \frac{x_{14}^2 x_{23}^2}{x_{13}^2 x_{24}^2} \, , \quad
w \bar{w} = \frac{x_{15}^2 x_{34}^2}{x_{13}^2 x_{54}^2}
\, , \quad
(1-w)(1-\bar{w}) = \frac{x_{14}^2 x_{35}^2}{x_{13}^2 x_{45}^2} \, .
\end{equation}
One can use the 2$n$-gon formula of~\eqref{eq:2n-gon} to compute
the integrability result for these graphs. The relevant
polygons have the edges
\begin{equation}
{\rm{Diagram}} \, \, a) \,= \{1, 5, 3, 4, 2, 4, 3, 5 \} \, , \quad
{\rm{Diagram}} \, \, b) \,= \{1, 2, 5, 4, 3, 4, 5 \} \, .
\end{equation}
Using the properties of the functions $m$ given in~\eqref{eq:midentities},
it is possible to show that indeed the graphs
of \figref{fig:FiveNon1E} give a zero
one-loop contribution, and the comparison
between integrability and the perturbative data
of~\cite{\Hexagonalizationtwo} is correct.

\begin{figure}
\centering
\includegraphics[scale=0.45]{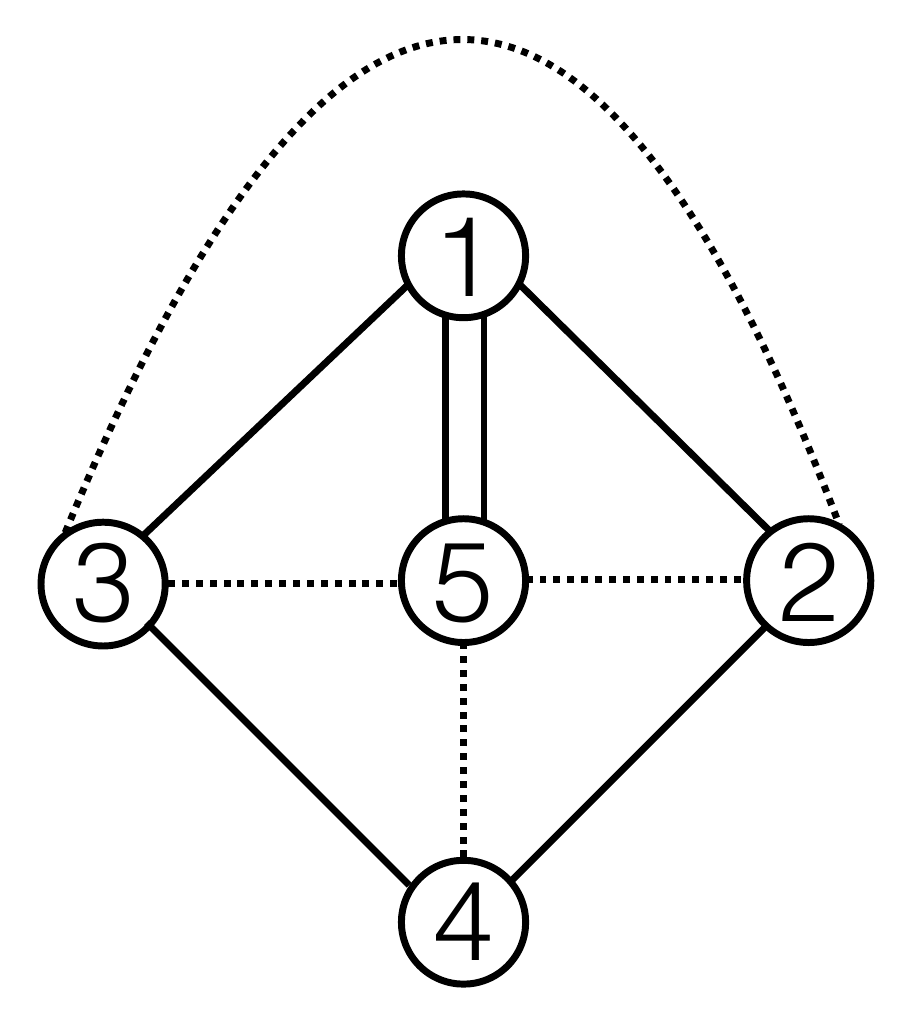}
\caption{One non-1EI graph
contributing to the five-point function of
four length-two and one length-four BPS operators.
Contrary to the original expectation, this graph gives a non-zero
contribution from the integrability calculation. This requires a refinement of the prescription
for summing over graphs of~\cite{\Hexagonalizationone}.}
\label{fig:FiveNon1ENonVanishing}
\end{figure}

In~\cite{\Hexagonalizationone}, the prescription for summing over graphs
was to not include non-1EI graphs in the summation,
because they were expected to vanish.
Using this prescription, the four-point functions
of arbitrary BPS operators
and some five-point functions were computed using
integrability, and the result agreed with perturbation theory.
Nevertheless, the general case for $n$-point functions
is more complicated even at one loop. In
\figref{fig:FiveNon1ENonVanishing}, we show
an example of a non-vanishing one-loop non-1EI graph
for the case of four length-two and one length-four BPS
operator, as one can see by computing the graph using
the $2n$-gon expression of~\eqref{eq:2n-gon} (it gives two times the one-particle contribution of the square).
This result contradicts the assumptions of the prescription
that has to the refined.
The correct prescription is to sum over all graphs
including both 1EI and non-1EI graphs. This gives
the correct result for arbitrary one-loop planar correlation functions of BPS operators, as argued in the previous
subsection using YM insertion lines.

\section{Contributions from Disconnected Graphs}
\label{app:disconnected}

\begin{figure}
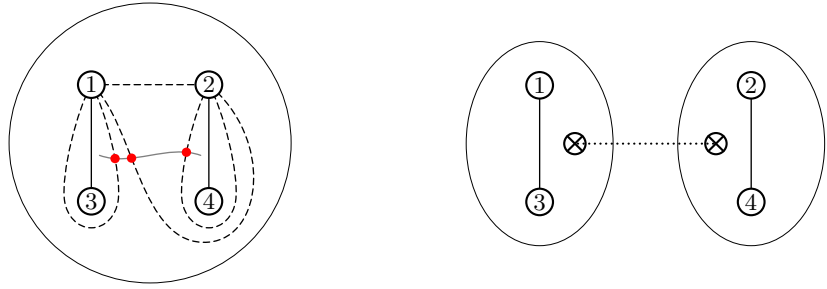

\centering
\includegraphicsbox{FigDisconnectedPlanar1}
\hspace{2cm}
\includegraphicsbox{FigDisconnectedPlanar2}
\caption{In the left figure, we depict a disconnected graph drawn on
a sphere, including its hexagonalization. The solid lines are
non-zero-length bridges, while the dashed lines denote zero-length bridges.
The red dots are a combination of magnons that can potentially
contribute at one loop. On the right, we drew a corresponding
subtraction graph which is given by two disconnected spheres, each
with one marked point.}
\label{fig:discsphere}
\end{figure}

In this appendix, we discuss disconnected planar graphs on the sphere, and
argue that their contribution to the planar four-point function vanishes
at one loop (in agreement with perturbation theory). In fact, there is
only one disconnected planar four-point graph; it is depicted in
\figref{fig:discsphere}, including its hexagonalization.
Much like the secretly planar graphs discussed in
the main text, this graph corresponds to a degenerate Riemann
surface, namely a sphere which splits into two connected components.
We therefore need to consider Dehn-twist identifications as well as
the subtraction of the degenerate case in
order to correctly evaluate its contribution.

As shown in the figure, the graph has a cycle formed by the
zero-length bridges, and one has to identify magnon configurations
that are related by Dehn twists performed on this cycle. As in the
case of secretly planar graphs discussed in the main text, we
conjecture that the net effect of the Dehn twist is to identify
configurations that include closed loops of magnon with the analogous
configurations without any loops. For the tessellation we chose, the
only configuration that does not contain a loop (and that ``feels''
all the four-operators) is the one depicted in
\figref{fig:discsphere} (on the left). The contribution from this configuration is
given by $\polygon(1,3,1,2,4,2)$, which evaluates to zero owing to the
pinching rule.

Having evaluated the contribution from the disconnected graph on the
sphere, the next task is to evaluate the subtraction, which comes from
two spheres, each with two operator insertions and a single marked
point (on the right in \figref{fig:discsphere}). As discussed in
\secref{sec:subtractions}, their contributions are related to the one
without marked points by a shift of the gauge group rank. Since the
(planar) two-point functions do not receive loop corrections, this
immediately shows that the contribution from the subtraction is zero
for our case.

Therefore, in summary, we have $(0-0)=0$, which shows that the
disconnected graphs do not contribute at one loop, as claimed in the
beginning of this section.

\pdfbookmark[1]{\refname}{references}
\bibliographystyle{nb}
\bibliography{references}

\end{document}